\documentclass[12pt,a4paper,twoside,sort&compress]{book}

\usepackage{epsfig}
\usepackage{amsfonts}
\usepackage{amssymb}
\usepackage{amsmath}
\usepackage{setspace}
\usepackage{lscape}
\usepackage[latin1]{inputenc}
\usepackage[TS1,T1]{fontenc}
\usepackage{lmodern}
\usepackage{amsthm}
\usepackage{latexsym}
\usepackage{psfrag}
\usepackage{graphicx}
\usepackage{rotating}
\usepackage{graphics}
\usepackage{subfigure}
\usepackage{mathenv}
\usepackage{amsthm}
\usepackage[titletoc]{appendix}
\usepackage{etoolbox}
\usepackage{tikz}
\newrobustcmd*{\mysquare}[1]{\tikz{\filldraw[draw=#1,fill=#1] (0,0)
rectangle (0.2cm,0.2cm);}}
\newrobustcmd*{\mycircle}[1]{\tikz{\filldraw[draw=#1,fill=#1] (0,0) circle [radius=0.1cm];}}
\newrobustcmd*{\mytriangle}[1]{\tikz{\filldraw[draw=#1,fill=#1] (0,0) --
(0.2cm,0) -- (0.1cm,0.2cm);}}
\usepackage{amssymb}
\usepackage{amsmath}
\usepackage{tcolorbox}
\usepackage{graphics}
\usepackage[final]{pdfpages}
\usepackage{slashed}
\usepackage{amssymb}
\usepackage{epstopdf}

\newcommand{\ignore}[1]{}
\usepackage[latin1]{inputenc}
\usepackage{cite}
\usepackage{subfigure}
\usepackage[small,bf]{caption}
\usepackage{color}
\usepackage[Conny]{fncychap}
\ChNameVar{\LARGE\sffamily\bfseries} \ChNumVar{\Huge} \ChTitleVar{\Huge\sffamily\bfseries}
\ChRuleWidth{0.85pt} \ChNameUpperCase \ChTitleUpperCase %\ChTitleUpperCase
\usepackage{fancyhdr}
\usepackage{amssymb}

\renewcommand{\vec}[1]{\mbox{\boldmath$#1$\unboldmath}}

\newcommand{\B}{\beta_{lm}}

\catcode`@=11
\def\seceqaa{\@addtoreset{equation}{section}
           \def\theequation{A\arabic{equation}}}
\def\seceqbb{\@addtoreset{equation}{section}
           \def\theequation{B\arabic{equation}}}
\def\seceqcc{\@addtoreset{equation}{section}
           \def\theequation{C\arabic{equation}}}
\def\seceqdd{\@addtoreset{equation}{section}
           \def\theequation{D\arabic{equation}}}
\def\seceqee{\@addtoreset{equation}{section}
           \def\theequation{E\arabic{equation}}}
\def\seceqff{\@addtoreset{equation}{section}
           \def\theequation{F\arabic{equation}}}
\def\seceqgg{\@addtoreset{equation}{section}
           \def\theequation{G\arabic{equation}}}
\def\seceqhh{\@addtoreset{equation}{section}
           \def\theequation{H\arabic{equation}}}
\def\seceqjj{\@addtoreset{equation}{section}
           \def\theequation{J\arabic{equation}}}
\def\seceqll{\@addtoreset{equation}{section}
           \def\theequation{L\arabic{equation}}}
\catcode`@=11

\begin{document}
\frontmatter
\pagestyle{plain}
%\input{decl}
    %%% Fancy Header %%%%%%%%%%%%%%%%%%%%%%%%%%%%%%%%%%%%%%%%%%%%%%%%%%%%%%%%%%%%%%%%%%
    % Fancy Header Style Options
    \pagestyle{fancy}                       % Sets fancy header and footer
    \fancyfoot{}                            % Delete current footer settings
    \renewcommand{\chaptermark}[1]{         % Lower Case Chapter marker style
      \markboth{\chaptername\ \thechapter.\ #1}{}} %
    \renewcommand{\sectionmark}[1]{         % Lower case Section marker style
      \markright{\thesection.\ #1}}         %
    \fancyhead[LE,RO]{\bfseries\thepage}    % Page number (boldface) in left on even
                                            % pages and right on odd pages
    \fancyhead[RE]{\bfseries\leftmark}      % Chapter in the right on even pages
    \fancyhead[LO]{\bfseries\rightmark}     % Section in the left on odd pages
    \renewcommand{\headrulewidth}{0.3pt}    % Width of head rule
    %%% Clear Header %%%%%%%%%%%%%%%%%%%%%%%%%%%%%%%%%%%%%%%%%%%%%%%%%%%%%%%%%%%%%%%%%%
    % Clear Header Style on the Last Empty Odd pages
    \makeatletter
    \def\cleardoublepage{\clearpage\if@twoside \ifodd\c@page\else%
        \hbox{}%
        \thispagestyle{empty}%              % Empty header styles
        \newpage%
        \if@twocolumn\hbox{}\newpage\fi\fi\fi}
    \makeatother
    %%%%%%%%%%%%%%%%%%%%%%%%%%%%%%%%%%%%%%%%%%%%%%%%%%%%%%%%%%%%%%%%%%%%%%%%%%%%%%%
    %\addcontentsline{toc}{chapter}{CANDIDATES DECLARATION}
%%\input{decl}
%\begin{titlepage}
%\title{}
\begin{center}
{\large \bf APPLICATIONS OF TOP-DOWN HOLOGRAPHIC THERMAL QCD AT FINITE COUPLING}
\footnote{Based on author's Ph.D. thesis defended on March 21, 2018}\\
Karunava Sil\footnote{e-mail: krusldph@iitr.ac.in}
 \\
Department of Physics, Indian Institute of Technology,
Roorkee - 247 667, Uttaranchal, India

 \date{\today}
\end{center}
\thispagestyle{empty}
%\begin{abstract}
{\bf Abstract}: Large-$N$ thermal QCD laboratories like strongly coupled QGP (sQGP) require not only a large t'Hooft coupling but also a finite gauge coupling \cite{Natsuume}. Unlike almost all top-down holographic models in the literature, holographic large-$N$ thermal QCD models based on this assumption, therefore necessarily require addressing this limit from M theory.

Using the UV-complete top-down type IIB holographic dual of large-$N$ thermal QCD as constructed in \cite{metrics} involving a fluxed resolved warped deformed conifold, its delocalized type IIA S(trominger)-Y(au)-Z(aslow) mirror as well as its  M-theory uplift constructed in \cite{MQGP}, in \cite{NPB}, the type IIB background of \cite{metrics} was shown to be thermodynamically stable. We also showed that the temperature dependence of DC electrical conductivity  mimics a one-dimensional Luttinger liquid, and the requirement of the Einstein relation (ratio of electrical conductivity and charge susceptibility equal to the diffusion constant) to be satisfied requires a specific dependence of the Ouyang embedding parameter on the horizon radius. Any strongly coupled medium behaves like a fluid with interesting transport properties. In \cite{EPJC-2}, we addressed these properties by looking at the scalar, vector and tensor modes of metric perturbations and solve Einstein's equation involving appropriate gauge-invariant combination of perturbations as constructed in \cite{klebanov quasinormal}. Due to finite string coupling, we obtained the speed of sound, the shear mode diffusion constant and the shear viscosity $\eta$ (and $\frac{\eta}{s}$) upto (N)ext to (L)eading (O)rder in $N$. The NLO terms in each of the coefficients serve as a the non-conformal corrections to the conformal results. Another interesting result for the temperature dependence of the thermal (and electrical) conductivity and the consequent deviation from the Wiedemann-Franz law, upon comparison with  \cite{WF}, was obtained at leading order in $N$. The results for the above qualitatively mimic a 1+1-dimensional Luttinger liquid with impurities.  Also  we obtained the QCD deconfinement temperature compatible with lattice results (a study that was in fact initiated in \cite{NPB}).

On the holographic phenomenology side,  in  \cite{glueball2017}, we computed the masses of the $0^{++}, 0^{-+},0^{--}, 1^{++}, 2^{++}$ `glueball' states corresponding to fluctuations in the dilaton or complexified two-forms or appropriate metric components in the same aforementioned backgrounds. All these calculations were done both for a thermal background with an IR cut-off $r_0$ and a black hole background with horizon radius $r_h$. We used WKB quantization conditions on one hand and imposed Neumann/Dirichlet boundary conditions at $r_0$/$r_h$ on the solutions to the equations of motion on the other. We found that the former technique produces results closer to the lattice results \cite{hep-lat/9804008},\cite{hep-lat/0510074}.
%\end{abstract}
%\end{titlepage}.
%\addcontentsline{toc}{chapter}{Abstract}
%\input{abstract}
%\chapter*{Abstract}

%\input{ack}
%\input{dedication}
%\newpage
%\clearpage
%\input{empty1}
%\addcontentsline{toc}{chapter}{List of Publications}
%\input{pubs}
%\newpage

%\clearpage
%\newpage
%\input{empty1}
\newpage
~\\
~\\

\thispagestyle{empty}
\cleardoublepage
\begin{spacing}{1.53}
\addcontentsline{toc}{chapter}{Table of Contents}\tableofcontents
\end{spacing}
\clearpage
%\input{empty1}
%\newpage
\newpage
~\\
~\\

\thispagestyle{empty}
\cleardoublepage
\addcontentsline{toc}{chapter}{List of Figures}
\begin{spacing}{1.34}
\listoffigures
\newpage
~\\
~\\

\thispagestyle{empty}
\cleardoublepage
\end{spacing}
\addcontentsline{toc}{chapter}{List of Tables}
\begin{spacing}{1.34}
\listoftables
\newpage
~\\
~\\

\thispagestyle{empty}
\cleardoublepage
\end{spacing}
\mainmatter
\pagestyle{fancy}
\setcounter{chapter}{0}
\setcounter{section}{0}
\setcounter{subsection}{0}
\setcounter{tocdepth}{3}

%\input{chap1}
%\input{chap2}
%\input{chap3}
%\input{chap4}
%\input{chap5}
%\input{chap6}

%\input{partA}
%\clearpage
%\input{empty}
%\input{Chapter1/chap1}
\chapter{Introduction}

\graphicspath{{Chapter1/}{Chapter1/}}

\section{Need for the gauge-gravity duality}
The duality between string theory and gauge theory has turned out to be a very useful approach in the study of strongly coupled quantum field theories. The AdS/CFT correspondence \cite{maldacena} - see \cite{Minwalla-1} for a summary of its applications - is the first explicit example of such duality between a particular known string theory and a gauge theory. The properties of these strongly coupled field theories at finite temperature, for example, transport coefficients have been extensively studied in recent years using this approach. Also the results of recent RHIC experiments motivate such theoretical studies in particular of the strongly coupled plasma phase of non-abelian gauge theories.
In RHIC experiment one collides two heavy nuclei such as Pb or Au. The name `heavy ion' is given due to the reason that before the collision these atoms are ionized as they are electrically neutral. After the collision a plasma state is formed at high temperature. Due to high temperature an expansion occurs in the plasma which decreases its temperature. As the temperature falls below the transition temperature $T_c$, the quarks get confined into hadrons. The temperature that is achieved so far in RHIC experiment is about $T=2T_c$.

As per the standard model of particle physics fundamental constituents of matter include quarks. Quarks come in six flavors and three colors. There exist strong interaction between the quarks. Due to this strong force quarks usually form bound state inside protons and neutrons. Quantum chromodynamics or QCD is the theory which describes the physics of strong interactions. It is a gauge theory with $SU(3)$ gauge group. The particles which mediate the strong force between quarks are the gluons. Unlike QED, where the force mediating particles, photons are charge neutral, in QCD the gluons also carry color charge. Quarks transform under the fundamental representation of $SU(3)$ gauge group, while the gluons transform under the adjoint of $SU(3)$.

QCD has a very interesting phase structure. At low temperature and low baryon chemical potential quarks are found to be in a confined state. In this phase QCD behaves as a strongly coupled theory. However, as temperature increases, the interactions between the quarks are weakened due to Debye screening. At sufficiently high temperatures, quarks and gluons are completely deconfined. This phase of QCD is known as the Quark Gluon Plasma phase or QGP. The transition from the confining phase to the deconfined QGP phase is estimated to occur at temperature $T_c=150-200$ Mev. Although in plasma phase the interaction strength between quarks and the gluons becomes weak at very high energy, it is quite strong at any intermediate stage. Specifically, upto temperature $T=2 T_c$ (as achieved by the RHIC experiments) the interaction is so strong that one cannot apply perturbative method. The lattice simulation suggest that ideal gas behavior of quarks and gluons can be achieved at extremely high temperature $T\sim 1000T_c$. The pressure of the ideal gas of quarks and gluons as obtained in the lattice calculation is given as: \cite{dmateosADSCFT}
\begin{equation}
P_{SB}=\frac{8\pi^2}{45}\left(1+\frac{21}{32}N_{f}\right)+\mathcal{O}(m^2_qT^2),
\end{equation}
where `SB' stands for Stefan-Boltzman, $N_f$ is the number of flavor and $m_q$ is the quark mass. However at RHIC temperature, the same pressure is obtained to be:
\begin{equation}
\frac{P}{P_{SB}}\Biggr|_{T\sim T_{RHIC}}\sim \left(0.75\right).
\end{equation}
Now, for the equilibrium properties of QCD such as thermodynamic properties, the weak coupling approximation is not that bad. Hence one can extrapolate the perturbative results for the intermediate coupling region. But the out-of-equilibrium phenomenon, such as the transport coefficients, depend strongly on the coupling. Hence for the out of equilibrium phenomenon, one can not trust the perturbative results for the intermediate region. Also, the low energy physics of QCD such as the computation of glueball spectrum is difficult using perturbative QCD technique. Theses problems are resolved by the gauge/gravity duality. Moreover, if one can find a weakly coupled gravity dual for the strongly coupled gauge theory then things can be handled in a better and easier way.
\section{The AdS/CFT Correspondence}
\subsection{History}
In its original version in the sixties, string theory was formulated as a theory of strong interaction. Soon after this in 1971 the asymptotic freedom was discovered and based on enough experimental evidence it was concluded that the theory of strongly interacting particles, quarks and gluons, are best described by QCD. So string theory was abandoned as a theory of strong interaction. Then in 1974, t'Hooft showed that a large-$N$ expansion in gauge theory, where $N$ represents the number of colors, looks like a string theory. Also around the same time it was realized that string theory also includes quantum gravity. Using lattice QCD it was observed that quarks in QCD can be confined by strings. After that the holographic principle was given by t'Hooft in 1993 followed by the discovery of D-branes by Polchinski in 1995. Finally in 1997, Juan Maldacena gave the AdS/CFT correspondence and in 1998 Witten made the connection with holographic principle. In the following subsections we will establish the AdS/CFT correspondence in steps and discuss only those aspects which are directly relevant to this thesis. We have closely followed \cite{HONGLIU} for this discussion.
\subsection{Large $N$ Gauge theory and String theory}
In usual perturbative expansion method we start with the free Lagrangian and expand around that free theory as a power series in a small parameter. In QCD there are no such small parameters. However, in 1974 t'Hooft proposed the idea of large $N$ expansion, $N$ being the number of colors of the gauge theory. The idea here is to treat $N$ as a parameter and then do a $\frac{1}{N}$-expansion in the limit $N\rightarrow \infty$. This $\frac{1}{N}$-expansion as we will show below is a string theory.
\begin{itemize}
\item{\it Large $N$ expansion in Gauge theory:}
\end{itemize}
Let's consider a large-$N$ $SU(N)$ gauge theory. Since each gluon field has one color and one anti-color index, they will be represented as $N\times N$ matrices. The large-$N$ limit is given as:
\begin{equation}
N\rightarrow \infty ~~\textrm{such that} ~~\lambda=g^2 N~~ \textit{is large but finite},
\end{equation}
where $g$ is the gauge coupling and $\lambda$ is the t'Hooft coupling. The vacuum energy of the theory will be given by the sum of all diagram without any external legs. To calculate the amplitude of any arbitrary vacuum diagram one has to sum over all the color indices of the gluon fields. As each contraction gives a factor of $N$ in the amplitude and at the same time in terms of the double line notation each loop corresponds to a single contraction, the $N$ counting will be given by the number of loops in the diagram. Now, as the gluon fields are represented as matrices and since matrices do not commute, the $N$ counting will depend on whether the contraction of indices is between two neighboring fields or not. Hence there will be two types of diagrams. The diagrams that can be drawn on a plane without crossing any two lines are called the planar diagrams. On the other hand the class of diagrams which can not be drawn on a plane without crossing lines are the non-planar diagrams. Also it can be shown that for the planar diagram the number of loops in the double line notation are equal to the number of disconnected regions as created by the usual Feynman diagrams on the plane. However, for the non-planar diagram we cannot count such disconnected regions as they cannot be drawn on a plane. Interestingly, it can be observed that these non-planar diagrams can be straightened out on non trivial topological surfaces such as a tours. Hence, we conclude that just like the planar diagram, the power of $N$ for non-planar diagrams are given by the number of faces in each diagrams after one straightened it out to a planar diagram. So each Feynman diagram is nothing but a partition of the surface on which it is drawn into polygons.
The amplitude of any vacuum diagram with $E$ number of propagator, $V$ number of vertices and $F$ number of faces can be written as,
\begin{equation}\begin{split}\label{vacua}
A & \sim  \left(g^2\right)^{E-V} N^{F}\\&=(\lambda)^{L-1}N^{\chi},
\end{split}
\end{equation}
where $L=E-V+1$ is the no of loops in a diagram and $\chi=F+V-E$ is the Euler characteristics. As any 2-dimensional surface is classified by the genus or handles $g$ that a surface has, two surfaces with the same number of $g$ are topologically equivalent. The Euler number $\chi$ is related to $g$ as $\chi=2-2g$. For example a genus zero surface is a sphere and $g=1$ is a torus. So for a genus-zero surface the vacuum amplitude would be,
\begin{equation}
\left(c_0+c_1 \lambda+c_2\lambda^2+....\right)N^2=f(\lambda)N^2.
\end{equation}
Also $g=0$ corresponds to the leading order result in $N$. Hence at leading order in $N$ only the planner diagram contribute to the vacuum energy. The partition function is given by the sum of all possible connected vacuum diagrams,
\begin{equation}
\label{partition}
\log{Z_{gauge}}=\sum^{\infty}_{g=0}N^{2-2g}f_{g}(\lambda).
\end{equation}
\begin{itemize}
\item{\it Implication of large-$N$ expansion in String theory:}
\end{itemize}
Before going into the details of what the large-$N$ expansion in gauge theory implies, let's first talk a little about the idea of string theory. As we all know that QFT is a theory of particle. In QFT we start with a Lagrangian and then quantize it to obtain the spectrum of particles. This is called the second quantization approach. However, in the first quantization instead of the Lagrangian one quantizes the motion of a given particle in spacetime. Let's consider the motion of a particle parameterized by a single parameter $\tau$. This motion can be mapped in the spacetime by it's coordinate as a function of the parameter as $x^{\mu}(\tau)$. Now, to quantize the particle we need to integrate over all possible paths of that particle or in other words we need to evaluate the following path integral:
\begin{equation}
\int Dx^{\mu}(\tau) e^{i S_{particle}},
\end{equation}
where $S_{particle}$ is the length of a given path.
String theory is formulated based on the first quantization approach. Here instead of a particle one quantizes the motion of a one dimensional string in spacetime. For the time being, let's consider the motion of a closed string only. This particular consideration is required for the diagrammatic analysis as discussed in the later part of this section. The motion of a string in spacetime will generate a two-dimensional surface, the worldsheet. A worldsheet is parameterized by two parameter: $\sigma$ and $\tau$ and the embedding of this two-dimensional surface in spacetime can be written in terms of the spacetime coordinate $x^{\mu}(\sigma,\tau)$. So to quantize a string, one needs to consider all possible embeddings of such two dimensional surfaces in spacetime that gives all kinds of string motion. The path integral for the string motion is given as:
\begin{equation}
\int Dx^{\mu}(\tau,\sigma) e^{i S_{string}},
\end{equation}
where $S_{string}$ is related to the area of the worldsheet and is given as: $S_{string}=T\int dA$, with $T=\frac{1}{2 \pi \alpha^{\prime}}$ is the string tension. Analogous to QFT, the sum of all the vacuum amplitudes is given as,
\begin{equation}\begin{split}
\mathcal{A} &=
  \sum_{\substack{\textrm{all}~ \textrm{closed}\\
                  2\textrm{d} ~\textrm{surfaces}}}
      e^{S_{string}}\\&
      =\sum^{\infty}_{g=0}\sum_{\substack{\textrm{all}~ \textrm{closed}\\
                  2\textrm{d} ~\textrm{surfaces}\\~\textrm{of}~\textrm{given}~g}}
      e^{-S_{string}}
      \end{split}
\end{equation}
In the next step one add a weight factor $e^{-\Lambda \chi}$ by hand in the right hand side of the above equation. The implication of this weight will be clear soon. The vacuum energy now becomes:
\begin{equation}\begin{split}\label{amplitude}
\mathcal{A} &=
  \sum^{\infty}_{g=0}e^{-\Lambda \chi}\sum_{\substack{\textrm{all}~ \textrm{closed}\\
                  2\textrm{d} ~\textrm{surfaces}\\~\textrm{of}~\textrm{given}~g}}
      e^{-S_{string}}
      \end{split}
\end{equation}
\begin{figure}
 \begin{center}
%\begin{center}
 \includegraphics[scale=0.5]
 %[height= 21cm,width=+15cm]
 {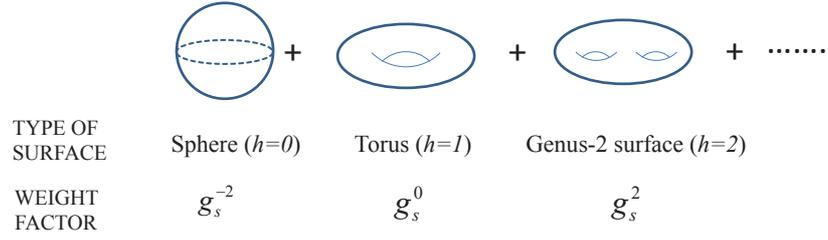}
 \end{center}
 \caption{Different genus Surfaces}
\end{figure}
\begin{figure}
 \begin{center}
%\begin{center}
 \includegraphics[scale=0.5]
 %[height= 21cm,width=+15cm]
 {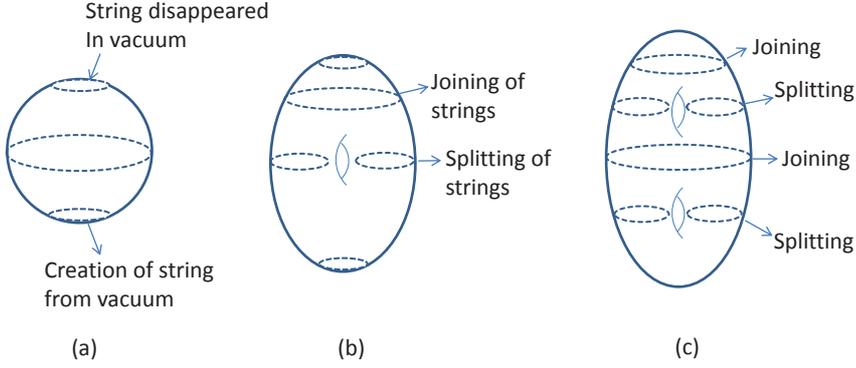}
 \end{center}
 \caption{Virtual propagation of closed string through different genus Surfaces}
\end{figure}

Defining $e^{\Lambda \chi}=g_s$, the above expression can be defined diagrammatically as given in {\bf Fig-1.1}.
Let's discuss more about each of the three diagrams of different $g$ in {\bf Fig-1.2}.

First consider the sphere ($g=0$). This can be generated by the virtual motion of a closed string with varying radius. Similarly, the torus diagram can be explained by the same virtual motion of a closed string but this time one can imagine a single splitting followed by a single joining of the closed string as shown in {\bf Fig-1.2(b)}. Also for the genus-2 surface there will be a total number of four alternative splitting and joining processes as depicted in {\bf fig-1.2(c)}. An important point to notice that in {\bf Fig-1.2}, any two conjugative diagrams differs from each other by two powers of $g_s$. Interestingly the total number of joining and splitting in a particular diagram also differs by two in between successive diagrams. Hence one concludes that each of the joining and splitting of closed string is equivalent to a multiplication by one $g_s$ in the vacuum amplitude. So the reason behind the inclusion of the factor $e^{-\Lambda \chi}$ in the sum is to assign a weight to each joining and splitting process. In other words, $g_s$ measure the strength of the string interaction. It is called the string coupling. Equation (\ref{amplitude}) can be rewritten as:
 \begin{equation}\begin{split}\label{amplitude1}
\mathcal{A} &=
  \sum^{\infty}_{g=0}g^{2g-2}_{s}F_{g}(\alpha^{\prime}),
      \end{split}
\end{equation}
where in the continuous limit $F_{g}(\alpha^{\prime})$ can be written as,
 \begin{equation}\label{continuous}
F_{g}(\alpha^{\prime})=\int\limits_{\substack{\textrm{genus}~g\\\textrm{surfaces}}}Dx(\sigma,\tau)e^{-S_{string}}.
\end{equation}
Interestingly the expression for the vacuum energy as given by equation (\ref{partition}) for the gauge theory and by (\ref{amplitude1}) for the string theory has the same mathematical structure.
Comparing these two equations the following two conclusions can be made:

(i) $\frac{1}{N}$ expansion in the gauge theory corresponds to the expansion in terms of $g_s$ in string theory side.

(ii) The sum over all Feynman diagram of genus $g$ in the gauge theory is equivalent to the sum over all possible string worldsheet of genus $g$.

Also note that the above equivalence is more prominent in the large $\lambda$ limit. This is because of the following reason:
First of all equation (\ref{continuous}) implies that in string theory a two dimensional surface of a given topology can be embedded in spacetime in infinitely many different ways and it's a continuous process. Where, in the gauge theory side a Feynman diagram actually discretizes the surface of a given genus on which it is drawn. Hence for simple Feynman diagrams the proper geometric structure of the surface would not be clear. But for complicated diagrams where the number of propagators and the number of vertices are infinitely large, the proper geometric picture will emerge. So in order to have a continuum limit just as the string theory side one has to consider complicated Feymman diagrams also along with the simpler one. Now, equation (\ref{vacua}) suggest that to include diagrams with large number of propagators one has to consider terms with large power of $\lambda$ and hence the 't Hooft coupling $\lambda$ has to be large in order to have a proper equivalence with string theory.

Hence it is clear that the large $N$ expansion is actually a string theory. Although this does not tell anything about the kind of string theory. Since the spacetime can be arbitrary, for different spacetimes one gets different action. So given a quantum field theory or a set of Feynman diagrams one has to look for some equivalent action which describes the motion of some surface in spacetime. This choice is in some sense is infinite and hence despite the above mathematical equivalence, giving an explicit example of this is not an easy task.
\begin{itemize}
\item{\it Hint from the Holographic Principle:}
\end{itemize}
 Let's consider a gauge theory in $3+1$ dimensional minkowski spacetime $\mathcal{M}_{3+1}$. A natural guess for the string theory would be the one in $3+1$ dimensional minkowski space. But string theory is consistent quantum mechanically in ten dimension. Since the equivalent string theory has to have the same amount of symmetry as the the gauge theory, the spacetime for the string theory has to have the form $\mathcal{M}_{3+1}\times \mathcal{N}$, where $\mathcal{N}$ is some compact manifold. Now, as $\mathcal{N}$ is a compact manifold, the above spacetime will only have $3+1$ dimensional Poincare symmetry just like the gauge theory. However there is still another problem. Gravity appeared naturally in string theory. Quantization of string theory gives massless spin two particle, graviton in it's spectrum. But from Weinberg Witten theorem \cite{ww} it is known that any $3+1$ dimensional relativistic QFT cannot have a spin two massless particle. This problem is resolved by the holographic principle. According to holographic principle, the degrees of freedom of any quantum gravity system is bounded by it's area. In other words, a quantum gravity system can be described by the degrees of freedom living on it's boundary. Hence one considers the non-compact part of the spacetime for string theory to be five dimensional and put the gauge theory on it's four dimensional boundary. This way a theory with gravity defined in the bulk can be described by a theory without gravity on it's boundary.
 \begin{itemize}
\item{\it Anti de Sitter space and Conformal field theory:}
\end{itemize}
The most general $4+1$ dimensional spacetime for the string theory with translation and lorentz symmetry is given as,
\begin{equation}\label{five}
ds^2=f^2(z)\left(dz^2+\eta_{\mu\nu}dx^{\mu}dx^{\nu}\right),
\end{equation}
where $z$ represents the fifth dimension. To get the exact form of $f(z)$, one has to consider some extra symmetry in the gauge theory side. For example one may take the field theory to be scale invariant. So the field theory must be invariant under the scaling of it's spacetime coordinate given by:
\begin{equation}
x^{\mu}\rightarrow \alpha x^{\mu},
\end{equation}
for some constant $\alpha$.
Now the metric (\ref{five}) must respect such scaling symmetry. This can be achieved if under the above transformation $z$ and $f(z)$ also transform as,
\begin{equation}\begin{split}
z&\rightarrow \alpha z\\f(z)&\rightarrow  f(z)\alpha.
\end{split}
\end{equation}
The above can be satisfied only if $f(z)=\frac{L}{z}$ with $L$ as a constant. So equation (\ref{five}) takes the form,
\begin{equation}\label{five1}
ds^2=\frac{L^2}{z^2}\left(dz^2+\eta_{\mu\nu}dx^{\mu}dx^{\nu}\right).
\end{equation}
This is precisely the AdS metric. So one concludes in general that a large $N$ conformal field theory in $d$-dimensional minkowski spacetime is equivalent to a string theory in $d+1$-dimensional AdS spacetime.
\subsection{Overview of String theory and D branes}
There are two types of strings in string theory, open strings and closed strings. A string has a tension $\mathrm{T}$ with the dimension of $[\textrm{Length}]^{-2}$. String tension is related to the string length $l_s$ as: $\mathrm{T}=\frac{1}{2\pi l^{2}_{s}}$. The fundamental string can oscillate in different modes and each oscillation mode corresponds to a spacetime particle. Consider the motion of a string in $d$ dimensional Minkowski spacetime. Quantum mechanically consistent quantization of open and closed Bosonic string requires the dimension of the Minkowski space to be $d=26$. The massless excitation of open and closed sector of the Bosonic string are:
\begin{itemize}
\item{Open string:} Photon $(A_{\mu})$
\item{Closed string:} Graviton $(h_{\mu\nu})$, Antisymmetric tensor $(B_{\mu\nu})$, Dilaton $(\phi)$.
\end{itemize}
Requiring the theory to be supersymmetric, the dimension of the target spacetime reduces to $10$. In $10$ dimensional superstring theory, depending on the periodic or antiperiodic boundary conditions on fermions one gets two types of string theories: type IIB and type IIA. The massless field content of $10$ dimensional type IIB and type IIA superstring theory at low energies are,
\begin{itemize}
\item{Type IIA:} ~~~$\underbrace{g_{\mu\nu}, B_{\mu\nu}, \phi}_{\textrm{NS} ~\textrm{Sector}}$~~~~~$\underbrace{A_{\mu}, C^{(3)}_{\mu\nu\lambda}}_{\textrm{RR}~ \textrm{Sector}}$
\item{Type IIB:} ~~~$\underbrace{g_{\mu\nu}, B_{\mu\nu}, \phi}_{\textrm{NS} ~\textrm{Sector}}$~~~~~$\underbrace{\chi, C^{(2)}_{\mu\nu}, C^{(4)}_{\mu\nu\lambda\rho}}_{\textrm{RR}~ \textrm{Sector}}$.
\end{itemize}

In the quantization process of open string one considers the Neumann boundary condition at both ends of the string for each direction of the spacetime. On the other hand imposing Dirichlet boundary condition constrains the motion of the end points to lie on some hypersurface within the spacetime. These hypersurfaces are called the $D$ branes. For example, a $Dp$-brane is a $p$ dimensional hypersurface. Let's consider a $Dp$-brane in d-dimensional Minkowski spacetime. This breaks the Lorentz and Poincare symmetry of the Minkowski space to a subgroup where along the world-volume of the brane the Poincare symmetry is still preserved. Also in the transverse directions there is a $SO(d-p-1)$ rotational symmetry. Now, quantization of an open string in Minkowski space using Neumann boundary condition and stack of coincident $D$-branes gives d-dimensional gauge fields $A_{\mu}$ as a massless excitations. With the introduction of $D$-branes, some of the gauge components now oscillate in the transverse direction to the brane world volume and hence effectively become scalar fields. Therefore the number of massless scalar fields are equal to the number of transverse directions to the $D$-brane. The Dirac-Born-Infeld action of a $Dp$-brane is given as,
\begin{equation}\label{DBI}
S_{DBI}=-T_{Dp}\int d^{p+1}x\sqrt{-det\left(g_{\alpha\beta}+2\pi\alpha^{\prime}F_{\alpha\beta}\right)},
\end{equation}
where $g_{\alpha\beta}$ is the induced metric on the $D$-brane in the full Minkowski space, $F_{\alpha\beta}$ is the field strength for the gauge field, $\phi$ is the scalar field and $T_{p}$ is the brane tension. In the low energy limit one can expand the above action as,
\begin{equation}\label{DBIEX}
S_{DBI}=-T_{Dp}\int d^{p+1}x\sqrt{-det{g_{\alpha\beta}}}\left(1+\frac{1}{4}F_{\alpha\beta}F^{\alpha\beta}+\frac{1}{2}\partial_{\alpha}\phi^{a}\partial^{\alpha}\phi^{a}+....\right).
\end{equation}
Here $\alpha$ and $a$ denotes the directions along the brane and transverse to the brane respectively. The tension of the $T_{Dp}$ of the D brane is related to the mass $M_{Dp}$ and volume $V_{Dp}$ of the D brane as $M_{Dp}=T_{Dp}V_{Dp}$.
Now for the time being lets put the gauge field to zero. Then from (\ref{DBIEX}), one gets,
\begin{equation}\label{DBIE}
S_{DBI}=\int dt\left(-M_{Dp}-\frac{1}{2}M_{Dp}\left(\dot{\phi}^{a}\right)^2+....\right).
\end{equation}
Equation (\ref{DBIE}) describe the motion of a massive object which can move in the spacetime with the field $\phi$ as the degrees of freedom describing it's motion. Let us discuss the importance of the above result. D branes are introduced at the beginning by some rigid boundary condition and hence they look like some non-dynamical object. But when the open string on the D branes are quantized, one realizes that the degrees of freedom on the $D$-branes correspond to their fluctuations. As these excitations on these $D$-branes vary coherently, they become a fully dynamical object.

Now, it can be shown that for a particular case of $N$ coincident $D3$-branes in flat Minkowski space one getss a four dimensional $SU(N)$ gauge theory where the open string excitations: a gauge field $A_{\alpha}$ and the scalar field $\phi^{a}$ transforms as the adjoint representation of the $SU(N)$ gauge group. Hence at low energies the gauge theory is a Yang-Mills theory with maximum supersymmetry and is scale invariant. The low energy effective action is given as,
\begin{equation}
S_{eff}=-\frac{1}{g^2_{YM}}\int d^{4}x Tr\left(-\frac{1}{4}F_{\alpha\beta}F^{\alpha\beta}-\frac{1}{2}\left(D_{\alpha}\phi^{a}\right)\left(D_{\alpha}\phi^{a}\right)+[\phi^{a},\phi^{b}]^{2}+...\right),
\end{equation}
where $g_{YM}$ is the Yang-Mills gauge coupling and it is related to the string coupling $g_s$ as $g^{2}_{YM}\sim g_{s}$. $D_{\alpha}$ is the covariant derivatives defined as $D_{\alpha}\phi^{a}=\partial_{\alpha}\phi^{a}-i[A_{\alpha},\phi^{a}]$.

Let's stop here for a moment and talk about another interpretation of D branes: they are non-perturbative charged objects. To see the non perturbative nature of a D-brane, one needs to calculate the tension of a D-brane. As discussed before, the mass of a p-dimensional D-brane is related to it's tension by the relation $M_{D_p}=T_{D_p}V_{D_p}$. Now, the mass of a D-brane is equal to it's energy when it is at ground state, i.e. when none of the open strings on it are excited. In other words, the mass $M_{D_p}$ is equal to the vacuum energy $E_{\rm vac}$ open strings living on it,
\begin{equation}\begin{split}
M_{D_p}=E_{\rm vac}& =\rm{sum ~over~ all~ vacuum~ diagrams~ of~open~ string}\\
&=\rm{sum ~over~ all~2~dimensional ~surfaces~with ~at~least ~one~boundary}
\end{split}
\end{equation}
\begin{figure}
 \begin{center}
%\begin{center}
 \includegraphics[scale=0.5]
 %[height= 21cm,width=+15cm]
 {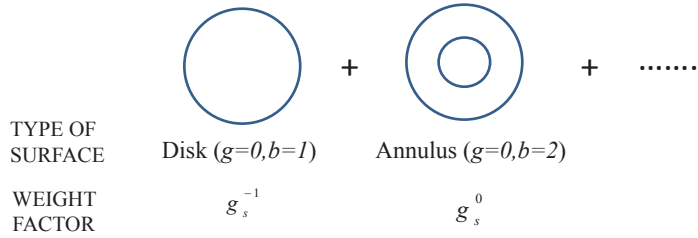}
 \end{center}
 \caption{Different two dimensional Surfaces}
\end{figure}
The first few diagrams can be diagrammatically presented as given in {\bf Fig-1.3}. Here each of the diagrams are weighted by the factor $g^{2g+b-2}_{s}$, with $g$ and $b$ defined as the number of genus and the number of boundary respectively. So, for example the weight factor for the disk with only one boundary is $g_{s}^{-1}$ and that for the annulus with two boundary is $g_{s}^{0}$. In weak coupling limit, $g_s\ll 1$ and hence the tension can ne approximated as $T_{D_p}\sim \frac{1}{g_{s}}$. Thus in the weak coupling region the tension of a D-brane is very large, which makes the D-brane a non-perturbative object.
Also, as discussed before, the RR sector of the spectrum of Type IIA and IIB superstring theory contains different antisymmetric gauge fields. D branes are the objects which couples to these antisymmetric fields and are charged either electrically or magnetically or both under these fields. This is basically the generalization of what we had in usual electromagnetic theory: the one form vector potential $A_{\mu}$ couples to the electron which has an electric charge. In Type IIB superstring theory the four form field $C^{(4)}$ couple to $D3$ branes. Since the field strength $F_5$ corresponding to this four form is self dual, the $D3$ brane has both electric and magnetic charge. These dual behaviour of $D$-branes: (i)a defect in spacetime where open string can end and (ii)a non-perturbative charged object under generalized gauge fields, is at the core of the establishment of AdS/CFT correspondence.
\subsection{Geometry of $D3$-branes in Minkowski space}
A stack of $N$ $D3$-branes extended along the coordinates ${t,x_1,x_2,x_3}$ within ten dimensional flat Minkowski space can be viewed as a point in it's transverse space $\mathbb{R}^6$. Hence the space surrounding the $D3$ brane is $S^5$ in $\mathbb{R}^6$. Based on this symmetry one writes down the metric as the following ansatz,
\begin{equation}\label{ADSmetric}
ds^2=f_{1}(r)\left(-dt^2+\sum^{3}_{i=1}dx^2_{i}\right)+f_{2}(r)\left(dr^2+r^2d\Omega^2_{5}\right).
\end{equation}
Since the $D3$ branes are charged under $C_4$ potential (with five form field strength $F_5$) one can write:
\begin{equation}\label{charge}
q=\int_{S^5}*F_{5}~~~~~~~~g=\int_{S^5}F_{5},
\end{equation}
where $q$ and $g$ are the electric and the magnetic charge respectively. As $F_5$ is self dual equation (\ref{charge}) gives $q=g$. Using Dirac quantization rule for $N$ $D3$ brane one getss $q=g=\sqrt{2\pi}N$.
Now considering type IIB supergravity action with only $F_5$ flux and solving Einstein's equation with the above ansatz one getss the following solution:
\begin{equation}\begin{split}\label{IIBsol}
& f_{1}(r)=H^{-1/2}(r),~~~~~~~~~~~~~~~f_{2}(r)=H^{1/2}(r),
\\\textrm{with} ~~& H(r)=\left(1+\frac{L^4}{r^4}\right),~~~~~~~~~~~~~~~L^4=4\pi g_s N (\alpha^{\prime})^2.
\end{split}
\end{equation}
The above solution has the following properties.
\begin{itemize}
\item In the limit $r\rightarrow 0$, $H(r)=1$ and one recovers the ten dimensional minkowski metric.
\item For $r\gg L$, $H(r)=1+\mathcal{O}(\frac{L^4}{r^4})$ and there is a long range Coulomb potential that varies as $\frac{1}{r^4}$ in $\mathbb{R}^6$.
\item Near $r\sim L$, the spacetime deforms from flat minkowski metric due to gravitational effects of the D branes and the curvature becomes $L^{-2}$
\end{itemize}
Now, in the classical gravity limit, one must have very small curvature of spacetime compared to $\frac{1}{\alpha^{\prime}}$. Also in the classical limit the quantum loop corrections can be ignored and hence the string coupling $g_s$ must be very small. So in the supergravity approximation the region of validity of the solution is given by the following limit:
\begin{equation}
g_s\ll 1~~~~~~~~~~\textrm{and}~~~~~~~~~~\alpha^{\prime}L^{-2}\ll 1.
\end{equation}
In the near horizon limit $(r\rightarrow 0)$, the metric has the form:
\begin{equation}\label{ADSmetric1}
ds^2=\frac{r^2}{L^2}\left(-dt^2+\sum^{3}_{i=1}\right)+\frac{L^2}{r^2}dr^2+L^2 d\Omega^2_5.
\end{equation}
From the above metric one realize that the proper distance $l$ of the $D3$ brane location in $\mathbb{R}^6$ can be calculated as:
\begin{equation}
\int \frac{dr^2}{r^2}=l\\
 \Rightarrow  l=\log{r}+\textrm{constant}.
\end{equation}
So as $r\rightarrow 0$, the proper distance $l$ goes to negative infinity. Therefore it is not possible to reach the location of $D3$ branes as it is infinite proper distance away. In other words, at $r=0$ the $D3$ branes disappear and the only thing left is the flux of $F_5$ through $S^5$. In trying to reach the point $r=0$, one finds an infinite `throat' and not the point itself. The cross section of the throat is $S^5$ with radius $L$. The metric of the throat region is that of $AdS^5\times S^5$. Of course at $r\rightarrow\infty$ it is still a flat Minkowski space.
\begin{itemize}
\item \it{Dual description of $D3$ branes}
\end{itemize}
The dual description of $D$ branes has already been discussed in the previous subsection. Here we will discuss the case of a stack of $D3$-branes in details.

(i) Description {\bf A}: In this case a $D3$-brane is an object in ten dimensional space extended along $\mathbb{R}^{1,3}$ where open strings can end. There are also closed strings in the same spacetime but outside the brane world volume which can interact with the open strings on the branes as well as with the other closed strings in the space.

(ii) Description {\bf B}: In this case the $D3$ branes are considered as massive charged objects that curve the spacetime. So here, as we just discussed, there are no $D3$ branes at all but a $AdS$ metric and $F_5$ flux through $S^5$. This means there are only closed strings in curved spacetime and no open strings.

These two descriptions must be equivalent as both of them describes the same configuration. An important point at this time is that, although the supergravity approximation was considered at the beginning to get the solution, the above equivalence in general can be extended for any $\alpha^{\prime}$ and $g_s$. In short the equivalence between description {\bf A} and {\bf B} is valid for all $\alpha^{\prime}$ and $g_s$. Now considering the low energy limit of each description has interesting consequences. The low energy limit corresponds to sending $\alpha^{\prime}E^2$ to zero, with $E$ defined as the energy. Let us discuss the low energy limit of the two descriptions below.

Low energy description of {\bf A}: one getss (i) $\mathcal{N}=4$ SYM theory in four dimensions with gauge group $U(N)$ from the open string sector. The gauge coupling $g_{YM}$ is related to string coupling $g_s$ as $g^2_{YM}=4\pi g_s$ and (ii) graviton, dilaton and other massless modes are part of the closed string sector. The interaction between the closed and open strings or between two closed strings is governed by the gravitational constant $G_N$ given by:
\begin{equation}
G_N\sim g^2_s(\alpha^{\prime})^4.
\end{equation}
Hence in the low energy limit the dimensionless quantity $G_N E^8$ goes to zero and the close/open string or two different close string do not interact in the low energy limit. So from picture {\bf A} one gets,
\begin{equation}
\nonumber \underbrace{\mathcal{N}=4 ~\textrm{SYM}~\textrm{theory}}_{\textrm{Open~ string~ sector}}+\underbrace{\textrm{Free}~ \textrm{graviton}}_{\textrm{Close~ string~ sector}}.
\end{equation}

Low energy limit of {\bf B}: Since the spacetime is curved, time is no more absolute and hence one has to consider energy at local proper time $\tau$, defined as $E_{\tau}$. The energy $E$ in description {\bf A} is at time $`t`$ which is the time at $r\rightarrow\infty$. The relation between $E_{\tau}$ and $E$ can be obtained from the metric (\ref{ADSmetric}) as $E_{\tau}=H^{1/4}E$.

In the region $r\gg L$, $H\approx 1$ and hence taking the low energy limit, all the massive modes are suppressed and one is left with the massless gravitons.
On the other hand the region $r\ll L$, $H\approx \frac{L^4}{r^4}$ and $E^2 \alpha^{\prime}\rightarrow 0$ gives $E_{\tau}\frac{r^2}{\sqrt{4\pi g_s N}}\rightarrow 0$. So in this case the low energy limit can be satisfied by sending $r$ to zero for any value of $E_{\tau}$. In other words, if one go deep enough inside the "throat" region, no matter what the proper energy is, it will always appear energetically low from the region $r\ll L$. Therefore in the low energy limit, the excitations of region $r\gg L$ decouples from that of $r\ll L$. Also as $E_{\tau}$ can be arbitrarily large, the excitations in $r\ll L$ region can also include massive modes. Hence from the low energy limit of description {\bf B} one gets:
\begin{equation}
\nonumber {\rm Full~ string~ theory~ in}~ AdS_{5}\times S^5 ~{\rm with ~fluxes}+{\rm Free~ graviton~ at}~ r\rightarrow\infty
\end{equation}
Finally from the equivalence of the two descriptions one conclude that:
\begin{itemize}
\item $\mathcal{N}=4$ SYM theory with $SU(N)$ gauge group on $\mathbb{R}^{1,3}$ is equivalent to type IIB string theory in $AdS_{5}\times S^5$.
\end{itemize}
This is the strongest form of the AdS/CFT correspondence. We will not discuss this any further here. For detailed discission on gauge/gravity duality see \cite{JEgaugegravity} \cite{JEgaugegravity2}.
\section{Generalization of AdS/CFT correspondence}
The AdS/CFT correspondence we just discussed describes a gauge theory which is maximally supersymmetric with $16$ supercharges and also conformal. This is very far from a realistic theory which is less supersymmetric and non-conformal. So a very natural and obvious question is whether the approach that one uses to establish the AdS/CFT correspondence, can be used to understand the Physics of non-conformal gauge theories. To be more specific, one has to generalize the AdS/CFT correspondence to find a supergravity description which is dual to QCD like theories. For this one has to try to
\begin{itemize}
\item reduce and in fact break the supersymmetry
\item break the conformal invariance
\end{itemize}
of the dual gauge theory.

Now the discussion of the previous section suggest that the gravity dual of non-conformal gauge theories should include non-AdS like geometry. So the recipe that works well for the AdS/CFT might not be useful at all for the present case. Another problem in establishing the non-AdS/non-CFT correspondence is the issue of decoupling. In AdS/CFT, there is a decoupling between the degrees of freedom of the open and closed sector in the low energy limit. This decoupling leads to the equivalence between the maximally supersymmetric gauge theory and closed superstring theory on anti-de Sitter space. However, in a particular limit of the dimensionless parameters of the two theories, the above correspondence can also be achieved in the supergravity approximation of the superstring theory without any stringy correction. For the non-conformal case, the complete decoupling of the gauge degrees of freedom from that of the closed string sector is not possible within the supergravity regime. In this case one must include the string states and hence has to go beyond the supergravity limit. It is not an easy task to check the duality beyond the supergravity limit. Although, one can still consider only the supergravity modes and try to study non-conformal field theories for some interesting properties as much as possible using the open/close strong duality. Hence, stepwise one requires to (i) consider D branes in particular geometry which is not maximally supersymmetric (ii) write down the DBI action for those $D$-branes and study their dynamics (iii) find a way to break conformal invariance and finally (iv) use the open/closed string duality. See \cite{S.Roy-1}, \cite{S.Roy-2}, \cite{S.Roy-3}, \cite{S.Roy-4}, \cite{S.Roy-5}, \cite{S.Roy-6},\cite{S.Roy-7} for more recent works on the aforementioned decoupling and extension of the AdS/CFT to non-supersymmetric cases.

There are different ways to achieve the above. But the one that would be relevant here is to choose a specific target spacetime along with a particular D brane configuration such that the conformal invariance and the supersymmetry are both broken from the very beginning. Some examples of such configurations are:
Regular D branes on orbifold and conifild geometries, Fractional D branes at orbifold or conifold fixed points, D branes wraping non trivial cycles of the Calabi Yau spaces, branes suspended between other branes etc.
Here we will discuss the D brane configurations in Calabi Yau spaces with conical singularities. Now, $D$-branes placed at any smooth point inside a CY manifold the same maximally supersymmetric $\mathcal{N}=4$ SYM theory as AdS/CFT. Instead if they are placed at the conical singularity, then the amount of supersymmetry can be broken. So let's discuss a specific configuration of regular $D3$ branes at the singular point of a conifold below.
\subsection{Regular $D3$ branes at conifold singularity}
A conifold is a Calabi Yau three fold and can be described as a cone over a five dimensional base. The metric of a conifold is given as \cite{Candelas} \cite{Knauf+Gwyn[2007]}:
\begin{equation}\label{ds6}
ds^{2}_{6}=dr^{2}+r^{2}ds^{2}_{T^{1,1}},
\end{equation}
where the base $T^{1,1}$ is a Sasaki-Einstein manifold and is given as,
\begin{equation}
ds^{2}_{T^{1,1}}=\frac{1}{9}\left[d\psi+\sum^{2}_{i=1}\cos\theta_{i}d\phi_{i}\right]^{2}+\frac{1}{6}\sum^{2}_{i=1}\left[d\theta^{2}_{i}
+\sin^{2}\theta_{i}d\phi^{2}_{i}\right].
\end{equation}
where $\theta_i\in[0,\pi],\ \phi_i\in[0,2\pi]$ and $\psi\in[0,4\pi]$; see \cite{Skenderis-4} for more recent applications of Sasaki-Einstein manifolds to type IIB string theory (trucation). The topology of $T^{1,1}$ is that of $S^{2}\times{S^{3}}$. At the tip of the conifold $\left(r=0\right)$, both the spheres shrinks to point and hence there is a singularity. Also $T^{1,1}$ has $SU(2)\times{SU(2)}\times{U(1)}$ symmetry. With this one embeds $N$ $D3$-branes in 10-dimensional spacetime with the metric
\begin{equation}
ds^{2}_{10}= -dt^{2}+\sum^{3}_{i=1}dx^{2}_{i}+ds^{2}_{6},
\end{equation}
where $dS^{2}_{6}$ is given in (\ref{ds6}). Here the $D3$-branes live along the four dimensional flat Minkowski spacetime and are fixed at the tip of the conifold. The resulting gauge theory on such $D3$-branes is a $\mathcal{N}=1$ supersymmetric theory with gauge group $SU(N)\times{SU(N)}$ coupled to complex matter fields $A_{i},B_{i},i=1,2$ transforming in the bi-fundamental representation. Since the dilation and the NS-NS $B_{2}$ are both constant, the gauge couplings of the two gauge groups do not run and the theory is conformal. Let's now analyze the string theory solution of the same configuration in the supergravity action with only self dual five from flux $F_{5}$, the metric is found to be
\begin{equation}\label{KWmetric}
 \begin{split}
 &ds^{2}=h^{-1/2}\left(-dt^{2}+\sum^{3}_{i=1}{dx^{2}_{i}}\right)+h^{1/2}dr^{2}+r^{2}+r^{2}h^{1/2} ds^{2}_{T^{1,1}}\\
 & F_{5}=\frac{1}{g}\left[d^{4}x\wedge dh^{-1}+\ast\left(d^{4}x\wedge dh^{-1}\right)\right]\\
 &\textrm{with}~ h=1+\frac{L^{4}}{r^{4}},~~~~~~~~~~ L^{4}=4\pi g_{s}N\alpha^{{\prime}^{2}}
 \end{split}
\end{equation}
From (\ref{KWmetric}) one see that the near-horizon geometry is that of $AdS_{5}\times{T^{1,1}}$. This is the Klebanov-Witten (KW) \cite{KW} model.
\subsection{Fractional $D3$ branes at conifold singularity}
To break the conformal invariance of the KW model, a stack of M $D5$ branes was introduced wrapping vanishing two-cycle of the conifold base and can be interpreted as fractional $D3$ branes at the tip of the conifold. These fractional branes changes the gauge group to $SU\left(M+N\right)\times{SU\left(N\right)}$. Addition of fractional branes preserves $\mathcal{N}=1$ supersymmetry with the matter fields $A_{i}/B_{i}$ transforming as under the $(\overline{M+N},N)$ representation of the gauge group.
The supergravity solution of this brane setup was obtained exactly by Klebanov and Tseytlin \cite{KT}. The warp factor $h(r)$ in (\ref{KWmetric}) is now given as:
\begin{equation}\label{hKT}
h\left(r\right)= \frac{L^{4}}{r^{r}}\left(1+\frac{3g_{s}M^{2}}{2\pi N}\log r\right).
\end{equation}
Unlike KW solution, in this case one has finite three forms flux sourced by the $D5$ branes along with the $F_{5}$ flux. Also the NS-NS $B$ field is not constant anymore and depends logarithmically on $r$. This non trivial $B_{2}$ field is responsible for the running of the gauge coupling and hence breaking the conformality.

The solution for the fluxes and the $B_{2}$ field is given as
\begin{equation}
 \begin{split}
 F_{3}=M w_{3}, ~~~~~~~~~~~~~~~~B_{2}=3g_{s}M w_{2}\log{\frac{r}{r_{0}}}.
\end{split}
\end{equation}
where $w_{3}\wedge{w_{2}}$ gives the volume of $T^{1,1}$. The self dual five form flux including the back reaction from $F_{3}$ is given as
\begin{equation}
\begin{split}
\widetilde{F_{5}}=F_{5}+\ast_{10}F_{5},~~~~
\text{with}~~ F_{5}=\left(M+\frac{3}{2\pi}g_{s}M^{2}\log{\frac{r}{r_{0}}}\right)\text{vol}(T^{1,1}).
\end{split}
\end{equation}
Where if one define the effective number of $D3$ branes as,
\begin{equation}\label{Neff}
N_{eff}\left(r\right)= N+\frac{3}{2\pi}g_{s}M^{2}\log{ \frac{r}{r_{0}}}.
\end{equation}
then $N_{eff}\left(r\right)$ deceases as $r\rightarrow{0}$, While the number of fractional $D3$ branes remains constant. Also notice from equation (\ref{hKT}) that the warp factor $h\left(r\right)$ also vanishes at some $r$ in the IR region. This makes the KT solution singular in the IR.
\subsection{Seiberg duality cascade}
From the expression of $N_{eff}$ as given in (\ref{Neff}), one can see that the decease of the number of $D3$ branes with $r$ is in units of $M$. This changes the group from  $SU\left(N+M\right)\times{SU\left(N\right)}$ to $SU\left(N\right)\times{SU\left(N-M\right)}$. This is known as the Seiberg duality. It was realized by Klebanov and Strasslar \cite{KS} that as one moves from UV into the deep IR region, the theory goes through a series of Seiberg dualities called the Seiberg duality cascade. If $N$ is an integer multiple of $M$ then in the far IR, after the duality cascade all of the $N$ $D3$ branes cascade away leaving behind only $M$ fractional $D3$ branes with gauge group $SU\left(M\right)$.
Now the issue of singularity of the KT solution in the IR regime can be fixed via strong IR dynamics. More precisely in the KS solution the $U\left(1\right)_{R}$ symmetry of the chiral field is broken at the quantum level to $\mathbb{Z}_{2M}$ in the presence of M fractional branes. This $\mathbb{Z}_{2M}$ symmetry is then spontaneously broken to $\mathbb{Z}_{2}$ in the IR due to gaugino condensation resulting in desingularizing the singular conifold into a deformed conifold.

So in the IR regime the KS solution realizes a deformed conifold metric. Also due to duality cascade the gauge theory in the IR is pure $SU(M)$ SYM theory. However in the UV the supergravity solution is still singular as the metric in the UV is that of a singular conifold. In that sense the KS model shows a geometric transition from UV to IR. The gauge theory in UV has an infinite number of massive states. The logarithmic gauge coupling of the KS model diverges in the UV. This necessitates modification of the UV sector of KS.
\subsection{Resolved conifold and $D7$ brane embedding}
The singularity of the KT solution in the deep IR region can also be removed by an $S^2$ blow-up at the tip of the conifold. This is precisely the resolved conifold geometry. The six dimensional metric of the resolved conifold is given as,
\begin{equation}\label{resmetric}
\begin{split}
ds^2_{res}&=\kappa(\rho)^{-1}d\rho^2+\frac{\kappa(\rho)}{9}\rho^2\left(d\psi+\cos\theta_1 d\phi_1+\cos\theta_2 d\phi_2\right)^2\\&+\frac{\rho^2}{6}\left(d\theta^2_1+\sin^{2}{\theta_1} d\phi^2_1\right)+\frac{\rho^2+6a^2}{6}\left(d\theta^2_2+\sin^2{\theta_2} d\phi^2_2\right),
\end{split}
\end{equation}
where $\kappa(\rho)=\frac{\rho^2+9 a^2}{\rho^2+6 a^2}$ and $a$ is called the Resolution parameter. Here $\rho$ is a newly defined radial coordinate and is related to $r$. In the limit $\rho\rightarrow 0$, one realize from (\ref{resmetric}) that the sphere parameterized by $\left(\theta_2,\phi_2\right)$ remains finite because of the finite resolution $a$. For $a\rightarrow 0$, one getss back the singular conifold metric.

In \cite{PT}, a configuration with $N$ regular $D3$ branes at the tip of a resolved conifold along with $M$ fractional $D3$ branes wrapping the blow-up two cycle of the same, was considered. The resulting supergravity solution has an RR three-form $F_3$, a NS-NS three-form $H_3$ and a self dual five form $F_5$ with a constant dilaton $\phi$. The expression of all these are given in \cite{PT} and they depends on the resolution parameter $a$. The three form $G_3$ defined as $F_3+\iota H_3$ is imaginary self dual. $G_3$ has both primitive $(2,1) $ part and non-primitive $(1,2)$ part. The non-primitive structure of $G_3$ breaks the supersymmetry of the PT solution.

To complete the story one needs to include the fundamental quarks also in the theory. This is done by the inclusion of flavor branes. The pioneer work on this was done by \cite{karch} where $D7$ brans were embedded in the AdS geometry in the probe limit to ignore the back reaction on the target space. A variety of interesting issues has been looked upon with flavor branes using gauge/gravity duality in \cite{arnabkundu1708.01775}\cite{arnabkundu1612.08624}\cite{arnabkundu1507.00818}\cite{arnabkundu1306.2178}. In the context of embedding flavor branes in conifold background, interesting work has been done by \cite{ouyang}, where a stack of $D7$ branes were considered in a singular conifold geometry via supersymmetry preserving holomorphic Ouyang embedding; see \cite{Nunez-2} for a discussion of studying flavor dynamics in the so-called Veneziano limit, and \cite{Nunez-3}, \cite{Nunez-4} in the Klebanov-Strassler/Tseytlin/Witten backgrounds, in the context of gauge-gravity duality. However, the particular configuration which is relevant to us is the one given in \cite{dasgupta d term}. In this case a stack of $N_f$ $D7$ branes were embedded in a non supersymmetric resolved conifold background we just discussed via the same Ouyang embedding. These $D7$ branes wraps a non trivial four cycle inside the resolved conifold geometry with the embedding equation given as,
\begin{equation}
 \label{eq:Ouyang_embedding_RC}
 \left(\rho^6 + 9 a^2\rho^4\right)^{\frac{1}{4}}e^{\frac{i}{2}(\psi-\phi_1-\phi_2)}\sin\frac{\theta_1}{2}\sin\frac{\theta_2}{2} = \mu,
 \end{equation}
with $\mu $ as the Ouyang embedding parameter.   Due to these $D7$ branes, the dilaton $\Phi$ is no longer a constant but runs with $r$:
\begin{equation}\label{dilaton1}
e^{-\Phi}=\frac{1}{g_s}-\frac{N_f}{8\pi}\log\left(r^6+9a^2r^4\right)-\frac{N_f}{2\pi}\log\left(\sin{\frac{\theta_1}{2}}\sin{\frac{\theta_2}{2}}\right).
\end{equation}
In \cite{dasgupta d term}, the Einstein's equation was solved in this non trivial dilaton background.

Now, as the holomorphic Ouyang embedding is supersymmetric, one must find a way to break supersymmetry. It turns out that the pullback of the non primitive $(1,2)$ flux of the $B_2$ field onto the four cycle that the $D7$ brane wraps, creates an additional D-term to the superpotential which is responsible for the breaking of supersymmetry. In the same paper this D-term was evaluated exactly using the solution in the non trivial dilaton background for the simple embedding with $\mu=0$.
\subsection{Finite temperature and the Dasgupta et al's set up}
In order to introduce temperature, one must consider non-extremal geometry in the dual supergravity side of the theory. In \cite{Buchel}, finite-temperature/non-extremal version of the abovementioned KT solution was considered with the proposition that the aforementioned KT singularity is cloaked behind $r=r_h$(horizon radius) making therefore Seiberg duality cascade, unnecessary. Unfortunately, the solution was not regular as the non-extremality/black hole function and the ten-dimensional warp factor vanished simultaneously at the horizon radius $r_h$. The authors of \cite{Gubser-et-al-finitetemp} were able to construct a supergravity dual of $SU(M+N)\times SU(N)$ gauge theory which approached the abovementioned KT solution asymptotically and possessed a well-defined horizon. The same was characterized by: modification of $T^{1,1}$ via a `squashing factor' of the $U(1)_\psi$ fiber, non-constancy of the dilaton and non self-duality of the fluxes. But it was valid only for large temperatures with no fundamental quark flavors.

In order to include fundamental quarks at non-zero temperature in the context of type IIB string theory, to the best of our knowledge, the following model proposed in \cite{metrics} (and various aspects also nicely explained in \cite{Keshav-more-1},\cite{Keshav-more-2}, \cite{Keshav-more-3}, \cite{Keshav-more-4}, \cite{Keshav-more-5}) is the closest to a UV complete holographic dual of large-$N$ thermal QCD. The KS (duality cascade) and QCD have similar IR behavior: $SU(M)$ gauge group and IR confinement; see \cite{Nunez-6} for confining ${\cal N}=1$ gauge theories arising from $NS5/D5$-branes wrapping two-cycles. However, they differ drastically in the UV as the former yields a logarithmically divergent gauge coupling (in the UV) - Landau pole. This is due to the presence of non zero three form flux which grows logarithmically to infinity in the deep UV region.  This necessitates modification of the UV sector of KS apart from inclusion of non-extremality factors. With this in mind and building up on all of the above, the type IIB holographic dual of
 \cite{metrics} was constructed. The brane construction of \cite{metrics} is summarized below.
\begin{itemize}
 \item
From a gauge-theory perspective, the authors of \cite{metrics} considered a stack of $N$ $D3$-branes and $M$ $D5$ branes wrapping the vanishing two cycle placed at the tip of six-dimensional conifold along with $M\ \overline{D5}$-branes around the antipodal point relative to the location of $M\ D5$ branes on the blown-up $S^2$ of the cone. In terms of the radial direction $r$, the deep IR is defined as the region with very small value of $r$ where the gauge theory is confining with the gauge group $SU(M)$ due to duality cascade. Now as $r$ increases, one can feel the presence of $\overline{D5}$ branes, and the three form flux starts decaying until at some point all it just goes to zero. This region is called the IR-UV interpolating region. Beyond this there are no three form flux and it is the UV region. If one Define the $D5/\overline{D5}$ separation as ${\cal R}_{D5/\overline{D5}}$, then this provides the boundary common to the outer UV-IR interpolating region and the UV region.
\item
 $N_f\ D7$-branes, via Ouyang embedding, are holomorphically embedded in the UV (asymptotically $AdS_5\times T^{1,1}$), the IR-UV interpolating region and dipping into the (confining) IR (up to a certain minimum value of $r$ corresponding to the lightest quark) and $N_f\ \overline{D7}$-branes present in the UV and the UV-IR interpolating (not the confining IR). This is to ensure turning off of three-form fluxes, constancy of the axion-dilaton modulus and hence conformality and absence of Landau poles in the UV. Further, the global  flavor group in the UV-IR interpolating and UV regions, due to presence of $N_f$ $D7$ and $N_f\ \overline{D7}$-branes, is $SU(N_f)\times SU(N_f)$, which is broken in the IR to $SU(N_f)$ as the IR has only $N_f$ $D7$-branes. The same kind of $D7-\overline{D7}$ brane configurations were also considered in Kuperstein-Sonnenschein model at finite temperature in \cite{arnabkundu1208.2663}.
 \end{itemize}
Due to the presence of $\overline{D5}$ and $\overline{D7}$ branes, one has $SU(N+M)\times SU(N+M)$ color gauge group (implying an asymptotic $AdS_5$) and $SU(N_f)\times SU(N_f)$ flavor gauge group, in the UV: $r\geq {\cal R}_{D5/\overline{D5}}$. It is expected that there will be a partial Higgsing of $SU(N+M)\times SU(N+M)$ to $SU(N+M)\times SU(N)$ at $r={\cal R}_{D5/\overline{D5}}$  \cite{K. Dasgupta et al [2012]}. The two gauge couplings, $g_{SU(N+M)}$ and $g_{SU(N)}$ flow  logarithmically  and oppositely in the IR:
\begin{equation}
\label{RG}
4\pi^2\left(\frac{1}{g_{SU(N+M)}^2} + \frac{1}{g_{SU(N)}^2}\right)e^\phi \sim \pi;\
 4\pi^2\left(\frac{1}{g_{SU(N+M)}^2} - \frac{1}{g_{SU(N)}^2}\right)e^\phi \sim \frac{1}{2\pi\alpha^\prime}\int_{S^2}B_2.
\end{equation}
Had it not been for $\int_{S^2}B_2$, in the UV, one could have set $g_{SU(M+N)}^2=g_{SU(N)}^2=g_{YM}^2\sim g_s\equiv$ constant (implying conformality) which is the reason for inclusion of $M$ $\overline{D5}$-branes at the common boundary of the UV-IR interpolating and the UV regions, to annul this contribution. In fact, the running also receives a contribution from the $N_f$ flavor $D7$-branes which needs to be annulled via $N_f\ \overline{D7}$-branes. The gauge coupling $g_{SU(N+M)}$ flows towards strong coupling and the $SU(N)$ gauge coupling flows towards weak coupling. Upon application of Seiberg duality, $SU(N+M)_{\rm strong}\stackrel{\rm Seiberg\ Dual}{\longrightarrow}SU(N-(M - N_f))_{\rm weak}$ in the IR; assuming after repeated Seiberg dualities or duality cascade, $N$ decreases to 0 and there is a finite $M$, one will be left with $SU(M)$ gauge theory with $N_f$ flavors that confines in the IR - the finite temperature version of the same is what was looked at by \cite{metrics}.
The resultant ten-dimensional geometry hence involves a resolved warped deformed conifold; see \cite{Nunez-1} for appearance of resolved warped deformed conifolds in type IIB solutions generated from type I involving a combination of Higgsing and Seiberg duality cascade. Back-reactions are included, e.g., in the ten-dimensional warp factor. Of course, the gravity dual, as in the KS construct, at the end of the Seiberg-duality cascade will have no D3-branes and the D5-branes are smeared/dissolved over the blown-up $S^{3}$ and thus replaced by fluxes in the IR.

Let's now concentrate on the supergravity solution in this resolved warped deformed conifold background with a black-hole. In \cite{metrics}, assuming an ansatz for the metric where there is a squashing between the two 2-spheres of the conifold base, the type IIB supergravity action was solved for the wrap factor and different fluxes which now will receive backreaction from the non-extremal geometry and also from the $D7$ brane. The supergravity limit in this case is given as,
\begin{equation}\begin{split}\label{limit}
\left(N,M,N_f,g_{s}N,g_{s}M\right)&\sim {\rm Large}\\
\left(g_{s},g_{s}M/N,g_{s}N_{f},M/N\right)&\sim {\rm Small}.\end{split}
\end{equation}
The final form of the metric is given by,
\begin{equation}
\label{metric}
ds^2 = \frac{1}{\sqrt{h}}
\left(-g_1 dt^2+dx_1^2+dx_2^2+dx_3^2\right)+\sqrt{h}\biggl[g_2^{-1}dr^2+r^2 d{\cal M}_5^2\biggr],
\end{equation}
where the black hole functions $g_i$ in the limit (\ref{limit}) for the introduction of temperature in the gravity side are of the form,
\begin{equation}
g_{1,2}(r,\theta_1,\theta_2)= 1-\frac{r_h^4}{r^4} + {\cal O}\left(\frac{g_sM^2}{N}\right),
\end{equation}
where $r_h$ is the horizon, and the ($\theta_1, \theta_2$) dependence come from the ${\cal O}\left(\frac{g_sM^2}{N}\right)$ corrections; see \cite{Minwalla-3} for a review on large-$N$ gauge theoretic description of black holes in $AdS_5\times S^5$ in backgrounds dual to  confining gauge theories.
The compact five dimensional metric in (\ref{metric}), is given as:
\begin{eqnarray}
\label{RWDC}
& & d{\cal M}_5^2 =  h_1 (d\psi + {\rm cos}~\theta_1~d\phi_1 + {\rm cos}~\theta_2~d\phi_2)^2 +
h_2 (d\theta_1^2 + {\rm sin}^2 \theta_1 ~d\phi_1^2) +   \nonumber\\
&&  + h_4 (h_3 d\theta_2^2 + {\rm sin}^2 \theta_2 ~d\phi_2^2) + h_5~{\rm cos}~\psi \left(d\theta_1 d\theta_2 -
{\rm sin}~\theta_1 {\rm sin}~\theta_2 d\phi_1 d\phi_2\right) + \nonumber\\
&&  + h_5 ~{\rm sin}~\psi \left({\rm sin}~\theta_1~d\theta_2 d\phi_1 +
{\rm sin}~\theta_2~d\theta_1 d\phi_2\right),
\end{eqnarray}
where in the UV/IR-UV interpolating region, $r\gg a, r \gg({\rm deformation\ parameter})^{\frac{2}{3}}$ and hence $h_5\sim\frac{({\rm deformation\ parameter})^2}{r^3}\ll  1$.  The $h_i$'s appearing in internal metric are not constant and up to linear order depend on $g_s, M, N_f$ are given as below:
\begin{eqnarray}
\label{h_i}
& & \hskip -0.45in h_1 = \frac{1}{9} + {\cal O}\left(\frac{g_sM^2}{N}\right),\  h_2 = \frac{1}{6} + {\cal O}\left(\frac{g_sM^2}{N}\right),\ h_4 = h_2 + \frac{a^2}{r^2},\nonumber\\
& & h_3 = 1 + {\cal O}\left(\frac{g_sM^2}{N}\right),\ h_5\neq0,\
%&& \hskip -0.45in M_{\rm eff}/N_{f}^{\rm eff} = M/N_f + \sum_{m\ge n} (a/b)_{mn} (g_sN_f)^m (g_sM)^n,\nonumber\\
 L=\left(4\pi g_s N\right)^{\frac{1}{4}}.
\end{eqnarray}
One sees from (\ref{RWDC}) and (\ref{h_i}) that one has a non-extremal resolved warped deformed conifold involving
an $S^2$-blowup (as $h_4 - h_2 = \frac{a^2}{r^2}$), an $S^3$-blowup (as $h_5\neq0$) and squashing of an $S^2$ (as $h_3$ is not strictly unity). The horizon (being at a finite $r=r_h$) is warped squashed $S^2\times S^3$. The resolution parameter $a$ is no longer a constant and depends on the horizon radius $r_h$ due to non-extremal geometry.
 The warp factor that includes the back-reaction is given as,
\begin{eqnarray}
\label{eq:h}
&& \hskip -0.45in h =\frac{L^4}{r^4}\Bigg[1+\frac{3g_sM_{\rm eff}^2}{2\pi N}{\rm log}r\left\{1+\frac{3g_sN^{\rm eff}_f}{2\pi}\left({\rm
log}r+\frac{1}{2}\right)+\frac{g_sN^{\rm eff}_f}{4\pi}{\rm log}\left({\rm sin}\frac{\theta_1}{2}
{\rm sin}\frac{\theta_2}{2}\right)\right\}\Biggr],
\end{eqnarray}
where, in principle, $M_{\rm eff}/N_f^{\rm eff}$ are not necessarily the same as $M/N_f$; we however will assume that up to ${\cal O}\left(\frac{g_sM^2}{N}\right)$, they are.
The three-forms fluxes, up to ${\cal O}(g_s N_f)$ (\ref{limit}) and setting $h_5=0$,  are as given as \cite{metrics},
\begin{eqnarray}
\label{three-form-fluxes}
& & \hskip -0.4in (a) {\widetilde F}_3  =  2M { A_1} \left(1 + \frac{3g_sN_f}{2\pi}~{\rm log}~r\right) ~e_\psi \wedge
\frac{1}{2}\left({\rm sin}~\theta_1~ d\theta_1 \wedge d\phi_1-{ B_1}~{\rm sin}~\theta_2~ d\theta_2 \wedge
d\phi_2\right)\nonumber\\
&& \hskip -0.3in -\frac{3g_s MN_f}{4\pi} { A_2}~\frac{dr}{r}\wedge e_\psi \wedge \left({\rm cot}~\frac{\theta_2}{2}~{\rm sin}~\theta_2 ~d\phi_2
- { B_2}~ {\rm cot}~\frac{\theta_1}{2}~{\rm sin}~\theta_1 ~d\phi_1\right)\nonumber \\
&& \hskip -0.3in -\frac{3g_s MN_f}{8\pi}{ A_3} ~{\rm sin}~\theta_1 ~{\rm sin}~\theta_2 \left(
{\rm cot}~\frac{\theta_2}{2}~d\theta_1 +
{ B_3}~ {\rm cot}~\frac{\theta_1}{2}~d\theta_2\right)\wedge d\phi_1 \wedge d\phi_2, \nonumber\\
& & \hskip -0.4in (b) H_3 =  {6g_s { A_4} M}\Biggl(1+\frac{9g_s N_f}{4\pi}~{\rm log}~r+\frac{g_s N_f}{2\pi}
~{\rm log}~{\rm sin}\frac{\theta_1}{2}~
{\rm sin}\frac{\theta_2}{2}\Biggr)\frac{dr}{r}\nonumber \\
&& \hskip -0.3in \wedge \frac{1}{2}\Biggl({\rm sin}~\theta_1~ d\theta_1 \wedge d\phi_1
- { B_4}~{\rm sin}~\theta_2~ d\theta_2 \wedge d\phi_2\Biggr)
+ \frac{3g^2_s M N_f}{8\pi} { A_5} \Biggl(\frac{dr}{r}\wedge e_\psi -\frac{1}{2}de_\psi \Biggr)\nonumber  \\
&&  \wedge \Biggl({\rm cot}~\frac{\theta_2}{2}~d\theta_2
-{ B_5}~{\rm cot}~\frac{\theta_1}{2} ~d\theta_1\Biggr). \nonumber\\
\end{eqnarray}
The asymmetry factors in (\ref{three-form-fluxes}) are given by \cite{metrics},
\begin{equation}\begin{split}
A_i&=1 +{\cal O}\left(\frac{a^2}{r^2}\ {\rm or}\ \frac{a^2\log r}{r}\ {\rm or}\ \frac{a^2\log r}{r^2}\right) + {\cal O}\left(\frac{{\rm deformation\ parameter }^2}{r^3}\right),  \\B_i& = 1 + {\cal O}\left(\frac{a^2\log r}{r}\ {\rm or}\ \frac{a^2\log r}{r^2}\ {\rm or}\ \frac{a^2\log r}{r^3}\right)+{\cal O}\left(\frac{({\rm deformation\ parameter})^2}{r^3}\right).
\end{split}
\end{equation}
As in the UV, $\frac{({\rm deformation\ parameter})^2}{r^3}\ll  \frac{({\rm resolution\ parameter})^2}{r^2}$, we will assume the same three-form fluxes for $h_5\neq0$.Further, to ensure UV conformality, it is important to ensure that the axion-dilaton modulus approaches a constant implying a vanishing beta function in the UV.
\subsection{The `MQGP Limit'}
In \cite{MQGP}, the authors had considered the following two limits,
\begin{eqnarray}
\label{limits_Dasguptaetal-i}
&   & \hskip -0.17in (i) {\rm weak}(g_s){\rm coupling-large\ t'Hooft\ coupling\ limit}:\nonumber\\
& & \hskip -0.17in g_s\ll  1, g_sN_f\ll  1, \frac{g_sM^2}{N}\ll  1, g_sM\gg1, g_sN\gg1\nonumber\\
& & \hskip -0.17in {\rm effected\ by}: g_s\sim\epsilon^{d}, M\sim\left({\cal O}(1)\epsilon\right)^{-\frac{3d}{2}}, N\sim\left({\cal O}(1)\epsilon\right)^{-19d}, \epsilon\ll  1, d>0
 \end{eqnarray}
(the limit in the first line  though not its realization in the second line, considered in \cite{metrics});
\begin{eqnarray}
\label{limits_Dasguptaetal-ii}
& & \hskip -0.17in (ii) {\rm MQGP\ limit}: \frac{g_sM^2}{N}\ll  1, g_sN\gg1, {\rm finite}\
 g_s, M\ \nonumber\\
& & \hskip -0.17in {\rm effected\ by}:  g_s\sim\epsilon^d, M\sim\left({\cal O}(1)\epsilon\right)^{-\frac{3d}{2}}, N\sim\left({\cal O}(1)\epsilon\right)^{-39d}, \epsilon\lesssim 1, d>0.
\end{eqnarray}
Let us now elaborate upon the motivation for considering the MQGP limit. There are principally two.
\begin{enumerate}
\item
Unlike the AdS/CFT limit wherein $g_{\rm YM}\rightarrow0, N\rightarrow\infty$ such that $g_{\rm YM}^2N$ is large, for strongly coupled thermal systems like sQGP, what is relevant is $g_{\rm YM}\sim{\cal O}(1)$ and $N_c=3$. In the IR after the Seiberg duality cascade, effectively $N_c=M$ which in the MQGP limit of (\ref{limits_Dasguptaetal-ii})  can be tuned to 3. Further, in the same limit, the string coupling $g_s\stackrel{<}{\sim}1$. The finiteness of the string coupling necessitates addressing the same from an M theory perspective; see \cite{A.Sinha-4} for holography at finite coupling. This is the reason for coining the name: `MQGP limit'. In fact this is the reason why one is required to first construct a type IIA mirror, which was done in \cite{MQGP} a la delocalized Strominger-Yau-Zaslow mirror symmetry, and then take its M-theory uplift.

\item
From the perspective of calculational simplification in supergravity, the following are examples of the same and constitute therefore the second set of reasons for looking at the MQGP limit of (\ref{limits_Dasguptaetal-ii}):
\begin{itemize}
\item
In the UV-IR interpolating region and the UV,
$(M_{\rm eff}, N_{\rm eff}, N_f^{\rm eff})\stackrel{\rm MQGP}{\approx}(M, N, N_f)$
\item
Asymmetry Factors $A_i, B_j$(in three-form fluxes)$\stackrel{MQGP}{\rightarrow}1$  in the UV-IR interpolating region and the UV.

\item
Simplification of ten-dimensional warp factor and non-extremality function in MQGP limit
\end{itemize}
\end{enumerate}
With ${\cal R}_{D5/\overline{D5}}$ denoting the boundary common to the UV-IR interpolating region and the UV region, $\tilde{F}_{lmn}, H_{lmn}=0$ for $r\geq {\cal R}_{D5/\overline{D5}}$ is required to ensure conformality in the UV.  Near the $\theta_1=\theta_2=0$-branch, assuming: $\theta_{1,2}\rightarrow0$ as $\epsilon^{\gamma_\theta>0}$ and $r\rightarrow {\cal R}_{\rm UV}\rightarrow\infty$ as $\epsilon^{-\gamma_r <0}, \lim_{r\rightarrow\infty}\tilde{F}_{lmn}=0$ and  $\lim_{r\rightarrow\infty}H_{lmn}=0$ for all components except $H_{\theta_1\theta_2\phi_{1,2}}$; in the MQGP limit and near $\theta_{1,2}=\pi/0$-branch, $H_{\theta_1\theta_2\phi_{1,2}}=0/\frac{3 g_s^2MN_f}{8\pi}\ll  1.$ So, the UV nature too is captured near $\theta_{1,2}=0$-branch in the MQGP limit. This mimics addition of $\overline{D5}$-branes in \cite{metrics} to ensure cancellation of $\tilde{F}_3$.
\subsection{Construction of  the Delocalized SYZ IIA Mirror and Its M-Theory Uplift in the MQGP Limit}
A central issue to \cite{MQGP,transport-coefficients} has been implementation of delocalized mirror symmetry via the Strominger Yau Zaslow prescription according to which the mirror of a Calabi-Yau can be constructed via three T dualities along a special Lagrangian $T^3$ fibered over a large base in the Calabi-Yau. This sub-section is a quick review of precisely this.

{ To implement the quantum mirror symmetry a la S(trominger)Y(au)Z(aslow) \cite{syz}, one needs a special Lagrangian (sLag) $T^3$ fibered over a large base (to nullify contributions from open-string disc instantons with boundaries as non-contractible one-cycles in the sLag). Defining delocalized T-duality coordinates, $(\phi_1,\phi_2,\psi)\rightarrow(x,y,z)$ valued in $T^3(x,y,z)$ \cite{MQGP}:
\begin{equation}
\label{xyz defs}
x = \sqrt{h_2}h^{\frac{1}{4}}sin\langle\theta_1\rangle\langle r\rangle \phi_1,\ y = \sqrt{h_4}h^{\frac{1}{4}}sin\langle\theta_2\rangle\langle r\rangle \phi_2,\ z=\sqrt{h_1}\langle r\rangle h^{\frac{1}{4}}\psi,
\end{equation}
using the results of \cite{M.Ionel and M.Min-OO (2008)} it can be shown \cite{transport-coefficients}that the following conditions are satisfied:
\begin{eqnarray}
\label{sLag-conditions}
& & i^* J \approx 0,\nonumber\\
& & \Im m\left( i^*\Omega\right) \approx 0,\nonumber\\
& & \Re e\left(i^*\Omega\right)\sim{\rm volume \ form}\left(T^3(x,y,z)\right),
\end{eqnarray}
separately for the $T^2$-invariant sLags of \cite{M.Ionel and M.Min-OO (2008)} for a resolved/deformed conifold implying thus: $\left.i^* J\right|_{RC/DC}\approx0, \Im m\left.\left( i^*\Omega\right)\right|_{RC/DC} \approx 0, \Re e\left.\left(i^*\Omega\right)\right|_{RC/DC}\sim{\rm volume \ form}\left(T^3(x,y,z)\right)$. Hence, if the resolved warped deformed conifold is predominantly either resolved or deformed, the local $T^3$ of (\ref{xyz defs}) is the required sLag to effect SYZ mirror construction.}

{Interestingly, in the `delocalized limit' \cite{M. Becker et al [2004]}  $\psi=\langle\psi\rangle$, under the coordinate transformation :
\begin{equation}
\label{transformation_psi}
\left(\begin{array}{c} sin\theta_2 d\phi_2 \\ d\theta_2\end{array} \right)\rightarrow \left(\begin{array}{cc} cos\langle\psi\rangle & sin\langle\psi\rangle \\
- sin\langle\psi\rangle & cos\langle\psi\rangle
\end{array}\right)\left(\begin{array}{c}
sin\theta_2 d\phi_2\\
d\theta_2
\end{array}
\right),
\end{equation}
and $\psi\rightarrow\psi - \cos\langle{\bar\theta}_2\rangle\phi_2 + \cos\langle\theta_2\rangle\phi_2 - \tan\langle\psi\rangle ln\sin{\bar\theta}_2$
the $h_5$ term becomes

$h_5\left[d\theta_1 d\theta_2 - sin\theta_1 sin\theta_2 d\phi_1d\phi_2\right]$, $e_\psi\rightarrow e_\psi$, i.e.,  one introduces an isometry along $\psi$ in addition to the isometries along $\phi_{1,2}$. This clearly is not valid globally - the deformed conifold does not possess a third global isometry}.

{To enable use of SYZ-mirror duality via three T dualities, one also needs to ensure a large base (implying large complex structures of the aforementioned two two-tori) of the $T^3(x,y,z)$ fibration. This is effected via \cite{F. Chen et al [2010]}:
\begin{eqnarray}
\label{SYZ-large base}
& & d\psi\rightarrow d\psi + f_1(\theta_1)\cos\theta_1 d\theta_1 + f_2(\theta_2)\cos\theta_2d\theta_2,\nonumber\\
& & d\phi_{1,2}\rightarrow d\phi_{1,2} - f_{1,2}(\theta_{1,2})d\theta_{1,2},
\end{eqnarray}
for appropriately chosen large values of $f_{1,2}(\theta_{1,2})$. The three-form fluxes
 remain invariant. The fact that one can choose such large values of $f_{1,2}(\theta_{1,2})$, was justified in \cite{MQGP}.  The guiding principle is that one requires that the metric obtained after SYZ-mirror transformation applied to the non-K\"{a}hler  resolved warped deformed conifold is like a non-K\"{a}hler warped resolved conifold at least locally. Then $G^{IIA}_{\theta_1\theta_2}$ needs to vanish \cite{MQGP}.}
{
The mirror type IIA metric after performing three T-dualities, first along $x$, then along $y$ and finally along $z$, utilizing the results of \cite{M. Becker et al [2004]} was worked out in \cite{MQGP}. We can get a one-form type IIA potential from the triple T-dual (along $x, y, z$) of the type IIB $F_{1,3,5}$ in \cite{MQGP} and using which the following eleven dimensional metric was obtained in \cite{MQGP}:
%%%\vskip -0.4in
\begin{eqnarray}
\label{M3}
& &\hskip -0.6in   ds^2_{11} =  e^{-\frac{2{\phi}_{IIA}}{3}} \Biggl[
\frac{1}{\sqrt{h}}\biggl\{-\left(1-\frac{r^4_{h}}{r^4}\right) dt^2+\sum^{3}_{i=1}dx^2_{i}\biggr\}+ \sqrt{h}\left(\frac{6a^2+r^2}{9a^2+r^2}\right)\frac{dr^2}{\left(1-\frac{r^4_{h}}{r^4}\right)} +  ds^2_{IIA}({\theta_{1,2},\phi_{1,2},\psi})\Biggr]\nonumber\\
 & & \hskip -0.2in + e^{\frac{4{\Phi}_{IIA}}{3}}\biggl(dx_{11} + A^{F_1}+A^{F_3}+A^{F_5}\biggr)^2,
\end{eqnarray}
where the warp factor $h$ is as given in (\ref{eq:h}) with $M^{eff}$ and $N_{f}^{eff}$ replaced simply by $M$ and $N_{f}$ in the `MQGP' limit. Also the dilaton factor is same as given in (\ref{dilaton1}).

Locally, the uplift (\ref{M3}) can hence be thought of as black $M3$-brane metric, which in the UV, can be thought of as black $M5$-branes wrapping a two cycle homologous to:
$n_1 S^2(\theta_1,x_{10}) + n_2 S^2(\theta_2,\phi_{1/2}) + m_1 S^2(\theta_1,\phi_{1/2}) + m_2 S^2(\theta_2,x_{10})$ for some large $n_{1,2},m_{1,2}\in\mathbb{Z}$ \cite{transport-coefficients}.  In the large-$r$ limit, the $D=11$ space-time is a warped product of $AdS_5(\mathbb{R}^{1,3}\times\mathbb{R}_{>0})$ and ${\cal M}_6(\theta_{1,2},\phi_{1,2},\psi,x_{10})$
\begin{equation}
\hskip -0.4in
\label{M_6}
\begin{array}{cc}
&{\cal M}_6(\theta_{1,2},\phi_{1,2},\psi,x_{10})   \longleftarrow   S^1(x_{10}) \\
&\downarrow  \\
{\cal M}_3(\phi_1,\phi_2,\psi) \hskip -0.4in & \longrightarrow  {\cal M}_5(\theta_{1,2},\phi_{1,2},\psi)   \\
&\downarrow  \\
 & \hskip 0.9in {\cal B}_2(\theta_1,\theta_2)  \longleftarrow  [0,1]_{\theta_1}  \\
 & \downarrow  \\
& [0,1]_{\theta_2}
\end{array}.
\end{equation}
In the `MQGP' limit we choose to work around a particular values of $\theta_1$ and $\theta_2$, given by,
\begin{equation}
\theta_{1}\sim\frac{\alpha_{\theta_1}}{N^{1/5}}~~~~~~~~~~~~~\theta_{2}\sim\frac{\alpha_{\theta_2}}{N^{3/10}},
\end{equation}
at which the five dimensional spacetime defined by $\{t, x_{1,2,3},r\}$ decouples from the six dimensional internal space defined by $\{\theta_{1,2},\phi_{1,2},\psi,x_{11}\}$.
Hence the five dimensional black $M3$ brane metric is given as,
\begin{equation}\label{Mtheorymetric}
ds_{5}^2=e^{-\frac{2{\Phi}_{IIA}}{3}} \Biggl(
-g_{tt} dt^2+g_{\mathbb{R}^3}\sum^{3}_{i=1}dx^2_{i}+ g_{rr}dr^2\Biggr),
\end{equation}
where at the above mentioned values of $\theta_1$ and $\theta_2$, the metric components and the dilaton factor are given as,
{\footnotesize
\begin{equation}\begin{split}\label{Mtheorymetriccomp}
g_{tt} &= {(r^4-r_h^4)\over r^2\sqrt{4\pi g_sN}} \Bigg\{1  +  {3g_s M^2\over 4\pi N} \left[1 + {3g_sN_f\over 2\pi}\left(\log~r + {1\over 2}\right)
+ {g_sN_f\over 4\pi}\log\left({\alpha_{\theta_1}\alpha_{\theta_2}\over 4\sqrt{N}}\right)\right] \log~ r \Bigg\}\\
g_{\mathbb{R}^3} &=  {r^2\over \sqrt{4\pi g_sN}}\Bigg\{1  - {3g_s M^2\over 4\pi N} \left[1 + {3g_sN_f\over 2\pi}\left(\log~r + {1\over 2}\right)
+ {g_sN_f\over 4\pi}\log\left({\alpha_{\theta_1}\alpha_{\theta_2}\over 4\sqrt{N}}\right)\right] \log~ r \Bigg\} \\
g_{rr}&= {r^2\sqrt{4\pi g_sN}\over r^4-r_h^4}\left({6a^2 + r^2\over 9a^2 + r^2}\right)\Bigg\{1  - {3g_s M^2\over 4\pi N} \left[1 + {3g_sN_f\over 2\pi}\left(\log~r + {1\over 2}\right)
+ {g_sN_f\over 4\pi}\log\left({\alpha_{\theta_1}\alpha_{\theta_2}\over 4\sqrt{N}}\right)\right] \log~ r \Bigg\},
\end{split}
\end{equation}}
and the type IIA dilaton $\phi_{IIA}$ being the triple T-dual of the type IIB dilaton $\Phi_{IIB}$:
\begin{equation}
e^{-\Phi_{IIB}} = \frac{1}{g_s} - \frac{N_f}{8\pi}\log(r^6 + a^2 r^4) - \frac{N_f}{2\pi}\log\left({\alpha_{\theta_1}\alpha_{\theta_2}\over 4\sqrt{N}}\right).
\end{equation}
All the calculations that were presented in the subsequent chapters are mostly done using the above metric.
\section{Non-equilibrium physics and Transport coefficients}
In this section we will mention in short about some basic idea of non-equilibrium physics and the procedure to obtain transport coefficients for a fluid medium. In this thesis we have evaluated different transport coefficients for a strongly coupled medium. Since any strongly coupled medium behaves almost like a fluid, the following discussion will be useful in order to have a clear understanding on this topic.

It is not possible to realize transport phenomenon of a physical system at equilibrium. What is important here is the reaction or the response of the system when it is perturbed by some external means. Here it is very important to note however that the changes as made by the external source are very small away from the equilibrium such that the equilibrium fluctuation dictates the non-equilibrium process. Thus the non-equilibrium processes discussed here are actually very close to the equilibrium so that the linear response theory can be used. {\it Linear response theory} - see \cite{Policastro+Son+Starinets-2}, \cite{son correlator} in the context of application of gauge-gravity duality ideas to hydrodynamics -  can determine the response from the microscopic consideration and it is given in terms of the retarded Green's function; see \cite{Minwalla-2},\cite{Skenderis-1} and references therein for going beyond linear order in the context of fluid-gravity correspondence. For a more recent discussion on  non-equilibrium field theory dynamics from gauge/gravity duality, see \cite{Skenderis-2}, \cite{Skenderis-3}, and non-conformal hydrodynamics, see \cite{Skenderis-5}, \cite{Skenderis-6}. In other words, the change in the ensemble average of some observable $\mathcal{O}$ from it's value at equilibrium due to an external source $\phi$ is given in momentum space as:
\begin{equation}
\label{response}
\delta\left<\mathcal{O}(k)\right>=-G^{\mathcal{O}\mathcal{O}}_{R}(k)\phi(k),
\end{equation}
with $k_{\mu}=(w,\textbf{\textit{q}})$.
The retarded greens function $G^{\mathcal{O}\mathcal{O}}_{R}(k)$ is given as:
 \begin{equation}
\label{retarded}
G^{\mathcal{O}\mathcal{O}}_{R}(k)=-i \int d^4x ~e^{i k x}~\theta(t)\left<[\mathcal{O}(t,\overrightarrow{x}),\mathcal{O}(0, \bf 0)]\right>.
\end{equation}
Here, in the perturbed action of the system the operator $\mathcal{O}$ couples with the source $\phi$. For example, in a charged system the chemical potential $\mu$ acts as a source and the response is the charge density $\rho$. In a fluid medium the response is given by the energy-momentum tensor $T^{\mu\nu}$ and it is sourced by the fluctuation in spacetime $h_{\mu\nu}$. Hence using (\ref{retarded}), one can write for the above two example:
\begin{equation}\begin{split}
\label{rhoT}
 \delta\left<\rho\right>&=-G^{\rho\rho}_{R}\mu\\
 \delta \left<T^{\mu\nu}\right>&=-G^{\mu\nu,\mu\nu}_{R}h_{\mu\nu}.
\end{split}
\end{equation}
So linear response theory tells us that to determine the response one has to compute the retarded Green's function. But is there any easier way to get the same? The answer to this is {\it Hydrodynamics}. Hydrodynamics is an effective field theory where instead of the effective action one starts with the equation of motion involving the conserved quantities. Contrary to the linear response theory, hydrodynamics is a macroscopic theory. This is because it involves the macroscopic variables of the system and gives their dynamics in the low momentum and large wavelength limit. Below we will discuss, in short, how hydrodynamics simplifies the whole computation by considering a viscous fluid as an example. The hydrodynamics and transport behaviour of a fluid with homogeneity and isotropy is studied in \cite{Obers1710.06885}

For any fluid medium the energy momentum tensor $T^{\mu\nu}$ is the macroscopic variable. The corresponding conservation law is:
\begin{equation}
\partial_{\mu}T^{\mu\nu}=0.
\end{equation}
To close the above equation one needs the {\it Constitutive Equation}. Constitutive equation express $T^{\mu\nu}$ in terms of the temperature $T$ and four velocity $u^{\mu}$ if the fluid. Moreover, $T^{\mu\nu}$ can be expanded as a power series in derivatives with respect to the spatial coordinates. As in a perfect fluid there exist no dissipation, we need to consider only the zeroth order term which is given as:
\begin{equation}
\label{zero}
T^{\mu\nu}=\left(\epsilon+P\right)u^{\mu}u^{\nu}+Pg^{\mu\nu},
\end{equation}
where, $\epsilon$ and $P$ are respectively the energy density and pressure. For a viscous fluid one must add derivative corrections. At the next order the correction is given as:
\begin{equation}
\label{first}
\sigma^{\mu\nu}\stackrel{RF}{=}-\eta\left(\partial_{i}u_{j}+\partial_{j}u_{i}-\frac{2}{3}\delta_{ij}\partial_{k}u^{k}\right)-\xi \delta_{ij}\partial_{k}u^{k},
\end{equation}
where, `RF' referred to the rest frame of the fluid motion and in the rest frame the four velocity is given as: $u^{\mu}=\left(1,0,0,0\right)$. The coefficients $\eta$ and $\xi$ are called the shear and the bulk viscosity respectively.
\begin{itemize}
\item {\it Kubo's formula for shear viscosity}
\end{itemize}
Kubo's formula provides a relationship between the transport coefficients and the retarded Green's function. Here we will give a sketchy derivation of the Kubo's formula for shear viscosity. For a detailed derivation see \cite{natsumeadscft}.
Let's consider a perturbation of the 4-dimensional spacetime where the field theory live, of the form: $g_{\mu\nu}=g^{(0)}_{\mu\nu}+h_{\mu\nu}$, where $g^{(0)}_{\mu\nu}$ is the unperturbed background. Now, as discussed before, the response to this perturbation will be given by the energy-momentum tensor $\delta\left< T^{\mu\nu}\right>$. For shear viscosity, we consider $h_{xy}$ as the only nonzero perturbation. Generalizing equation (\ref{zero}), (\ref{first}) for the curved spacetime and substituting the perturbation one can easily find the response in the momentum space as,
\begin{equation}
\label{responsesigma}
\delta\left<\sigma^{xy}(w,\textbf{\textit{q}}=0)\right>=i w\eta h_{xy}.
\end{equation}
Comparing the above (\ref{responsesigma}) and the second relation in equation (\ref{response}), one getss:
\begin{equation}
\label{kuboeta}
\eta=-\lim_{w\to 0} \frac{1}{w}\textrm{Im}~ G^{xy,xy}_{R}(w, \textbf{\textit{q}}=0).
\end{equation}
This is the {\it Kubo's formula} for shear viscosity.

Let's discuss the implications of the above two results. Equation (\ref{responsesigma}) tells us that the computation of the transport coefficient is enough to know the response. On the ether hand to get the transport coefficient one requires only the $O(w)$ term in the retarded greens function (Kubo formula) and not the the full computation as in the linear response theory. This is the advantage provided by the hydrodynamics.

However, this is not the end of the story. It turns out that the transport coefficients can also be obtained from the pole structure of the retarded Green's function. To see this let's take a charged system and consider the diffusion of charges. Here $j^{\mu}$ defines the conserved current and $\rho=j^{0}$ is the conserved charge density. Writing the current as $j^{\mu}=\left(\rho,j^{i}\right)$, the conservation law is given by:
\begin{equation}
\label{conservedj}
\partial_{\mu}j^{\mu}=0.
\end{equation}
The constitutive equation in this case is given by the Fick's law,
\begin{equation}
\label{ficks}
j^{i}=-D\partial^{i}\rho,
\end{equation}
where $D$ is called the diffusion constant. Combining equation (\ref{conservedj}) and (\ref{ficks}), one gets the following equation,
\begin{equation}
\label{ficks1}
\partial_{0}\rho-D\partial^{2}_{i}\rho=0.
\end{equation}
To solve the above equation one consider a Fourier transformation in space and Laplace transform in time with the boundary condition $\rho(t=0,x)=\delta(x)$ to get in the momentum space, \cite{natsumeadscft}
\begin{equation}
\label{rhotilde}
\tilde{\rho}(w,q)=\frac{1}{-iw+Dq^2}.
\end{equation}
Hence the information about the diffusion constant $D$ can be obtained from the pole of $\tilde{\rho}(w,q)$. But from linear response theory we have seen that in equation (\ref{response}), the response $\delta\left<\rho\right>$ is proportional to the retarded Green's function $G^{\rho\rho}_{R}$. Therefore the retarded Green's function must also have the same pole structure.
So it is concluded that the transport coefficient can be calculated in two different ways:
\begin{itemize}
\item from the coefficient of $O(w)$ term in retarded Green's function (Kubo's formula)
\item from the pole of the retarded Green's function.
\end{itemize}
An important point to note here that the Green's function that we need to consider in order to get the transport coefficient from Kubo's formula is different from the one with the required pole structure.

Similarly, for the viscous fluid it is possible to get the dispersion relations without considering the spacetime fluctuation and simply considering the linear fluctuations of the hydrodynamic variables. This is called linearized hydrodynamics. Considering the perturbations to be along a particular spatial direction, say along $x$, one can write:
\begin{equation}\begin{split}
u^{i}=u^{i}(t,x)~~~~~~~~~~~~~~P=P(t,x).
\end{split}
\end{equation}
So in the flat spacetime where the gauge theory lives, there will be a $SO(2)$ rotational symmetry in the $y-z$ plane. Therefore one can decompose the current $T^{\mu\nu}$ according to it's transformation under $SO(2)$ as:

\begin{equation}
\begin{split}
 \textrm{Tensor mode}&:T^{yz}, T^{yy}=- T^{zz}\\
 \textrm{Vector mode}&:T^{ty}, T^{tz}, T^{xz}, T^{xy}\\
 \textrm{Scalar mode}&:T^{tt}, T^{tx}, T^{xx}, T^{yy}=T^{zz}.
\end{split}
\end{equation}
We consider the perturbation to have the form: $u^{i}=e^{-i wt+iqx}$. Substituting this perturbation in the constitutive equation, and using the conservation law one gets the following dispersion relations for the vector and the scalar mode components of $T^{\mu\nu}$ as,
\begin{equation}\begin{split}
\textrm{Vector mode:}~w&=-i \frac{\eta}{\epsilon+P}q^2\\
\textrm{Scalar mode:}~w&=\pm v_{s}q-\frac{i}{2}\Gamma q^2+O(q^3),
\end{split}
\end{equation}
where $v_s$ is the speed of sound and $\Gamma$ is the attenuation constant. In {\bf Chapter 3} we have obtained the corresponding transport coefficients. Hence the only task left is to calculate the retarded Green's function. This retarded Green's function or the two point correlation function can be calculated using the AdS/CFT correspondence.
\subsection{Recipe to find Minkowski Correlators}
Following \cite{son correlator} we briefly review the prescription to find the thermal correlator in Minkowski signature. According to AdS/CFT correspondence, there exists an operator $\mathcal{O}$ in the field theory side dual to a field $\phi$ defined in the bulk of AdS geometry such that on the boundary of the anti-de Sitter space $\phi$ tends to a value $\phi_{0}$ which acts as a source for the operator $\mathcal{O}$. We are interested in calculating the retarded Green's function $G^R$ of the operator $\mathcal{O}$ in space-times with Minkowski signature.

Our working background (Type $IIB$ or it's M-theory uplift) can be expressed as the following $5d$ metric,
\begin{eqnarray}
\label{5dmetric inu}
ds^2=-g_{tt}(u)dt^2+g_{xx}(u)\Biggl(dx^2+dy^2+dz^2\Biggr)+g_{uu}(u)du^2.
\end{eqnarray}
Here $u$ is the new coordinate defined as $u=r_h/r$ so that $u=0$ is the boundary and $u=1$ is the horizon of the AdS space. A solution of the linearized field equation for any field $\phi(u,x)$ choosing $q^\mu=(w,q,0,0)$ is given as,
 \begin{equation}
 \phi(u,x)=\int\frac{{d^4}q}{(2\pi)^4}e^{-i wt + i qx}f_{q}(u)\phi_{0}(q)
 \end{equation}
where $ f_{q}(u)$ is any function which depends only on the radial variable $u$ and is normalized to 1 at the boundary and satisfies the incoming wave boundary condition at $u=1$, and $\phi_{0}(q)$ is determined by,
 \begin{equation}
 \phi(u=0,x)=\int\frac{{d^4}q}{(2\pi)^4}e^{-i wt + i qx}\phi_{0}(q).
\end{equation}
If the kinetic term for $\phi(u,x)$ is given by: $\frac{1}{2}\int {d^4}x du A(u)\left(\partial_u\phi(x,u)\right)^2$, then using the equation of motion for $\phi$ it is possible to reduce an on-shell action to the surface terms as,
\begin{equation}
S=\int\frac{{d^4}q}{(2\pi)^4}\phi_{0}(-q)\mathcal{F}(q,u)\phi_{0}(q)|^{u=1}_{u=0}
\end{equation}
where the function
\begin{equation}
\label{F}
\mathcal{F}(q,u) = A(u) f_{\pm q}(u)\partial_{u}f_{\pm q}(u).
\end{equation}
 Finally, the retarded Green's function is given by the formula:
\begin{equation}
G^{R}(q)=-2\mathcal{F}(q,u)|_{u=0}.
\end{equation}

The different retarded Green's functions are defined as
 \begin{equation}
 G^{R\ T}_{\mu\nu,\rho\sigma}(q)=-i\int {d^4}x e^{-i wt + i qx}\theta(t) \langle[T_{\mu\nu}(x),T_{\rho\sigma}(0)]\rangle,
\end{equation}
with $\langle[T_{\mu\nu},T_{\rho\sigma}]\rangle\sim\frac{\delta^2S}{\delta h_{\mu\nu}\delta h_{\rho\sigma}}$ and
 \begin{equation}
G^{R\ J}_{\mu\nu}(q)=-i\int {d^4}x e^{-i wt + i qx}\theta(t) \langle[J_{\mu}(x),J_{\nu}(0)]\rangle
\end{equation}
with $\langle[J_{\mu}(x),J_{\nu}(0)]\rangle\sim\frac{\delta^2S}{\delta A_\mu \delta A_\nu}$,
as the energy-momentum tensor $T_{\mu\nu}(x)$ and the current $J_{\mu}(x)$ couple respectively to the metric and gauge field.
\section{Summary}
Let us now summarise the rest of the thesis chapter wise below.

In {\bf Chapter 2}, we first evaluated the DBI action for $N_f$ flavor $D7$ branes in the presence of a $U(1)$ gauge field (assuming it to have only a non-zero temporal component with only a radial dependence, corresponding to a baryon chemical potential) by {\it first} evaluating in the MQGP limit, the angular integrals exactly and then taking the UV limit of the (incomplete) elliptic integrals so obtained. Demanding square integrability of the aforementioned $U(1)$ gauge field and using the Dirichlet boundary condition at an IR cut-off and demanding a mass parameter appearing in the solution to be related to the mass of the lightest known vector meson mass, we related the mass of the lightest vector meson to the IR cut-off. The computation of the QCD deconfinement transition temperature or equivalently the critical temperature $T_c$ corresponding to the first order Hawking-Page phase transition between a thermal and a black hole backgrounds, is then carried out from five-dimensional Einstein-Hilbert and Gibbons-Hawking-York actions with the angular portions decoupling in the delocalized MQGP limit. It has been shown that:
\begin{itemize}
\item
 it is possible to obtain the QCD deconfinement temperature  consistent with lattice results for $N_f$ equal to three, ensuring at the same time the thermodynamical stability of the type $IIB$ background;
 \item
  the Ouyang embedding parameter required to be dialed in to reproduce $T_c$ is happily exactly what also reproduces the mass scale of the first generation (light) quarks;
 \item
 $T_c$ decreases with increase in $N_f$ in accordance with lattice computations.
\end{itemize}
Also, using the aforementioned $U(1)$ background, we then looked at both $U(1)$ and $SU(2)$ (for $N_f=2$) gauge fluctuations. By looking at two-point correlation functions of either the former or the diagonal sector of the latter, we calculated the DC electrical conductivity and the temperature dependence of the same (above $T_c$), and found:
\begin{itemize}
\item
 demanding the Einstein relation (ratio of electrical conductivity and charge susceptibility to equal the diffusion constant) to be satisfied within linear perturbation theory, requires a non-trival dependence of the Ouyang embedding paramter on the horizon radius;

\item
a prediction that the temperature dependence of the DC electrical conductivity above $T_c$, curiously mimics a one-dimensional Luttinger liquid with an appropriately tuned interaction parameter.

\end{itemize}
{\bf Chapter 3} is entirely dedicated to the transport properties of strongly coupled QGP medium. Due to the `MQGP' limit, the string coupling $g_s$ is small but finite which necessitates the transport coefficients to be evaluated upto NLO in $N$. Here we start by considering a linear perturbation of the five dimensional black $M_3$ brane metric. Based on the spin of different metric perturbations under rotation, the sam are categorized into Scalar, Vector and Tensor modes. Then we solve the linearized Einstein's equations separately with scalar, vector and tensor modes of the metric perturbations to get respectively the speed of sound $(v_s)$, the diffusion constant $(D)$ and the shear viscosity $(\eta)$ (also the shear viscosity to entropy density ratio $\frac{\eta}{s}$). The Einstein's equations as obtained for the above mentioned three modes are coupled and are difficult to solve. Following \cite{klebanov quasinormal}, we construct the gauge invariant combination of different perturbations ($Z_{s}$ for scalar mode, $Z_{v}$ for vector mode, $Z_{t}$ for tensor modes) and were able to write down the coupled equations as a single equation involving the corresponding gauge invariant variable which is then solved for the quasinormal frequencies with pure incoming wave boundary condition at the black hole horizon and Dirichlet boundary condition at spacial infinity. For the metric fluctuations in the sound channel the corresponding quasinormal frequency is given by $w=\pm v_s q-i\Gamma_s q^2$ with $v_s$ defined as the speed of sound and $\Gamma_s$ as the damping constant of the sound mode. Again for the sound channel the pole of the correlations of longitudinal momentum density gives the same dispersion relation. The quasinormal frequency for the vector modes of black brane metric fluctuation reads $\omega=-iD q^2$, where $D$ is the shear mode diffusion constant. This dispersion relation also follows from the pole structure of the correlations of transverse momentum density.
The results for the NLO (in $N$) corrections of $v_s$, $D$ and $\frac{\eta}{s}$ are particularly important as they suggest a scale dependance to the above mentioned quantities and hence leads to a non-conformal nature of the field theory in the IR. We have also make a comparison of the result for $\frac{\eta}{s}$ with the RHIC data of \cite{eta-over-s-RHIC}.

In this chapter we also compute the temperature dependance of thermal (electrical) conductivity via Kubo's formula at finite temperature and finite baryon density up to LO in $N$. For this we turn on simultaneously gauge and vector modes of metric fluctuations, and evaluate the thermal
($\kappa_T$) and electrical ($\sigma$) conductivities, and the Wiedemann-Franz law ($\frac{\kappa_T}{T\sigma}$). {The new insight gained is that for $\mu\equiv$(Ouyang embedding parameter)$\sim r_h^\alpha,\ \alpha\leq0$, the temperature dependence of $\kappa_T, \sigma$ and the consequent deviation from the Wiedemann-Franz law, all point to the remarkable similarity with $D=1+1$ Luttinger liquid with impurities at `$\frac{1}{3}$-doping'}; for $\alpha=\frac{5}{2}$ one is able to reproduce the expected linear large-$T$ variation of DC electrical conductivity for most strongly coupled gauge theories with five-dimensional gravity duals with a black hole \cite{SJain_sigma+kappa}.

In {\bf Chapter 4} using a large-$N$ top-down holographic dual of thermal QCD, we obtain the spin $2^{++}, 1^{++},$ $0^{++}, 0^{--}, 0^{-+}$ glueball spectrum explicitly for QCD$_{3}$  from type IIB, type IIA and M theory perspectives. For each of the above computations, we consider two different scenarios in the background geometry. These different backgrounds corresponds to two classical solution to the gravitational action. In one solution there exists a black hole in the background while the other solution has no notion of black hole and is known as the thermal background, where in the later case the singularity is removed by an infrared cut-off. An important point to remember at this stage is that the IR cut-off at $r=r_0$ is not put by hand but is a consequence of the $D7$ embedding. From a top-down perspective this IR cut-off will in fact be proportional to two-third power of the Ouyang embedding parameter obtained from the minimum radial distance (corresponding to the lightest quarks) requiring one to be at the South Poles in the $\theta_{1,2}$ coordinates, in the holomorphic Ouyang embedding of flavor $D7$-branes. In the detailed calculation of different glueball given below, we refer the solution with a black hole as `Background with a black hole' and the solution with a thermal background as `Background with an IR cut-off'. In the spirit of \cite{witten}, the time direction for both cases will be compact with fermions obeying anti-periodic boundary conditions along this compact direction, and hence we will be evaluating three-dimensional glueball masses.
\chapter{Deconfinement Temperature and Hints of  a $D=1+1$ Luttinger Liquid}
\chaptermark{Deconfinement Temperature and\\ Hints of  a $D=1+1$ Luttinger Liquid}
\graphicspath{{Chapter2/}{Chapter2/}}
\section{Introduction and Motivation}
In chapter {\bf 1}, we have presented a reasonably detailed discussion on AdS/CFT correspondence, but that discussion is valid strictly at zero temperature. However one interesting generalization of AdS/CFT correspondence is the introduction of temperature. In usual AdS/CFT correspondence, a normalizable mode in string theory on AdS spacetime corresponds to some states in the dual field theory. For example, in the absence of any excitations, a vacuum state in the field theory side is dual to pure AdS in the bulk. So starting with a pure $AdS_5$, as one starts to excite the normalizable modes then the field theory also goes to an excited state from the vacuum. One such excited state is the finite temperature or thermal state. Now the obvious question is: what does this thermal state in the field theory correspond to in the gravity picture?

Before answering this question one must note that whatever the solution is in the gravity side, it has to satisfy the following conditions:
\begin{itemize}
\item it has to be asymptotically $AdS_5$,
\item it must have the notion of temperature,
\item it must have all the symmetries of the thermal system such as translation symmetry, rotational symmetry.
\end{itemize}
Now there are two such candidates which follow the above conditions:
\begin{enumerate}
\item the thermal AdS background,
\item a black hole in AdS geometry.
\end{enumerate}
Let us talk about the thermal AdS background first. To get the thermal AdS geometry one needs to go to the Euclidean signature first and then the Euclidean
time is identified periodically with the inverse temperature $\beta$. The 5-dimensional thermal AdS metric is given as:
\begin{equation}
ds^2=\frac{r^2}{L^2}\left(dt^2_{E}+\sum^{3}_{i=1}dx^2_{i}\right)+\frac{L^2}{r^2}dr^2,
\end{equation}
with the euclidean time defined as $t_{E}\equiv t_{E}+\beta$. The above metric tells us that the measure of the size of the Euclidean time circle is given as $\frac{r^2}{L^2}$ and hence as $r\rightarrow 0$, i.e., deep into the interior of the AdS space, the size of the circle goes to zero giving a singular solution.

The other candidate is the AdS-Black Hole solution with the correct symmetry. More precisely to ensure the translation symmetry the black hole background has to have a black brane metric or in other words, the black hole must have a planar horizon. The 5-dimensional ansatz which respects all the symmetries is given by:
\begin{equation}\label{bhansatz}
ds^2=\frac{r^2}{L^2}\left(-f(r)dt^2+\sum^{3}_{i=1}dx^2_{i}\right)+\frac{L^2}{r^2}g(r)dr^2,
\end{equation}
where by solving Einstein's equations the functions, $f(r)$ and $g(r)$ can be obtained as $f(r)=g(r)=1-\frac{r^4_h}{r^4}$. Here the horizon is at $r=r_h$ and the space at the horizon is indeed $\mathbb{R}^3$, i.e., planar. Again, one must go to the Euclidean signature and introduce a periodicity in the Euclidean time. The inverse of this period is the temperature of the black hole. To calculate the temperature one have to impose regularity of the solution at the horizon $r_h$.

The singularity that arises for the thermal background can be removed by introducing a cut-off in the radial coordinate at $r=r_0$ such that the region for which $r<r_0$ is not accessible any more. In other words the coordinate $r$ never reaches zero. This is known as the `Hard wall' model. There is also a `Soft wall' model where the cut-off is provided by some particular $r$ dependant dilaton profile and not that abruptly as in the hard wall model.
The background that we are using is not an AdS geometry in general but asymptotically it is indeed an AdS space as required. The singularity at $r=0$ is fixed by a thermal IR cut-off provided by the Ouyang embedding parameter $\mu$ \cite{ouyang} for the flavor $D7$ brane embedding. More specifically, the IR cut-off $r_0$ in \cite{EPJC-2} is taken to be related to the embedding parameter $\mu$ as $|\mu|^{2/3}=\delta  r_0$, where $\delta$ is a positive constant and is greater than one. The thermal metric in our set up is given as:
\begin{equation}\label{thermalmet}
ds_{TH}^2 = A(r) \left(dt^2 + \sum_{i = 1}^3 dx_i^2\right) + B(r)dr^2 ,
\end{equation}
where the components $A(r)$ and $B(r)$ are given as,
\begin{equation}\begin{split}
A(r)&={e^{-\frac{2\Phi}{3}}r^2\over \sqrt{4\pi g_sN}}\Bigg\{1  - {3g_s M^2\over 4\pi N} \left[1 + {3g_sN_f\over 2\pi}\left(\log~r + {1\over 2}\right)
+ {g_sN_f\over 4\pi}\log\left({\alpha_{\theta_1}\alpha_{\theta_2}\over 4\sqrt{N}}\right)\right] \log~ r \Bigg\}\\
B(r)&={e^{-\frac{2\Phi}{3}} \sqrt{4\pi g_sN}\over r^2}\Bigg\{1  - {3g_s M^2\over 4\pi N} \left[1 + {3g_sN_f\over 2\pi}\left(\log~r + {1\over 2}\right)
+ {g_sN_f\over 4\pi}\log\left({\alpha_{\theta_1}\alpha_{\theta_2}\over 4\sqrt{N}}\right)\right] \log~ r \Bigg\}.
\end{split}
\end{equation}
The metric for the black hole background in our set up is given in {\bf Chapter 1} (\ref{Mtheorymetric}) with the components in (\ref{Mtheorymetriccomp}). We rewrite the same black hole background metric in Euclidean signature here for the convenience of the reader but using a slightly different notation,
\begin{equation}\label{blackmet}
ds^2_{BH}=G^{\cal M}_{tt}dt^2+G^{\cal M}_{\mathbb{R}^3}\sum_{i=1}^{3}dx_{i}^2+G^{\cal M}_{rr}dr^2,
\end{equation}
where we have defined $G^{\cal M}_{tt}=e^{-\frac{2\Phi}{3}}g_{tt}$, $G^{\cal M}_{\mathbb{R}^3}=e^{-\frac{2\Phi}{3}}g_{\mathbb{R}^3}$ and $G^{\cal M}_{rr}=e^{-\frac{2\Phi}{3}}g_{rr}$ with $g_{tt}$, $g_{\mathbb{R}^3}$ and $g_{rr}$ as given in (\ref{Mtheorymetriccomp}).

Using the above one obtains the following expression for the black hole temperature up to $\mathcal{O}(a^2)$,
\begin{eqnarray}
\label{T-RC}
& & T_{h} = \frac{\partial_rG^{\cal M}_{tt}}{4\pi\sqrt{G^{\cal M}_{tt}G^{\cal M}_{rr}}}\nonumber\\
& & = {r_h} \left(\frac{1}{2 \pi ^{3/2} \sqrt{{g_s} N}}-\frac{3 {g_s}^{\frac{3}{2}} M^2 {N_f} \log {r_h} \left(\frac{8 \pi }{g_s N_f}-\log {N}+12 \log {r_h}- \log {16}+6\right)}{64 \pi ^{7/2} N^{3/2}} \right)\nonumber\\
   & & + a^2 \left(\frac{3}{4 \pi ^{3/2} \sqrt{{g_s}} \sqrt{N} {r_h}}-\frac{9 {g_s}^{3/2} M^2 {N_f} \log ({r_h})
   \left(\frac{8 \pi }{g_s N_f}-\log {N}+12 \log {r_h}- \log {16}+6\right)}{128 \pi ^{7/2} N^{3/2}
   {r_h}}\right),\nonumber\\
   & &
\end{eqnarray}
where the resolution parameter $a$ is taken to be \cite{EPJC-2},
\begin{equation}
a=r_h\left(b+c_1\frac{g_s M^2}{N}+c_2\frac{g_s M^2}{N}\log{r_h}\right),
\end{equation}
with $b$, $c_1$ and $c_2$ as positive constants.

The domain of integration with respect to the non compact radial direction is partitioned differently for thermal and black hole backgrounds. Let us discuss this separately for the two different backgrounds below.
\begin{itemize}
\item For the thermal background:
\begin{enumerate}
\item $r=r_0$ is the thermal IR cut-off; this is the point from where the radial direction starts,
\item $r=|\mu|^{2/3}=\delta r_0$ is the end point of the IR-UV interpolating region or this is where the UV region begins,
\item $r\rightarrow \infty$ is the far UV region.

From the above one realizes that:
\begin{enumerate}
\item the region $r_0\leq r\leq |\mu|^{2/3}$ is the IR/IR-UV interpolating region where we have non zero three form fluxes and also non trivial running dilaton profile.
\item the region $|\mu|^{2/3}<r<\infty$ is the UV region where the three form fluxes goes to zero and the dilaton is a constant.
\end{enumerate}
\end{enumerate}
\end{itemize}
\begin{itemize}
\item For the black hole background
\begin{enumerate}
\item $r=r_h$ is the horizon and it is the starting point of the radial direction.
\item $r={\cal R}_{D5/\overline{D5}}=\sqrt{3}a$ is taken to be the point where IR-UV interpolating region ends or the UV region starts,
\item $r\rightarrow \infty$ is again the far UV region.

As for the thermal background:
\begin{enumerate}
\item the region $r_h\leq r\leq {\cal R}_{D5/\overline{D5}}$ is the IR/IR-UV interpolating region where we have a running dilaton profile.
\item the region ${\cal R}_{D5/\overline{D5}}<r<\infty$ is the UV region where the dilaton is a constant.
\end{enumerate}
\end{enumerate}
\end{itemize}
The dilaton profile for the two backgrounds is given as:
\begin{eqnarray}
\label{dilaton}
& (a) & {\rm Black ~hole~ background}:\nonumber\\
& & e^{-\Phi} = \frac{1}{g_s} - \frac{N_f}{8\pi}\log(r^6 + a^2 r^4) - \frac{N_f}{2\pi}\log\left(\frac{\alpha_{\theta_1}\alpha_{\theta_2}}{ 4\sqrt{N}}\right),\ r<{\cal R}_{D5/\overline{D5}},\nonumber\\
& & e^{-\Phi} = \frac{1}{g_s},\ r>{\cal R}_{D5/\overline{D5}};\nonumber\\
& (b) & {\rm Thermal ~background}:\nonumber\\
& & e^{-\Phi} = \frac{1}{g_s} - \frac{3 N_f}{4\pi}\log r - \frac{N_f}{2\pi}\log\left(\frac{\alpha_{\theta_1}\alpha_{\theta_2}}{ 4\sqrt{N}}\right),\ r<\left|\mu\right|^{\frac{2}{3}},\nonumber\\
& & e^{-\Phi} = \frac{1}{g_s},\ r>\left|\mu\right|^{\frac{2}{3}}.
\end{eqnarray}
In this chapter, using the top-down holographic thermal QCD model of \cite{metrics}, we have discussed the following QCD-related properties at finite temperature:\footnote {Interesting work has been done in the context of large $N$ gauge theories at finite temperature with quarks in electric and magnetic field in \cite{arnabkundu0709.1547}\cite{arnabkundu0709.1554}}
\begin{itemize}
 \item
  evaluation of lattice-compatible $T_c$ for the right number and masses of light quarks,
  \item
  demonstrating the thermodynamical stability of \cite{metrics},
  \item
  obtaining the temperature dependence of electrical conductivity $\sigma$, charge susceptibility $\chi$ and hence seeing the constraints which the Einstein's law (relating $\frac{\sigma}{\chi}$ to the diffusion constant) imposes on the holomorphic Ouyang embedding of $D7$-branes into the resolved warped deformed conifold geometry of \cite{metrics};
  \end{itemize}

A black hole with temperature T can radiate energy due to quantum fluctuations and become unstable. A black hole is unstable in an asymptotically flat space time due to its negative specific heat. However stability can be achieved at high temperature in asymptotically AdS black-hole background, while at low temperature the (thermal) AdS solution is preferred. There exists a first order phase transition between these two regimes at a temperature $T_c$, known as the Hawking-Page phase transition \cite{Hawking-Page_1983}. In the dual gauge theory this corresponds to the confinement/deconfinement phase transition. Using the setup as discussed in {\bf Chapter 1}, one of the things we do here is to calculate the QCD deconfinement temperature.
This  is motivated by the following query. From a holographic dual of thermal QCD, at a finite baryon chemical potential, is it possible to simultaneously (within the same holographic dual):
\begin{itemize}
\item
obtain a $T_c$ compatible with lattice QCD results for the right number of light quark flavors,

\item
obtain the mass scale of the light quarks,

\item
incorporate the right mass of the lightest vector meson,

\item
obtain a $T_c$ which increases with decrease of $N_f$ (as required by lattice computations \cite{dTcoverdNfnegative}),

\item
ensure thermodynamical stability?

\end{itemize}
Needless to say, if a proposed holographic dual of thermal QCD  is able to satisfy all the above requirements (in addition to the requirements of UV conformality, IR confinement, etc.), it could be treated as a viable dual.
A particularly interesting issue in this context is the incorporation of $N_f$ $D7$ branes in the resolved warped deformed conifold background geometry. The inclusion of quark matter, is achieved by these $D7$-brane probes. Now at finite baryon density, as provided by the $D7$ branes, we show that the confinement/deconfinement phase transition occurs at a temperature around $175 MeV$, which is consistent with the
lattice QCD result. In deriving the deconfinement temperature we use the mass $m_{\rho}$ of the lightest vector boson as
an input which is around $760 MeV$ from lattice QCD results. Also the consistency of the result demands the number
of light flavors $N_f$  to be equal to 2 or 3 with their masses around $5.6 MeV$, not far from the actual value of the first generation quark masses.

Considering a non-abelian gauge field fluctuations using the gauge-gravity duality prescription, we obtain the $SU(2)$ EOM for $N_f=2$ and investigate the temperature dependence of the electrical conductivity as well as charge susceptibility along with the Einstein relation relating their ratio to the diffusion constant, and show that the Ouyang embedding parameter is required to have a non-trivial dependence on the horizon radius.
Further, we will see that the temperature dependence of electrical conductivity resembles a one-dimensional Luttinger liquid for appropriately tuned Luttinger interaction parameter.
\section{Baryon Chemical Potential and $T_c$ Consistent with Lattice Results and First Generation Quark Masses}
In this section we discuss the evaluation of the QCD confinement-deconfinement transition temperature $T_c$ in the presence of a finite baryon chemical potential/charge density. The evaluation of $T_c$ will be mostly influenced by \cite{witten}, \cite{Herzog-Tc}; see \cite{Minwalla-4} for calculations of $T_c$ in large-$N$ gauge theories on spheres. Particularly interesting works with non-zero chemical potential are the investigation of jet jet quenching of virtual gluons and thermalization of a strongly-coupled plasma via the gauge/gravity duality as done in \cite{arnabkundu1212.5728}\cite{arnabkundu1208.6368}.

Here is first, an outline of how the calculations in this section will proceed.
\begin{enumerate}
\item
For starters, we revisit our calculation of \cite{MQGP} of the baryon chemical potential  generated via $D7$-brane gauge fields in the background of \cite{metrics}. The temporal component of bulk $U(1)$ field on the $D7$-brane world-volume is related to chemical potential which is defined in a gauge-invariant manner as follows:  $\mu_C=\int_{r_h}^\infty dr F_{rt}$. The field strength's only non-zero component, $F_{rt}$, can be evaluated by solving the Euler-Lagrange equation of motion for DBI Action. Instead of taking the UV-limit of the DBI action for $D7$-branes before performing the angular $\theta_{1,2}$ integrals therein as was done in \cite{MQGP}, we will first perform the angular integral exactly and then take the UV limit of the resultant (incomplete) elliptic integrals, in this section.
\item
Using the sum of the five-dimensional Einstein-Hilbert and Gibbons-Hawking-York action and the $A_t(r)$ from step 1., the Hawking-Page transition or QCD deconfinement temperature $T_c$ is obtained.

\end{enumerate}
We will assume $i\mu\in\mathbb{R}$ in Ouyang's embedding: $r^{\frac{3}{2}}e^{\frac{i}{2}(\psi-\phi_1-\phi_2)}\sin\frac{\theta_1}{2} \sin\frac{\theta_2}{2}=i|\mu|$,  which could be satisfied for $\psi=\phi_1+\phi_2+\pi$ and $r^{\frac{3}{2}}\sin\frac{\theta_1}{2} \sin\frac{\theta_2}{2}=|\mu|$. Using the same, one obtains the following metric for a space-time-filling wrapped $D7$-brane embedded in the resolved warped deformed conifold:
\begin{equation}
\label{i*g}
ds^2 = \frac{1}{\sqrt{h}}
\left(-\left(1-\frac{r_h^4}{r^4}\right) dt^2+dx^2+dy^2+dz^2\right)+\sqrt{h}\Big[\frac{dr^2}{\left(1-\frac{r_h^4}{r^4}\right)}+r^2 d{\cal M}_3^2\Big],
\end{equation}
where
{\small
\begin{eqnarray}
\label{eq:metric_D7}
& & d{\cal M}_3^2 = {h_1} \left[{d\phi_2} (\cos ({\theta_2})+1)+{d\phi_1} \left(2-\frac{2 |\mu| ^2 \csc
   ^2\left(\frac{{\theta_2}}{2}\right)}{r^3}\right)\right]^2+ \nonumber\\
   & & {h_2} \left[\left\{1-\left(1-\frac{2 |\mu| ^2
   \csc ^2\left(\frac{{\theta_2}}{2}\right)}{r^3}\right)^2\right\} {d\phi_1}^2+\frac{|\mu| ^2 \left(\frac{3
   {dr}}{r}+{d\theta_2} \cot \left(\frac{{\theta_2}}{2}\right)\right)^2}{r^3 \left(\sin
   ^2\left(\frac{{\theta_2}}{2}\right)-\frac{|\mu| ^2}{r^3}\right)}\right]\nonumber\\
   & & +{h_5} \cos
   ({\phi_1}+{\phi_2}) \left(-\frac{{d\theta_2} |\mu|  \left(\frac{3 {dr}}{r}+{d\theta_2} \cot
   \left(\frac{{\theta_2}}{2}\right)\right)}{r^{3/2} \sqrt{\sin
   ^2\left(\frac{{\theta_2}}{2}\right)-\frac{|\mu| ^2}{r^3}}}-{d\phi_1} {d\phi_2} \sqrt{1-\left(1-\frac{2
   |\mu| ^2 \csc ^2\left(\frac{{\theta_2}}{2}\right)}{r^3}\right)^2} \sin ({\theta_2})\right)\nonumber\\
   & & +{h_5} \sin
   ({\phi_1}+{\phi_2}) \left(-\frac{|\mu|  \left(\frac{3 {dr}}{r}+{d\theta_2} \cot
   \left(\frac{{\theta_2}}{2}\right)\right) \sin ({\theta_2}) {d\phi_2}}{r^{3/2} \sqrt{\sin
   ^2\left(\frac{{\theta_2}}{2}\right)-\frac{|\mu| ^2}{r^3}}}+{d\phi_1}{d\phi_2} \sqrt{1-\left(1-\frac{2 |\mu| ^2 \csc
   ^2\left(\frac{{\theta_2}}{2}\right)}{r^3}\right)^2} \right)\nonumber\\
   & & +{h_4} \left({h_3}
   {d\theta_2}^2+{d\phi_2}^2 \sin ^2({\theta_2})\right).
\end{eqnarray}}
From (\ref{three-form-fluxes}), using the Ouyang embedding which implies \cite{MQGP}
\begin{equation}\begin{split}
d\psi & = d\phi_1 + d\phi_2\\
d\theta_1 &= - \tan\left(\frac{\theta_1}{2}\right)\left(3 \frac{dr}{r} + \cot\left(\frac{\theta_2}{2}\right)d\theta_2\right)
\end{split}
\end{equation}
one get,
\begin{equation}\begin{split}
\label{B_Ouyang}
 B_2& = - \frac{3}{r}\tan\frac{\theta_1}{2}\left(B_{\theta_1\phi_1} + B_{\theta_1\psi}\right)dr\wedge d\phi_1  \\
 & - \frac{3}{r}\tan\frac{\theta_1}{2}\left(B_{\theta_1\phi_2} + B_{\theta_1\psi}\right)dr\wedge d\phi_2\\&+
 \left[B_{\theta_2\phi_1}- \tan\frac{\theta_1}{2}\cot\frac{\theta_2}{2}\left(B_{\theta_1\phi_1} + B_{\theta_1\psi}\right)\right]d\theta_2\wedge d\phi_1
\\&+\left[ B_{\theta_2\phi_2} - \tan\frac{\theta_1}{2}\cot\frac{\theta_2}{2}\left(B_{\theta_1\phi_2} + B_{\theta_1\psi}\right)\right]d\theta_2\wedge d\phi_2.
\end{split}
\end{equation}
Now the $D7$ brane DBI action is given as,
\begin{equation}
\label{SDBI-arb-mu}
 S_{\rm D7}=\int d^8 x\sqrt{det\left(i^*(g + B) + F\right)},
\end{equation}
where $i^*g$ denoting the pulled-back metric as given in (\ref{i*g}) and (\ref{eq:metric_D7}), and $i^*B$ denoting the pulled-back NS-NS $B$ as given in (\ref{B_Ouyang}).

In the MQGP limit, taking the large-$r$ limit after angular integration in (\ref{SDBI-arb-mu}) one obtains:
\begin{equation}
\label{S_aftertheta2UV}
S_{D7}\sim\int_{r=r_h}^\infty dr\left[\sqrt{|\mu|}r^{\frac{9}{4}}\sqrt{1 - F_{rt}^2} + {\cal O}\left(r^{\frac{3}{2}},(1,h_5,\frac{a^2}{r^2})\left[\frac{1}{\sqrt{g_sN}},\frac{g_sM^2}{N}\right]\right)\right].
\end{equation}
With $e^{-\phi}\approx \frac{1}{g_s} - \frac{N_f}{2\pi} \log{\mu}$ in the MQGP limit, one obtains:
\begin{equation}\begin{split}
\label{At}
A_t &= r \ _2F_1\left(\frac{2}{9}, \frac{1}{2}, \frac{11}{9}, -\frac{r^{\frac{9}{2}} \left(\frac{1}g_{s} - \frac{N_f \log{ \mu}}{2 \pi}\right)^2}{C^2}\right)\\&\approx \left(\frac{2^{4/9}
   \Gamma \left(\frac{5}{18}\right) \Gamma \left(\frac{11}{9}\right) (C
   {g_s})^{4/9}}{\pi^{1/18} ({g_s} N_f \log (\mu )-2 \pi )^{4/9}}\right)-\left(\frac{36 \pi
    \Gamma \left(\frac{11}{9}\right)C g_{s}}{5 \Gamma
   \left(\frac{2}{9}\right) ({g_s} N_f \log (\mu )-2 \pi )}\right)\left(\frac{1}{r^{5/4}}\right) \\&+\left(\frac{72 \pi^3 \Gamma
   \left(\frac{11}{9}\right)\left(C g_{s}\right)^3 }{23 \Gamma \left(\frac{2}{9}\right) ({g_s}
   N_f \log (\mu )-2 \pi )^3}\right)\left(\frac{1}{r^{23/4}}\right) \\& \equiv \gamma_1 - \frac{\gamma_2}{r^{\frac{5}{4}}} + \frac{\gamma_3}{r^{\frac{23}{4}}}.
   \end{split}
   \end{equation}
   Now, using (\ref{At}) the chemical potential is given as,
   \begin{equation}\begin{split}
   \label{muC}
    \mu_C &= \int_{r_h}^\infty dr F_{rt}\\
   & = \frac{2^{4/9}  \Gamma \left(\frac{5}{18}\right) \Gamma \left(\frac{11}{9}\right)\left(C g_s\right)^{4/9} }{\pi^{1/18} \left(2 \pi -g_s N_{f} \log (\mu )\right)^{\frac{4}{9}}}-{r_h} \,
   _2F_1\left(\frac{2}{9},\frac{1}{2};\frac{11}{9};-\frac{r_{h}^{9/2} (g_s N_{f} \log (\mu )-2 \pi )^2}{4\pi ^2 C^2 g_s^2 }\right).
   \end{split}
   \end{equation}
We Choose a quantity $\gamma$ such that:
 \begin{equation}
 \int_{r_h}^{r_\Lambda}\sqrt{g}\left(A_t - \gamma\right)^2\sim\int_{r_h}^{r_\Lambda}r^3\left(A_t - \gamma\right)^2<\infty,
 \end{equation}
for some UV cut-off scale $r_{\Lambda}$, then the following equation,
 \begin{equation}
 \frac{8}{11} {\gamma_2} r_{\Lambda}^{11/4} (\gamma -\gamma_{1})+\frac{1}{4}
   r_{\Lambda}^4 (\gamma -\gamma_{1})^2+\frac{2}{3} \gamma_{2}^2
   r_{\Lambda}^{3/2} = 0,
 \end{equation}
 can be solved for $\gamma$ as,
 \begin{equation}
 \gamma = \frac{\gamma_{3}}{r_{\Lambda}^{23/4}}+\frac{1}{33} \gamma_{2}
   \left(-\frac{33}{r_{\Lambda}^{5/4}}+\frac{2 \left(24+5 i
   \sqrt{6}\right)}{r_{\Lambda}^{5/4}}\right).
 \end{equation}
 Utilizing that dimensionally $[C]=[r^{\frac{9}{4}}]$, this implies that one can impose a Dirichlet boundary condition at the IR cut-off $r_0: A_t(r_0) - \gamma=0$ where the cut-off $r_{0}$ is given by:
 \begin{equation}
 \label{cut off}
 \frac{C g_s \pi}{r_0^{9/4}\left(g_s N_f \log{ \mu}- 2 \pi\right)} = \pm\left(\frac{23}{10}\right)^{1/2}.
 \end{equation}
 As $e^{-\phi}\approx \frac{1}{g_s} - \frac{N_f \log{\mu}}{2\pi}>0$ we choose the minus sign in (\ref{cut off}).
 Writing $C\equiv m_\rho^{\frac{9}{4}}$ on dimensional grounds, where $m_\rho$ provides the mass scale of the lightest vector boson, one obtains:
 \begin{equation}
 \label{mrho}
 m_\rho = \left(\frac{23}{10}\right)^{2/9}\frac{ r_{0} \left(2 \pi -{g_s} {N_f} \log
   (|\mu|)\right)^{4/9}}{(g_{s}\pi) ^{4/9}}.
 \end{equation}
 If $m_\rho=760$ MeV the cut-off $r_0$ in units of $MeV$, from (\ref{mrho}), is given by:
\begin{equation}
\label{r0}
 r_0 = 760\times \left(\frac{10}{23}\right)^{2/9}\frac{(g_s \pi) ^{4/9}}{\left(2 \pi -{g_s} {N_f}   \log (|\mu| )\right)^{4/9}}.
 \end{equation}
Our next task would be to establish a relationship between the QCD deconfinement temperature and $r_0$, incorporating thereby the effects of non-zero baryon chemical potential and charge density, and in the process working out the dependence of $T_c$ on $N_f$.

We consider the Einstein-Hilbert (EH) action along with the Gibbons-Hawking York surface term of the form
\begin{equation}
V=-\frac{1}{2\kappa^2}\int_{M} d^5x \sqrt{g}e^{-2\phi}\left(R-2\lambda\right)-\frac{1}{\kappa^2}\int_{\partial M} d^4x \sqrt{g_B}e^{-2\phi}K.
\end{equation}
where $g_B$ is the metric at the boundary and $K$ is the extrinsic curvature of the boundary.
Using the two metric corresponding to the thermal background (\ref{thermalmet}) and that of the black hole background (\ref{blackmet}), we first need to calculate the total action for the same two backgrounds. In the black hole case the periodicity of $t$ is given as $0\leq t\leq T_h$, while for thermal background it is not constrained. Now for the regularity of the action at the boundary for both the solution, we integrate up to a UV cut-off $r=r_{\Lambda}$ but will take the limit of $r_{\Lambda}\rightarrow\infty$ at the end.
Setting the Newton's constant to unity, the regularized action for thermal AdS background is given by:
\begin{equation}\begin{split}
\label{V_1}
 V_1
 & = \frac{3  r_{\Lambda}^4}{2\sqrt{2}N^{5/4}  \pi^{5/4} g_{s}^{13/4}} -\frac{1}{32 \left(\sqrt{2} \pi^{9/4} g_{s}^{13/4}\right)}\Biggl\{\frac{1}{{N^{5/4}}} \Biggl(4
   \pi  g_{s}^2 (\log N)^2 \left({|\mu|^{\frac{8}{3}}}-{r_0}^4\right)
   \\
    & - {g_s} \log N
   \Biggl[\left({r_0}^4-{|\mu|^{\frac{8}{3}}}\right) \{-12 {g_s} {N_f} \log ({|\mu|^{\frac{2}{3}}})+{g_s} {N_f} (3+16 \pi  \log {4})+16 \pi \}\\
    & -12
   {g_s} {N_f} {r_0}^4 \log \left(\frac{{r_0}}{{|\mu|^{\frac{2}{3}}}}\right)\Biggr]-16 \pi  \left({|\mu|^{\frac{8}{3}}}-2
   {r_h}^4\right)\Biggr)\Biggr\}
+ {\cal O}\left(\frac{1}{{r_\Lambda}^2}\right).
\end{split}
\end{equation}
Similarly, for the black hole background, for which $r\in[r_h,r_\Lambda]$ one obtains:
\begin{equation}\begin{split}
\label{V_2}
 V_2
 & = \frac{3 r^4_{\Lambda}}{2 \sqrt{2}N^{5/4} \pi^{5/4} g_{s}^{13/4}}+\frac{9 a^2
   r_{\Lambda}^2}{4 \sqrt{2} N^{5/4} \pi^{5/4} g_{s}^{13/4}}\\
   &  +\frac{1}{32 \sqrt{2} \pi ^{9/4} g_{s}^{13/4}
   {{\cal R}^2_{D5/\overline{D5}}}}\Biggl\{\frac{1}{N^{5/4}} \Biggl[6 \pi  a^2 g_{s}^2 \left(\log
   {N}\right)^2 \left(r_{h}^4-{{\cal R}^4_{D5/\overline{D5}}}\right)\\
   & - 3 a^2 g_{s} \log {N} \Biggl(6 g_{s} {N_f}
   \left(r_{h}^4-{{\cal R}^4_{D5/\overline{D5}}}\right) \log ({{\cal R}_{D5/\overline{D5}}})\\
   &  +\left({{\cal R}^2_{D5/\overline{D5}}}-{r_h}^2\right) \Biggl[{g_s} {N_f} \left({{\cal R}^2_{D5/\overline{D5}}} (8 \pi
    \log {4}-9)+r_{h}^2 (8 \pi  \log {4}-3)\right)\\
    &  +8 \pi  \left({{\cal R}^2_{D5/\overline{D5}}}+r_{h}^2\right)\Biggr]\Biggr)+8 \pi  \left(3 a^2
   \left({{\cal R}^4_{D5/\overline{D5}}}-r_{h}^4\right)-4 {{\cal R}^2_{D5/\overline{D5}}} r_{h}^4\right)\Biggr]\Biggr\}  + {\cal O}\left(\frac{1}{{r_\Lambda^2}}\right).
\end{split}
\end{equation}
Now, in the $r_\Lambda\rightarrow\infty$-limit, realizing:
\begin{eqnarray}
\label{CT-i}
& & \left.\sqrt{-g^{\rm Thermal}}\right|_{r=r_\Lambda} = \frac{{r_\Lambda}^4}{4 \pi  {g_s} N},\nonumber\\
& & \left.\sqrt{-g^{\rm BH}}\right|_{r=r_\Lambda} = \frac{{r_\Lambda}^4-3 a^2 {r_\Lambda}^2}{8 \sqrt{2} \pi ^{3/4} {g_s}^{3/4} N^{3/4}},
\end{eqnarray}
one sees that the required counter term required to be added to $V_2-V_1$ (required later) is:
\begin{eqnarray}
\label{CT-ii}
& & \int_{r=r_\Lambda}\left(-\frac{3 \left( (N\pi g_s)^{1/4}   \sqrt{-g^{\rm Thermal}}-2 \sqrt{2}
   \sqrt{-g^{\rm BH}}\right)}{\sqrt{2}\sqrt{N} \sqrt{\pi} g_{s}^{5/2}}\right).
\end{eqnarray}
Therefore, comparing the UV finite part of the two action we get,
\begin{eqnarray}
\label{V_2-V_1-UVfinite}
& & \hskip -0.3in (V_2 - V_1)^{\rm UV-finite} = \nonumber\\
& & \frac{1}{32 \sqrt{2}N^{5/4} \pi^{9/4} g_{s}^{13/4}} \Biggl(3 g_{s}^2 \log{ N} {N_f} \left(9 a^4-r_{h}^4\right) (2 \log {a}+\log {3})\nonumber\\
& & -9 a^4
   \left(g_{s}^2 \log{N} \{2 \pi  \log{ N}+{N_f} (8 \pi  \log {4}-9)\}+8 \pi  g_{s} \log {N}-8 \pi \right)-18 a^2
   g_{s}^2  N_{f} r_{h}^2 \log {N}\nonumber\\
   & & +4 \pi  g_{s}^2 (\log{ N})^2 \left(|\mu|^{8/3}-r_{0}^4\right)-g_{s} \log {N}
   \Biggl[\left(r_{0}^4-|\mu|^{8/3}\right) \{-8 g_{s} N_{f} \log{(|\mu|)}\nonumber\\
   & & +g_{s} N_{f} (3+16 \pi  \log {4})+16 \pi\} -12
   g_{s} N_{f} r_{0}^4 \log \left(\frac{r_{0}}{|\mu|^{2/3}}\right)\Biggr]-16 \pi  \left(|\mu|^{8/3}-2
   r_{h}^4\right)\nonumber\\
   & & -r_{h}^4 \left(g_{s}^2 \log{ N} \{N_{f} (3-8 \pi  \log {4})-2 \pi  \log {N}\} -8 \pi  g_{s}
   \log {N}+40 \pi \right)\Biggr).
\end{eqnarray}
Now assuming
${\cal R}_{D5/\overline{D5}} = \sqrt{3}a$ (to be justified by $0^{++}$ glueball mass calculation via the WKB quantization method in {\bf Chapter 4}), $|\mu|^{\frac{2}{3}}=\delta r_0$, equating $(V_2 - V_1)^{\rm UV-finite}$ to zero one get the following equation:
\begin{equation}\begin{split}
&2 \pi  {g_s}^2 {\log N}^2 \left\{\left(1-9 b^4\right) r_{h}^4+2 \left(\delta ^{8/3}-1\right) {r_0}^4\right\}-8 \pi  \left\{\left(1-9 b ^4\right) r_{h}^4+2 \delta ^{8/3} r{0}^4\right\}\\&+6 g_{s}^2
   {\log N} N_{f} \left\{\left(9 \alpha ^4-1\right) r_{h}^4 \log {r{h}}-2 \left(\delta ^{8/3}-1\right) r_{0}^4 \log
   r_{0}\right\}= 0,
   \end{split}
   \end{equation}
The above equation is solved for $r_{0}$. The solution of $r_0$ involve a `ProductLog' function and it's exact expression is rather long. In the large $N$ limit the solution is given as,
\begin{equation}
\label{r0rh}
r_0  =  r_h\left(\left|\frac{9 b^4-1}{2(\delta^{\frac{8}{3}}-1)}\right|\right)^{1/4} + {\cal O}\left(\frac{1}{\log N}\right).
\end{equation}
Now in the UV limit one can consider a constant dilaton so that $b=0$ and due to no $D7$ brane in the UV we must have $\delta=0$. Also the black hole temperature as given in (\ref{T-RC}) can be modified as $T_h= \left(\frac{r_h}{2 \pi ^{3/2} \sqrt{{g_s} N}}\right)$ so that the transition temperature is obtained as,
\begin{equation}
\label{Tc}
T_c=\frac{2^{1/4}r_0}{2\pi^{3/2}(g_sN)^{1/2}}
\end{equation}
So, from (\ref{r0}) and (\ref{Tc}), one obtains:
  \begin{equation}
 \label{Nf_1}
 N_f \sim \frac{1}{23 \log (|\mu|)}\left(\frac{46 \pi }{{g_s}} \pm \frac{144167}{\pi^{19/8} g_{s}^{9/8} N^{9/8} T_{c}^{9/4}}\right);
 \end{equation}
 we choose the plus sign as, in accordance with lattice calculations, $T_c$ must decrease with $N_f$ \cite{dTcoverdNfnegative}.
 In the MQGP limit taking $g_s=0.8$   in (\ref{Nf_1}), one obtains:
 \begin{equation}
 \label{Nf_2}
 N_f=\frac{7.85398 + \frac{2.94676}{{T_c}^{9/4}}}{\log (|\mu|)}.
 \end{equation}
Hence, for   $\mu = 13.7 i, N_f = 3$, one obtains the QCD deconfinement temperature $T_c=175-190$ MeV, consistent with lattice calculations \cite{Lattice_Tc} and the correct number of light quark flavors.

Now, dimensionally, $[\mu]=[r^{\frac{3}{2}}]$ and using the  AdS/CFT dictionary, hence mass dimensions of 3/2. Curiously, if one set $\sqrt{|\mu|} = m_q^{\frac{3}{4}}$, one would obtain,  in units of $MeV$, $m_q\approx 5.6$ - exactly the mass scale of the first generation light quarks. In the context of quark masses and non-local operators generating the same, at strong coupling, in the Sakai-Sugimoto model \cite{Sakai-Sugimoto} involving $D4, D8, \overline{D8}$-branes, see \cite{A.Sinha-5}.

The thermodynamical stability conditions are governed by inequalities imposed on certain thermodynamical quantities such as $\Delta S<0, \Delta E>0 ~{\rm and} ~\Delta H>0$ (which measure deviations from
equilibrium values implied). Considering that $\Delta E(S,V,N)$ and $\partial^2 E(S,V,N) >0$  and expanding $\partial^2 E(S,V,N)$ around equilibrium values of $(S_0, V_0, N_0)$ leads to three conditions  $C_v>0, \left.\frac{\partial \mu_C}{\partial T}\right|_{N_f}<0, \left.\frac{\partial\mu_C}{\partial N_f}\right|_{T}>0$ for the system to be in stable thermodynamic equilibrium at constant value of S, V and N \cite{Bruno}. From (\ref{muC}), one sees that for $g_s=0.8,N_f=3,\mu=13.7i$:
   \begin{eqnarray}
   \label{th-stab}
   & &  \left.\frac{\partial \mu_C}{\partial T}\right|_{N_f}= - \frac{\partial S}{\partial N_f}\Biggr|_T = \pi \sqrt{4\pi g_sN}\left.\frac{\partial \mu_C}{\partial r_h}\right|_{N_f}=\pi\sqrt{4\pi g_sN}\left(-\frac{1}{\sqrt{\frac{{r_h}^{9/2} (g_s {Nf} \log (|\mu| )-2 \pi )^2}{4 \pi ^2 C^2 g_s^2}+1}}\right)<0;\nonumber\\
   & &  \left.\frac{\partial \mu_C}{\partial N_f}\right|_{T}= \frac{4\ 2^{4/9} \Gamma \left(\frac{5}{18}\right) \Gamma \left(\frac{11}{9}\right) \log (|\mu| )}{9 \sqrt[18]{\pi } C \left(\frac{({g_s} {Nf} \log (|\mu| )-2
   \pi )^2}{C^2 {g_s}^2}\right)^{13/18}}\nonumber\\
   & &  -\frac{4 {g_s} {r_h} \log (|\mu| ) \left(\frac{1}{\sqrt{\frac{{r_h}^{9/2} (g_s {Nf} \log (|\mu| )-2
   \pi )^2}{4 \pi ^2 C^2 g_s^2}+1}}-\, _2F_1\left(\frac{2}{9},\frac{1}{2};\frac{11}{9};-\frac{{r_h}^{9/2} (g_s {Nf} \log (|\mu| )-2 \pi )^2}{4
   C^2 {g_s}^2 \pi ^2}\right)\right)}{9 ({g_s} {N_f} \log (|\mu| )-2 \pi )}>0,\nonumber\\
   & &
\end{eqnarray}
which demonstrates the thermodynamical stability of the type IIB background of \cite{metrics}.

Hence, ensuring thermodynamical stability and with the lightest vector meson mass as an input,  for an appropriate imaginary Ouyang embedding parameter, it is possible to obtain the QCD deconfinement temperature consistent with lattice results for the right number of light quark flavors, in the MQGP limit from the type IIB background of \cite{metrics}  in such a way that the modulus of the Ouyang  embedding parameter gives the correct first generation quark mass scale.
\section{$N_f=2$ Gauge Field Fluctuations}
Within the framework of linear response theory, the Einstein's relation according to which the ratio of the DC electrical conductivity and charge susceptibility yields the diffusion constant, must be satisfied. The main result of this section is that imposing the Einstein's relation requires the Ouyang embedding parameter corresponding to the holomorphic embedding of $N_f$ $D7$-branes in the non-extremal resolved warped deformed conifold, to have a specific dependence on the horizon radius $r_h$.

We first discuss the EOMs and their solutions for non-abelian gauge field fluctuations for $N_f=2$ using the formalism of \cite{Erdmenger_et_al} and then calculate the DC electrical conductivity and the charge susceptibility. Finally, we comment on the Einstein relation relating their ratio to the diffusion constant.

Considering a chemical potential with $SU(2)$ flavor structure the general action is given by:
\begin{equation}
S=-T_rT_{D7}\int d^8\xi \sqrt{\det(g+\hat{F})}
\end{equation}
where the group-theoretic factor $T_r=\frac{1}{2}$ for $SU(2)$ and the field strength tensor is given as:
\begin{equation}
\hat{F}_{\mu\nu}=\sigma^{a}(2\partial_{[{\mu}}\hat{A}^{a}_{\nu]}+\frac{r^2_h}{2\pi\alpha^{'}}f^{abc}\hat{A}^{b}_{\mu}\hat{A}^{c}_{\nu}),
\end{equation}
$\sigma^{a}$ are the Pauli matrices and $\hat{A}$ is given by
\begin{equation}
\hat{A}_{\mu}=\delta^{0}_{\mu}\tilde{A}_{0}+A_{\mu}
\end{equation}
with the $SU(2)$ background gauge field
\begin{equation}
\tilde{A}^{3}_{0}\sigma^{3}=\tilde{A}_{0}\left(
                                           \begin{array}{cc}
                                             1 & 0 \\
                                             0 & -1 \\
                                           \end{array}
                                         \right).
\end{equation}
Now collecting the induced metric $g$ and the background field tensor $\tilde{F}$ as another background tensor $G=g+\tilde{F}$ we get equation of motion for gauge field fluctuation $A^{a}_{\mu}$ on $D7$-brane from the action quadratic in the same gauge fluctuation as in \cite{Erdmenger_et_al}:
\begin{equation}
\partial_\kappa[\sqrt{{\rm det}\ G}(G^{\nu\kappa}G^{\sigma\mu}-G^{\nu\sigma}G^{\kappa\mu})\widehat{F^a_{\mu\nu}}]=\sqrt{{\rm det}\ G}\frac{r^2_h}{2\pi{\alpha^\prime}}
\tilde{A}^3_0f^{ab3}(G^{\nu t}G^{\sigma\mu}-G^{\nu\sigma}G^{t\mu})\widehat{F^b_{\mu\nu}}.
\end{equation}
This simplifies to yield:
\begin{eqnarray}
\label{eom}
& & -2\partial_u[\sqrt{{\rm det}\ G}(G^{uu}G^{yy})(2\partial_u A^a_y)]-2\partial_t[\sqrt{{\rm det}\ G}G^{yy}G^{tt}(2\partial_tA^a_y)+\sqrt{{\rm det}\ G}G^{yy}G^{tt}f^{ab3}\tilde{A}^3_0\frac{r^2_h}{2\pi{\alpha^\prime}} A^b_y]\nonumber\\
& & = -2\sqrt{{\rm det}\ G}\frac{r^2_h}{2\pi{\alpha^\prime}}\tilde{A}^3_0f^{ab3}G^{yy}G^{tt}(2\partial_t A^b_y)-2\sqrt{{\rm det}\ G}\frac{r^2_h}{2\pi{\alpha^\prime}}\tilde{A}^3_0G^{yy}G^{tt}f^{ab3}f^{bc3}
\tilde{A}^3_0\frac{r^2_h}{2\pi{\alpha^\prime}}A^c_y.
\end{eqnarray}
Now, choosing the momentum four-vector in $\mathbb{R}^{1,3}$ as $q^\mu = (w, q, 0, 0)$, and with a slight abuse of notation, writing $A_\mu^a(x,u)=\int d^4q e^{- i w t + i q x}A^a_\mu(q,u)$, the simplification of (\ref{eom}) and rewriting in terms of the gauge-invariant variables or electric field components $E^a_T = \omega A^a_y, a=1,2,3$ as well as a further simplification using $X \equiv E^1 + i E^2, Y \equiv E^1 - i E^2$, in the $q=0$-limit, their solutions up to linear order in $w$, are presented in Appendix {\bf A.2}.

In the same appendix, for the purpose of evaluation of DC electrical conductivity, the on-shell action too is worked out.  As shown in \cite{Erdmenger_et_al}, the on-shell action is given by:
\begin{equation}
%\label{S-on-shell-1}
S_{\rm on-shell}\sim T_r T_{D7}\int d^4x\sqrt{{\rm det}\ G}\left.\left(G^{\nu u}G^{\nu'\mu}-G^{\nu \nu'}G^{u \mu}\right)A^a_{\nu'}\widehat{F^a_{\mu\nu}}\right|_{u=0}.
\end{equation}
Working in the gauge $A^a_u=0$, in appendix {\bf A.2}, the following on-shell action's integrand is worked out:
\begin{eqnarray}
\label{DBI-EOM}
& & \sqrt{{\rm det}\ G}\left[\frac{4G^{uu}G^{xx}(G^{ut}G^{ut}-G^{uu}G^{tt})}{q^2(G^{uu}G^{xx})+w^2(G^{tt}G^{uu}-G^{ut}G^{ut})}E^a_x(\partial_uE^a_x)-\frac{4}{w^2}G^{uu}G^{\alpha \alpha}E^a_{\alpha}(\partial_uE^a_{\alpha})+...\right]_{u=0}\nonumber\\
& & = 4\left(\frac{r_h u(u^4-1)}{w^2(\frac{r_h}{u})^{3/4}\sqrt{\frac{r^4_h\sqrt{\frac{r_h}{u}}}{r^4_h\sqrt{\frac{r_h}{u}}+c^2e^{2\phi}u^4}}}E^a_x(\partial_uE^a_x) + \frac{r_h u(u^4-1)}{w^2(\frac{r_h}{u})^{3/4}\sqrt{\frac{r^4_h\sqrt{\frac{r_h}{u}}}{r^4_h\sqrt{\frac{r_h}{u}}+c^2e^{2\phi}u^4}}}E^a_{\alpha}(\partial_uE^a_{\alpha})+...
\right)_{u=0}\nonumber\\
& & \sim \left.\frac{r_h^{\frac{1}{4}}u^{\frac{7}{4}}}{w^2}\left(E^a_x(\partial_uE^a_x) + E^a_{\alpha}(\partial_uE^a_{\alpha})\right)+....\right|_{u\rightarrow0},
\end{eqnarray}
where the dots include the flavor anti-symmetric terms.

Defining the longitudinal electric field as $E_{x}(q,u)= E_{0}(q) \frac{E_{q}(u)}{E_{q}(u=0)}$, the flux factor as defined in \cite{[10]} in the zero momentum limit, using (\ref{F}) and (\ref{DBI-EOM}) will hence be given as:
\begin{eqnarray}
{\cal F}(q,u)=  -\frac{ e^{-\phi(u)}  r_h^{\frac{1}{4}}u^{\frac{7}{4}} }{w^2 }  \frac{E_{-q}(u)\partial_{u}E_{q}(u)}{ E_{-q}(u=0)E_{q}(u=0)},
\end{eqnarray}
and  the retarded Green's function for $E_{x}$, using the prescription of \cite{[10]}, will be given by: ${\cal G}(q,u)= -2 {\cal F}(q,u)$. The retarded Green function for $A_{x}$ is $w^2$ times above expression and for $q=0$, it gives
\begin{eqnarray}
{\cal G}_{x x} = \left.2 e^{-\phi(u)}   r_h^{\frac{1}{4}}u^{\frac{7}{4}} \frac{ \partial_{u}E_{q}(u)}{ E_{q}(u)}\right|_{u=0}.
\end{eqnarray}
The spectral functions in zero momentum limit will be given as:
\begin{equation}
\label{correlator_sigma}
{\cal X}_{x x}(w,q=0)= -2 Im {\cal G}_{x x}(w,0) = { e^{-\phi(u)}  r_h^{\frac{1}{4}} }Im\left[u^{\frac{7}{4}} \frac{ \partial_{u}E_{q}(u)}{ E_{q}(u)}\right]_{u=0}.
\end{equation}
The DC conductivity is given by the following expression \cite{Mateos}, \cite{[10]}:
\begin{equation}
\label{conductivity}
\sigma = \lim_{w\rightarrow0}\frac{{\cal X}_{x x}(w,q=0)}{w}=\lim_{u\rightarrow0,w\rightarrow0}\frac{r_h^{\frac{1}{4}}u^{\frac{7}{4}}\Im m\left(\frac{E'(u)}{E(u)}\right)}{w}.
\end{equation}
%If one takes $u\rightarrow0$ limit in (\ref{conductivity}) instead of demanding finiteness of $E^3(u)$ or $E(u)$ at $u=0$, then (\ref{conductivity}) yields the required expression for the DC conductivity.
The final result for the DC conductivity $\sigma$ is given as under:
\begin{equation}
\label{sigma-DC}
\sigma=\frac{r_h^{\frac{1}{4}}}{\pi T}\Im m\left(\frac{c_2\left(\frac{i}{16}(-)^{\frac{3}{4}}c_1 + \frac{\gamma_0}{4}c_2\right) - c_3\frac{c_1\gamma_0}{4}}{c_2^2}\right)\sim\left(g_sN\right)^{\frac{1}{8}}T^{-\frac{3}{4}}\frac{c_1}{c_2}.
\end{equation}
Interestingly, this mimics a one-dimensional interacting system - Luttinger liquid - on a lattice for appropriately tuned Luttinger parameter \cite{fraccond-Luttinger-1d}.
 \footnote{We wishe to thank S. Mukerjee for pointing out this fact as well as \cite{fraccond-Luttinger-1d}.}

Another physically relevant quantity  is the charge susceptibility $\chi$, which is thermodynamically defined as response of the charge  density to the change in chemical potential, is given by the following expression  \cite{J. Mas et al [2008]}:
\begin{eqnarray}
\label{chi-a}
& & \chi=\left.\frac{\partial n_q}{\partial \mu_C}\right|_{T},
\end{eqnarray}
 where $n_q = \frac{\delta S_{DBI}}{\delta F_{rt}}$, and the chemical potential $\mu_C$ is defined as
 $\mu_C=\int_{r_h}^{r_B} { F_{rt}} dr$. The charge density will be given as:
\begin{equation}
\label{nq-ii}
n_q = \frac{\delta S_{\rm DBI}}{\delta F_{rt}} \sim \frac{F_{rt}\sqrt{|\mu|}r^{\frac{9}{4}}}{\sqrt{1-F_{rt}^2}},
\end{equation}
and using (\ref{chi-a}), one gets the following charge susceptibility:
\begin{eqnarray}
\label{chi-ii}
& & \frac{1}{\chi} = \int_{r_h}^\infty dr \frac{dF_{rt}}{dn_q}  = \int_{r_h}^\infty dr \frac{r^{\frac{9}{2}}}{\sqrt{|\mu|}\left( \frac{C^2}{\left(\frac{1}{g_s} - \frac{N_f}{2\pi}\log|\mu|\right)^2} + r^{\frac{9}{2}}\right)^{\frac{3}{2}}}\nonumber\\
& & = \frac{1}{45 \sqrt{\mu
   } {r_h}^{5/4} \left(  \frac{C^2}{\left(\frac{1}{g_s} - \frac{N_f}{2\pi}\log|\mu|\right)^2}+{r_h}^{9/2}\right)}\Biggl\{414 {r_h}^{9/2}\ _2F_1 \left(-\frac{1}{2},\frac{5}{18};\frac{23}{18};-\frac{
    \frac{C^2}{\left(\frac{1}{g_s} - \frac{N_f}{2\pi}\log|\mu|\right)^2}}{{r_h}^{9/2}}\right)\nonumber\\
    & & +\left(4  \frac{C^2}{\left(\frac{1}{g_s} - \frac{N_f}{2\pi}\log|\mu|\right)^2}-5 {r_h}^{9/2}\right)
  \  _2F_1 \left(\frac{5}{18},\frac{1}{2};\frac{23}{18};-\frac{  \frac{C^2}{\left(\frac{1}{g_s} - \frac{N_f}{2\pi}\log|\mu|\right)^2}}{{r_h}^{9/2}}\right)\Biggr\}\nonumber\\
   & & = \frac{4}{5 \sqrt{|\mu| }\left(4\pi g_sN\right)^{\frac{5}{8}} {T}^{5/4}} + {\cal O}\left(\frac{1}{\left( g_sN\right)^{\frac{23}{8}}}\right).
\end{eqnarray}
Hence, the charge susceptibility is given by:
\begin{equation}
\chi\sim\sqrt{|\mu| }\left( g_sN\right)^{\frac{5}{8}} {T}^{5/4}.
\end{equation}
Given that one is in the regime of linear response theory, one expects the Einstein's relation: $\frac{\sigma}{\chi}=D\sim\frac{1}{T}$, to hold\footnote{We would like to thank V.B.Shenoy and S. Mukerjee for clarifications on this point.}. However, a naive application yields $\frac{\sigma}{\chi}\sim \frac{c_1}{c_2}\frac{1}
{\sqrt{|\mu|g_sN}}\frac{1}{T^2}$. One expects the Ouyang embedding parameter to be related to the deformation parameter if there were supersymmetry. In the MQGP limit, there is approximate supersymmetry. The resolution parameter possesses an $r_h$-dependence.
If one assumes that $|\mu|\sim\frac{1}{r_h^2}$ (in $\alpha^\prime=1$-units), then the Einstein's relation is preserved.

The fact that the Ouyang embedding parameter turns out to be dependent on the horizon radius is reminiscent of the fact that the resolution parameter too turns out to be dependent on the horizon radius \cite{K. Dasgupta  et al [2012]}, and serves as an important constraint while studying Ouyang embeddings. Further, the 1+1-dimensional subspace singled out in the plane wave basis of the Fourier modes of the gauge field fluctuations, via the evaluation of the electrical conductivity, provides an important prediction that the theory mimicks a 1+1-dimensional Luttinger liquid for appropriately tuned interaction parameter.
\section{Significance of Results Obtained}
Systems like QGP are expected to be strongly coupled. In fact, apart from having a large t'Hooft coupling, it is believed that the same must also be characterized by finite gauge coupling. It is hence important to have a framework in the spirit of gauge-gravity duality, to be able to address this regime in string theory. Finite gauge coupling would under this duality translate to finite string coupling hence necessitating addressing the same from $M$ theory perspective.
\begin{itemize}
 \item Being able to reproduce the confinement-deconfinement temperature compatible with lattice results, serves as a non-trivial check for a proposed holographic dual of large-$N$ thermal QCD. In this respect, this result is very significant as it is able to successfully incorporate in a self-consistent way, a lattice-compatible $T_c$ for the right number of light quark flavors and light quark masses, thermodynamical stability, the right lightest vector mass for the number of quark colors $N_c$ given in in the IR (relevant to a low value of $T_c$) by $M$ which can be tuned to equal 3 (as one ends up with an $SUN(M)$ gauge theory at finite temperature in the IR at the end of the Seiberg duality cascade).

\item
Given that one is working within linear perturbation/response theory, one expects the Einstein relation relating the ratio of the DC electrical conductivity and charge susceptibility to the diffusion constant, to hold. This necessitates taking the Ouyang embedding parameter, analogous to the resolution parameter \cite{K. Dasgupta  et al [2012]}, to be dependent on the horizon radius with a specific form of dependence.
\item
  The temperature dependence at temperatures above $T_c$, i.e., the deconfined phase curiously  mimics a one-dimensional Luttinger liquid for a specific choice of the Luttinger parameter. The one-dimensional identification could be due to the $(t,x)$ singled out in the plane-wave basis of the Fourier modes of the gauge field fluctuations upon the choice of the dual $q^\mu=(w,q,0,0)$.

\end{itemize}
%\begin{subappendices}
%\input{AppendixA}
%\end{subappendices}
%\input{empty}
%\input{Chapter3/chap3}
\chapter{Speed of Sound, Diffusion Coefficient, $\frac{\eta}{s}$, Wiedemann-Franz law and $D=1+1$ Luttinger Liquid}
\chaptermark{Speed of Sound, Diffusion Coefficient, $\frac{\eta}{s}$, \\ Wiedemann-Franz law and $D=1+1$ Luttinger Liquid}
\graphicspath{{Chapter3/}{Chapter3/}}
\section{Introduction}
QCD has an interesting phase structure. The phase diagram of QCD indicates a confining phase at low temperature, while beyond some temperature, $T\gg T_c$, where $T_c$ is the only scale that we have here, it is a non-confining theory. It possesses a phase transition from confining phase to a deconfined plasma phase at $T=T_c$. At sufficiently high temperatures i.e., at $T\gg T_c$, 't Hooft coupling is much less than unity and hence the theory is weakly coupled. However, the plasma as obtained in RHIC experiment at temperature which is about $2T_c$, is not weak enough. In particular, to explore the physics of QCD at temperature close to $T_c$, we have to take a look at the strongly coupled regime where the 't Hooft coupling is around unity. This is called the strongly coupled plasma phase of QCD and the medium is referred as the `sQGP'; see \cite{A.Sinha-1}.

Due to the strong nature of the coupling, the perturbative method is not quite applicable in this case. In lattice gauge theory using numerical simulations the equilibrium properties of the strongly coupled hot QCD can be explored. But interesting non-equilibrium properties such as hydrodynamic behavior or the real time dynamics cannot be seen from the equilibrium correlation functions. So the lack of non-perturbative methods to study hot QCD, forces us to look for either a different theory/model or a different limit of a known theory/model. At finite temperature the equilibrium or non-equilibrium properties of the Euclidean theory are studied requiring time to have periodicity $\beta\sim\frac{1}{T}$. Thus, at non-zero temperature, the Euclidean space-time looks like a cylinder with the topology $\mathbb{R}^3\times S^1$. The AdS/CFT correspondence tells us that at $T=0$ the $4d$ SYM  theory defined on $\mathbb{R}^4$ is dual to string theory on $5d$ AdS space with $\mathbb{R}^4$ as the boundary of the same. So at zero temperature we can think of the field theory as living on the boundary of AdS space. However, the prime interest is to investigate the finite temperature aspects of the dual field theory from the physics of supergravity. Hence at finite temperatures, the space-time of the gravitational description somehow has to be changed such that one gets a geometry of the boundary which is equivalent to $\mathbb{R}^3\times S^1$ and not $\mathbb{R}^4$. In other words one needs to find some bulk geometry which has a boundary with the topology $\mathbb{R}^3\times S^1$. One possible answer is the AdS-BH space-time with the following metric sometimes called black-brane metric given as:
\begin{equation}
ds^2=a(r)\Biggl(-g(r)dt^2+d\vec{x}^2\Biggr)+b(r)dr^2
\end{equation}
with Minkowskian signature. Here $r$ is the radial coordinate and $g(r)$, dependent on the horizon radius $r_h$, is a `black-hole function'. By construction, the time coordinate is defined to be periodic with period $\beta$ which is inverse of temperature and is related to the horizon radius $r_h$. Now, in the Klebanov-Strasslar model where the temperature is turned on in the field theory side is effected by introducing a black hole in the dual geometry. Interestingly the KS background with a black hole has the geometry equivalent to the AdS-BH spacetime in the large $r$ limit. Moreover, the embedding of $D7$-branes in KS model via the holomorphic Ouyang embedding \cite{ouyang} and finally the M-theory uplift of the whole set up keeps the background geometry as required provided we consider some limiting values of the parameters in the theory. The details about this, based on \cite{metrics}, \cite{ouyang}, has already discussed in {\bf Chapter 1}.

Most of the large-$N$ holographic models cater to the large 't Hooft-coupling limit while keeping the gauge coupling vanishingly small. However, in systems such as sQGP, it is believed that not only should the 't Hooft coupling be large, but even { the gauge/string coupling should also be finite} \cite{Natsuume}. A finite gauge coupling would imply a finite string coupling which necessitates addressing the limit from an M-theory point of view. In the context of top-down holographic models of large-$N$ thermal QCD {\it at finite gauge coupling}, in this chapter we evaluate the non-conformal corrections to some transport coefficients such as the shear viscosity $\eta$ (as well as the shear-viscosity-entropy-density ratio $\frac{\eta}{s}$), diffusion constant $D$ and the speed of sound $v_s$. These {non-conformal corrections at finite gauge coupling are particularly relevant in the IR and in fact also encode the scale-dependence of aforementioned physical quantities, and hence are extremely important to be determined for  making direct contact with sQGP. { The main non-trivial insight gained via such computations is the realization that at NLO in $N$ there is a partial universality in these corrections determined by $N_f$ and $M$ apart from $N$.} Also using the type IIB gravity dual at leading order in $N$, we reproduce the expected linear large-$T$ variation of DC electrical conductivity characteristic of most strongly coupled gauge theories with five-dimensional gravity duals with a black hole \cite{SJain_sigma+kappa}.

Analogous to \cite{KS}, the non-conformality in \cite{metrics} is introduced via $M$ number of fractional $D3$-branes, the latter appearing explicitly in $B_2, H_3$ and after construction of a delocalized SYZ type IIA mirror (resulting in mixing of $B_2$ with the metric components after taking a triple T-dual of \cite{metrics}) as well as its local M-theory uplift, also in the metric. {In the context of a (local) M-theory uplift of a top-down holographic thermal QCD dual such as that of \cite{metrics} {\it at finite gauge coupling}, we estimate for the first time, the non-conformal  corrections appearing at the NLO in $N$ to the speed of sound $v_s$, shear mode diffusion constant  $D$, the shear viscosity $\eta$ and the shear viscosity - entropy density ratio $\frac{\eta}{s}$.} { The main new insight gained by this set of results is  that the non-conformal corrections in all the aforementioned quantities are found to display a partial universality in the sense that at the NLO in $N$ the same are always determined by $\left(\frac{(g_s M^2)(g_s N_f)}{N}\right)$}, $N_f$ being the number of flavor $D7$-branes. Thus, {we see that the same are determined by the product of the very small $\frac{g_sM^2}{N}\ll1$ - part of the MQGP limit (\ref{limits_Dasguptaetal-ii}) - and the finite $g_s N_f\sim {\cal O}(1)$ (also part of (\ref{limits_Dasguptaetal-ii})).}  Of course,  the leading order conformal contributions though at vanishing string coupling and large t'Hooft coupling were (in)directly known in the literature. {It is interesting to see the conformal limit of our results at finite $g_s$ obtained by turning off of $M$ - which encodes the non-conformal contributions -  reduce to the known conformal results for vanishing $g_s$}.

Now, in {\bf Chapter 1} we have seen that the response of a physical system to an external source is given by the retarded Green's function as per the linear response theory. Although hydrodynamics tells us that it is enough to compute the transport coefficients to know the response. Also there are two different ways to get the transport coefficients. First, from the coefficient of $\mathcal{O}(w)$ term in the retarded greens function (Kubo formula) and second, from the pole of the retarded greens function. So it is not necessary to calculate the full retarded greens function but an appropriate part of it does the job. This is precisely done by the AdS/CFT correspondence using the dual gravitational background. Another way of getting the pole of retarded greens function is to solve for the quasinormal modes of the bulk gravity fields. Quasinormal modes are defined for a non-conservative system with the same analogy as the usual normal modes. The frequency of the quasinormal modes are complex valued and the dissipation is given by it's imaginary part.
To get the quasinormal frequency we must solve for the linearized fluctuations around an uniform black brane with the boundary condition which is normalizable at the spatial infinity and purely incoming at the horizon.
It was shown in \cite{0302026} that the quasinormal frequency associated with the quasinormal modes defined above in an asymptotically AdS spacetime exactly matches with the pole of the two point correlation function involving operators in the field theory dual to different metric perturbations. Hence evaluating the quasinormal frequency $\omega$ as a function of the special momentum $q$, gives the thermodynamic and hydrodynamic behavior of the plasma. Interestingly, though not addressed in this thesis, but in the context of holographic thermoelectric, piezoelectric and flexoelectric behaviors, response coefficients of charged black branes/holes, exhibiting the same, was discussed in \cite{Obers-2}, \cite{Erdmenger-4}.
\subsection{Perturbations of the background and the gauge invariant combinations}
We consider a small linear fluctuation of the background metric of (\ref{5dmetric inu}) as:
\begin{eqnarray}
\label{metric perturbation}
g_{\mu\nu}=g^0_{\mu\nu}+h_{\mu\nu},
\end{eqnarray}
where $g^0_{\mu\nu}$ denotes the background metric. The inverse metric is defined as(up to second order in perturbation)
\begin{equation}
\label{inverse_fluc}
g^{\mu\nu}=g^{(0)\mu\nu}-h^{\mu\nu}+h^{\mu l}h_{l}^{~\nu}.
\end{equation}
Assuming the momenta to be along the $x$-direction, the metric fluctuations can be written as the following fourier decomposed form:
\begin{eqnarray}
h_{\mu\nu}( x,t,u)=\int \frac{\varepsilon^{4}q}{(2\pi)^4 }e^{-iwt+iqx} h_{\mu\nu}(q,w,u).
\end{eqnarray}
We will work in the gauges where $h_{u\mu}$ is zero for all $\mu$ including $u$.
Based on the the spin of different metric perturbations under $SO(2)$ rotation in $(y,z)$ plane, the same can be classified into three types as follows:
\begin{enumerate}
\item[(i)]
 vector modes: $h_{x y}, h_{t y}\neq 0$ or $h_{x z},  h_{t z}\neq 0$, with all other $h_{\mu \nu}=0$.
\item[(ii)]
Scalar modes: $h_{x x}=h_{y y}=h_{z z}=h_{tt}\neq 0$, $h_{x t}\neq 0$, with all other $h_{\mu \nu}=0$.
\item[(iii)]
Tensor modes: $h_{y z}\neq 0$, with all other $h_{\mu \nu}=0$.
\end{enumerate}
The EOMs for the scalar and vector type metric perturbations are all coupled to each other and hence they are not easy to solve. However following \cite{klebanov quasinormal} one can construct a particular combination of different perturbations which is gauge invariant and all the coupled EOMs can be replaced by a single equation involving the gauge invariant variable. This combination which is invariant under diffeomorphisms: $h_{\mu\nu}\rightarrow h_{\mu\nu}-\nabla_{(\mu}\xi_{\nu)}$ is given as \cite{klebanov quasinormal}:
\begin{equation}\begin{split}
\label{Z-scalar mode}
{\rm Scalar\ type}: Z_s&=-q^2(1-u^4)H_{tt}+2wqH_{xt}+w^2H_{xx}\\&
+q^2(1-u^4)\left(1+\frac{g_{xx}(-4u^3)}{g^{\prime}_{xx}(1-u^4)}-\frac{w^2}{q^2(1-u^4)}\right)H_{yy}
\end{split}
\end{equation}
\begin{equation}\begin{split}
\label{Z-vector mode}
 &{\rm Vector\ type}: Z_v = q H_{ty} + w H_{xy}~~~~~~~~~~~~~~~~~~~~~~~~~~~~~~~~~~~~~~~~~~
\end{split}
\end{equation}
\begin{equation}\begin{split}
\label{Z-tensor mode}
&{\rm Tensor\ type}: Z_t=H_{yz},~~~~~~~~~~~~~~~~~~~~~~~~~~~~~~~~~~~~~~~~~~~~~~~~
\end{split}
\end{equation}
where $H_{tt}=-g^{tt}h_{tt}$, $H_{xx}=g^{xx}h_{xx}$, $H_{yy}=g^{xx}h_{yy}$, $H_{xt}=g^{xx}h_{xt}$, $H_{xy}=g^{xx}h_{xy}$ and the prime denotes the derivative with respect to $u$. The two second order differential equations corresponding to the EOMs of $Z_v$, $Z_s$ and $Z_t$ are solved and the required quasinormal modes are obtained by imposing Dirichlet  boundary conditions at $u=0$ \cite{klebanov quasinormal}.
\section{Scalar Metric Perturbation Modes and Speed of Sound in MQGP Limit}
In this section, by considering scalar modes of metric perturbations, we will evaluate the speed of sound, first up to leading order in $N$ four ways: (i) (subsection {\bf 3.2.1.1}) the poles appearing in the common denominator of the solutions to the individual scalar modes of metric perturbations (the pure gauge solutions and the incoming-wave solutions); (ii) (subsection {\bf 3.2.1.1}) the poles appearing in the coefficient of the asymptotic value of the square of the time-time component of the scalar metric perturbation in the on-shell surface action; (iii) (subsection {\bf 3.2.2.1}) the dispersion relation obtained via a Dirichlet boundary condition imposed on an appropriate single gauge-invariant metric perturbation - using the prescription of \cite{klebanov quasinormal} - at the asymptotic boundary; (iv) (subsection {\bf 3.2.2.2}) the poles appearing in the coefficient of the asymptotic value of the square of the time-time component of the scalar metric perturbation in the on-shell surface action written out in terms of the same single gauge-invariant metric perturbation. The third approach is then extended to include the non-conformal corrections to the metric and obtain an estimate of the corrections to $v_s$ up to NLO in $N$.

Using the black $M_3$ brane metric, up to leading order in $N$ and considering the non-zero scalar modes of metric perturbations, we get a set of seven differential equations from the Einstein's equation. Defining the dimensionless energy and momentum,
\begin{eqnarray}
 \label{four momentum}
 \omega_3=\frac{w}{\pi T},~~~~~~~~~~~~~~~~~\ q_3=\frac{q}{\pi T},
 \end{eqnarray}
the set of seven equations are given as:
 \begin{eqnarray}
 \label{7scalar_EOMs}
& &  H_{tt}^{\prime\prime} + \frac{1}{u} \left(-\frac{6}{g} + 5\right) H_{tt}^\prime +
  H_s^{\prime\prime} + \frac{1}{u} \left(-\frac{2}{g} + 1\right) H_s^\prime = 0,\nonumber\\
    & & H_{tt}^{\prime\prime} + \frac{2}{u} \left(-\frac{3}{g} + 1\right) H_{tt}^\prime + \frac{1}{u} \left(-\frac{2}{g} + 1\right) H_s^\prime - \frac{q_3^2}{g} H_{tt}
 + \frac{\omega_3^2}{g^2} H_s + 2 \frac{q_3 \omega_3}{g^2} H_{{x}t} = 0,\nonumber\\
& &   H_s^{\prime\prime} - \frac{3}{u} H_{tt}^\prime - \frac{2}{u} \left(1 + \frac{2}{g}\right) H_s^\prime - \frac{q_3^2}{g} H_{tt} + \frac{\omega_3^2}{g^2} H_s -\frac{4q_3^2}{g} H_{{y}{y}} +\frac{2 \omega_3 q_3}{g^2} H_{{x}t} = 0,\nonumber\\
& &  H_{{y}{y}}^{\prime\prime} - \frac{H_{tt}^\prime}{u} - \frac{H_s^\prime}{u} + \frac{1}{u} \left(-\frac{4}{g} + 1\right) H_{{y}{y}}^\prime + \frac{1}{g^2} \left(\omega_3^2 - g q_3^2\right) H_{{y}{y}} =  0,\nonumber\\
& &  H_{{x}t}^{\prime\prime} - \frac{3}{u} H_{{x}t}^\prime + \frac{2 q_3 \omega_3}{g} H_{{y}{y}} = 0,\nonumber\\
& &  q_3 \left(-g H_{tt}^\prime + 2 u^3 H_{tt}\right) - 2 q_3 g H_{{y}{y}}^\prime +  \omega_3 H_{{x}t}^\prime= 0,\nonumber\\
& &   \omega_3 \left(g H_s^\prime + 2 u^3 H_s\right) +
  q_3 \left(g H_{{x}t}^\prime + 4 u^3 H_{{x}t}\right) = 0
\end{eqnarray}
where we define $H_{tt}=\left(\frac{g^{2/3}_{s}u^2 L^2}{r^{2}_{h} g}\right)h_{tt}
$, $H_{{x}{x}}=\left(\frac{g^{2/3}_{s}u^2L^2}{r^{2}_{h}                                                                                 }\right)h_{{x}{x}}$, $H_{{y}{y}}=H_{{z}{z}}=\left(\frac{g_s^{2/3}u^2L^2}{r_h^2g}\right)h_{{y}{y}}$, and $H_s=H_{{x}{x}}+2H_{{y}{y}}$.
The above system of equations can be reduced to the following linearly independent set of four equations
\begin{eqnarray}
\label{4scalar_EOMs}
 H^\prime_{{x}{x}} &=&\frac{3\omega_3^2-2q_3^2u^4}{q_3^2\left(u^4-3\right)}H^\prime_{tt}+\frac{2 u \left(q_3^4 \left(1-u^4\right)^2-\omega_3^2 \left(-2 u^6+6 u^2+\omega_3^2\right)\right)}{q_3^2 \left(u^4-3\right)
   \left(1-u^4\right)^2}H_{{y}{y}}\nonumber\\
    & & +\frac{u \omega_3^2 \left(q_3^2 \left(u^4-1\right)+2 u^6-6 u^2-\omega_3^2\right)}{q_3^2 \left(1-u^4\right)^2 \left(u^4-3\right)}H_{{x}{x}}+\frac{2 u w3 \left(q3^2 \left(u^4-1\right)+2 u^6-6 u^2-w3^2\right)}{q3 \left(1-u^4\right)^2 \left(u^4-3\right)}H_{{x}t}\nonumber\\
     & & +\frac{u \left(q_3^2 \left(u^4-1\right)+2 u^6-6 u^2-\omega_3^2\right)}{\left(u^4-3\right) \left(u^4-1\right)}H_{tt}\nonumber\\
     H^\prime_{{y}{y}}&=&-\frac{q3^2 \left(u^4-3\right)+3 \omega_3^2}{2 q_3^2 \left(u^4-3\right)}H^\prime_{tt}+\frac{u \omega_3^2 \left(q3^2 \left(u^4-1\right)-2 u^6+6 u^2+\omega_3^2\right)}{q_3^2 \left(1-u^4\right)^2 \left(u^4-3\right)}H_{{y}{y}}\nonumber\\
     & & +\frac{u \omega_3^2 \left(-2 u^6+6 u^2+\omega_3^2\right)}{2 q_3^2 \left(1-u^4\right)^2 \left(u^4-3\right)}H_{{x}{x}}+\frac{u \omega_3 \left(-2 u^6+6 u^2+\omega_3^2\right)}{q_3 \left(1-u^4\right)^2 \left(u^4-3\right)}H_{{x}t}\nonumber\\
      & & +\frac{u \left(q_3^2 \left(u^4-1\right)+2 u^6-6 u^2-\omega_3^2\right)}{\left(u^4-3\right) \left(u^4-1\right)}H_{tt}\nonumber\\
       H_{{x}t}^\prime&=&\frac{3 \left(u^4-1\right) \omega_3}{q_3 \left(u^4-3\right)}H_{tt}^\prime-\frac{2 u \omega_3 \left(q_3^2 \left(u^4-1\right)-2 u^6+6 u^2+\omega_3^2\right)}{q_3 \left(u^4-3\right) \left(u^4-1\right)}H_{{y}{y}}\nonumber\\
        & & -\frac{u \omega_3 \left(-2 u^6+6 u^2+\omega_3^2\right)}{q_3 \left(u^4-3\right) \left(u^4-1\right)}H_{{x}{x}}+\frac{2 u \left(2 u^6-6 u^2-\omega_3^2\right)}{\left(u^4-3\right) \left(u^4-1\right)}H_{{x}t}\nonumber\\
         & & -\frac{u q_3  \omega_3}{u^4-3}H_{tt}\nonumber\\
          H_{tt}^{\prime\prime}&=&\frac{u^8+2 u^4+9}{u \left(u^4-3\right) \left(u^4-1\right)}H_{tt}^\prime-\frac{2 \left(q_3^2 \left(u^4+1\right)+2 \omega_3^2\right)}{\left(u^4-3\right) \left(u^4-1\right)}H_{{y}{y}}\nonumber\\  & & -\frac{2 \omega_3^2}{\left(u^4-3\right) \left(u^4-1\right)}H_{{x}{x}}-\frac{4 q_3 \omega_3}{\left(u^4-3\right) \left(u^4-1\right)}H_{{x}t} -\frac{2 q_3^2}{u^4-3}H_{tt}.
\end{eqnarray}
To solve the system of equation (\ref{4scalar_EOMs}) we look for the behavior of the solution near $u=1$. Hence for time being we reconsider equation (\ref{7scalar_EOMs}) and write them as the following system of six first order differential equations
\begin{equation}\begin{split}
&  H^\prime_{tt}=\frac{1}{g}P_{tt}\\
&  H^\prime_{yy}=-\frac{1}{2 g}P_{tt}+\frac{  u^3}{ g}H_{tt}+\frac{ \omega_3}{2 q_3g }P_{xt}\\
&  H^\prime_{s} =-\frac{2 u^3 }{g}H_s-\frac{4 q_3 u^3}{w_{3} g}H_{xt}-\frac{q_3 }{\omega_3}P_{xt}\\
&  H^\prime_{xt}=P_{xt}\\
&  P^\prime_{xt}=\frac{3}{u}P_{xt}-\frac{2q_3 \omega_3 }{g}H_{yy}\\
&  P^\prime_{xt}=-\frac{2(u^4-2)}{u g}P_{tt}+q_3^2H_{tt}-\frac{q_3(u^4+1)}{u \omega_3}P_{xt}-\frac{2u^2+2u^6+\omega_3^2}{g}
 \left(H_s+\frac{2q_3}{\omega_3}H_{xt}\right).
 \end{split}
\end{equation}
In matrix form the above equation can be written as
\begin{eqnarray}
\label{one}
X^\prime=A(u)X
\end{eqnarray}
where $A$ is a $6\times6$ matrix and is singular for all values of $u$. Equation (\ref{one}) can be solved by substituting the ansatz $X=(1-u)^r F(u)$ into the same, where the exponent $r$ can be evaluated from the eigenvalues of the matrix $(1-u)A(u)$ near $u=1$. They are given by $r_1=r_2=0,r_3=-1/2,r_4=i\omega_3/4,r_5=-i\omega_3/4$ and $r_6=1/2$. Two of the eigenvalues namely $r=\mp i\omega_3/4$ represent the incoming/outgiong wave.
\subsection{The Longer Route up to Leading Order in $N$ - Via Solutions of EOMs}
In this subsection, we describe the evaluation of $v_s$, first from the solutions to the EOMs for the scalar metric perturbation modes and then from two-point correlation function of energy momentum tensor: $\langle T_{tt}T_{tt}\rangle$. We limit ourselves, in this subsection, to the leading order in $N$.
\subsubsection{From the Pole Structure of Solutions to $H_{ab}(u)$}
Based on \cite{PSS-scalar}, we give below a discussion on three gauge transformations that preserve $h_{\mu u}=0$, for the black $M3$-brane metric. This is then utilized to obtain solutions to the scalar metric perturbation modes' equations of motion (\ref{7scalar_EOMs}) near $u=0$ and thereafter the speed of sound. We verify the result for the speed of sound by also calculating the same from an two-point energy-momentum correlation function.
Demanding that infinitesimal diffeomorphism which is given as:
\begin{equation}\begin{split}
x^\mu &\rightarrow x^\mu + \xi^\mu\\
g_{\mu\nu} & \rightarrow g_{\mu\nu} - \nabla_{(\mu}\xi_{\nu)},
\end{split}
\end{equation}
preserves the gauge condition $h_{\mu u}=0$ implies imposing \cite{PSS-scalar}:
\begin{equation}
\partial_{(\mu}\xi_{u)} - 2 \Gamma^\rho_{\mu u}\xi_\rho = 0,
\end{equation}
wherein $\Gamma^\rho_{\mu u}$ is calculated w.r.t. $g_{\mu\nu} = g_{\mu\nu}^{(0)} + h_{\mu\nu}$. There are three residual gauge transformations under which the system of differential equations (\ref{7scalar_EOMs}) remains invariant. They are given in (\ref{GTI}), (\ref{GTII}) and (\ref{GT-III}).

Choosing $C_{u}, \tilde{C}_{{x},u}:  \left(C_{u},\frac{\tilde{C}_{t,{x}}}{i}\right)\frac{g_s^{\frac{2}{3}}}{L^2}=1$, the non-zero pure gauge solutions gauge equivalent to $H_{ab}=0$ ($H_{ab}=0, \xi_a=0$), near $u=0$, are given by:
\begin{equation}
\begin{split}
\label{H_ab-0_u-0}
 H_{xx}^{(I)}(0)& = - 2 q_3
 \\H_{xt}^{(I)}(0)& = \omega_3 \\\\
 H_{tt}^{(II)}(0) &= 2 \omega_3\\
H_{xt}^{(II)}(0) &= q_3
\\\\H_{xx}^{(III)}(0)& = 2
\end{split}
\end{equation}

Writing $H^{\rm inc}_{ab}(u)$ as the incoming solution to the differential equations, the general solution can be written as the following form,
\begin{equation}
\label{Hcomponents}
H_{ab}(u) = a H^{(I)}_{ab}(u) + b H^{(II)}_{ab}(u) + c H^{(III)}_{ab}(u) + d H^{\rm inc}_{ab}(u).
\end{equation}
To determine $H^{\rm inc}_{ab}(u)$, we Solve (\ref{7scalar_EOMs}) near the horizon $u=1$ (this enables solving the fourth, fifth and sixth equations of (\ref{7scalar_EOMs}) independent of the first, second, third and seventh equations), where we have already shown that the same is a regular singular point with exponent of the indicial equation corresponding to the incoming solution given by $-\frac{i\omega_3}{4}$, implying that $H_{ab}^{\rm inc}(u)=(1-u)^{-\frac{i\omega_3}{4}}{\cal H}_{ab}(u)$.  Making double perturbative ansatze:
\begin{equation}
{\cal H}_{ab}(u) = \sum_{m=0}^\infty\sum_{n=0}^\infty {\cal H}_{ab}^{(m,n)}(u)q_3^m\omega_3^n,
\end{equation}
one obtains near u=0 the solutions given in (\ref{hab}).

Upon using $H_{tt}(0)=H_{t}^{(0)}, H_{xt}(0)=H_{xt}^{(0)}, H_s(0)=H_s^{(0)}$ and solving for $a, b, c$ and $d$, the following is the common denominator:
\begin{eqnarray}
\label{pole-speed_s}
 \Omega(\omega_3,q_3) & \equiv & \alpha_{yy}^{(0,0)} + \alpha_{yy}^{(1,0)} q_3 + C_{1yy}^{(2,0)} q_3^2 +
 \alpha_{yy}^{(1,0)} \omega_3 + \left(-\frac{i}{4} + C_{2yy}^{(1,1)} - \frac{2}{9} C_{1yy}^{(1,1)} e^3\right) q_3 \omega_3
 \nonumber\\
& & + \left(C_{1yy}^{(0,2)} +
     C_{2yy}^{(0,2)} + \frac{i}{4} \Sigma_{2yy}^{(0,1)}\right) \omega_3^2,
\end{eqnarray}
where $\alpha_{yy}^{(m,n)}, C_{ayy}^{(m,n)}, a,b=1,2$ are constants appearing in the solutions to ${\cal H}_{ab}^{(m,n)}(u)$ in (\ref{hab}).
Now, (\ref{pole-speed_s}) can be solved for $\omega_3$ and the solution is given in (\ref{pole-speed_s_ii}) in appendix {\bf B.4}.

Assuming $\alpha_{yy}^{(0,0)}\ll1, |\Sigma_{2\ yy}^{(0,1)}|\gg1(i \Sigma_{2\ yy}^{(0,1)}\in\mathbb{R}): \alpha_{yy}^{(0,0)}\Sigma_{2\ yy}^{(0,1)}<1; \alpha_{yy}^{(1,0)} = - |\alpha_{yy}^{(1,0)}|$, consistent with the constraints such as (\ref{constraints_I}) and (\ref{pole-speed_s_ii}) of appendix {\bf B.4}, implies the roots (\ref{root1-i}) and (\ref{root2-i}) as given in appendix {\bf B.4}. In the same appendix, it is shown that:
\begin{eqnarray}
 \label{pole-speed_s_iii}
 \omega_3\approx \pm q_3\left(1 + i \frac{\alpha_{yy}^{(00)}\Sigma_{2\ yy}^{(0,1)}}{2\left(\alpha_{yy}^{(1,0)}\right)^2}\right)\equiv\pm v_s q_3.
 \end{eqnarray}
One can show that one can consistently choose $ \frac{\alpha_{yy}^{(00)}\left(i\Sigma_{2\ yy}^{(0,1)}\right)}{2\left(\alpha_{yy}^{(1,0)}\right)^2} = \frac{1}{\sqrt{3}} - 1$ to yield $v_s=\frac{1}{\sqrt{3}}$.
\subsubsection{Via Two-Point Correlation Function $\langle T_{tt} T_{tt}\rangle$ using ON-Shell Action and LO EOM's Solutions}
We will now concentrate on the evaluation of the two-point correlation function $\langle T_{tt}T_{tt}\rangle$ from the on-shell action.
On-shellness dictates that:
$R^{(0)}=\frac{10}{3}\Lambda$ under the metric perturbation given in (\ref{metric perturbation}). The pure gravitational part of the $5d$ action along with the Gibbons-Hawking York surface term \cite{Liu+Tseytlin} and a counter term (required to regularize the action) is given by (without worrying about the overall constant term):
\begin{eqnarray}
\label{full action}
S\sim\int_0^1 du \int {d^4}x \sqrt{-g}(R - 2 \Lambda)+\int {d^4}x\sqrt{-g_B}~2K+a\int {d^4}x \sqrt{-g_B}
\end{eqnarray}
where $\Lambda$ is a cosmological constant term and is given as: $\lambda=-\frac{6 g^{2/3}_{s}}{L^2}$. $g_B^{\mu\nu}$ is the pull-back metric on the boundary of AdS space and $K$ is the extrinsic curvature. We choose $a=-\frac{6 g^{1/3}_{s}}{L}$ to make the action in equation (\ref{full action}) finite. On-shell, the bilinear part of the above action, in the limit $q_3\rightarrow0, \omega_3\rightarrow0$, reduces to the following surface term:
\begin{eqnarray}
\label{on-shell-surface-action}
S\sim & & \int {d^4}x\Biggl[\frac{1}{4}\Biggl(H_{tt}^2+8H_{{x}t}^2+2H_{{x}{x}}H_{tt}+4H_{{y}{y}}H_{tt}+4H_{{x}{x}}H_{{y}{y}}-H_{{x}{x}}^2\Biggr)
\nonumber\\
& & -\frac{1}{2\epsilon^3}\Biggl(H_{{x}t}^2+H_{{y}{y}}^2+H_{{x}{x}}H_{tt}+H_{{y}{y}}H_{tt}+2H_{{x}{x}}H_{{y}{y}}\Biggr)^\prime~\Biggr].
\end{eqnarray}
The equations of motion imply that $H_{tt}^\prime(u=0) = H_s^\prime(u=0) = H_{xt}^\prime(u=0) = H_{yy}^\prime(u=0) = 0$, and we will further assume that
\begin{eqnarray}
& & \left(\begin{array}{c}H_{yy}(u=0)\\ H_{s}(u=0)\end{array}\right) = \left(\begin{array}{cc} -\beta_{yt} & -\beta_{yx} \\ -\beta_{st} & -\beta_{sx} \end{array}\right)\left(\begin{array}{c} H_{tt}(u=0)\\ H_{xx}(u=0)\end{array}\right).
\end{eqnarray}
So, the relevant two-point correlation function involving $T_{00}$ will require finding out the coefficient of $\left(H^{(0)}_{tt}\right)^2$ upon substitution of (\ref{GTI}) - (\ref{GTIII}) and (\ref{hab}) along with the values of $a, b, c, d$ with the common denominator $\Omega(\omega_3,q_3)$ of (\ref{pole-speed_s}). As the generic form of this two-point function in the hydrodynamical limit \cite{hyrdodynamical_limit} : $\omega_3\rightarrow0, q_3\rightarrow 0: \frac{\omega_3}{q_3}=\alpha\equiv$ constant - is expected to be of the form: $\frac{q_3^2}{\omega_3^2 - v_s^2 q_3^2}$, we isolate these terms and work up to leading order in $\Sigma_{2yy}^{(0,1)}$. We find from (\ref{on-shell-surface-action}) the following  coefficients of $(H^{(0)}_{tt})^2$ coming from the $H^2$-like terms and $HH^{\prime}$-like terms:
\begin{equation}\begin{split}
\label{T00T00}
& H^2\ {\rm terms}:
    -\frac{i \Sigma_{2yy}^{01}}{16 \left(\alpha
   ^2-1\right)}\left\{\alpha^4 \left(\beta_{st}^2+\beta_{st} (2-8 \beta_{yt})+12 \beta_{yt}^2-1\right)-32 \beta_{yt} (\pi \beta_{yt}-2)\right.\\& \left.-\alpha^2\left(\beta_{st}^2+\beta_{st} (56 \beta_{yt}+2)+12 \beta_{yt}^2-1\right)\right\};\\
& \left.\frac{\left(HH^\prime\right)^{{\cal O}(u^0)}}{u^3}\right|_{u=\epsilon} =-\frac{i \alpha ^2  \Sigma_{2yy}^{01}\beta_{yt}}{16 \left(\alpha ^2-1\right)}
\left\{\alpha ^2 ((8+\pi ) \beta_{st}-2 ((\pi -6) \beta_{yt}+1))\right.\\&\left.+(16+\pi ) \beta_{st}-12
   \beta_{yt}+\pi ^2 \beta_{yt}+14 \pi  \beta_{yt}-2 \pi -22\right\} ;\\&\left.\frac{\left(HH^\prime\right)^{{\cal O}(u)}}{u^3}\right|_{u=\epsilon}=-\frac{1}{16 \left(\alpha
   ^2-1\right)}\Biggl\{i \alpha ^2 \Sigma_{2yy}^{01} \Biggl(\alpha ^2 \beta_{yt} ((16+\pi ) \beta_{st}-(\pi -20)\beta_{yt}+2)\\
 & -2 \alpha  (2 \beta_{st}+\pi
   \beta_{yt}-2)+\beta_{yt} \left((\pi -24)\beta_{st}+\left(-20-3 \pi +\pi ^2\right) \beta_{yt}-2 \pi +6\right)\Biggr)\Biggr\};\\
     &\left.\frac{\left(HH^\prime\right)^{{\cal O}(u^2)}}{u^3}\right|_{u=\epsilon}=-\frac{1}{32 \left(\alpha ^2-1\right)}\Biggr\{i \alpha ^2  \Sigma_{2yy}^{01}\beta_{yt} \Biggl(2 \alpha ^2 ((\pi -24) \beta_{st}+10\beta_{yt})\\&-(\pi -36) \alpha  (2  \beta_{st}+\pi
   \beta_{yt}-2)
   +2 \left((\pi -24)  \beta_{st}+\left(-10-24 \pi +\pi ^2\right) \beta_{yt}-2 \pi +48\right)\Biggr)\Biggl\};\\&
     \left.\frac{\left(HH^\prime\right)^{{\cal O}(u^3)}}{u^3}\right|_{u=\epsilon}=
    \frac{i \Sigma_{2yy}^{01} }{160 \left(\alpha ^2-1\right)^2}\Biggl\{5 \alpha ^6 \left(24 \beta_{st}^2+ \beta_{st} (8-2 (\pi -2) \beta_{yt})+\beta_{yt} (2-\pi
   \beta_{yt})\right)\\& +2 \alpha ^4 \left(60  \beta_{st}^2+ \beta_{st} (6 (15 \pi
   -8) \beta_{yt}-200)-5 \left((\pi -11) \pi \beta_{yt}^2+(22-4 \pi ) \beta_{yt}+4\right)\right) \\&+\alpha ^2 \left(2  \beta_{st} ((38+35 \pi )\beta_{yt}-60)+\pi  (70 \pi -233) \beta_{yt}^2+(466-280 \pi )
   \beta_{yt}+280\right)\\&-40 \alpha ^5 \beta_{yt} (2  \beta_{st}+\pi \beta_{yt}-2)+40 \alpha ^3\beta_{yt} (2 \beta_{st}+\pi
   \beta_{yt}-2)+128 \beta_{yt} (\pi  \beta_{yt}-2)\Biggr\}.
   \end{split}
   \end{equation}
   From (\ref{T00T00}), we see that for $\beta_{yt}=0,\beta_{st}=1$, the first line in (\ref{on-shell-surface-action}) yields
   a contribution: $i\alpha^2\Sigma_{2yy}^{01}\frac{q_3^2}{\left(\omega_3^2 - v_s^2 q_3^2\right)}$ and
   from the second line in (\ref{on-shell-surface-action}), there is no required contribution from $\frac{\left(HH^\prime\right)^{{\cal O}(u^{0,1,2})}}{u^3}$ and $\frac{\left(HH^\prime\right)^{{\cal O}(u^3)}}{u^3}$ terms yield:
   $i \alpha^2\Sigma_{2yy}^{01}\frac{q_3^2}{\left(\omega_3^2 - v_s^2 q_3^2\right)}$.
%\begin{tcolorbox}
%\end{tcolorbox}
\subsection{The Shorter Route - Use of Gauge-Invariant Variable}
In this subsection, we carry on the same calculation as we did in the last subsection for the speed of sound up to leading order in $N$ via a different approach. This time following \cite{klebanov quasinormal}, we first obtain the EOM for appropriate gauge-invariant variable corresponding to the non zero scalar modes of metric perturbations as defined in (\ref{Z-scalar mode}) and then compute the quasinormal modes, hence the speed of sound $v_s$ by solving that EOM in the hydrodynamic approximation. we have also calculate the two point correlation function of energy momentum tensor using the above solution for the gauge invariant variable. Latter following the same approach we compute the next to leading order correction to speed of sound by using the M-theory metric as given in (\ref{Mtheorymetric}) and (\ref{Mtheorymetriccomp}).
\subsubsection{From the solution of Gauge Invariant Variable up to Leading Order in $N$}
Going back to (\ref{4scalar_EOMs}) we see that the four linearly independent equations using the following gauge invariant combination of perturbations namely,
\begin{eqnarray}
\label{gauge invariant}
Z_s(u)=2 q_3 \omega_3 H_{{x}t}+\omega_3^2 H_{{x}{x}}+H_{{y}{y}}\left[q_3^2
   \left(u^4+1\right)-\omega_3^2\right]-q_3^2 \left(1-u^4\right) H_{tt},
\end{eqnarray}
can be written as a single second order differential equation involving $Z_s(u)$:
\begin{eqnarray}
\label{single EOM}
& &  Z_s^{\prime\prime}(u)-\frac{q_3^2 \left(7 u^8-8 u^4+9\right)-3 \left(u^4+3\right) \omega_3^2}{u \left(u^4-1\right) \left(q_3^2 \left(u^4-3\right)+3
  \omega_3^2\right)}Z_s^\prime(u)\nonumber\\
  & & +\frac{q_3^4 \left(u^8-4 u^4+3\right)+2 q_3^2 \left(8 u^{10}-8 u^6+2 u^4 \omega_3^2-3 \omega_3^2\right)+3 \omega_3^4}{\left(1-u^4\right)^2
   \left(q_3^2 \left(u^4-3\right)+3 \omega_3^2\right)}Z_s(u)=0.
\end{eqnarray}
The above equation can be solve by considering an ansatz $Z_s(u)=(1-u)^rF(u)$ where $F(u)$ is regular near the horizon $u=1$. We have already obtained the value of exponent $r$ at the end of section {\bf 5} and it is given by $\pm \frac{i\omega_3}{4}$. we choose the negative sign here as it represents an incoming wave. The evaluation of the function $F(u)$ can be done perturbatively using hydrodynamic approximation, given as: $\omega_3\ll1$, $q_3\ll1$. For analytic solution the momentum has to be light-like, means $\omega_3$ and $q_3$ would be of the same order. Hence we can rescale $\omega_3$ and $q_3$ by a same parameter $\lambda$ as: $\omega_3\rightarrow\lambda \omega_3$, $q_3\rightarrow\lambda q_3$ and expand equation (\ref{single EOM}) to first order in $\lambda$, where the limit $\lambda\ll 1$ ensure that we are working in the hydrodynamic regime.
We choose the following series expansion of $F(u)$ for small frequency and momentum as:
\begin{equation}
\label{regular F}
F(u)=F_0(u)+\omega_3F_1(u)+\mathcal{O}(\omega_3^2,q_3^2,\omega_3q_3).
\end{equation}
Plugging in the equation (\ref{regular F}) into the equation (\ref{single EOM}) one can get an equation involving $F_0(u)$ only:
\begin{equation}\begin{split}
 &u \left(u^4-1\right) \left(q_3^2 \left(u^4-3\right)+3 \omega_3^2\right) F_0^{\prime\prime}+ \left(q_3^2 \left(-7 u^8+8
   u^4-9\right)+3 \left(u^4+3\right) \omega_3^2\right)F_0^\prime\\&+16 q_3^2 u^7 F_0=0.
   \end{split}
\end{equation}
A solution to the above equation is given by,
\begin{eqnarray}
& & F_0(u)=\frac{c_1 \left(q_3^2 \left(u^4+1\right)-3 \omega_3^2\right)}{17 q_3^2-3 \omega_3^2}\\& & \nonumber +\frac{c_2 \left(q_3^2 \left(u^4+1\right)-3
   \omega_3^2\right) \left(-\frac{2 q_3^2-3 \omega_3^2}{q_3^2 \left(u^4+1\right)-3 \omega_3^2}-\frac{1}{4} \log
   \left(u^4-1\right)\right)}{17 q_3^2-3 \omega_3^2}.
\end{eqnarray}
For the regularity of $F_0(u)$ near the horizon $u=1$, we choose the constant $c_2$ to be equal to zero.
Using this solution for $F_0(u)$, another equation for $F_1(u)$ can be found from (\ref{single EOM}),
\begin{eqnarray}
& & u \left(u^4-1\right)\left\{17 q_3^4
   \left(u^4-3\right)-3 q_3^2\omega_3^2 \left(u^4-20\right) -9 \omega_3^4\right\}F_1^{\prime\prime}\nonumber\\
    & & + \left\{-17 q_3^4 \left(7 u^8-8
   u^4+9\right)+3 q_3^2 \omega_3^2\left(7 u^8+9 u^4+60\right) -9 \omega_3^4 \left(u^4+3\right)\right\}F_1^{\prime}\nonumber\\
   & & +16  u^7q_3^2 \left(17
  q_3^2-3 \omega_3^2\right) F_1+16 i  u^7 q_3^2\left(2 q_3^2-3 \omega_3^2\right)c_1=0.
\end{eqnarray}
A general solution is given as:
\begin{eqnarray}
& & F_1(u)=-\frac{c_1 i \left(2 q_3^2-3 \omega_3^2\right)}{17 q_3^2-3 \omega_3^2}+\frac{c_2 \left(q_3^2 \left(u^4+1\right)-3
  w3^2\right)}{17 q_3^2-3 \omega_3^2}\nonumber\\
  & & +\frac{c_3 \left(q_3^2 \left(u^4+1\right)-3 \omega_3^2\right) \left(-\frac{2 q_3^2-3
   \omega_3^2}{q_3^2 \left(u^4+1\right)-3 \omega_3^2}-\frac{1}{4} \log \left(u^4-1\right)\right)}{17 q_3^2-3 \omega_3^2}.
\end{eqnarray}
Again demanding the regularity of the above solution near the horizon, we put $c_3$ to zero. Also imposing a boundary condition $F_1(u=1)=0$, we determine the constant $c_2$ to be equal to $ic_1$. With this the final expression of $Z_s(u)$ is given as:
\begin{eqnarray}
\label{solution single eom}
Z_s(u)=c_1(1-u^4)^{-i\omega_3/4}\left(\frac{q_3^2 \left(u^4+1\right)-3 \omega_3^2}{17 q_3^2-3 \omega_3^2}-\frac{i q_3^2\omega_3 \left(1-u^4\right)}{17 q_3^2-3 \omega_3^2}\right).
\end{eqnarray}
Imposing the Dirichlet boundary condition $Z(u=0)=0$ we get the quasinormal frequency,
\begin{eqnarray}
\omega_3=\pm\frac{q_3}{\sqrt{3}}-\frac{i q_3^2}{6}+\mathcal{O}.
\end{eqnarray}
Using (\ref{four momentum}), we get the following dispersion relation:
\begin{eqnarray}
w=\pm\frac{q}{\sqrt{3}}-\frac{i q^2}{6\pi T}.
\end{eqnarray}
Comparing this with the dispersion relation corresponding to the sound wave mode,
\begin{eqnarray}
w=\pm q v_s - i\Gamma_s q^2
\end{eqnarray}
where $v_s$ is the speed of sound and $\Gamma_s$ is the attenuation constant, we get their exact values.
\subsubsection{Via Two-Point Correlation Function $\langle T_{00} T_{00}\rangle$- Using the Solution of EOM involving Gauge Invariant Variable}
The relevant part of the bilinear surface term of the full action (\ref{full action}) as given in (\ref{on-shell-surface-action}) can be rewritten in terms of the gauge invariant variable $Z_s(u)$ as:
\begin{eqnarray}
\label{bilinear action2}
S^{(2)}_{\epsilon}=\lim_{u\rightarrow 0}\int \frac{dw dq}{2\pi^2} A(\omega_3,q_3,u)Z_s^\prime(u,q)Z_s(u,-q).
\end{eqnarray}
Using the equations of motion (\ref{4scalar_EOMs}) along with (\ref{on-shell-surface-action}), we find the function $A(\omega_3,q_3,u)$ as:
\begin{eqnarray}
A(\omega_3,q_3,u)=\frac{3}{u^3 \left(q_3^2 \left(u^4-3\right)+3 \omega_3^2\right)^2}
\end{eqnarray}
For the computation of two point function we need the solution of equation (\ref{single EOM}) as given in equation (\ref{solution single eom}), where the constant $c_1$ is determined by the boundary condition
\begin{eqnarray}
Z_s(u=0)=-H_{tt}^0 q_3^2+2 H_{{x}t}^0 q_3 \omega_3+H_{{x}{x}}^0 \omega_3^2+H_{{y}{y}}^0 \left(q_3^2-\omega_3^2\right),
\end{eqnarray}where we define $H_{ab}(u=0)=H_{ab}^0$.
Now putting the above expression of $A(\omega_3,q_3,u)$ and the solution $Z_s(u)$ back in equation (\ref{bilinear action2}) one get the two point correlator $G_{tt,tt}$ as:
\begin{eqnarray}
\nonumber G_{tt,tt}=\frac{\delta^2S_{\epsilon}^{(2)}}{\delta H_{tt}^{(0)}(k)\delta H_{tt}^{(0)}(-k)}\\
=\frac{8 q^6}{3 \left(q^2-3 w^2\right) \left(q^2-w^2\right)^2}
\end{eqnarray}
Hence the pole structure of the Green's function gives the correct value of the speed of sound wave, $v_s=\frac{1}{\sqrt{3}}$ propagating through hot plasma. The above value of speed of sound also matches exactly with the value that we have already got from the solution of hydrodynamic equations, thus provides a quantitative checks of the validity of Gauge/Gravity duality.
\subsection{From the solution of Gauge Invariant Variable - Going up to NLO in $N$ in the MQGP Limit}
Considering the Next-to-Leading Order corrections in $N$ of the metric components as given in (\ref{Mtheorymetric}), and using the gauge invariant combination given in (\ref{Z-vector mode}), (\ref{Z-scalar mode}) and (\ref{Z-tensor mode}), the Einstein equation can be expressed in terms of a single equation of the form $Z_i^{\prime\prime}(u) = m_i(u) Z_i^\prime(u) + l_i(u) Z_i(u)$, where, $i=s({\rm calar})$, $v({\rm ector})$, $t({\rm ensor })$.

In {\bf 3.2.3.1}, we first evaluate $v_s$ including the non-conformal contribution to the M-theory metric evaluated at a finite $r$ and large $N$, i.e., $\log\left(\frac{r}{\sqrt{\alpha^\prime}}\right)<\log N$, thereby dropping $\log r \log N$  as compared to $\left(\log N\right)^2$. Then, in {\bf 3.2.3.2}, we attempt a full-blown non-conformal estimate of $v_s$ up to NLO in $N$ by working at an $r: \log\left(\frac{r}{\sqrt{\alpha^\prime}}\right)\sim \log N$. It turns out, unlike the former, the horizon becomes an irregular singular point for the latter.  We set $\alpha^\prime$ to unity throughout. Given that in both, {\bf 3.2.3.1} and {\bf 3.2.3.2}, we are interested in numerics, exact numerical factors in all expressions will be replaced by their decimal equivalents.
\subsubsection{Dropping $\log r \log N$ As Compared to $\left(\log N\right)^2$}
Including the NLO terms, the EOM for the gauge invariant variable $Z_s(u)$ - given by (\ref{Z-scalar mode}) -  can be rewritten as:
\begin{equation}
\label{EOM_vs_i}
(u-1)^2Z_s^{\prime\prime}(u) + (u-1)P(u-1) Z_s^\prime(u) + Q(u-1) Z_s(u) = 0,
\end{equation}
in which $P(u-1) = \sum_{n=0}^\infty p_n(u-1)^n$ and $Q(u-1) = \sum_{m=0}^\infty q_n (u-1)^n$  wherein, up to ${\cal O}\left(\frac{1}{N}\right)$, $p_n, q_n$ are worked out in (\ref{pn+qn_up_to_2nd_order}).  The horizon $u=1$ being a regular singular point of (\ref{EOM_vs_i}), the Frobenius method then dictates that the incoming-wave solution is given by:
\begin{equation}
\label{solution-i}
Z_s(u) = \left(1 - u \right)^{\frac{3 {g_s}^2 M^2 {N_f}  \log (N) \left(8 {q_3}^2 {\omega_3}^2 \log (N)+\left({\omega_3}^2+4\right) \left(10 {q_3}^2-27
   {\omega_3}^2\right)\right)}{2048 \pi ^2 N q_3^2 \omega_3 \left(-1\right)^{3/2}}-\frac{i {\omega_3}}{4}}\left(1 + \sum_{m=1}a_m (u - 1)^m\right),
\end{equation}
where $a_{1,2}$ are given in (\ref{a1a2-sound}).  Following \cite{klebanov quasinormal},  imposing Dirichlet boundary condition $Z_s(u=0)=0$ and going up to second order in powers of $(u-1)$ in (\ref{solution-i}) and considering in the hydrodynamical limit $\omega_3^nq_3^m:m+n=2$ one obtains:
\begin{equation}
\label{vs+Gammas-second_order}
\omega_3 = -\frac{2 {q_3}}{\sqrt{3}}-\frac{9 i {q_3}^2}{32},
\end{equation}
which yields a result for the speed of sound similar to, though not identical to, (\ref{v_s 1})  for $n=0,1$.

To get the LO or conformal result for the speed of sound $v_s = \frac{1}{\sqrt{3}}$, let us go to the fourth order in (\ref{solution-i}). For this, up to ${\cal O}\left(\frac{1}{N}\right)$, $p_n, q_n$ are worked out in (\ref{pn+qn_up_to_fourth_order}).

We will not quote the expressions for $a_3$ and $a_4$ because they are too cumbersome. Substituting the expressions for $a_{1,2,3,4}$ into $Z_s(u)$ and implementing the Dirichlet boundary condition: $Z_s(u=0)=0$, in the hydrodynamical limit, going up to ${\cal O}(\omega_3^4)$ one sees that one can write the Dirichlet boundary condition as a quartic: $a \omega_3^4 + b \omega_3^3 + c \omega_3^2 + f \omega_3 + g = 0$ where $a, b, c, d, f, g$ are given in (\ref{a+b+c+f+g}).
One of the four roots yields:
\begin{equation}
\label{dispersion}
\omega_3 \approx 0.46 q_3 - 0.31 i q_3^2,
\end{equation}
with no ${\cal O}\left(\frac{1}{N}\right)$-corrections! The coefficient of $q_3$ is not too different from the conformal value of $\frac{1}{\sqrt{3}}\approx 0.58$. We expect the leading order term in the coefficient of $q_3$ to converge to $\frac{1}{\sqrt{3}}$. Also, the coefficient of $q^2$ term turns out to be $\frac{0.31}{\pi}$ which is not terribly far from the conformal result of $\frac{0.17}{\pi}$. We are certain that the inclusion of higher order terms in (\ref{solution-i}) will ensure that we get a perfect match with the conformal result. The reason we do obtain the NLO non-conformal contribution to $v_s$ is that at the very outset, we have neglected the non-conformal $\log r$-contributions by working at a large but finite $r$, but such that $\frac{\frac{r}{\sqrt{\alpha^\prime}}}{N}\ll1.$ We will see how to obtain the non-conformal contribution with the inclusion of the same in {\bf 3.2.3.2} below.
\subsubsection{Retaining $\log r \log N$ and $\left(\log N\right)^2$ Terms}
Constructing a $Z_s(u)$ given by (\ref{Z-scalar mode}) and retaining the non-conformal $\log r \log N$-contribution as well as $\left(\log N\right)^2$ terms, one sees one obtains (\ref{Z-EOM}) as the equation of motion for $Z_s(u)$.
The horizon $u=1$ due to inclusion of the non-conformal corrections to the metric,  becomes an irregular singular point. One then tries the ansatz: $Z_s(u) = e^{S(u)}$ near $u=1$ \cite{Bender_Orzag}. Assuming that $\left(S^{\prime}\right)^2\gg S^{\prime\prime}(u)$ near $u=1$ the differential equation (\ref{Z-EOM}), which could written as $Z_s^{\prime\prime}(u) = m(u)Z_s^\prime + l(u) Z_s(u)$ can be approximated by:
\begin{equation}
\label{S_EOM-text}
\left(S^\prime\right)^2 - m(u) S^\prime(u) - l(u) \approx 0.
\end{equation}
A solution to (\ref{S_EOM-text}) is given in (\ref{S-solution}). Taking first the MQGP limit, integrating with respect to $u$, the solution (\ref{S-solution})  will reflect the singular nature of $Z_s(u)$'s equation of motion (\ref{S-solution}) via
\begin{equation}
\label{pole-soln-Z}
Z_s(u)\sim \left(1 - u \right)^{-\frac{1}{2} + \frac{15 {g_s}^2 M^2 {N_f} {\omega_3}^2 \log \left(\frac{1}{N}\right)}{256 \pi ^2 N \left(2 {q_3}^2-3{\omega_3}^2\right)}}F(u),
\end{equation}
 where $F(u)$ is regular in $u$ and its equation of motion, around $u=0$, is given by (\ref{F-EOM}) whose solution is given in (\ref{F-eom-solution}). One notes from (\ref{F-eom-solution}) that $F(u\sim0) = c_1$. This needs to be improved upon by including the sub-leading terms in $u$ in $F'(u)$ which is discussed in detail in Appendix {\bf B.3}.

For $Z_s(u=0)=0$ to obtain $\omega = \omega(q)$ to determine the speed of sound, one requires $F(u=0)=0$. From (\ref{solution-improved-F-EOM}), this can be effected by requiring
\begin{equation}
\hskip -0.5in\frac{225 {g_s}^4 {N_f}^2 {\omega_3}^2 \log ^2(N) M^4+4800 {g_s}^2 N
   {N_f} \pi ^2 \left(4 {q_3}^2-5 {\omega_3}^2\right) \log (N) M^2+139264 N^2 \pi ^4 \left(2 {q_3}^2-3 {\omega_3}^2\right)}{128 N \pi ^2
   \left(15 {g_s}^2 {N_f} \left(8 {q_3}^2-11 {\omega_3}^2\right) \log (N) M^2+896 N \pi ^2 \left(2 {q_3}^2-3
   {\omega_3}^2\right)\right)} = - n\in\mathbb{Z}^-
\end{equation}
   or
\begin{equation}
   \label{v_s 1}
\omega = {q_3} \left(\frac{\sqrt{14 {n}+17}}{\sqrt{21 {n}+\frac{51}{2}}}+\frac{5 (2 {n}+5) {g_s}^2 M^2 {N_f} \log N}{128 \pi ^2 \sqrt{14 {n}+17} \sqrt{84 {n}+102} N}\right),
\end{equation}
implying the following estimate of the speed of sound:
\begin{equation}
\label{v_s 2}
v_s\approx  \frac{\sqrt{14 {n}+17}}{\sqrt{21 {n}+\frac{51}{2}}}+\frac{5 (2 {n}+5) {g_s}^2 M^2 {N_f} \log
   N}{128 \pi ^2 \sqrt{14 {n}+17} \sqrt{84 {n}+102} N}.
\end{equation}
Given that (\ref{S-solution}) is an approximate solution to (\ref{S_EOM-text}), one expects to obtain an expression for $v_s$ from an $M3$-brane uplift\footnote{For a $p$-brane solution, to LO in $N$, one expects $v_s=\frac{1}{\sqrt{p}}$ \cite{Herzog-vs}.}, to be of the form $v_s \approx \frac{{\cal O}(1)}{\sqrt{3}} + {\cal O}\left(\frac{g_sM^2}{N}\right)$,
and (\ref{v_s 1}) is exactly of this form for $n=0,1$.
\section{Vector Mode Perturbations and Shear Mode Diffusion Constant  up to NLO in $N$ in the MQGP Limit}
The equations of motion for the vector perturbation modes upto next-to-leading order in $N$, can be reduced to the following single equation of motion in terms of a gauge-invariant variable $Z_v(u)$ (given by (\ref{Z-vector mode})):
\begin{equation}
\label{vector-modes-ZEOM-text}
Z_v^{\prime\prime}(u) - m_v(w_3,q_3,u) Z_v^\prime(u) - l_v(w_3,q_3,u) Z_v(u) = 0,
\end{equation}
where $m_v(w_3,q_3,u), l_v(w_3,q_3,u)$ are given in (\ref{m+l-vec-definitions}).
The horizon $u=1$ is a regular singular point of (\ref{vector-modes-ZEOM-text}) and the root of the indicial equation corresponding to the incoming-wave solution is given by:
\begin{equation}
\label{root-solution-incoming-wave}
-\frac{i {\omega_3}}{4} + \frac{3 i  \omega_3 g^2_s M^2 N_{f} \log ^2{N}}{256 \pi ^2 N}.
\end{equation}
Next we will solve equation (\ref{vector-modes-ZEOM-text}), for the appropriate dispersion relation using the Frobenius method. We took the solution about $u=1$ to be
\begin{equation}
\label{solution1-text}
Z_v(u) = (1 - u)^{-\frac{i {\omega_3}}{4} + \frac{3 i \omega_3 g^2_{s} M^2 N_{f}  \log ^2{N}}{256 \pi ^2 N}}\left(1 + \sum_{n=1}^\infty a_n (u - 1)^n\right),
\end{equation}
\begin{itemize}
\item {\it Considering terms up to ${\cal O}((u-1)^2)$}
\end{itemize}
Truncating the infinite series in (\ref{solution1-text}) to ${\cal O}((u-1)^2)$ one obtains in (\ref{a1a2}), the values for $a_1, a_2$.

In the hydrodynamical limit retaining terms only up to ${\cal O}(\omega_3^mq_3^n):\ m+n=4$, we get from the Dirichlet boundary condition $Z(u=0)=0$,
\begin{equation}
a \omega_3^4 + b \omega_3^3 + c \omega_3^2 + f \omega_3 + g = 0
\end{equation}
 where $a, b, c, d, f, g$ are given in (\ref{a b c d f g-i}). Analogous to the calculation in the previous section, once again as we are interested in numerics, exact numerical factors in all expressions will be replaced by their decimal equivalents for most part of this section.

One of the four roots of  $Z_v(u=0)=0$ is:
\begin{equation}
\label{root-at-second-order}
\omega_3 = -8.18 i + \frac{0.14 i g_s^2 M^2 N_f(\log N)^2}{N} + \left(-0.005 i - \frac{0.002 i g_s^2 M^2 N_f (\log N)^2}{N}\right)q_3^2 + {\cal O}(q_3^3).
\end{equation}
\begin{itemize}
\item {\it Considering terms up to ${\cal O}((u-1)^3)$}
\end{itemize}
Going up to ${\cal O}((u-1)^3)$ in (\ref{solution1-text}), one obtains in (\ref{NLOa3})values of $a_3$.

The Dirichlet condition $Z_v(u=0)=0$ in the hydrodynamic limit reduces to
\begin{equation}
a \omega_3^4 + b \omega_3^3 + c \omega_3^2 + f \omega_3 + g = 0
\end{equation}
where $a, b, c, d, f, g$ are given in (\ref{a b c d f g-ii}).
One of the four roots of the quartic in $\omega_3$ is:
\begin{equation}
\label{root-at-third-order}
\omega_3 =  \left(- 0.73 i + \frac{0.003 i g^2_s M^2 N_f (\log N)^2}{N}\right)q^2_3 + {\cal O}(q^3_3).
\end{equation}
The leading order coefficient of $q_3^2$ is not terribly far off the correct value $-\frac{i}{4}$ already at the third order in the infinite series (\ref{solution1-text}).
\begin{itemize}
\item {\it Considering terms up to ${\cal O}((u-1)^4)$}
\end{itemize}
Let us look at (\ref{solution1-text}) up to the fourth order. One finds in (\ref{a_4}) the value of $a_4$.
In the hydrodynamical limit the Dirichlet boundary condition $Z_v(u=0)=0$ reduces to
\begin{equation}
a \omega_3^4 + b \omega_3^3 + c \omega_3^2 + f \omega_3 + g = 0
\end{equation}
where $a, b, c, d, f, g$ are given in (\ref{a b c d f g-iii}).
Incredibly, one of the roots of the quartic equation in $\omega_3$ is:
\begin{eqnarray}
\label{root-at-fourth-order}
& & \omega_3 = \left( - \frac{i}{4} + \frac{3 i g^2_s M^2 N_f \log N\left(5 + 2 \log N\right)}{512 \pi^2 N}\right)q^2_3 + {\cal O}\left(q^3_3\right).
\end{eqnarray}
Hence, the leading order (in $N$) yields a diffusion constant of the shear mode $D = \frac{1}{4\pi T}$, exactly the conformal result! Including the non-conformal corrections which appear at NLO in $N$, one obtains:
\begin{equation}
\label{D}
D = \frac{1}{\pi T}\left(\frac{1}{4} - \frac{3  g_s^2 M^2 N_f \log N\left(5 + 2 \log N\right)}{512 \pi^2 N}\right).
\end{equation}
We conjecture that all terms in (\ref{solution1-text}) at fifth order and higher, do not contribute to the Dirichlet boundary condition up to the required order in the hydrodynamical limit.

\begin{figure}
 \begin{center}
%\begin{center}
 \includegraphics[scale=0.8]
 %[height= 21cm,width=+15cm]
 {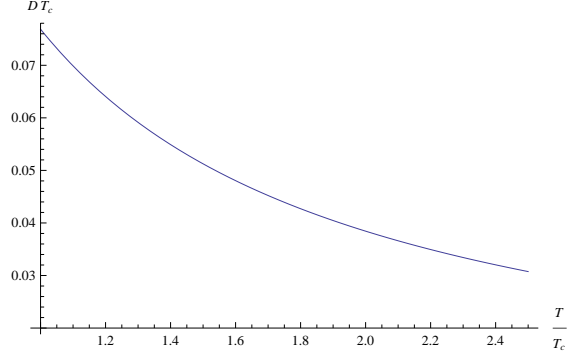}
 \end{center}
 \caption{$D T_c$ vs. $\frac{T}{T_c}$ for $T\geq T_c$}
\end{figure}
The variation of the shear mode diffusion constant with temperature is shown in Figure {\bf 3.1} for $N_f=3$, $M=3$, $g_s=0.9$, $N=100$. As the lowest order conformal result we obtain exactly $\frac{1}{4\pi T}$ as obtained in \cite{Starinets-Quasinormal spectrum}, for the black brane metric of the form (\ref{5dmetric inu}).
\section{NLO Corrections in N to $\eta$ and $\frac{\eta}{s}$}
In this section we will evaluate the non-conformal $\frac{g_s M^2}{N}$-corrections to the shear viscosity $\eta$ by considering the EOM for the tensor mode of metric fluctuations up to NLO in $N$,  and also estimate the same for the shear viscosity - entropy density ratio $\frac{\eta}{s}$; see \cite{A.Sinha-2}, \cite{A.Sinha-6}, \cite{A.Sinha-3}, \cite{S.Trivedi-1} and \cite{S.Trivedi-2} for a summary, higher derivative corrections to and violations of the $\frac{\eta}{s}=\frac{1}{4\pi}$-bound.
The EOM for the tensor mode of metric fluctuation, using (\ref{Z-tensor mode}),  is given as under:
\begin{eqnarray}
\label{EOM-tensor}
& & Z_t^{\prime\prime}(u) + Z_t^\prime(u) \left(-\frac{15 g^2_s M^2 {N_f} \log {N}}{64 \pi^2 N u}+\frac{u^4+3}{u \left(u^4-1\right)}\right)+Z_t(u) \Biggl(\frac{{q^2_3}
   \left(u^4-1\right)+{\omega_3}^2}{\left(u^4-1\right)^2}\nonumber\\
   & & -\frac{3 \left(q^2_3 u^4-{q_3}^2+{\omega_3}^2\right) \left({g^2_s} M^2 {N_f} \log
   ^2{N}+2 {g_s}^2 M^2 {N_f} \log {N} \log {r_h}\right)}{32 \pi
   ^2 N \left(u^4-1\right)^2}\Biggr)=0.\nonumber\\
   & &
   \end{eqnarray}
Realizing the horizon is a regular singular point, one makes the following double perturbative ansatz in $\omega_3$ and $q_3$,
\begin{equation}
\label{ansatz-solution-tensor}
\hskip -0.4in Z_t(u) = \left(1 - u\right)^{-i {\omega_3} \left(\frac{1}{4}-\frac{3  {g^2_s} M^2 {N_f} \log (N) \log r_h}{128 \pi ^2 N}\right)}\left\{z_{00}(u) + \omega_3 z_{01}(u) + q_3 z_{10}(u) + {\cal O}(q^2_3\omega^2_3)\right\}.
\end{equation}
Using equations (\ref{w3}) - (\ref{constant}), we obtain the following solusion,
\begin{eqnarray}
\label{Phi}
& & Z_t(u) = -\frac{i}{3072 \pi^2 N}(1-u)^{-i {\omega_3} \left(\frac{1}{4}-\frac{3  {g^2_s} M^2 {N_f}\log {r_h} \log {N} }{128 \pi ^2 N}\right)}\nonumber\\
    & & \times\Biggl[-3 {g^2_s} M^2 {N_f} u \log {N} \left\{4 c_2 \left(2 u^2+3 u+6\right) {\omega_3} \log{r_h}+15 i c_5 {q_3} u^3 (1-4 \log {u})\right\}\nonumber\\
   & & -6 c_2 {g_s}^2 M^2 {N_f} u \left(2 u^2+3 u+6\right) {\omega_3} \left(\log
   {N}\right)^2 +128 \pi ^2 N \biggl(6 i \left(c_5 {q_3} u^4+4 c_3 {q_3}+4 c_4 {\omega_3}\right)\nonumber\\
   & & +c_2 \left(2 u^3 {\omega_3}+3 u^2 {\omega_3}+6 u
   {\omega_3}+24 i\right)\biggr)\Biggr].
\end{eqnarray}
Setting $q_3=0$ one obtains (\ref{Phi'overPhi}) wherein the ${\cal O}(u^3\omega_3)$ term, without worrying about overall numerical multiplicative constants, is given by:
\begin{equation}
   \label{w3u3}
   \frac{i}{4}-\frac{3 i {g^2_s} M^2 {N_f}  \log r_h\log {N}}{128 \pi ^2 N}
   \end{equation}
Using arguments of \cite{transport-coefficients},  setting $\kappa_{11}^2=1$, the coefficient of the kinetic term of $Z_t(u)$ near $u=0$ and near   $\theta_1=\frac{\alpha_{\theta_1}}{N^{\frac{1}{5}}}$ (whereat an explicit local $SU(3)$-structure of the type IIA mirror and an explicit local $G_2$-structure of the M-theory uplift was obtained in \cite{NPB}) is
 \begin{equation}
 \label{kinetic-term-fluctuation}
 \frac{r_h^4}{g_s^2u^3}\int d\theta_1\cot ^3\theta_1 \sin\theta_1 f_1(\theta_1)\delta\left(\theta_1 - \frac{\alpha_{\theta_1}}{N^{\frac{1}{5}}}\right)\sim \frac{r_h^4}{g_s^2u^3}\frac{N^{\frac{2}{5}}}{\alpha_N\alpha_{\theta_1}^2},
 \end{equation}
 where $f_1(\theta_1) = \frac{\cot\theta_1}{\alpha_N},\ f_2(\theta_2) = -\alpha_N\cot\theta_2$ \cite{NPB}.

 The exact result for the temperature, assuming the resolution to be larger than the deformation in the resolved warped deformed conifold in the type IIB background of \cite{metrics} has already been calculated in (\ref{T-RC}). One can hence calculate the shear viscosity $\eta$:
\begin{eqnarray}
\label{eta}
& &\hskip -0.8in \eta = \Upsilon\frac{N^{\frac{2}{5}}}{g_s^2\alpha_N\alpha_{\theta_1}^2}\lim_{\omega_3\rightarrow0}\left(\frac{1}{\omega_3 T}\lim_{u\rightarrow0}\left[\frac{r_h^4}{u^3}\Im m \left(\frac{Z_t^\prime(u)}{Z_t(u)}\right)\right]\right).
\end{eqnarray}
where $\Upsilon$ is an overall multiplicative constant.

For the purpose of comparison of $\frac{\eta}{s}$ with lattice/RHIC data for QGP and consequently be able to express
$r_h$ in terms of $\tilde{t}\equiv \frac{T}{T_c}-1$, we need to use the relation between the IR cut-off $r_0$ and the horizon radius $r_h$ as obtained in {\bf Chapter 1} regarding the calculation of transition temperature $T_c$.
\begin{equation}
\label{r0rh}
r_0  =  r_h\sqrt[4]{\left|\frac{9b^4-1}{2(\delta^{\frac{8}{3}}-1)}\right|} + {\cal O}\left(\frac{1}{\log N}\right).
\end{equation}

Now, the lightest $0^{++}$ scalar glueball mass as will be obtained in {\bf Chapter 4} is given by:
\begin{equation}
\label{glueball-mass-i}
m_{\rm glueball} \approx \frac{4 r_0}{L^2}.
\end{equation}
Lattice calculations for $0^{++}$ scalar glueball masses \cite{SU3_lattice_glueball_masses}, yield the lightest mass to be around $1,700 MeV$. From (\ref{glueball-mass-i}), replacing $\frac{r_0}{L^2}$ by $\frac{m_{\rm glueball}}{4}$ we obtain:
\begin{equation}
\label{Tc}
T_c = \left.\frac{m_{\rm glueball}\left(1 + \frac{3b^2}{2}\right)}{2^{\frac{7}{4}}\pi\sqrt[4]{\left|\frac{9 b ^4-1}{2(\delta^{\frac{8}{3}}-1)}\right|}}\right|_{b=0.6,\delta=1.02}
=179 MeV.
\end{equation}

As the expressions in the following will become very cumbersome to deal with and to type, specially for the purpose of comparison with RHIC QGP data, we will henceforth deal only with numerical expressions setting $g_s=0.9, N=100, M=3, N_f=2, b=0.6, \delta=1.02$.

We now discuss the $\frac{1}{N}$-corrections to the entropy density $s$ by estimating the same from the $D=11$ supergravity action result of \cite{MQGP}, and hence work out the $\frac{1}{N}$ corrections to $\frac{\eta}{s}$.
The UV-finite part of the $D=11$ supergravity action, given by the Gibbons-Hawking-York (GHY) surface action $S_{\rm GHY}$ from \cite{MQGP} (without worrying about overall multiplicative constants) is \cite{MQGP}:
\begin{equation}
\label{K_MQGP_i}
S_{GHY}\sim\left.\int_{r=r_{\Lambda}}K\sqrt{\gamma}\right|_{\theta_{1,2}\sim0}\sim\frac{\cot^2\theta_1 f_2(\theta_2)}{g_s^{\frac{11}{4}}N^{\frac{3}{4}}\left(\sin^2\theta_1 + \sin^2\theta_2\right)}\left(\frac{1}{T}\right),
\end{equation}
 where,$K$ being the extrinsic curvature and $\gamma$ being the determinant of the pull-back of the $D=11$ metric on to $r={\cal R}_{\rm UV}$ (UV cut-off). Further, assume that what appears in (\ref{K_MQGP_i}) is $f_1(\theta_1)$.

 Now, unlike the scaling given in (\ref{limits_Dasguptaetal-ii}) used in \cite{MQGP}, we will be using:
 $\theta_{1,2}\rightarrow0$ as $\theta_1=\frac{\alpha_{\theta_1}}{N^{\frac{1}{5}}},\ \theta_2=\frac{\alpha_{\theta_2}}{N^{\frac{3}{10}}} (N\sim10^8)$ - as used in \cite{NPB} (to discuss a local $SU(3)$ structure of the type IIA delocalized SYZ mirror and a local $G_2$ structure of its M-theory uplift), as well as this thesis. This can be used to evaluate $ S^{\rm UV-finite}_{GHY}$ and the entropy density: $s =  - T \frac{\partial S^{\rm UV-finite}_{\rm GHY}}{\partial T} - S^{\rm UV-finite}_{\rm GHY}$. This yields:
%where $\alpha_{\rm GHY}\equiv\frac{N^{3/4}}{{\alpha_N} \alpha_{\theta_1}^5 {g_s}^{11/4}}$.
%$\alpha_{\theta_1}\sim{\cal O}(1)$:
\begin{eqnarray}
\label{eta_over_s-i}
& & \hskip -0.8in \frac{\eta}{s} =  {\cal O}(1)\times\nonumber\\
& & \hskip -0.8in \left[\frac{\frac{1}{4 \pi }-0.00051 \log {r_h}}{1 - 0.064 c_2 + 0.004 c_2^2 + \sum_{n=1}^4 a_n(c_1,c_2)\log^n r_h + \frac{\sum_{n=0}^4 b_n(c_1,c_2)\log^n r_h}{\sum_{n=0}^2c_n(c_1,c_2)\log^nr_h}}\right],
\end{eqnarray}
\noindent where $a_n,b_n,c_n$ are known functions of $c_1$ and $c_2$, and there is freedom to choose the ${\cal O}(1)$ constant. We will impose two conditions, as per RHIC QGP data, on $c_1$ and $c_2$ and the ${\cal O}(1)$ constant to get: $\left.\frac{\eta}{s}\right|_{T=T_c}=0.1,$ and
$\left.\frac{d\left(\frac{\eta}{s}\right)}{d\tilde{t}}\right|_{\tilde{t}>0}>0$.

Numerically, one sees that setting $(c_1,c_2)=(4,4)$ and consequently $r_h=\frac{35546.9 ({\tilde{t}}+1)}{{\cal PL}(2706.3 ({\tilde{t}}+1))}$ where ${\cal P L}$ is the ``ProductLog" function, and the ${\cal O}(1)$ constant equal to $5.8$, fits the bill. Hence,
{\footnotesize
\begin{eqnarray}
\label{eta_over_s-ii}
& & \hskip -0.7in\frac{\eta}{s} = 5.8\Biggl[\frac{{9.18\times 10^{-8}} \log ^3({r_h})-1.6\times 10^{-5} \log
   ^2({r_h})+2.7\times 10^{-4} \log ({r_h})+1.7\times 10^{-3}}{-{2.5\times 10^{-7}} \log
   ^6({r_h})+\frac{9 \log ^5({r_h})}{10^6}-\frac{\log ^4({r_h})}{10^4}+3.1\times 10^{-4}
   \log ^3({r_h})+0.002 \log^2({r_h})+3.6\times 10^{-3} \log ({r_h})+0.047}\Biggr].\nonumber\\
   & &
\end{eqnarray}}
The graphical variation of $\frac{\eta}{s}\left(N_f=3,M=3,g_s=0.9,N=100\right)$ vs. $\tilde{t}=\frac{T - T_c}{T_c}$ is shown in the following graph in Figure {\bf 3.2}, and the RHIC data plot from \cite{eta-over-s-RHIC}\footnote{One of us (KS) thanks R. Lacey to permit us to reproduce the graph in Figure {\bf 3.3} from their paper \cite{eta-over-s-RHIC}.}, is shown in Figure {\bf 3.3}.

\begin{figure}
 \begin{center}
%\begin{center}
 \includegraphics[scale=0.8]
 %[height= 21cm,width=+15cm]
 {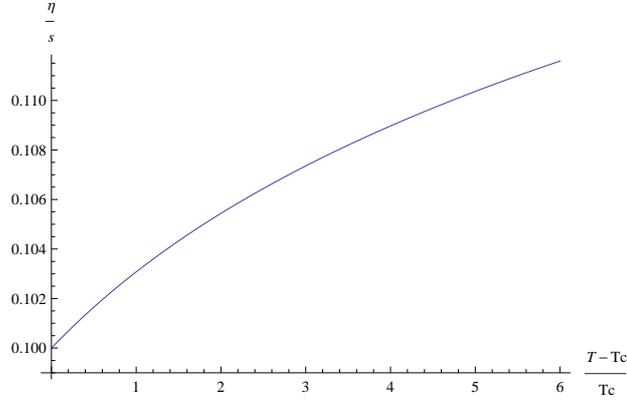}
 \end{center}
 \caption{$\frac{\eta}{s}$ vs. $\frac{T-T_c}{T_c}$ for $T\geq T_c$ assuming $\left.\frac{\eta}{s}\right|_{T=T_c}=0.1$}
\end{figure}

\begin{figure}
 \begin{center}
%\begin{center}
 \includegraphics[scale=0.5]
 %[height= 21cm,width=+15cm]
 {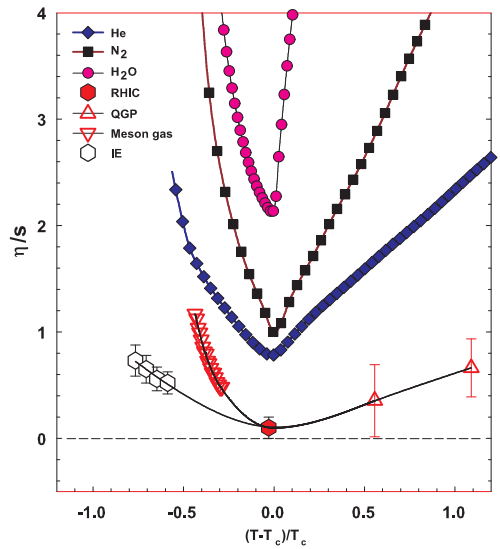}
 \end{center}
 \caption{$\frac{\eta}{s}$ vs. $\frac{T-T_c}{T_c}$ reproduced from \cite{eta-over-s-RHIC}.}
\end{figure}

We draw a third graph in which the plots of Figures {\bf 3.2} and {\bf 3.3} are drawn on the same graph.
\begin{figure}
 \begin{center}
%\begin{center}
 \includegraphics[scale=0.8]
 %[height= 21cm,width=+15cm]
 {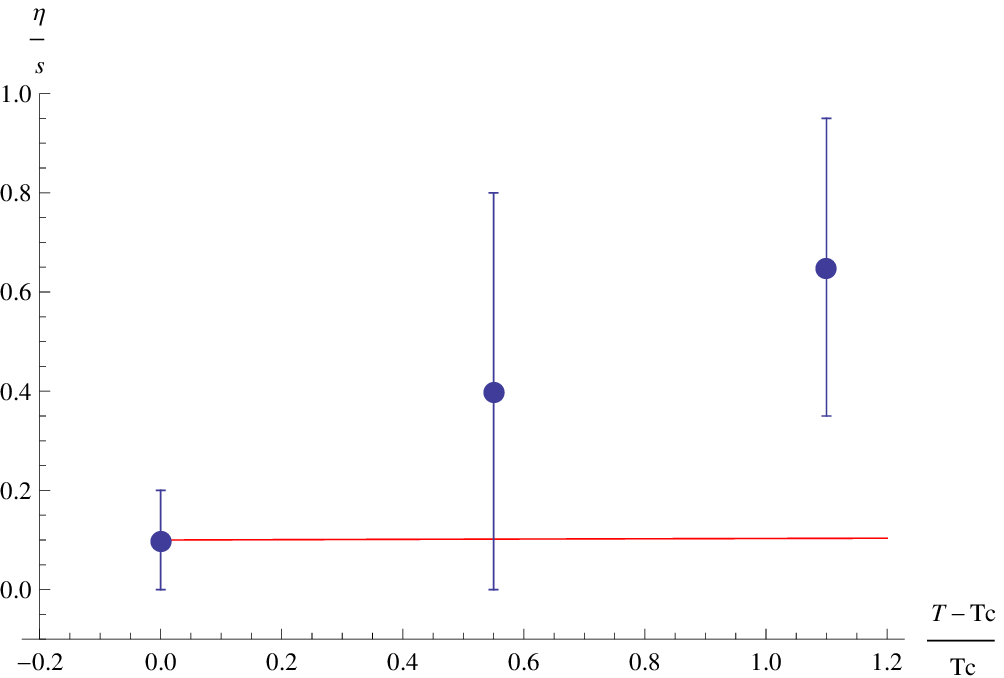}
 \end{center}
 \caption{Combined plots of Figures {\bf 3.2} and {\bf 3.3}: the graph in red is from Figure {\bf 3.2} (our calculations) and the set of three points with error bars are from Figure {\bf 3.3} (RHIC QGP data from \cite{eta-over-s-RHIC}).}
\end{figure}
The combined plots in Figure {\bf 3.4} make the comparison of our results with those of RHIC data in \cite{eta-over-s-RHIC}, very clear. We conclude the following:
 \begin{itemize}
 \item
 $\frac{\eta}{s}(T=T_c)= 0.1$, and $\left.\frac{d\left(\frac{\eta}{s}\right)}{dT}\right|_{T>T_c}>0$ - this is clear from Figure {\bf 3.2}.

 \item
 The numerical values, unlike \cite{eta-over-s-RHIC}, remain close to the value at $T=T_c$. In other words, unlike Figure {\bf 3.3} quoted from \cite{eta-over-s-RHIC}, in Figure {\bf 3.4}, $\frac{\eta}{s}$ is found to be a much more slowly varying function of $\tilde{t}=\frac{T - T_c}{T_c}$.  Also, $\frac{d^2\left(\frac{\eta}{s}\right)}{d\tilde{t}^2}<0$ in Figure {\bf 3.2}  and $\frac{d^2\left(\frac{\eta}{s}\right)}{d\tilde{t}^2}>0$ in Figure {\bf 3.3}. The error bars appearing in Figure {\bf 3.3} as shown more clearly in Figure {\bf 3.4},  for $\frac{T - T_c}{T_c}\in[0,1.1]$ - the range covered in \cite{eta-over-s-RHIC} - permit our deviations from \cite{eta-over-s-RHIC} at least for $\frac{T - T_c}{T_c}\in[0,0.6]$.

\end{itemize}
\section{Thermal (Electrical) Conductivity, Deviation from the Wiedemann-Franz Law and $D=1+1$ Luttinger Liquids up to LO in $N$}
In this section we compute the temperature dependance of thermal ($\kappa_T$) and electrical ($\sigma$) conductivities for a gauge theory at finite temperature and density, and hence explore deviation from the Wiedemann-Franz law. Remarkably, we find that the results qualitatively mimic those of a $D=1+1$ Luttinger liquid with impurities.

Finite temperature in the  gauge theory is effected by placing a black hole in the dual bulk gravitational background. To get the finite density in the boundary we consider the theory at non-zero chemical potential. The embedding of $N_f$ $D7$ branes in the background geometry introduces $N_f$ no of flavor fields, all in fundamental representation of the gauge group $U(N_f)$. The $U(1)_B$ subgroup of $U(N_f)$ is identified as the baryon number. Hence the $D7$ brane puts the boundary field theory at finite baryon density or equivalently at finite chemical potential $\mu_C$; see \cite{Erdmenger-5} for study of thermodynamics of a thermal field theory in presence of  a baryon and isospin chemical potential. For this we consider a probe of several D7-branes embedded in the AdS-Schwarzschild black hole background. In the supergravity description one have a $U(1)$ gauge field $A_{\mu}$ in the worldvolume of the $D7$ brane, dual to the current operator $j_{\mu}$ in the boundary. The nonzero time component $A_t$ of the gauge field has to be turned on to get a finite baryon density $<j_t>$ in the field theory. Here we will consider the $5d$ Einstein-Hilbert action and the $D7$ brane DBI action together, of course after integrating over the three angular directions of the later.

The $D7$ brane DBI action in presence of a $U(1)$ gauge field is given as:
\begin{equation}
\label{DBI}
S_{D7}= T_{D7}\int d^8\xi e^{-\Phi}\sqrt{-\det(g+B+\hat{F})}
\end{equation}
where $g$ is the induced metric on $D7$ brane and $\hat{F}$ is the gauge field strength with the only nonzero component given by $\hat{F_{rt}}=\frac{ce^{\Phi}}{\sqrt{c^2e^{2\Phi}+r^{9/2}}}$ \cite{NPB}, where $\Phi$ is the dilaton and $c$ is a constant.

Now the finite chemical potential or equivalently the finite charge density will mix the heat (energy) current and the electric current together. According to the AdS/CFT correspondence for every operator in the boundary field theory, there is a bulk field in the dual gravity theory. Heat current is sourced by the energy momentum tensor $T_{\mu\nu}$ in the boundary and the corresponding field in the gravitational description is the bulk metric $g_{\mu\nu}$. Similarly, as already mentioned, the electric current sourced by the current operator $j_{\mu}$ corresponds to the $U(1)$ bulk gauge field $A_{\mu}$. Hence for the computation of thermal conductivity we consider the following linear fluctuations of both the background metric $g^{(0)}_{\mu\nu}$ and the gauge field $A^{(0)}_{\mu}$ as,\begin{equation}g_{\mu\nu}=g^{(0)}_{\mu\nu}+h_{\mu\nu}  ~~~~~~~~~~~~~~~~  A_{\mu}=A^{(0)}_{\mu}+\mathcal{A_{\mu}},\end{equation}
where $h_{\mu\nu}$ and $\mathcal{A_{\mu}}$ represents the metric and the gauge field fluctuations respectively.
Considering the $y$-component of the gauge field as the only perturbation, it can be shown that only the $(ty)$ and the $(xy)$ component of the metric gets perturbed.
Assuming that the above perturbations depends only on the radial coordinate $u$, time $t$ and spatial coordinate $x$, can be decomposed as the following way,
\begin{equation}
h_{ty}=g^{(0)}_{xx}H_{ty}(u)e^{-iwt+iqx} ~~~~~~~~~~~~~  h_{xy}=g^{(0)}_{xx}H_{xy}(u)e^{-iwt+iqx} ~~~~~~~~~~~~~  \mathcal{A}_{y}=\phi(u)e^{-iwt+iqx}
\end{equation}
Now including the above fluctuations in the DBI action, we perform the three angular integrations on $\phi_1$, $\phi_2$ and $\theta_2$. The integration over two of the three angular variable namely $\phi_1$ and $\phi_2$ gives constant factors.
To perform the $\theta_2$ integration, we first expand the DBI action in (\ref{DBI}) (Taking into account the fluctuations) up to quadratic order in fluctuating fields to get,
\begin{equation}\begin{split}
&\sqrt{-\det(g+h+B+\hat{F}+F)}\\&=\sqrt a_1 \left(1 + \frac{a_2 H^2_{ty}(u) + a_3 H_{ty}(u) \phi^{'}(u) + a_4 H^2_{xy}(u) + a_5 \phi^{2}(u) + a_6 \phi^{\prime}(u)^2}{2 a_1}\right).
\end{split}
\end{equation}
where $h$ and $F$ represents the fluctuations of the two fields. The coefficients $a_1,a_2,a_3,a_4,a_5,a_6$ are given as,
\begin{eqnarray}
\label{a-coefficients}
& & a_1 = -\frac{\cot ^2(\frac{\theta_2}{2}) \csc ^4(\frac{\theta_2}{2})}{1296 \left(c^2 e^{2 \Phi }+r^{9/2}\right)}r^{9/2}\left\{\Biggl(r^3+2 (5 \mu ^2-2 r^3) \cos
   \theta_2+14 \mu^2+3 r^3 \cos 2 \theta_2\Biggr)\right\}\nonumber\\
& &  \times \Biggl\{\left(8 \mu^2-4 r^3\right) \cos \theta_2+r^3 (\cos 2
   \theta_2+3)\Biggr\}\nonumber\\
   & &  a_2 = a_1\frac{r^{-\frac{1}{2}}\left(c^2 e^{2 \Phi }+r^{9/2}\right)}{\left(r^4-r_h^4\right)}e^{2i(qx-tw)}\nonumber\\
       & &  a_3 = 2a_1\frac{ ce^{\Phi}\sqrt{\left(c^2 e^{2 \Phi }+r^{9/2}\right)}}{r^{\frac{9}{2}}}e^{2i(qx-tw)}\nonumber\\
    & &       a_4 = a_1e^{2i(qx-tw)}\nonumber\\
    & & a_5 =4 a_1\frac{(g_s N\pi)}{r^{\frac{9}{2}}\left(r^4-r_h^4\right)} \left\{iw^2 c^2 e^{2 \Phi }+\sqrt{r} \left(iq^2 \left(r_h^4-r^4\right)+iw^2
   r^4\right)\right\}e^{2i(qx-tw)}\nonumber\\
   & & a_6 = a_1\frac{\left(c^2 e^{2 \Phi }+r^{9/2}\right)}{r^{\frac{9}{2}}}\left(1 - \frac{r_h^4}{r^4}\right)e^{2i(qx-tw)},
\end{eqnarray}
where the coordinate $r$ is related to $u$ as $u=\frac{r_h}{r}$. Upon changing the variable from $r$ to $u$ to the above mentioned variables one see that the coefficients $a_2,a_3,a_4,a_5,a_6$ each after the division by $a_1$ are independent of $\theta_2$ and only depends on $u$. The integration of $\sqrt{a_1}$ over $\theta_2$ gives some function of $u$ say $\mathcal{M}(u)$ given by,
\begin{eqnarray}
\mathcal{M}(u)=\sqrt{\mu}\left(\frac{r_h}{u}\right)^{9/2}\sqrt{\frac{1}{c^2e^{2\Phi}+\left(\frac{r_h}{u}\right)^{9/2}}}
\end{eqnarray}
 In this way reducing the dimension from eight to five the DBI action takes the following form:
\begin{eqnarray}
\label{kinetic-fluctuations}
\nonumber & & S_{D7}=\left(\frac{a_{DBI}T_{D_7}}{g_s}\right)\\& &
\nonumber \int du~ d^4x~ \mathcal{M}(u)\left(1+\frac{a_2 H^2_{ty}(u) + a_3 H_{ty}(u) \phi^{'}(u) + a_4 H^2_{xy}(u) + a_5 \phi^{2}(u) + a_6 \phi^{'}(u)^2}{2 a_1}\right)\\
& &
\end{eqnarray}
where $a_{DBI}$ includes all the constant terms resulting after the angular integrations; $T_{D7}$ is the tension on the $D7$ brane. We will henceforth be working in a hydrodynamical approximation wherein we will approximate the plane-wave exponentials by unity.

Finally taking into account the Einstein-Hilbert action given as,
\begin{eqnarray}
S_{EH}=a_{EH}\int du d^4x\sqrt{-g_{(5)}}(R-2\lambda),
\end{eqnarray}
where $g_{(5)}$ is the determinant of the $5d$ metric, the total action is given by:
$S_{tot}=S_{\rm EH}+S_{D7}$.

The type $IIB$ metric satisfying the above action $S_{tot}$ has the form:
\begin{equation}
ds^2=g_{tt}dt^2+g_{xx}(dx^2+dy^2+dz^2)+g_{uu}du^2,
\end{equation}
where  the different background metric components, in the UV (as the gauge fluctuation will be solved for, near the UV $u=0$) and to LO in $N$, are given as,
\begin{eqnarray}
\nonumber g_{tt}=\frac{\left(u^4-1\right)r_h^2}{2 u^2 \sqrt{\pi g_s   N }}\\
\nonumber g_{xx}=g_{yy}=g_{zz}=\frac{r_h^2}{2 u^2 \sqrt{\pi g_s   N }}\\
g_{uu}=\frac{2 u^2 \sqrt{\pi g_s   N }}{\left(1-u^4\right)r_h^2}.
\end{eqnarray}
Now from the total action defined above, we can write down the EOMs in the hydrodynamical limit for $H_{ty}$, $H_{xy}$~and $\phi$ and they are given as:
\vskip 0.1in
\noindent {\bf $H_{ty}(u)$ EOM:}
\begin{eqnarray}
& & a_{EH}\sqrt{-g_{(5)}}\left(\mathcal{R}^{(1)}_{ty}-\frac{2}{3}\lambda g_{xx}H_{ty}(u)\right)+\left(\frac{a_{DBI}T_{D_7}}{g_s}\right)\mathcal{M}(u)\left(\frac{a_2(u)}{a_1(u)}H_{ty}(u)
+ \frac{a_3(u)}{2a_1(u)}\phi^{'}(u)\right)=0;\nonumber\\
& &
\end{eqnarray}
{\bf $H_{xy}(u)$ EOM:}
\begin{eqnarray}
& & a_{EH}\sqrt{-g_{(5)}}\left(\mathcal{R}^{(1)}_{xy}-\frac{2}{3}\lambda g_{xx}H_{xy}(u)\right)+\left(\frac{a_{DBI}T_{D_7}}{g_s}\right)\mathcal{M}(u)\left(\frac{a_4}{a_1(u)}H_{xy}(u)\right)=0;\nonumber\\
& &
\end{eqnarray}
{\bf $\phi(u)$ EOM:}
\begin{eqnarray}
\nonumber & & \frac{d}{du}\left(\frac{a_3(u)}{2 a_1(u)}\mathcal{M}(u)H_{ty}(u)\right) + \frac{d}{du}\left(\frac{a_6(u)}{a_1(u)}\mathcal{M}(u)\right)\phi^{'}(u)
\\&&+\left(\frac{a_6(u)}{a_1(u)}\mathcal{M}(u)\right)\phi^{''}(u)-\left(\frac{a_5(u)}{a_1(u)}\mathcal{M}(u)\right)\phi(u)=0,
\nonumber\\
& &
\end{eqnarray}
where $\mathcal{R}^{(1)}_{\mu\nu}$ is the linear ordered perturbation of the Ricci tensor. Now substituting the exact form of $\mathcal{M}(u)$ as well as all of the six coefficients $a_1(u), a_2(u), a_3(u), a_4, a_5(u), a_6(u)$, the above three equations regarding $H_{ty},H_{xy}$ and $\phi$ can be rewritten as (\ref{EOM-Hty}) - (\ref{EOM-phi}) in appendix {\bf B.1} which also contains their solutions.

As the pre-factor multiplying $\frac{\phi'(u)}{\phi(u)}$ from (\ref{kinetic-fluctuations}), the $A(u)$ in (\ref{F}) - the coefficient of the kinetic term of $\phi(u)$ - that will appear in the current-current correlator is $\left(\frac{\sqrt{\mu}r_h^{\frac{13}{4}}}{72u^{\frac{17}{4}}}\right)$, to obtain
   a finite $\left\{\lim_{{u}\rightarrow0}\frac{1}{u^{\frac{17}{4}}}\frac{\phi'(u)}{\phi(u)}\right\}$, one needs $\phi(u)\sim e^{{\rm constant}\ u^{\frac{21}{4}}}$. Expanding (\ref{phiu0solution_i}) about $u=0$:
\begin{eqnarray}
\label{large-N-small-u-rh_TN-small-w}
 \phi(u\sim0;q=0) & =  & \frac{\left(i g_s N \pi\right)^{7/8}\omega^{7/4}c_2^\Phi u^{\frac{21}{4}}}{33^{\frac{3}{4}}r_h^{\frac{7}{2}}} -\frac{4cg_sc_1^\Phi\Gamma(\frac{13}{24})u^{\frac{13}{4}}}{63r_h^{\frac{9}{4}}\Gamma(\frac{37}{24})}+\frac{4cg_sc_1^\Phi\Gamma(-\frac{1}{3})u^{\frac{13}{4}}}{63r_h^{\frac{9}{4}}\Gamma(\frac{2}{3})}
\nonumber\\
& & + c_1^\Phi + \frac{c}{r_h^{\frac{9}{4}}}{\cal O}(u^6).
\end{eqnarray}
Now, in terms of a dimensionless ratio: $\kappa\equiv\frac{C}{r_h^{\frac{9}{4}}}$ and choosing $C$ to be $m_{\rho}$ and $r_h$ in units of GeV implying $\kappa\ll1$ {\cite{NPB}}. Therefore,
\begin{eqnarray}
\label{large-N-small-u-rh_TN-small-w k-small}
& & \phi(u\sim0;q=0)=\frac{(0.08+0.39\ i)\left(g_s N\right)^{\frac{7}{8}}w^{\frac{7}{4}}c_2^\Phi u^{\frac{21}{4}}}{r_h^{\frac{7}{2}}}+c_1^\Phi+\frac{c}{r_h^{\frac{9}{4}}}{\cal O}(u^6)\nonumber \\
& & \approx c_1^\Phi e^{\frac{(0.08+0.39\ i)g_s^{\frac{7}{8}}N^{\frac{7}{8}}w^{\frac{7}{4}}c_2^\Phi u^{\frac{21}{4}}}{c_1r_h^{\frac{7}{2}}}}.
\end{eqnarray}
 Next we require to calculate some of the thermodynamic parameters like pressure, energy density, entropy density etc. In particular, pressure and energy density follows from the thermodynamic relations as given by $s=\frac{\partial P}{\partial T}$ and $\epsilon=-P+Ts+\mu_{C}n_q$, where $s$ is called the entropy density and is given as,
\begin{eqnarray}
& &  s = \mathcal{O}(1)r_h^3 = \mathcal{O}(1)\pi^3\left(4\pi g_s N\right)^{3/2}T^3.
\end{eqnarray}
%The pressure and energy density is given by,
%\begin{eqnarray}
%P=\mathcal{O}(1)\frac{\pi^3}{4}\left(4\pi g_s N\right)^{3/2}T^4,\\
%\epsilon=\mathcal{O}(1)\frac{3}{4}\pi^3\left(4\pi g_s N\right)^{3/2}T^4.
%\end{eqnarray}
Now the density of Gibbs potential $\Omega$ which is equal to the pressure with a minus sign can be used to find the charge density $n_q$ using the relation $n_q=\frac{\partial\Omega}{\partial\mu_C}$, where $\mu_c$ being the chemical potential is given by
\begin{eqnarray}
\label{mu_C_dimensionless}
& & \mu_C = \frac{\left(2 {\kappa} {g_s}\right)^{\frac{4}{9}} {r_h} \Gamma \left(\frac{5}{18}\right) \Gamma \left(\frac{11}{9}\right)}{\sqrt[18]{\pi } (2
   \pi -{g_s} {N_f} \log |\mu|)^{4/9}}-{r_h}\ _2F_1 \left(\frac{11}{9};-\frac{(2 \pi -{g_s} {N_f} \log |\mu| )^2}{4
   {\kappa}^2 {g_s}^2 \pi ^2}\right)\nonumber\\
   & & = \frac{36 \pi  {\kappa} {g_s} {r_h} \Gamma \left(\frac{11}{9}\right)}{5 \Gamma \left(\frac{2}{9}\right) (2 \pi -{g_s} {N_f}
   \log |\mu| )} + {\cal O}\left(\kappa^{\frac{19}{9}}\right),
\end{eqnarray}
from which we get
\begin{equation}
\label{T-func-mu}
 T=\left(\frac{8}{5}\right)^{4/5}\left(\frac{g_s^{\frac{3}{10}}C^{\frac{4}{5}}}{\left(2\pi - g_s N_f \log |\mu|\right)^{4/5}}\right)\left(\frac{\mu^{-\frac{4}{5}}_C}{2\pi^{\frac{7}{10}}\sqrt{N}}\right).
 \end{equation}
Substituting the above result for $T$ in the expression for Gibbs potential and differentiating w.r.t $\mu_C$ we get charge density as
\begin{eqnarray}
n_q=\left(\frac{8}{5}\right)^{\frac{16}{5}}\left(\frac{2}{5}\right)\left(\frac{g_s^{\frac{27}{10}}\pi^{\frac{17}{10}}C^{\frac{16}{5}}\mu_C^{-\frac{21}{5}}}{\sqrt{N}\left(2\pi - g_s N_f \log |\mu|\right)^{16/5}}\right).
\end{eqnarray}
Hence,
\begin{eqnarray}
\label{kappaT}
& & \hskip -0.3in \kappa_T = \frac{\left(\epsilon+P\right)^2\sigma}{n_q^2T} = -\left(\frac{\epsilon+P}{n_q}\right)^2\left(\frac{\sqrt{|\mu|}r_h^{\frac{13}{4}}}{72 T u^{\frac{17}{4}}}\right)\lim_{{\omega}\rightarrow0}\frac{1}{{\omega}}
\Im m\left.\frac{\phi^\prime(u)}{\phi(u)}\right|_{u=0}\nonumber\\
&& \hskip -0.3in = \frac{9}{200\sqrt{2}}\frac{g_s^{\frac{3}{4}}C^2}{ N^{\frac{5}{4}}\pi^{\frac{7}{4}}\left(2\pi - g_s N_f \log |\mu|\right)^2T^{\frac{7}{2}}}\sqrt{|\mu|}\left(T\pi \sqrt{4\pi g_s N}\right)^{13/4}\nonumber \\ && \hskip -0.3in \times \lim_{{\omega}\rightarrow0}\frac{(0.39i)g_s^{7/8}N^{7/8}{\omega}^{3/4}c_2^\Phi}{\left(T\pi\sqrt{4\pi g_s N}\right)^{7/2}},\nonumber\\
& &
\end{eqnarray}
which for $c_2^\Phi\sim -i {\omega}^{-\frac{3}{4}}$ implies:
\begin{eqnarray}
\label{WF}
& & \sigma=(0.39)\frac{\sqrt{|\mu|}\left(g_s N\right)^{\frac{3}{4}}}{2^{\frac{1}{4}}T^{\frac{1}{4}}\pi^{\frac{3}{8}}},\nonumber\\
& &  \kappa_T = \frac{9\times 0.39}{200\times2^{3/4}}\frac{\sqrt{|\mu|}g_s^{\frac{3}{2}}C^2}{\sqrt{N}\pi^{\frac{17}{8}}T^{\frac{15}{4}}\left(2\pi - g_s N_f \log |\mu|\right)^2};\nonumber\\
& &  {\rm Wiedemann-Franz\ law}: \frac{\kappa_T}{\sigma T}=\frac{9}{200\sqrt{2}}\frac{g_s^{\frac{3}{4}}C^2}{N^{\frac{5}{4}}\pi^{\frac{7}{2}}\left(2\pi - g_s N_f \log |\mu|\right)^2T^{\frac{9}{2}}}.
\end{eqnarray}

(a) Assuming the Ouyang embedding parameter to depend on the temperature via the horizon radius as $|\mu|\sim r_h^\alpha,\ \alpha\leq0$. Then, the temperature dependence of $\sigma, \kappa_T$ and the temperature dependences of the Wiedemann-Franz law in (\ref{WF}), upon comparison with Table 2  of \cite{WF}, qualitatively mimic a $D=1+1$ Luttinger liquid with impurities/doping (close to `$\frac{1}{3}$-filling') in the following sense.

With
\begin{itemize}
\item
$v_i, K_i, i=c$(harge), $s$(pin) being the parameters appearing in the Luttinger liquid Hamiltonian as $\sum_{i=c,s}v_i\left[K_i \left(\partial_x\theta_i\right)^2 + \frac{1}{K_i}\left(\partial_x\phi_i\right)^2\right]$ wherein the spin ($s$) and charge ($c$) densities are $\phi_{s,c}$ and their canonically conjugate fields are $\partial_x\theta_i$,
\item
 $n_s=0,1$ for even and odd $n_c$ respectively where $n_{c,s}$ along with $g,a$ appear in the Umklapp scattering Hamiltonian $\frac{g}{\left(2\pi a\right)^{n_c}}\int \left(e^{i\sqrt{2}\left(n_c\phi_c + n_s\phi_s\right) - i \Delta k x} + {\rm h.c.}\right)$,
\item
 $D$ as a parameter appearing in the two-point correlation function of the impurity field $\eta(x)$ via $\langle \eta(x)\eta(x^\prime)\rangle = D \delta(x-x^\prime)$ with $\eta(x)$ appearing in the back-scattering Hamiltonian due to disorder $\frac{1}{\pi a}\int dx \eta(x) \left[e^{i\sqrt{2}\phi_c}\cos\left(\sqrt{2}\phi_s\right) + {\rm h.c.}\right]$,
\end{itemize}
  the authors of \cite{WF} define the following dimensionless parameters: $\tilde{D} \equiv\frac{\rm Impurity\ scattering\ rate}{\rm Umklapp\ scattering\ rate}=\frac{D a^{2n_c-3}}{g^2\left(\frac{a T}{v_c}\right)^\gamma}, \tilde{\delta}\equiv\frac{\delta}{\tilde{D}^{\frac{1}{\gamma}}}$ where
$\delta\equiv\frac{v_c\Delta k}{\pi T}, \gamma\equiv (n_c^2-1)K_c + (n_s^2-1)K_s - 1$ and dimensionless temperature:
$\tilde{T}\equiv\frac{T}{T_D}$ where $T_D\equiv\frac{v_c}{a}\left(\frac{D a^{2n_c-3}}{g^2}\right)^{\frac{1}{\gamma}}$. One then notes that for $\tilde{\delta}=10, 20$ and for $T>T_D$, $\frac{d\sigma}{dT}, \frac{d\kappa_T}{dT}, \frac{d\left(\frac{\kappa_T}{T \sigma}\right)}{dT} < 0$ which is also reflected in (\ref{WF}).  In $\alpha^\prime=1$-units $[T] = [C^{\frac{4}{9}}]$, where $[..]$ denotes that canonical dimension.  To ensure a constant finite value of
$\frac{\kappa_T}{T\sigma}$ for small temperatures as per \cite{WF}, we assume, in the MQGP limit,  for $T :  \frac{T}{C^{\frac{4}{9}}}<1$, i.e., $T\sim C^{\frac{4}{9}}\epsilon^{\alpha_T>0}, 0<\epsilon<1$ and $N\sim\beta_N\epsilon^{-\alpha_N}$, so that if $0<\frac{9\alpha_T}{2} - \frac{5\alpha_N}{4}\ll1$ then $\lim_{T\rightarrow0}\left(\frac{\kappa_T}{T\sigma}\sim\frac{g_s^{\frac{3}{4}}C^2}{N^{\frac{5}{4}}T^{\frac{9}{2}}\left(2\pi - g_s N_f\log\mu\right)^2}\right)\sim\frac{g_s^{\frac{3}{4}}}{\epsilon^{-\frac{5\alpha_N}{4} + \frac{9\alpha_T}{2}}\left(2\pi - g_sN_f\left\{\frac{\alpha_N}{4} + \alpha_T\right\}\log\epsilon + \frac{g_sN_f}{4}\log(\beta_Ng_s)\right)^2}\neq0$.

(b) For $\alpha$(figuring in $|\mu|\sim r_h^\alpha$)$>0$, interestingly for the specific choice of
$\alpha=\frac{5}{2}$ one reproduces the large-$T$ (as $T>C^{\frac{4}{9}}=m_\rho=760 MeV$(\cite{NPB})$>T_c=175 MeV$, is considered large) linear behavior of DC electrical conductivity $\sigma\sim T$ characteristic of most strongly coupled gauge theories with a five-dimensional gravity dual with a black hole \cite{SJain_sigma+kappa}. As $\frac{C^2}{T^{\frac{9}{2}}}$ is dimensionless, this yields dimensionally $\kappa_T\sim ({\rm temperature})^2$, though $\kappa_T\sim T^{\frac{5}{2}}$ in the aforementioned large-$T$ limit.

\section{Summary and New Insights into (Transport) Properties of Large $N$ Thermal QCD at Finite Gauge Coupling}

A realistic computation pertaining to  thermal QCD systems such as sQGP, require a finite gauge coupling and not just a large t'Hooft coupling \cite{Natsuume}, and the number of colors $N_c$ equal to three. Further, computations quantifying the IR non-conformality in physical quantities pertaining  to large-$N$ thermal QCD at finite gauge coupling that appear at the NLO in $N$ in the corresponding holographic description in string \cite{metrics}/M-theory \cite{MQGP}. In this work, at finite gauge coupling with $N_c=M_{r\in{\rm IR}}=3$ as part of the MQGP limit (\ref{limits_Dasguptaetal-ii}), we have addressed a Math issue and {obtained new insights into some transport properties at LO in $N$, and  non-conformal corrections appearing at the NLO in $N$ in a variety of hydrodynamical quantities crucial to characterizing thermal QCD - like systems at finite gauge coupling such as sQGP}.

The following provides a summary of the new results discussed here as well as the new insights into the Physics of strongly coupled thermal QCD laboratories like sQGP gained therefrom.

\begin{itemize}

\item
{\bf $\kappa_T, \sigma$, Wiedemann-Franz law at LO in $N$ and  $D=1+1$ Luttinger Liquid with impurities}: As gauge fluctuations are tied  to vector modes of metric fluctuations, by solving the coupled set of equations for both, we obtained the temperature dependence of the thermal and electrical conductivities as well as looked at whether the Wiedemann-Franz law was satisfied. {This revealed a remarkable insight into the properties of large-$N$ thermal QCD at finite gauge coupling} namely that { the type IIB holographic dual of large-$N$ thermal QCD with a temperature-dependent Ouyang embedding parameter: $|\mu|\sim r_h^{\alpha\leq0}$, effectively  qualitatively mimicked a $D=1+1$ Luttinger liquid with impurities/doping}. It will be extremely interesting to explore this unexpected duality, further. For $\alpha=\frac{5}{2}$, one is able to reproduce the usual linear large-temperature dependence of DC electrical conductivity for most strongly coupled systems with five-dimensional gravity duals with  a black hole \cite{SJain_sigma+kappa}.

\item
{\bf The non-conformal/NLO-in-$N$ corrections to Transport Coefficients}: For ease of readability and convenience of the reader, the main results pertaining to obtaining the non-conformal temperature-dependent ${\cal O}\left(\frac{(g_sM^2)(g_sN_f)}{N}\right)$ corrections to $v_s$ (the speed of sound), $D$ (shear mode diffusion constant ), $\eta$ (shear viscosity) and $\frac{\eta}{s}$ (shear-viscosity-entropy density ratio) are summarized in Table 1 below.
%\begin{center}
\begin{table}[h]
\begin{tabular}{|c|c|c|}\hline
S. No. & Quantity & Expression up to ${\cal O}\left(\frac{(g_sM^2)(g_s N_f)}{N}\right)$  \\ \hline
1 & $v_s$ & $v_s\approx  \frac{\sqrt{14 {n}+17}}{\sqrt{21 {n}+\frac{51}{2}}}+\frac{5 (2 {n}+5) {g_s}^2 M^2 {N_f} \log
   N}{128 \pi ^2 \sqrt{14 {n}+17} \sqrt{84 {n}+102} N}, n\in\mathbb{Z}^+\cup{0}$ \\
   & & $\downarrow\ n=0,1$ \\
   & & $\frac{{\cal O}(1)}{\sqrt{3}} + {\cal O}\left(\frac{g_sM^2}{N}\right)$  \\ \hline
   2 & $D$ & $\frac{1}{\pi T}\left(\frac{1}{4} - \frac{3  g_s^2 M^2 N_f \log N\left(5 + 2 \log N\right)}{512 \pi^2 N}\right)$
   \\  \hline
   3 & $\frac{\eta}{s}$ & See (\ref{eta_over_s-i}) and (\ref{eta_over_s-ii}) \\ \hline
   \end{tabular}
   \caption{Summary of local non-nonformal ${\cal O}\left(\frac{g_sM^2}{N}\right)$ corrections   to $v_s, D, \frac{\eta}{s}$}
   \end{table}
%\end{center}

We showed that in the spirit of  gauge/gravity duality, the leading order result of speed of sound from the quasinormal modes can be reproduced from (a) the pole of the common denominator that appears in the solutions to the scalar modes of metric perturbations, (b) the pole of the retarded Green's function corresponding to the energy momentum tensor two-point correlation function $\langle T_{tt}T_{tt}\rangle$ using the on-shell surface action written in terms of the metric perturbation modes, (c) imposing Dirichlet boundary condition on the solution to the EOM of an appropriate single gauge-invariant perturbation and (d) $\langle T_{tt}T_{tt}\rangle$-computation using the on-shell surface action written in terms of this gauge-invariant perturbation. The leading order result for the diffusion constant of the shear mode as well as the ratio of shear viscosity-to-entropy density ratio were already discussed in \cite{transport-coefficients}.

{The non-trivial insight thus gained at LO in $N$ into the transport properties of holographic large-$N$ thermal QCD at finite gauge coupling is that the LO-in-$N$ conformal result for finite $g_s$ as obtained in this thesis, matches the LO-in-$N$ conformal result for vanishing $g_s$ as is expected/known in the literature for a $p$-brane for $p=3$.}

 { The non-conformal corrections in all the aforementioned quantities, start appearing at  ${\cal O}\left(\frac{(g_s M^2)(g_s N_f)}{N}\right)$}, $N_f$ being the number of flavor $D7$-branes. Thus, at NLO in $N$, { the new insight gained is that  there is a partial universality in the non-conformal corrections to the transport coefficients in the sense that the same are determined by the product of the very small $\frac{g_sM^2}{N}\ll1$ - part of the MQGP limit (\ref{limits_Dasguptaetal-ii}) - and the finite $g_s N_f\sim {\cal O}(1)$ (also part of (\ref{limits_Dasguptaetal-ii})).} The NLO-corrected results in this thesis reflect the non-conformality of the field theory in the IR. As discussed in {\bf Chapter 1} that in the Klebanov-Strassler backgroud \cite{KS} the number of $D3$ branes $N$ decreases with  decreasing (the non-compact radial coordinate) $r$, which according to AdS/CFT dictionary, behaves as an energy scale. This decrease in $N$ is due to a series of repeated Seiberg dualities, where in the extreme IR, at the end of this duality cascade the number of fractional $D3$ branes $M$ which is taken to be finite in the 'MQGP Limit' gets identified with the number of colors in the theory. In other words, the number of $D3$ branes $N$ exhibits a scale dependance due to the duality cascade. Hence from the NLO-corrected expressions of the shear mode diffusion constant and the viscosity, we conclude that these quantities also exhibit a scale dependance through $N$; the appearance of $M$ in the NLO-in-$N$ corrections to the transport coefficients appearing as $\frac{(g_s M^2)(g_s N_f)}{N}$ signals the non-conformality of the field theory in the IR. This is because of the following reason. In the KS picture the presence of finite number $M$ of fractional $D3$ branes makes the field theory non-conformal in the IR while in the UV the presence of $\overline{D5}$ branes cancels the effects of the  $D5$-branes and restore the conformality in the UV. Now at large $r$ the effective number of $D3$ branes are so large that the NLO term can be neglected and we will be left with the leading order conformal results. But in the IR region the NLO terms have to be considered due to small value of $N_{\rm eff}$ - this is rather nicely captured, e.g., by the non-conformal/NLO corrections to $\eta$ (See e.g. Table 1.)

\end{itemize}

We compared our results for $\frac{\eta}{s}$ with the QGP-related RHIC data for $T\geq T_c$ in Section {\bf 3.4}. Let us also make some remarks as regard comparison of some of our results with some well-cited bottom-up holographic QCD models like \cite{Kiritsis_et_al} (as well as references therein) and the more recent \cite{Veneziano-i} based on the Veneziano's QCD model. As regard the speed of sound, like \cite{Kiritsis_et_al}, for $T>T_c$ (which is the temperature range in which we calculated the speed of sound in Section {\bf 5}) the speed of sound approaches a constant value; the difference however is that the NLO non-conformal corrections in our results pushes the value to slightly above $\frac{1}{\sqrt{3}}$ - our LO result and the saturation value in \cite{Kiritsis_et_al}. Upon comparison with some of the results of \cite{Veneziano-ii} which works with the finite temperature version of \cite{Veneziano-i}, one sees that the authors of the same work in the limit: $N_f\rightarrow\infty, N_c\rightarrow\infty: \frac{N_f}{N_c}\equiv$ fixed and $g_{\rm YM}^2 N_c\equiv$ fixed, which is very different from the MQGP limit of (\ref{limits_Dasguptaetal-ii}). A similarity however pertaining to the QCD phase diagram in the same and our results of \cite{NPB} is that $\mu_C(T=T_c)$(for $N_f=2$)$\approx0$.

%\begin{subappendices}
%\input{AppendixB}
%\end{subappendices}
%\input{empty}
%\input{partB}
%\input{Chapter4/chap4}
\chapter{Glueball Spectrum}

\graphicspath{{Chapter4/}{Chapter4/}}

\section{Introduction}
In recent years the computation of the spectrum of glueballs which are the bound states of gluons, has become quite important and significant mainly because the masses of these bound states will be very useful to identify them in the modern ongoing experiments. Also, for a better understanding of QCD, especially the non-perturbative aspects of QCD one needs to have a careful look at glueballs. The glueball state is represented by quantum numbers $J^{PC}$, where $J$, $P$ and $C$ correspond to total angular momentum, parity and charge conjugation, respectively. Now, to capture such physics of strong interaction using the concept of the AdS/CFT correspondence \cite{maldacena}, it is required to break the conformal invariance. In other words, the AdS/CFT correspondence has to be generalized.

Different generalized versions of the AdS/CFT correspondence has thus far been proposed to study non-supersymmetric field theories with a running gauge coupling constant. The original proposal was given by Witten to obtain a gravity dual for non-supersymmetric field theories. Starting with a $4$-dimensional superconformal theory on a stack of parallel $D3$ branes it is possible to construct a 3-dimensional non-supersymmetric theory or $QCD_3$ by compactifying the theory on a circle and imposing anti-periodic boundary condition for the fermions around the same. on the otherhand, a $4$-dimensional non-supersymmetric gauge theory or $QCD_4$ can be achieved by compactifying a six dimensional superconformal theory living on a stack of $M5$ branes on two circles while imposing anti-periodic boundary condition on fermions around one of the two circles. In both the cases the compactifications leads to introducing a black hole in AdS geometry. The glueball masses on these supergravity backgrounds were studied in detail by \cite{Czaki_et_al-0-+},\cite{Mathur et al}. In \cite{Minahan}, glueball masses were obtained for supergravity Duals of QCD Models. In these cases the authors have used the WKB approximation to solve the supergravity equation of motion. In \cite{Boschi+Braga_AdS_BH_AdS_slice}, the authors have considered a model where an AdS slice with an IR cut-off is approximately dual to a confining gauge theory to estimate the mass of the glueball states. Various holographic bottom up approach such as soft-wall model, hard wall model, modified soft wall model, etc. have been used to obtain the glueball's spectra. In \cite {Colangelo,Nicotri} a soft wall holographic model was used to study the glueball spectrum. Both hard wall and soft wall holographic models were considered \cite {Forkel,ForkelStructure} to obtain the glueball correlation functions to study the dynamics of QCD. In \cite{Nunez-5}, ${\cal N}=1$ SYM glueballs from wrapped branes in type IIA, IIB, were studied. In \cite {Gordeli} holographic glueball spectrum was obtained in the singlet sector of ${\cal N}$=1 supersymmetric Klebanov-Strassler model.  States containing the bifundamental $A_i$ and $B_i$ fields were not considered. Comparison with the lattice data showed a nice agreement for $1^{+-}$ and $1^{--}$ states while $0^{+-}$ results were different because of its fermionic component.

Glueball masses can be obtained by evaluating the correlation functions of gauge invariant local operator. The first step to obtain the glueball spectrum in QCD$_3$ is to identify the operators in the gauge theory that have quantum number corresponding to the glueballs of interest. According to
the gauge/gravity duality each supergravity mode corresponds to a gauge theory operator. This operator couples to the supergravity mode at the boundary of the AdS space, for example,
the lowest dimension operator with quantum numbers $J^{PC}=0^{++}$ is Tr$F^2=$Tr$F_{\mu\nu}F^{\mu\nu}$ and this operator couples to the dilaton mode on the boundary. To calculate $0^{++}$ glueball mass we need to evaluate the correlator $\left< TrF^{2}(x)TrF^2(y)\right>$ =$\Sigma_{i} c_{i} e^{-m{i}|x-y|}$, where
$m_{i}$ give the value for glueball mass. However the masses can also be obtained by solving the wave equations for supergravity modes which couples to the gauge theory operators on the
boundary. The latter approach is used here.

The 11D metric obtained as the uplift of the delocalized SYZ type IIA metric, up to LO in $N$, can be interpreted as a black $M5$-brane wrapping a two-cycle, i.e. a black $M3$-brane \cite{transport-coefficients,NPB}. Taking this as the starting point, compactifying again along the M-theory circle, we land up at the type IIA metric and then compactifying again along the periodic temporal circle (with the radius given by the reciprocal of the temperature), one obtains ${\rm QCD}_3$ corresponding to the three non-compact directions of the black $M3$-brane world volume. The Type IIB background of \cite{metrics}, in principle, involves $M_4\times$ RWDC($\equiv$ Resolved Warped Deformed Conifold); asymptotically the same becomes $AdS_5\times T^{1,1}$. To determine the gauge theory fields that would couple to appropriate supergravity fields a la gauge-gravity duality, ideally one should work the same out for the $M_4\times$ RWDC background (which would also involve solving for the Laplace equation for the internal RWDC). We do not attempt to do the same here. Motivated however by, e.g.,
\noindent {\bf (a)} asymptotically the type IIB background of \cite{metrics} and its delocalized type IIA mirror of \cite{MQGP} consist of $AdS_5$ and
\noindent {\bf (b)} terms of the type $Tr(F^2(AB)^k), (F^4(AB)^k)$ where $F^2 = F_{\mu\nu}F^{\mu\nu}, F^4 = F_{\mu_1}^{\ \mu_2}F_{\mu_2}^{\ \mu_3}F_{\mu_3}^{\ \mu_4}F_{\mu_4}^{\ \mu_1} - \frac{1}{4}\left(F_{\mu_1}^{\ \mu_2}F_{\mu_2}^{\ \mu_1}\right)^2$, $A, B$ being the bifundamental fields that appear in the gauge theory superpotential corresponding to $AdS_5\times T^{1,1}$ in \cite{KW}, form part of the gauge theory operators corresponding to the solution to the Laplace equation on $T^{1,1}$ \cite{Gubser-Tpq} (the operator $Tr F^2$ which shares the quantum numbers of the $0^{++}$ glueball couples to the dilaton and $Tr F^4$ which also shares the quantum numbers of the $0^{++}$ glueball couples to trace of metric fluctuations and the four-form potential, both in the internal angular directions),
ere we calculate:\\
$\bullet$ type IIB dilaton fluctuations, which we refer to as $0^{++}$ glueball\\
 $\bullet$ type IIB complexified two-form fluctuations that couple to
 $d^{abc}Tr(F_{\mu\rho}^aF^{b\ \rho}_{\lambda}F^{c\ \lambda}_{\ \ \ \ \ \nu})$, which we refer to as $0^{--}$ glueball \\
$\bullet$ type IIA one-form fluctuations that couple to $Tr(F\wedge F)$, which we refer to as $0^{-+}$ glueball \\
$\bullet$  M-theory metric's scalar fluctuations which we refer to as another (lighter) $0^{++}$ glueball \\
$\bullet$ M-theory metric's vector fluctuations which we refer to as $1^{++}$ glueball, \\
and \\
$\bullet$ M-theory metric's tensor fluctuations which we refer to as $2^{++}$ glueball.

Now, for the glueball mass computation we have solved the supergravity equations by two different method: (i) using WKB approximation, (ii) imposing Neumann/Diriclet boundary condition at $r_h/r_0$. We will discuss the details of these two approaches below with the different energy regions of our set up as discussed in details in {\bf Chapter 2} for the two backgrounds.
\paragraph{WKB method:} To obtain the mass spectrum for different glueballs we need to solve the differential equations involving appropriate field perturbation. For example, assuming a particular perturbation of the form $H(r)=\tilde{H}(r)e^{ikx}$ with $k^2=-m^2$, $m$ being the mass of the corresponding glueball, the equation has the following general form,
\begin{equation}
\label{geneom}
\tilde{H}^{\prime\prime}+f_1(r)\tilde{H}^{\prime}+m^2f_2(r)\tilde{H}=0.
\end{equation}
Next, following the redefinition of \cite{Minahan}, we introduce new variables as given below,
\begin{itemize}
\item{\it{Background with a black hole}}: $r\rightarrow \sqrt{y}$, $r_h\rightarrow \sqrt{y_h}$ and then $y\rightarrow y_h\left(1+e^z\right)$
\end{itemize}
\begin{itemize}
\item{\it{Background with a black hole}}: $r\rightarrow \sqrt{y}$, $r_0\rightarrow \sqrt{y_0}$ and then $y\rightarrow y_0\left(1+e^z\right)$.
\end{itemize}
In terms of these newly defined variables equation (\ref{geneom}) can be written as the following form,
\begin{equation}
\label{geneominz}
\partial_z\left(f_3(z)\partial_z\tilde{H}\right)+m^2f_4(z)\tilde{H}=0.
\end{equation}
With a field redefinition: $\tilde{H}$ as $\psi(z)=\sqrt{f_3(z)}\tilde{H}(z)$, the above equation reduces to the following Schrodinger like form:
\begin{equation}\label{pote}
\left(\frac{d^2}{dz^2} + V(z)\right)\psi(z)=0,
\end{equation}
where $V(z)$ is the potential term. Once we get the potential, the mass can be found from the WKB quantization condition:
\begin{equation}
\int^{z_2}_{z_1}\sqrt{V(z)}dz=\left(n+\frac{1}{2}\right)\pi
\end{equation},
where $z_1$ and $z_2$ are the turning points obtained by solving for the roots, the equation $V(z)=0$: $V(z)>0$ for $z\epsilon [z_1,z_2]$.

In the present work, we have considered the two regions namely, IR and IR-UV interpolating/UV region separately. Moreover, the potential, the corresponding turning points and finally the spectrum are calculated for each of the above two regions.
%In one scenario, there exist a black hole with horizon radius $r_h$, while for the second case we consider an IR cut-off at a finite radial distance $r_0$ with no black hole in the background. The exact analytical calculation is not quite obvious and hence we had to consider the WKB approximation to get the final result. Although in some cases, mainly for the geometry with IR cut-off, we manage to solve exactly with Diriclet/Neumann boundary condition at $r_0$.
%In the following we first present a detailed discussion to get the mass spectrum using WKB approximation and also imposing Neumann/Diriclet boundary condition for the $0^{++}$ glueball. For the other glueball masses we will not write down all the intermediate steps but have tried to give the explicit expressions of important terms/parameters.
\section{$0^{++}$ Glueball spectrum from type IIB supergravity background}
In this section we compute the $0^{++}$ glueball spectrum in the type IIB background of \cite{metrics}. The $10-d$ metric with small but finite resolution $a$ is given as,
 \begin{equation}
\label{IIB metric1}
ds^2=\frac{1}{\sqrt{h}}\left(-g(r)dt^2+dx_1^2+dx_2^2+dx_3^2\right)+\frac{\sqrt{h}}{g(r)}\left({r^2 + 6a^2\over r^2 + 9a^2}\right)dr^2+\sqrt{h}~r^2 d{\cal M}_5^2,
\end{equation}
with $g(r)=1-\frac{rh^4}{r^4}$. As before we choose to work around a particular value of $\theta_1$ and $\theta_2$, namely at $\{\theta_1=\frac{\alpha_{\theta_1}}{N^{1/5}}$ and $\theta_2=\frac{\alpha_{\theta_1}}{N^{3/10}}\}$, where this time we take both $\alpha_{\theta_1}$ and $\alpha_{\theta_1}$ to be equal to one for convenience. Also for small resolution parameter, we series expand the $\{rr\}$ component of the metric (\ref{IIB metric1}) in $a$ and choose to keep terms only upto quadratic order. At the above mentioned values of $\theta_1$ and $\theta_2$, the internal five dimensional part ${\cal M}_5$ of the metric decouples to get,
\begin{equation}
\label{IIB metric1}
ds^2=\frac{r^2 \left(1-\frac{{\cal B}(r)}{2 N}\right)}{L^2}\left(-g(r)dt^2+\sum^{3}_{i=1}dx^2_{i}\right)+\frac{L^2\left(r^2-3a^2\right)\left(1+\frac{{\cal B}(r)}{2 N}\right)}{r^4 g(r)}dr^2,
\end{equation}
where, $L=\sqrt{4\pi g_s N}$ and the expression for ${\cal B}(r)$ is given as,
\begin{equation}
\label{B}
{\cal B}(r)=\frac{3g_s M^2  \log (r)} {16 \pi ^2}\Biggl\{8 \pi+g_s N_f\Biggl(6-2\log (4)+12 \log (r) - \log (N)\Biggr) \Biggr\}.
\end{equation}

The resolution parameter $a$ depends on the horizon radius. In fact it is proportional to $r_h$, such that
\begin{eqnarray}
\label{a}
a=r_h\left(b+c1\left(\frac{g_s M^2}{N}\right)+c_2\left(\frac{g_s M^2}{N}\right)\log (r_h)\right),
\end{eqnarray}
with $b$, $c_1$ and $c_2$ as some positive constants. Hence while computing the masses in the background with no horizon, we must put $r_h$ and $a$ to zero in the above metric.

Now, for the spectrum of $0^{++}$ glueball, we need to solve for the eigenvalues of the dilaton wave equation in the above background. The background dilaton profile at the particular values of $\theta_1$ and $\theta_2$ with ($r_h\neq 0$) and without ($r_h=0$) the black hole is discussed in {\bf Chapter 2}.
The dilaton equation that has to be solved is given as:
\begin{equation}
\label{EOM0++}
\begin{split}
\partial_{\mu}\left(e^{-2\Phi}\sqrt{g}g^{\mu\nu}\partial_{\nu}\phi\right)&=0.
\end{split}
\end{equation}
\subsection{WKB approximation}
\begin{itemize}
\item{\it{Background with a black hole}}
\end{itemize}
Assuming $\phi$ to be of the form $\phi=e^{i k.x}\tilde{\phi}(r)$ with $k^2=-m^2$, equation (\ref{EOM0++}) can be written as,
\begin{equation}
\begin{split}
\label{EOM0++1}
\partial_{r}\left(A(r)\partial_{r}\tilde{\phi}\right)+m^2B(r)\tilde{\phi}=0,
\end{split}
\end{equation}
where
\begin{equation}
A(r)=\sqrt{g}e^{-2\Phi}g^{rr},~~~~~~~~~~~~~~~~~~~~~~~B(r)=\sqrt{g}e^{-2\Phi}g^{x_1x_1}.
\end{equation}
Now, to simplify the calculations we impose here the large $N$ limit. Due to this limit, the term ${\cal B}(r)$ in the above metric simply goes to zero.
Following the redefinition of \cite{Minahan} as described before, the above equation (\ref{EOM0++1}) reduces to,
\begin{equation}\label{ee}
\partial_{z}(E_{z}\partial_{z}\tilde{\phi})+y^2_{h}F_{z}m^2\tilde{\phi}=0,
\end{equation}
where at leading order in $N$, $E_z$ and $F_z$ are given with $L=(4\pi g_s N)^{1/4}$ in (\ref{EF}).

 Now, transforming the wave function $\tilde{\phi}$ as $\psi(z)=\sqrt{E_{z}}\tilde{\phi}(z)$ equation (\ref{ee}) reduces to a Schr\"{o}dinger-like equation,
\begin{equation}\label{pote}
\left(\frac{d^2}{dz^2} + V(z)\right)\psi(z)=0
\end{equation}
 where the potential $V(z)$ is a rather cumbersome expression which we will not explicitly write out.

The WKB quantization condition becomes,
\begin{equation}
 \int_{z_1}^{z_2}\sqrt{V(z)} = \left(n + \frac{1}{2}\right)\pi,
 \end{equation}
 where $z_{1,2}$ are the turning points of $V(z)$.
Here, we will work below with a dimensionless glueball mass $\tilde{m}$ assumed to be large defined via: $m = \tilde{m} \frac{r_h}{L^2}=\tilde{m} \frac{\sqrt{y_h}}{L^2}$.

The next task is to determine the turning points $(z_1,z_2)$ in the IR and IR-UV interpolating/UV region separately. Also note that in the large $N$ limit $a$ is approximated by $b~r_h$.

In the IR, we have to take the limit $z\rightarrow-\infty$.  Now in the large $\tilde{m}$ and large $\log{N}$ limit this potential at small $z$ can be shown to be given as:
\begin{eqnarray}
\label{Vsmall}
& & \hskip -1in V_{IR}(z)= \frac{1}{8} \left(1-3 b^2\right) {e^z} {\tilde{m}}^2 + {\cal O}\left(e^{2z},\left(\frac{1}{{\tilde{m}}}\right)^2,\frac{1}{\log N}\right)\nonumber\\
   & &
\end{eqnarray}
For the value of $b$ as obtained in the previous chapter equal to $0.6$, the potential is negative. Hence, there are no turning points in the IR.

Now, in the UV, apart from taking the large $z$ limit we also have to take $N_f=0$, to get:
\begin{eqnarray}
\label{Vlarger}
& & V_{UV}(z) = -\frac{3 \left(b^2+1\right) \left({y_h}{\tilde{m}}^2+3\right)}{4 {y_h}{e^{2z}}}+\frac{3 b^2+{y_h}\tilde{m}^2+6}{4 {y_h}{e^z}}-1
   + {\cal O}(e^{-3z}).
\end{eqnarray}

The turning points of (\ref{Vlarger}) in the large $\tilde{m}$ limit are given as
\begin{equation}
 z_1=(3 + 3 b^2) + {\cal O}\left(\frac{1}{\tilde{m}^2}\right),~~~~~~~~~~~~ z_2=\frac{\tilde{m}^2}{4} - \frac{3(2 + 3 b^2)}{4} + {\cal O}\left(\frac{1}{\tilde{m}^2}\right).
 \end{equation}
To obtain a real spectrum, one  first notes:
\begin{eqnarray}
\label{Vlargezlargemtilde}
& & \sqrt{V_{UV}(z)} \sim \sqrt{\left(\frac{ e^{-z}}{4}- e^{-2 z}\right)}~\tilde{m} + {\cal O}\left(e^{-3z},\frac{1}{\tilde{m}^2}\right).
\end{eqnarray}
and therefore with $b=0.6$
\begin{eqnarray}
\label{intsqrtVWKB}
& & \int_{z_1}^{z_2}\sqrt{V_{UV}(z)} = 0.39 \tilde{m} - 2 = \left(n + \frac{1}{2}\right)\pi,
\end{eqnarray}
yielding:
\begin{equation}
\label{mn0++T>0}
m_n^{0^{++}} = 9.18 + 8.08 n.
\end{equation}
\begin{itemize}
\item{\it{Background with an IR cut-off}}
\end{itemize}
In this case we have worked out the spectrum with the full metric. In other words, this time we have included the (N)ext to (L)eading (O)rder term in $N$ in the metric components. Also as mentioned before, we must take the limit $r_h\rightarrow 0$ and hence $a\rightarrow 0$ as now there is no horizon.

Again following the same redefinition as before we get the equation for the dilaton as,
\begin{equation}\label{ff}
\partial_{z}(C_{z}\partial_{z}\tilde{\phi})+y^2_{h}D_{z}m^2\tilde{\phi}=0,
\end{equation}
where $C_{z}$ and $D_{z}$ are given up to NLO in $N$ in (\ref{CD}).

Defining a new function $\psi(z)$ such that: $\psi(z)=\sqrt{C_z}\tilde{\phi}(z)$, the above equation can be converted into a Schr\"{o}dinger-like equation,
\begin{equation}\label{poten}
\left(\frac{d^2}{dz^2} + V(z)\right)\psi(z)=0.
\end{equation}
The turning points in the large $\tilde{m}$ limit are given as
\begin{equation}
z_1=\log\left(\frac{3 {g_s}^2 M^2 {N_f} \log {N} \log {y_0}}{64 \pi ^2 \tilde{m} N}+\frac{1}{\tilde{m}}\right),~~~~~~~~~~~z_2=\log\left(\delta^2-1\right),
\end{equation}
where $\mu^{\frac{2}{3}} = \delta \sqrt{y_0}$ and the proportionality constant $\delta$ could be determined by matching with lattice calculations and turns out to be ${\cal O}(1)$.

Expanding $\sqrt{V}$ in $\tilde{m}$ and then integrating over the above mentioned domain, one gets the following quantization condition,
\begin{eqnarray}
\label{WKB-integral-ii}
& & \int_{z_1}^{z_2}\sqrt{V(z)}dz\nonumber\\
 & & = \Biggl\{\frac{1}{2} \left(\delta ^2-1\right) \tilde{m} \left(1+\frac{3 M^2g^2_s N_f \log{ N}\log {y_0}}{64 \pi ^2 N}\right) - 0.75\Biggr\} + {\cal O}\left(\frac{1}{\tilde{m}N},\frac{1}{\tilde{m}^2}\right) \nonumber\\& &= \left(n + \frac{1}{2}\right)\pi,
\end{eqnarray}
 yielding:
\begin{equation}
\label{WKB-result0++_T=0}
m_n^{0^{++}}=\frac{(6.28 n+4.64)}{\delta ^2-1} \left(1-\frac{0.004~ g^2_s  M^2 {N_f}\log {N} \log {y_0}}{N}\right).
\end{equation}
\subsection{Neumann boundary condition at $r_0$}
Following \cite{WKB-i}, we redefine the radial coordinate as $z=\frac{1}{r}$. With this change of variable, the radial cut-off now maps to $z=z_0$, with $z_0=\frac{1}{r_0}$. The dilaton equation using the full metric and the dilaton background in the limit $(r_h,a)\rightarrow0$ is given as:
\begin{equation}
\begin{split}
e^{2U}\partial_z\left(e^{-2U}\partial_z\tilde{\phi}\right)+\left(\frac{4g_s N\pi({\cal B}(z)+1)}{(1-{\cal B}(z))}\right)m^2\tilde{\phi}=0,
\end{split}
\end{equation}
where upto NLO in $N$ we have,
\begin{equation}
\begin{split}
e^U & =\frac{8 \times 2^{1/4} g_s^{13/8}\pi^{13/8}N^{5/8} z^{3/2}}{4\pi+4 \pi g_s N_f \log {4}  +2 \pi g_s N_f  \log {N}+3 g_s N_f  \log {z}}
\\ & -\frac{15
   M^2 g_s^{21/8} z^{3/2}  \log {z} \left(8\pi+6 N_f g_s-g_s N_f \log {16} -12g_s N_f \log {z}-g_s N_f  \log {N}\right)}{8\times 2^{3/4} \pi ^{3/8} N^{3/8}
   \left(4\pi+4 \pi g_s N_f  \log {4}+3 N_f g_s \log {z}+2 \pi  N_f g_s \log {N}  \right)}.
\end{split}
\end{equation}
Now to convert the above equation in a one-dimensional schrodinger like form we introduce a new field variable $\psi(z)$ as: $\psi(z)=e^{-U}\tilde{\phi}(z)$.

With this one can write the equation in the following schrodinger like form,
\begin{equation}
\begin{split}\label{pot}
\frac{\partial^2\psi(z)}{\partial z^2} & = V(z)\psi(z),
\end{split}
\end{equation}
where, the full expression for the potential $V(z)$ is too large to solve analytically. This potential can be simplified at large $N$ and large $\log{N}$ and is given as:
\begin{eqnarray}
\label{VV}
& & \hskip -0.6in V(z)= 4 \pi  {g_s} m^2 N+\frac{6}{\pi  z^2 \log (N)}-\frac{15}{4 z^2} + {\cal O}\left(\frac{1}{(\log N)^2},\frac{g_sM^2}{N}\right).
\end{eqnarray}
Hence, the Schr\"{o}dinger equation becomes:
\begin{equation}
\label{Schroedinger}
\psi''(z)+\psi(z) \left(4 \pi  {g_s} m^2 N+\frac{6}{\pi  z^2 \log (N)}-\frac{15}{4 z^2}\right)=0,
\end{equation}
whose solution is given as under:
\begin{equation}\begin{split}
\label{Schroedinger-solution}
\psi(z) = c_1 \sqrt{z}~ J_{\sqrt{\frac{2\left(2 \pi  \log {N}-3\right)}{\pi\log {N}}}}\left(2 \sqrt{{g_s}} m \sqrt{N} \sqrt{\pi } z\right)+c_2 \sqrt{z}~
   Y_{\sqrt{\frac{2\left(2 \pi  \log {N}-3\right)}{\pi\log {N}}}}\left(2 \sqrt{{g_s}} m \sqrt{N} \sqrt{\pi } z\right).
\end{split}
\end{equation}
Requiring finiteness of $\psi(z)$ at $z=0$ requires setting $c_2=0$.

Then imposing Neumann boundary condition on $\tilde{\phi}(z)$ at $z=z_0$ implies, in the large-$N$ large-$z$ (as the Neumann boundary condition will be implemented in the IR) limit:
\begin{eqnarray}
\label{Neumann_bc}
& & \frac{1}{2} x_0 J_{\sqrt{4-\frac{6}{\pi  \log (N)}}-1}(x_0)-\frac{1}{2} x_0 J_{\sqrt{4-\frac{6}{\pi  \log (N)}}+1}(x_0)+2 J_{\sqrt{4-\frac{6}{\pi  \log (N)}}}(x_0)=0,
\end{eqnarray}
where $x_0\equiv 2\sqrt{g_s N \pi}m z_0$. The graphical solution points out that the ground state has a zero mass and the lightest (first excited state) glueball mass is approximately given by $3.71 \frac{r_0}{L^2}$.
\begin{figure}
\begin{center}
 \includegraphics[scale=0.8]
 %[height= 21cm,width=+15cm]
 {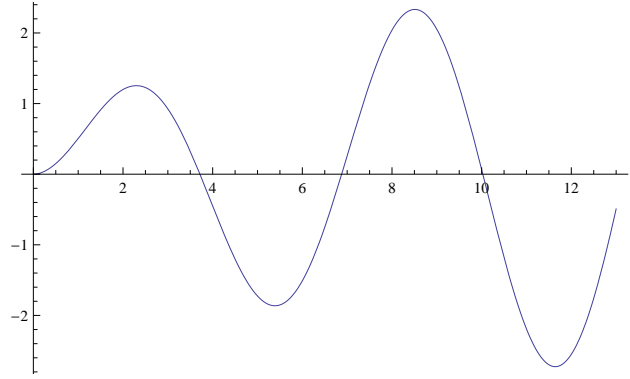}
 \end{center}
 \caption{$0^{++*,++**,++***}$ Masses obtained by graphical solution of the Neumann boundary condition in the $r_h=0$ Limit}
\end{figure}
\section{Scalar Glueball ($0^{-+}$) Masses}
The Wess-Zumino term for the type IIA $D4-$brane is given by,
\begin{equation}
\int_{\Sigma_{4,1}}A\wedge F\wedge F,
\end{equation}
where the operator $Tr(F\tilde{F})$ with $P=-,C=+$ couples to the gauge field $A_{\mu}$.

Hence to obtain the glueball mass in this case we consider the EOM involving the one form gauge field in the type IIA background which is given as:
\begin{equation}
\label{0-+-i}
\partial_\nu\left(\sqrt{g^{\rm IIA}}g_{\rm IIA}^{\mu\sigma}g_{\rm IIA}^{\nu\rho}\left(\partial_{[\sigma}A_{\rho]}\right)\right)=0,
\end{equation}
where the indices $\mu,\nu,...=\{0,1,2,.....,9\}$. Also all the non zero components of the IIA gauge field is given in \cite{MQGP}.

Following \cite{Czaki_et_al-0-+}, we consider the linear perturbation of the type IIA gauge field $A_{\mu}$ about the background value $A^0_{\mu}$ of the form,
\begin{equation}
A_{\mu}=A^0_{\mu}+\mathcal{A}_{\mu}.
\end{equation}

Assuming the perturbation of the form $\mathcal{A}_\mu=\delta^{\theta_2}_\mu a_{\theta_2}(r)e^{i k\cdot x}$ with $k^2=-m^2$, the $0^{-+}$ EOM (\ref{0-+-i}) reduces to,
\begin{equation}
\label{0-+-ii}
 \sqrt{g^{\rm IIA}}g_{\rm IIA}^{\theta_2\theta_2}g_{\rm IIA}^{rr}a_{\theta_2}^{\prime\prime}(r) + \partial_r\left(\sqrt{g^{\rm IIA}}g_{\rm IIA}^{\theta_2\theta_2}g_{\rm IIA}^{rr}\right)a_{\theta_2}^\prime(r) + \sqrt{h}m^2\sqrt{g^{\rm IIA}}g_{\rm IIA}^{\theta_2\theta_2}a_{\theta_2}(r)= 0,
\end{equation}
where the different type IIA metric components are given in (\ref{Mtheorymetriccomp}).
\subsubsection{WKB approximation}
\begin{itemize}
\item{\it{Background with a black hole}}
\end{itemize}
The potential corresponding to the Schr\"{o}dinger-like equation, substituting $m = \tilde{m} \frac{\sqrt{y_h}}{L^2}$, is given by:
\begin{eqnarray}
\label{V_0-+-Minahan-i}
& & V = \frac{e^z \left(4\times 3^{2/3} \tilde{m}^2 \left(3 e^z+e^{2 z}+2\right)-64 e^z-108 e^{2 z}-25 e^{3 z}+96\right)}{16
   \left(e^z+1\right)^2 \left(e^z+2\right)^2}.
\end{eqnarray}
In the IR $(z\ll 0)$, the above potential is given as,
\begin{eqnarray}
\label{V_0-+Minahan-ii}
& & V_{IR}(z)= \left(-\frac{3}{16} 3^{2/3} \tilde{m}^2-\frac{11}{2}\right) e^{2 z}+\left(\frac{1}{8} 3^{2/3}
   \tilde{m}^2+\frac{3}{2}\right) e^z + {\cal O}(e^{-3z}),
\end{eqnarray}
The turning points in this region are obtained as,
\begin{equation}
 z_1=-\infty,~~~~~~~~~z_2=-2.526,
\end{equation}
to get,
\begin{equation}
\label{WKB-Minahan-0-+_i}
\int_{z_1}^{z_2}\sqrt{V_{IR}}dz = 0.283 \tilde{m} = \left(n + \frac{1}{2}\right)\pi,
\end{equation}
which obtains:
\begin{equation}
\label{WKB-Minahan-0-+_ii}
 m_n^{0^{-+}} = 5.56 (1 + 2n)\frac{r_h}{L^2}.
\end{equation}

Similarly, in the UV $(z\gg1)$:
\begin{equation}
\label{V0-+_T_UV}
V_{UV}(z) = \left({\frac{1}{4} 3^{2/3} \tilde{m}^2+\frac{21}{8}}\right)e^{-z}+\left(\frac{9}{16}-\frac{3}{4} 3^{2/3}
   \tilde{m}^2\right) e^{-2 z}-\frac{25}{16} + {\cal O}(e^{-3z}),
\end{equation}
whose turning points in the large $\tilde{m}$ limit are,
\begin{equation}
z_1=\log(3 + {\cal O}\left(\frac{1}{\tilde{m}^2}\right)),~~~~~z_2=\log \left(0.33 \tilde{m}^2 - 1.32 + {\cal O}\left(\frac{1}{\tilde{m}^2}\right)\right).
\end{equation}
Again in the large $\tilde{m}$ limit we have,
   $\sqrt{V_{UV}(z)} = \frac{1}{2} \sqrt[3]{3} \tilde{m} e^{-z} \sqrt{e^z-3} + {\cal O}\left(\frac{1}{\tilde{m}}\right)$.

Therefore,
   \begin{equation}
   \label{WKB0-+_UV_T}
   \int_{z_1}^{z_2}\sqrt{V_{UV}(z)}dz = 0.654 \tilde{m} - 2.5 = \left(n + \frac{1}{2}\right)\pi,
   \end{equation}
   which obtains:
\begin{equation}
\label{mn0-+_UV_T}
m_n^{0^{-+}} = \left(6.225 + 4.804 n\right)\frac{r_h}{L^2}.
\end{equation}
\begin{itemize}
\item{\it{Background with an IR  cut-off}}
\end{itemize}
The `potential' with $m = \tilde{m} \frac{\sqrt{y_h}}{L^2}$ at leading order in $N$ is given by:
\begin{equation}
\label{V0-+_T=0}
V(z) = \frac{4\times 3^{2/3} \tilde{m}^2 e^{2 z}+\frac{e^{2 z} \left(e^z+1\right) \left(5 e^z+12\right)^2}{\left(e^z+2\right)^2}-2\left(e^z+1\right)
\left(14 e^z+25 e^{2 z}+4\right)}{16 \left(e^z+1\right)^3} + {\cal O}\left(\frac{g_sM^2}{N}\right).
\end{equation}
Therefore, in the IR $(z\ll1)$ the above can be approximated as,
\begin{equation}
\label{V0-+_IR-T=0}
V_{IR}(z) = -\frac{1}{2} - \frac{3}{4}e^z + \left(\frac{9}{8} + \frac{3^{\frac{2}{3}}}{4}\tilde{m}^2\right)e^{2z} + {\cal O}(e^{3z}).
\end{equation}
The turning points are: $\left\{z_1=\log\left(\frac{\sqrt{2}}{3^{\frac{1}{3}}\tilde{m}}\right),~~ z_2=\log(\delta^2-1)\right\}$, giving,
\begin{equation}
\int_{z_1}^{z_2}\sqrt{V_{IR}(z)} dz= \frac{3^{\frac{1}{3}}(\delta^2-1)}{2} \tilde{m} - 1.1126,
\end{equation}
yielding:
\begin{equation}
\label{mn0-+_T=0_IR}
m_n^{0^{-+}} = \frac{\left(3.72 + 4.36 n\right)}{\left(\delta^2-1\right)}\frac{r_0}{L^2}.
\end{equation}
Also, in the UV $(z\gg0)$:
\begin{equation}
\label{V0-+_UV-T=0}
V_{UV}(z) = \left(-\frac{3}{4}\times 3^{2/3} \tilde{m}^2-\frac{103}{16}\right) e^{-2 z}+\left(\frac{1}{4}\times 3^{2/3} \tilde{m}^2+\frac{21}{8}\right)
   e^{-z}-\frac{25}{16},
\end{equation}
whose turning points are: $\left\{z_1=3 + {\cal O}\left(\frac{1}{\tilde{m}^2}\right),~~~z_2=\frac{4}{25}\times 3^{\frac{2}{3}}\tilde{m}^2 - \frac{33}{25}\right\}$, yielding,
\begin{equation}
\int_{z_1}^{z_2}\sqrt{V_{UV}(z)}~dz
=\frac{\pi}{4\times 3^{\frac{1}{6}}} = \left(n + \frac{1}{2}\right)\pi
\end{equation}
which obtains:
\begin{equation}
\label{m0-+_T=0}
m_n^{0^{-+}} = 4.804\left(n + \frac{1}{2}\right)\frac{r_0}{L^2}.
\end{equation}
\subsubsection{Neumann/Dirichlet boundary condition}
\begin{itemize}
\item{\it{Background with a black hole}}
\end{itemize}
Taking the large-N limit one can show that the equation of motion (\ref{0-+-ii}) yields::
\begin{equation}
\label{0-+-vi}
\frac{8 \pi  \left(r^4-{r^4_h}\right) }{\sqrt{3}}a_{\theta_2}^{\prime\prime}(r)-\frac{32 \pi  \left(r^4-2 {r_h}^4\right)}{\sqrt{3} ~r}a^{\prime}_{\theta_2}(r)+\left(32\times
   3^{1/6} \pi^2 {g_s} m^2 N\right){a_{\theta_2}}(r)=0.
\end{equation}
Working near $r=r_h$, the above equation is approximated by,
\begin{equation}
\label{0-+-vii}
\frac{32 \pi  {r^3_h} (r-{r_h}) }{\sqrt{3}}{a^{\prime\prime}_{\theta_2}}(r)-\left(\frac{160 \pi  {r^2_h} (r-{r_h})}{\sqrt{3}}-\frac{32 \pi
   {r_h}^3}{\sqrt{3}}\right) {a^{\prime}_{\theta_2}}(r)+\left(32 \times{3^{1/6}} \pi ^2 {g_s} m^2 N\right){a_{\theta_2}}(r)=0,
\end{equation}
whose solution is given as under,
\begin{eqnarray}
\label{0-+-viii}
a_{\theta_2}(r) = c_1 U\left(-\frac{3^{2/3} {g_s} m^2 N \pi }{5 {r_h}^2},1,\frac{5 r}{{r_h}}-5\right)+c_2 L_{\frac{3^{2/3} \pi  {g_s} m^2 N}{5
   {r_h}^2}}\left(\frac{5 r}{{r_h}}-5\right).
\end{eqnarray}
Now utilizing,
\begin{eqnarray}
\label{0-+-ix}
& & U\left(1-\frac{3^{2/3} \pi  {g_s} m^2 N}{5 |{r_h}|^2},2,\frac{5 r}{{r_h}}-5\right) = \nonumber\\
& &  -\frac{{r_h}^3}{3^{2/3} \pi  {g_s} m^2 N (r-{r_h}) \Gamma \left(-\frac{3^{2/3} {g_s} m^2 N \pi }{5 {r_h}^2}\right)}+\frac{\left(1-\frac{3^{2/3} {g_s} m^2 N \pi }{5 {r_h}^2}\right)+\log \left(\frac{5 r}{{r_h}}-5\right)+2 \gamma -1}{\Gamma
   \left(-\frac{3^{2/3} {g_s} m^2 N \pi }{5 {r_h}^2}\right)}\nonumber\\
   & & +\frac{(r-{r_h}) \left(5 {r_h}^2-3^{2/3} \pi  {g_s} m^2 N\right) \left(2
   \left(2-\frac{3^{2/3} {g_s} m^2 N \pi }{5 {r_h}^2}\right)+2 \log (r-{r_h})+2 \log \left(\frac{5}{{r_h}}\right)+4 \gamma
   -5\right)}{4 {r_h}^3 \Gamma \left(-\frac{3^{2/3} {g_s} m^2 N \pi }{5 {r_h}^2}\right)}\nonumber\\
   & & +{\cal O}\left((r-{r_h})^2\right),
\end{eqnarray}
with Neumann boundary condition: $a_{\theta_2}^\prime(r=r_h)=0$, one requires $c_2=0$ and at the same time using $r_h=T\sqrt{4\pi g_s N}$ at leading order in $N$ one get the following condition,
 \begin{eqnarray}
 \label{0-+-x}
 \frac{3^{2/3} m^2}{20 T^2}=n.
 \end{eqnarray}
Therefore the mass is given as,
 \begin{eqnarray}
 \label{0-+-xi}
 m_n^{0^{-+}} = \frac{2 \sqrt{5} \sqrt{n} T}{\sqrt[3]{3}}.
 \end{eqnarray}
 One can show that imposing Dirichlet boundary condition $a_{\theta_2}(r=r_h)=0$, yields the same spectrum as (\ref{0-+-xi}).
If the temperature $T$ gets identified with $a$ of \cite{Czaki_et_al-0-+}, then the ground state, unlike \cite{Czaki_et_al-0-+}, is massless; the excited states for lower $n$'s are closer to $a=0$ and the higher excited states are closer to $a\rightarrow\infty$ in \cite{Czaki_et_al-0-+}.
\begin{itemize}
\item{\it{Background with an IR  cut-off}}
\end{itemize}
The $r_h=0$ limit of (\ref{0-+-vi}) gives:
\begin{equation}
\label{0-+_EOM_T=0}
\sqrt{3} r^4 {a^{\prime\prime}_{\theta_2}}(r)-4 \sqrt{3} r^3 {a^{\prime}_{\theta_2}}(r)+3 \sqrt[6]{3} {\tilde{m}}^2 r_0^2 {a_{\theta_2}}(r) = 0,
\end{equation}
which near $r=r_0$ yields:
\begin{equation}
\label{0-+_EOM_solution}
a_{\theta_2}(r) = (4 r-3 \tilde{m})^{5/4} \left(c_1 U\left(\frac{5}{4}-\frac{\tilde{m}^2}{4 \sqrt[3]{3}},\frac{9}{4},\frac{3 r}{\tilde{m}}-\frac{9}{4}\right)+c_2
   L_{\frac{1}{12} \left(3^{2/3} \tilde{m}^2-15\right)}^{\frac{5}{4}}\left(\frac{3 r}{\tilde{m}}-\frac{9}{4}\right)\right).
\end{equation}
Imposing Neumann boundary condition on (\ref{0-+_EOM_solution}) yields:
\begin{eqnarray}
\label{0-+_T=0_spectrum}
& & m^{0^{-+}}(r_h=0) = 0\nonumber\\
& & m^{0^{-+*}}(r_h=0) \approx 3.4 \frac{r_0}{L^2}\nonumber\\
& & m^{0^{-+**}}(r_h=0) \approx 4.35 \frac{r_0}{L^2}.
\end{eqnarray}
One can similarly show that imposing Dirichlet boundary condition on (\ref{0-+_EOM_solution}) for $c_2=0$ yields:
\begin{eqnarray}
\label{0-+_T=0_spectrum-ii}
& & m^{0^{-+}}(r_h=0) = 0\nonumber\\
& & m^{0^{-+*}}(r_h=0) \approx 3.06 \frac{r_0}{L^2}\nonumber\\
& & m^{0^{-+**}}(r_h=0) \approx 4.81 \frac{r_0}{L^2}.
\end{eqnarray}
\section{Glueball ($0^{--}$) Masses}
Given the Weiss-Zumino term $A^{\mu\nu}d^{abc}{\rm Tr}\left(F_{\mu\rho}^aF^{b\ \rho}_\lambda F^{c\ \lambda}_\nu\right)$ and the two-form potential $A_{\mu\nu}$ is dual to a pseudo-scalar, for $r_h\neq0$, corresponding to ${\rm QCD}_3$, one writes down the EOM for the fluctuation $\delta A^{23}$. The $B_{MN}, C_{MN}$ EOMs are:
\begin{eqnarray}
\label{B_C_EOMs}
& & D^M H_{MNP} = \frac{2}{3}F_{NPQRS}F^{QRS},\nonumber\\
& & D^M F_{MNP} = - \frac{2}{3}F_{NPQRS}H^{QRS},
\end{eqnarray}
or defining $A_{MN} = B_{MN} + i C_{MN}$, (\ref{B_C_EOMs}) can be rewritten as:
\begin{equation}
\label{A_EOM}
D^M\partial_{[M}A_{NP]} = - \frac{2i}{3}F_{NPQRS}\partial^{[Q}A^{RS]}.
\end{equation}
Now, we consider the perturbation of the two form $A_{\mu\nu}$ as,
\begin{equation}
A_{MN}=A^{(0)}_{MN} + \delta A_{MN},
\end{equation}
with $\delta A^{MN} = \delta^M_2 \delta^N_3 \delta A_{23}$.

The EOM satisfied by $\delta A_{23}(x^{0,1,2,3},r) = \int \frac{d^4k}{\left(2\pi\right)^4}e^{i k\cdot x}g_{22} G(r)$ reduces to:
\begin{equation}
\label{A_EOM_ii}
\partial_\mu\left(\sqrt{-g}g^{22}g^{33}g^{\mu\nu}\partial_\nu \delta A_{23}\right) = 0.
\end{equation}
Now considering the NLO term in $a$ (\ref{a}), with $b=0.6$, $c_1=c_2=4$ (as obtained in chapter $\bf{3}$, $k^2=-m^2$, and defining ${\cal G}(r)\equiv g_{22} G(r)$ the EOM for ${\cal G}(r)$ is:
\begin{eqnarray}
\label{0--i}
& &  {\cal G}''(r)+ D_1(r){\cal G}'(r)+D_2(r){\cal G}(r)=0,
\end{eqnarray}
where $D_1$ and $D_2$ at NLO in $N$ are given in (\ref{D1D2}).
\subsection{WKB approximation}
\begin{itemize}
\item{\it{Background with a black hole}}
\end{itemize}
The potential in the IR $(z\ll0)$ with $m=\tilde{m}\frac{\sqrt{y_h}}{L^2}$ at leading order in $N$ is given as:
\begin{eqnarray}
\label{V0--_IR_i}
& & V_{IR}(z) = \left(6-0.01 \tilde{m}^2\right) e^z+\left(0.15 \tilde{m}^2-16.18\right) e^{2 z} + {\cal O}\left(e^{3z}\right),
\end{eqnarray}
where in the `large' $\tilde{m}$-limit the turning points are: $\{z_1=-\infty,~~~z_2=\log{0.067}\}$, to get:
\begin{eqnarray}
\int_{z_1}^{z_2}\sqrt{V_{IR}(z)}\approx 0,
\end{eqnarray}
implying there is no contribution to the WKB quantization condition in the IR.

In the UV $(z\gg1)$, the potential is approximated as,
\begin{equation}
\label{V0--_UV}
V_{UV}(z) = \left(-1.02 \tilde{m}^2-22.5\right) e^{-2 z}+\left(0.25 \tilde{m}^2+8.25\right) e^{-z}-1 + {\cal O}\left(e^{-3z}\right).
\end{equation}
Again, in the large $\tilde{m}$ limit, the turning points are:

 $\{z_1=\log {4.08}+{\cal O}\left(\frac{1}{\tilde{m}^2}\right),~~~z_2=\log(0.25 \tilde{m}^2 + 4.17)+{\cal O}\left(\frac{1}{\tilde{m}^2}\right)\}$,
to get in the large $\tilde{m}$ limit:
\begin{eqnarray}
\label{WKB-0--_UV}
   & & \int_{z_1}^{z_2}\sqrt{V_{UV}(z)} \\&&=\int_{z_1}^{z_2} e^{-z}\left(0.25 e^z - 1.02\right)^{1/2} + {\cal O}\left(\frac{1}{\tilde{m}}\right)\nonumber\\
   & & = 0.389 \tilde{m} - 2 + {\cal O}\left(\frac{1}{\tilde{m}}\right) = \left(n + \frac{1}{2}\right)\pi.
\end{eqnarray}
Hence one obtains isospectrality with $0^{++}$; for large $n$, there is also isospectrality with $2^{++}$.

\begin{itemize}
\item{\it{Background with an IR  cut-off}}
\end{itemize}
In the IR $(z\ll0)$, the WKB `potential' at leading order in $N$ can be shown to be given by:
\begin{equation}
\label{WKB-integ_T=0}
V_{IR}(z) = -\frac{1}{4} + \frac{1}{4}\left(-3 + \tilde{m}^2\right)e^{2z} + {\cal O}(e^{3z}),
\end{equation}
where $m=\tilde{m}\frac{\sqrt{y_h}}{L^2}$.
The turning points are: $\{z_1=-\log(m_0),~~~z_2=\log({\delta^2-1})\}$.
Further dropping ${\cal O}\left(\frac{1}{\tilde{m}^3}\right)$ terms,
\begin{equation}
\int_{z_1}^{z_2}\sqrt{V_{IR}(z)} = \frac{\left(\delta^2-1\right)}{2} \tilde{m} - 0.785 = \left(n + \frac{1}{2}\right)\pi
\end{equation}
yielding:
\begin{equation}
\label{QKB0--_IR_T=0}
m_n^{0^{--}} =\left(\frac{4.71 + 6.28 n}{\delta^2-1}\right)\frac{r_0}{L^2}.
\end{equation}

In the UV $(z\gg1)$ the potential with $m=\tilde{m}\frac{\sqrt{y_h}}{L^2}$ at leading order in $N$ is:
\begin{eqnarray}
\label{V0--_UV_T=0}
& & V_{UV}(z) = -\frac{3}{4} \left({\tilde{m}}^2+3\right) e^{-2 z}+\frac{1}{4} \left({\tilde{m}}^2+6\right) e^{-z}-1,
\end{eqnarray}
with turning points: $\{z_1=\log{3} + {\cal O}\left(\frac{1}{\tilde{m}^2}\right),~~z_2=\log\left(\frac{\tilde{m}^2}{4} - \frac{3}{2}\right)+ {\cal O}\left(\frac{1}{\tilde{m}^2}\right)\}$.
In the large $\tilde{m}$ limit we have:
 $\sqrt{V_{UV}(z)} = \frac{e^{-z}}{2}\sqrt{e^z - 3}~\tilde{m} + {\cal O}\left(\frac{1}{\tilde{m}}\right)$. Hence,
\begin{eqnarray}
\label{WKB_0--_UV_T=0}
\int_{z_1}^{z_2}\frac{e^{-z}}{2}\sqrt{e^z - 3}\tilde{m}~dz = \left(n + \frac{1}{2}\right)\pi,
\end{eqnarray}
implying:
\begin{equation}
\label{mn0--_UV}
m_n^{0^{--}} =  (7.87 + 6.93 n)\frac{r_0}{L^2}.
\end{equation}

\subsubsection{NLO (in $N$)/Non-Conformal Corrections}

Up to NLO in $N$, in the IR, the potential is given by:
{\footnotesize
\begin{eqnarray}
\label{V2++_NLO_T=0}
& &V^{NLO}_{IR}(z) =  \frac{1}{256 \pi ^2 N}\Biggl\{e^{2 z} \Biggl(-{g_s}^2 M^2 {N_f} (6 {\log N}-72+\log (16777216))+36 {g_s}^2 M^2 \tilde{m}^2 {N_f} \log ^2({y_0})+{g_s}
   M^2 \log ({y_0})\nonumber\\
    & & \times\left({g_s} {N_f} \left(72-\tilde{m}^2 (6 {\log N}-36+\log (16777216))\right)+48 \pi  \tilde{m}^2\right) +48 \pi
    {g_s} M^2+64 \pi ^2 \left(\tilde{m}^2-3\right) N\Biggr)\Biggr\}-\frac{1}{4} + {\cal O}(e^{-3z}).\nonumber\\
    & &
\end{eqnarray}}
The turning points of (\ref{V2++_NLO_T=0}) up NLO in $N$ are given by:\\
$\Biggl[\log\left(\frac{1}{\tilde{m}}\left[1-\frac{{g_s} M^2 \log ({y_0}) (-{g_s} {N_f} (6 {\log N}-36+\log (16777216))+36 {g_s} {N_f} \log ({y_0})+48 \pi )}{128 \pi ^2
    N}\right]\right),\log(\delta^2-1)\Biggr]$. After evaluation of the integral of $\sqrt{V(IR,r_h=0)}$ between the aforementioned turning points, in the large-$\tilde{m}$-limit, one obtains the following quantization condition:
    \begin{eqnarray}
    \label{WKB_NLO_N+0++_T=0}
    & & \left(\frac{(\delta^2-1)}{2}-\frac{3(\delta^2-1)g_sM^2(g_sN_f)\log N\ \log r_0}{64\pi^2N}\right)\tilde{m} -\frac{\pi}{4} = \left(n + \frac{1}{2}\right)\pi,
    \end{eqnarray}
    which yields:
    \begin{eqnarray}
    \label{mn0--_NLON_T=0}
    & & m_n^{0^{--}}(r_h=0) = \frac{6.28319 n+4.71239}{\delta ^2-1}\left(1 + \frac{0.01 {g_s}^2 {\log N} M^2 {N_f} \log ({r_0})}{ N}\right).
    \end{eqnarray}

\subsection{Neumann/Dirichlet Boundary Conditions}
\begin{itemize}
\item{\it{Background with a black hole}}
\end{itemize}
Equation (\ref{0--i}), near $r=r_h$ can be approximated as:
\begin{equation}
\label{0--ii}
 {\cal G}''(r)+\left({b_1} + \frac{1}{r-{r_h}}\right) {\cal G}'(r) + {\cal G}(r) \left(\frac{{a_2}}{r-{r_h}}+{b_2}\right)=0,
\end{equation}
where $b_1$, $b_2$, $a_2$ are given in (\ref{b1b2a2}).

The solution to the above equation (\ref{0--ii}) is given by:
\begin{equation}\begin{split}
\label{0--iv}
{\cal G}(r)& =  e^{\frac{1}{2} r \left(-\sqrt{{b_1}^2-4 {b_2}}-{b_1}\right)}\Biggl\{ c_1 U\left(-\frac{2 {a_2}-{b_1}-\sqrt{{b_1}^2-4 {b_2}}}{2
   \sqrt{{b_1}^2-4 {b_2}}},1,\sqrt{{b_1}^2-4 {b_2}} ~r-\sqrt{{b_1}^2-4 {b_2}}~ {r_h}\right)\\&
+c_2  L_{\frac{2 {a_2}-\sqrt{{b_1}^2-4 {b_2}}-{b_1}}{2 \sqrt{{b_1}^2-4 {b_2}}}}\left(r
   \sqrt{{b_1}^2-4 {b_2}}-{r_h} \sqrt{{b_1}^2-4 {b_2}}\right)\Biggr\},
   \end{split}
\end{equation}
implying:
\begin{equation}
\label{0--v}
G^\prime(r) = \left(\frac{{\cal G}(r)}{g_{22}}\right)^\prime = \frac{1}{\Gamma \left(\frac{-2 {a_2}+{b_1}+\sqrt{{b_1}^2-4 {b_2}}}{2 \sqrt{{b_1}^2-4 {b_2}}}\right)}\sum_{n=-1}^\infty a_n(N,M,N_f,g_s,r_h)(r-r_h)^n.
\end{equation}
Assuming $c_2=0$, the Neumann boundary condition at $r=r_h$ can be satisfied by setting the argument of the gamma function to a negative integer $n$. It runs out setting $\frac{-2 {a_2}+{b_1}+\sqrt{{b_1}^2-4 {b_2}}}{2 \sqrt{{b_1}^2-4 {b_2}}}=-n\in\mathbb{Z}^-\cup\left\{0\right\}$ produces a negligible ground state $0^{--}$ mass. Hence, we consider $\frac{-2 {a_2}+{b_1}+3\sqrt{{b_1}^2-4 {b_2}}}{2 \sqrt{{b_1}^2-4 {b_2}}}=-n\in\mathbb{Z}^-\cup\left\{0\right\}$, which gives a finite ground state mass.  This condition up to LO in $N$ yields:
\begin{equation}
\label{0--vii}
\frac{T^2 \left\{\frac{3}{2T} \left(1.82\times 10^{12}~ T^2-1.66\times 10^9 ~m^2\right)^{1/2}-675867\right\}+265.15 m^2}{T \left(1.82\times 10^{12} ~ T^2-1.66\times 10^9 ~m^2\right)^{1/2}} = - n\in\mathbb{Z}^-\cup\{0\},
\end{equation}
the solution to which are given below:
\begin{eqnarray}
\label{0--viii}
& & m_{0^{--}} = 32.461 T\nonumber\\
& & m_{0^{--}}^* = 32.88 T \nonumber\\
& & m_{0^{--}}^{**} = 32.989 T \nonumber\\
& & m_{0^{--}}^{***} = 33.033 T \nonumber\\
& & m_{0^{--}}^{****} = 33.055 T.
\end{eqnarray}
One can show that one obtains the same spectrum as in (\ref{0--viii}) after imposing Dirichlet boundary condition $G(r=r_h)=0$.
\begin{itemize}
\item{\it{Background with a black hole}}
\end{itemize}
Considering the limit $r_h\rightarrow 0$, and hence $a\rightarrow 0$ in (\ref{0--ii}), the same reduces to:
\begin{equation}\begin{split}
\label{0--T=0-i}
 {\cal G}''(r)+\frac{5}{ r} \left\{1-\frac{3 {g_s} M^2 }{320\pi ^2 N}\left(8 \pi+24 {g_s} {N_f} \log (r)+{g_s} {N_f} \left(6-\log {16}-\log {N}\right)
   \right)\right\}{\cal G}'(r)\\
  +\frac{{g_s} m^2}{4 \pi  r^4}  \Biggl\{16 \pi ^2 N-8 \pi +3 {g^2_s} M^2 N_f\log {r}
  \left(6-\log {16}-\log {N}+12 \log{r}\right)\Biggr\}{\cal G}(r) = 0.
\end{split}
\end{equation}
Equaion (\ref{0--T=0-i}) near $r=r_0$ reduces to:
\begin{equation}
\label{0--T=0-ii}
{\cal G}''(r) + (\alpha_1 + \beta_1(r - r_0)){\cal G}'(r) + (\alpha_2 + \beta_2(r - r_0)){\cal G}(r) = 0
\end{equation}
where
\begin{eqnarray}
\label{0--T=0-iii}
& & \alpha_1 = \frac{5}{r_0}\left(1+\frac{3 {g_s}^2  M^2 {N_f}\log {N}}{320\pi ^2 N}\right),\nonumber\\
& & \beta_1 = -\frac{3 {g_s}^2 M^2 {N_f}}{64 \pi ^2 N {r_0}^2}-\frac{5}{{r_0}^2},\nonumber\\
& & \alpha_2 = \frac{4 \pi  {g_s} m^2 N}{{r_0}^4}-\frac{3 {g_s}^3 m^2 M^2 {N_f} \log {r_0}}{4 \pi  {r_0}^4},\nonumber\\
& & \beta_2 = \frac{3 {g_s}^3 {\log N} m^2 M^2 {N_f} (4 \log {r_0}-1)}{4 \pi  r^5_0}-\frac{16 \pi  {g_s} m^2
   N}{r^5_0}.
\end{eqnarray}
The solution to (\ref{0--T=0-ii}) is given by:
\begin{equation}\begin{split}
\label{0--T=0-iv}
 {\cal G}(r)& = e^{-{\alpha_1} r+\frac{{\beta_2} r}{{\beta_1}}-\frac{{\beta_1} r^2}{2}+{\beta_1} r r_0} \\&\Biggl\{c_2 \,
   _1F_1\left(\frac{{\beta_1}^3-{\alpha_2} {\beta_1}^2+{\alpha_1} {\beta_2} {\beta_1}-{\beta_2}^2}{2
   {\beta_1}^3};\frac{1}{2};\frac{\left((r-{r_0}) {\beta_1}^2+{\alpha_1} {\beta_1}-2 {\beta_2}\right)^2}{2
   {\beta_1}^3}\right)\\&
   +c_1 H_{\frac{-{\alpha_1} {\beta_1} {\beta_2}+{\alpha_2}
   {\beta_1}^2-{\beta_1}^3+{\beta_2}^2}{{\beta_1}^3}}\left(\frac{{\alpha_1} {\beta_1}+{\beta_1}^2 (r-{r_0})-2
   {\beta_2}}{\sqrt{2} {\beta_1}^{3/2}}\right)\Biggr\}.
   \end{split}
\end{equation}
One can then work out  $G'(r=r_0)=\left.\left(\frac{{\cal G}(r)}{g_{22}}\right)^\prime\right|_{r=r_0}$.
Now, setting $c_2=0$, defining $\tilde{m}$ via: $m = \tilde{m} \frac{r_0}{L^2}$, and using the large $\tilde{m}$-limit of Hermite functions:
\begin{eqnarray}
\label{0--T=0-vi}
& &   H_{\frac{-{\alpha_1} {\beta_1} {\beta_2}+{\alpha_2}
   {\beta_1}^2-(1\ {\rm or}\ 2){\beta_1}^3+{\beta_2}^2}{{\beta_1}^3}}\left(\frac{{\alpha_1} {\beta_1}-2 {\beta_2}}{\sqrt{2}
   {\beta_1}^{3/2}}\right)\longrightarrow H_{-\frac{16 \tilde{m}^4}{125}}\left(\frac{2^{\frac{5}{2}}}{5^{\frac{3}{2}}}\tilde{m}^2\right),
\end{eqnarray}
and
\begin{equation}
\label{Hn[x]_large-n}
H_n(x)\stackrel{n\gg1}{\longrightarrow}\frac{2^{\frac{n}{2}+\frac{1}{2}} e^{\frac{x^2}{2}} \left(\frac{n}{e}\right)^{n/2} \cos
   \left(\frac{\pi  n}{2}-x \sqrt{2 n-\frac{x^2}{3}+1}\right)}{\sqrt[4]{1-\frac{x^2}{2 n}}},
\end{equation}
one can show that the Neumann(/Dirichlet: $G(r=r_0)=0$) boundary condition at $r=r_0$ is equivalent to the condition:
\begin{equation}
\label{0--T=0-Neumann_+bc}
\frac{8}{375} \left(\sqrt{6} \tilde{m}^2 \sqrt{375-64 \tilde{m}^4}-6 i \pi  \tilde{m}^4\right)=i \pi  (2 n+1),
\end{equation}
yielding:
\begin{equation}
\label{0--T=0-ix}
m_n^{0^{--}} = \frac{1}{2} 5^{3/4} \left(\frac{-2 \left(\sqrt{6} \sqrt{\pi ^2 \left(16 n^2+22 n+7\right)+6}+6\right)-3 \pi ^2 (2 n+1)}{3 \pi ^2-32}\right)^{1/4}\frac{r_0}{L^2}.
\end{equation}
\section{ Glueball Masses from M theory}
The glueball spectrum for spin $0^{++}, 1^{++}$ and $2^{++}$ is calculated in this section from the M-theory perspective.
The $11$ dimensional M-theory action is given as:
\begin{equation}\label{Mtheoryaction}
S_{M}=\frac{1}{2k^{2}_{11}}\int_{M}d^{11}x\sqrt{G^{\cal M}}\left( R-\frac{1}{2} G_4\wedge *_{11}G_4\right),
\end{equation}
where $G_4=d C_3 + A_1 \wedge d B_2 + dx_{10}\wedge dB_2$, and $C_{\mu \nu 10}^M = B_{\mu \nu}^{IIA}, C_{\mu \nu \rho}^M = C_{\mu \nu \rho}^{IIA}$.

 Now, as shown in \cite{MQGP}, no $F_4^{IIA}$ (to be obtained via a triple
T-dual of type IIB $F_{1,3,5}$ where $F_1\sim F_{x/y/z}, F_3\sim F_{xy r/\theta_1/\theta_2}, F_{xz r/\theta_1/\theta_2},
 F_{yz r/\theta_1/\theta_2}$ and $F_5\sim F_{xyz \beta_1\beta_2}$ where $\beta_i=r/\theta_i$) can be generated,
%\footnote{Consider $T_x$ followed by $T_y$ followed by $T_z$ where $T_i$ means T-dualizing along i-th direction. As an example, $T_x %F_x^{IIB}\rightarrow
%{\rm non-dynamical\ 0-form\ field\ strength}^{IIA}\cite{kiritsis-book}, T_y T_x F_x^{IIB} \rightarrow F_y^{IIB}$,\\
% $T_z F_y^{IIB} \rightarrow F_{yz}^{IIA}$ implying one can never generate
%$F_4^{IIA}$ from $F_1^{IIB}$.
%    As also an example consider $T_x F_{xy\beta_i}^{IIB}\rightarrow F_{y \beta_i}^{IIA}, T_yF_{y \beta_i}^{IIA}\rightarrow F_{\beta_i}^{IIB}, T_z F_{\beta_i}^{IIB}
%\rightarrow F_{\beta_i z}^{IIA}$ again not generating $F_4^{IIA}$;
%   $ T_x F_{xyz \beta_1 \beta_2}^{IIB}\rightarrow F_{yz \beta_1 \beta_2}^{IIA}$,
%    $T_y F_{yz \beta_1 \beta_2}^{IIA}\rightarrow F_{z \beta_1 \beta_2}^{IIB},
%T_z F_{z \beta_1 \beta_2}^{IIB}\rightarrow F_{\beta_1 \beta_2}^{IIB}$; thus one can not generate $F_4^{IIA}$.}
the four-form flux $G_4$ can be obtained as,
\begin{equation}
\begin{split}
G_4&=d\left(C_{\mu\nu10}dx^\mu\wedge dx^\nu\wedge dx_{10}\right)
 + \left(A^{F_1}_1 + A^{F_3}_1 + A^{F_5}_1\right)\wedge H_3\\&=H_3\wedge dx_{10} + A\wedge H_3,
\end{split}
\end{equation}
where $C_{\mu\nu10}\equiv B_{\mu\nu}$, implying
\begin{eqnarray}
\label{flux_action_D=11-ii}
& & \int G_4\wedge *_{11}G_4 = \int \left(H_3\wedge dx_{10} + A\wedge H_3\right)\wedge *_{11}\left(H_3\wedge dx_{10} + A\wedge H_3\right).
\end{eqnarray}
Now, $H_3\wedge dx_{10}\wedge *_{11}\left(H_3\wedge A\right)=0$ as neither $H_3$ nor $A$ has support along $x_{10}$. Hence,
\begin{eqnarray}\begin{split}
\label{flux_action_D=11-iii}
& H_3\wedge dx_{10}\wedge *_{11}\left(H_3\wedge dx_{10}\right)
\\&=\sqrt{G}H_{\mu\nu\rho10}G^{\mu\mu_1}G^{\nu\nu_1}G^{\rho\rho_1}G^{10\lambda_1}H_{\mu_1\nu_1\rho_1\lambda_1}
dt\wedge...dx_{10}\\
& = \sqrt{G} H_{\mu\nu\rho10}\left(-G^{\mu10}G^{\nu\nu_1}G^{\rho\rho_1}G^{10\lambda_1}H_{\nu_1\rho_1\lambda_1} + G^{\mu\mu_1}G^{\nu10}G^{\rho\rho_1}G^{10\lambda_1}H_{\mu_1\rho_1\lambda_1} \right.\\
& \left.- G^{\mu\mu_1}G^{\nu\nu_1}G^{\rho10}G^{10\lambda_1}H_{\mu_1\nu_1\lambda_1} + G^{\mu\mu_1}G^{\nu\nu_1}G^{\rho\rho_1}G^{10\ 10}H_{\mu_1\nu_1\rho_1}\right)dt\wedge...dx_{10},
\end{split}
\end{eqnarray}
where $H_{\mu\nu\rho10}=H_{\mu\nu\rho}$, and
\begin{eqnarray}
\label{flux_action_D=11-iv}
& & \left(H\wedge A\right)\wedge *_{11}\left(H\wedge A\right)=\sqrt{G} H_{[\mu\nu\rho}A_{\lambda]}G^{\mu\mu_1}G^{\nu\nu_1}G^{\lambda\lambda_1}H_{[\mu_1\nu_1\rho_1}A_{\lambda_1]},
\end{eqnarray}
with
$H_{[\mu_1\mu_2\mu_3}A_{\mu_4]}\equiv H_{\mu_1\mu_2\mu_3}A_{\mu_4} - \left(H_{\mu_2\mu_3\mu_4}A_{\mu_1} - H_{\mu_3\mu_4\mu_1}A_{\mu_2} + H_{\mu_4\mu_1\mu_2}A_{\mu_3}\right)$.

Now, working near $\theta_1=N^{-1/5}$ and $\theta_2=N^{-3/10}$ and using the results of \cite{MQGP} for $H_3$, $A_1$ and the inverse of $11$-dimensional metric we calculate at leading order in $N$, the following flux-generated cosmological constant term:
\begin{equation}
\label{cc-G4squaredoversqrtGM}
G_{MNPQ}G^{MNPQ} =\frac{\mathcal{A}(r)}{N^{7/10}},
\end{equation}
 where $\mathcal{A}(r)$ is given in (\ref{cc}).
After performing the integration on the six compact coordinated the M-theory action (\ref{Mtheoryaction}) reduces to $5$-dimensions:
\begin{equation}\label{Mtheory5daction}
S_{M}\sim\int_{M}d^{5}x~\sqrt{G^{5}}\left( R+Cosmological ~constant ~term\right),
\end{equation}
\begin{itemize}
\item{\bf{Metric Fluctuations:}}
The background metric $g^{(0)}_{\mu\nu}$ is linearly perturbed as $g_{\mu\nu}=g^{(0)}_{\mu\nu}+h_{\mu\nu}$. Now we assume the perturbation to have the form: $h_{\mu\nu}=\epsilon_{\mu\nu}(r)e^{ikx_1}$. Clearly there is a $SO(2)$ rotational symmetry in the $x_2-x_3$ plane which allow us to classify different perturbations into three categories, namely tensor, vector and scalar type of metric perturbations.
\end{itemize}
\subsection{$0^{++}$ Glueball spectrum}

The $0^{++}$ glueball in M-theory corresponds to scalar metric perturbations \cite{Brower}:
\begin{eqnarray}
\label{0++_M_i}
& & h_{tt} = g_{tt} e^{i q x_1} q_1(r),\nonumber\\
& & h_{x_1r} = h_{rx_1} = i q~ g_{x_1x_1} e^{i q x_1} q_3(r),\nonumber\\
& & h_{rr} = g_{rr} e^{i q x_1} q_2(r),
\end{eqnarray}
where $g_{tt}$, $g_{x_ix_i}$ and $g_{rr}$ are the M-theory metric components, where we work only with the leading order terms in each.
\subsubsection{WKB approximation}
\begin{itemize}
\item{\it{Background with a black hole}}
\end{itemize}
Taking into account the above perturbation, we get the following differential equation for $q_3(r)$,
\begin{eqnarray}
\label{q3-EOM}
{q_3}''(r)+G(r){q_3}'(r)+H(r){q_3}(r)=0,
\end{eqnarray}
where the expression for $G(r)$ and $H(r)$ upto leading order in $N$ is given in (\ref{GH}).

Writing the above equation (\ref{q3-EOM}) as a Schrodinger like form one can read off the potential term with $m=\tilde{m}\frac{\sqrt{y_h}}{L^2}$ which in the IR region $(z\ll 0)$ is given as,
\begin{equation}
\label{V0++-WKB-IR}
V_{IR}(z)\sim{e^z} \left(-0.006 \tilde{m}^4-0.07\tilde{m}^2\right)+0.002 \tilde{m}^4+0.003 \tilde{m}^2-0.25 + {\cal O}\left(e^{2z}\right),
\end{equation}
where to simplify calculation we have used $a=0.6 ~r_h$ and set $g_s=0.9, N\sim(g_s)^{-39}\sim100, N_f=2$ - in the MQGP limit of \cite{MQGP}.

The potential (\ref{V0++-WKB-IR}) is found to be positive for the domain $z\in(-\infty,-2.526]$ and the same yields:
\begin{equation}
\label{WKB0++-IR}
\int_{-\infty}^{-2.62}\sqrt{V_{IR}(z)} = 0.09 \tilde{m}^2 \log (\tilde{m})-0.04 \tilde{m}^2=\pi  \left(n+\frac{1}{2}\right),
\end{equation}
to get,
\begin{equation}
\label{mn0++-WKB}
m_n^{0^{++}}= \frac{\sqrt{70 n+35}}{\sqrt{{\cal PL}(26.30 n+13.15)}}.
\end{equation}

In the UV region, one can show that,
\begin{equation}
\label{VUV}
V_{UV}(z)\sim \frac{-0.007\tilde{m}^4+0.3 \tilde{m}^2+1.62}{{e^{2z}}}-\frac{0.08 \tilde{m}^2+0.54}{{e^z}}-0.25,
\end{equation}
which remains negative for all values of $z$, implying no turning points in the UV.
\begin{itemize}
\item{\it{Background with an IR cut-off}}
\end{itemize}
The equation to be considered for this case can be obtained by taking the limit $r_h\rightarrow 0$ and hence $a\rightarrow 0$ in (\ref{q3-EOM}), to get,
\begin{equation}
\label{0++EOM_T=0}
q_3''(r) + q_3'(r) \left(\frac{4 \pi  {g_s} m^2 N}{3 r^3}-\frac{r^4}{4 \pi  {g_s} N}+\frac{9}{r}\right)+q_3(r) \left(\frac{8 \pi  {g_s} m^2 N}{3
   r^4}-\frac{5 r^3}{4 \pi  {g_s} N}+\frac{15}{r^2}\right)=0.
\end{equation}

Again, converting the above equation into a Schrodinger like equation one the potential function which in the large-$N$ limit is negative and hence has no turning points. The WKB method does not work in this case.
\subsubsection{Neumann/Diritchlet boundary condition}
\begin{itemize}
\item{\it{Background with a black hole}}
\end{itemize}
Using $a=0.6 r_h$, equation (\ref{q3-EOM}) in the large $N$ limit, near $r=r_h$ (writing $m = \tilde{m} \frac{r_h}{L^2}$) simplifies to:
\begin{equation}
\label{EOM_r=rh-i}
{q_3}''(r) + \frac{\left(2 -0.0067 \tilde{m}^2\right) {q_3}'(r)}{r-{r_h}}+\frac{\left(0.76 -0.1 \tilde{m}^2\right)
   {q_3}(r)}{{r_h} (r-{r_h})}=0.
\end{equation}
Lets write the above equation of the following form,
\begin{equation}
\label{EOM_r=rh-ii}
h''(r) +  \frac{p}{r-{r_h}}h'(r) + \frac{s}{r-{r_h}}h(r) =0,
\end{equation}
where we have, $p=\left(2 -0.0067 \tilde{m}^2\right)$,~~$s=\frac{\left(0.76 -0.1\tilde{m}^2\right)
   }{{r_h}}$.

The solution to (\ref{EOM_r=rh-ii}) is given by:
\begin{equation}\begin{split}
\label{EOM_r=rh-iii}
h(r) &= c_1 (2 r-2 {r_h})^{p/2} (r-{r_h})^{-p/2} (-s (r-{r_h}))^{\frac{1}{2}-\frac{p}{2}} I_{p-1}\left(2 \sqrt{-s (r-{r_h})}\right)\\& + (-1)^{1-p} c_2
   (2 r-2 {r_h})^{p/2} (r-{r_h})^{-p/2} (-s (r-{r_h}))^{\frac{1}{2}-\frac{p}{2}} K_{p-1}\left(2 \sqrt{-s (r-{r_h})}\right).
   \end{split}
\end{equation}
Setting $c_2=0$, one can verify that one satisfy the Neumann boundary condition: $h^\prime(r=r_h)=0$ provided:
\begin{equation}
\label{EOM_r=rh-iv}
p = - n\in\mathbb{Z}^-\cup\{0\},
\end{equation}
implying:
\begin{equation}
\label{EOM_r=rh-v}
\tilde{m} = 12.25\sqrt{2+n}.
\end{equation}
One can similarly show that by imposing Dirichlet boundary condition: $h(r=r_h)=0$:
\begin{equation}
\label{EOM_r=rh-vi}
\tilde{m} = 12.25\sqrt{1+n}.
\end{equation}
\begin{itemize}
\item{\it{Background with an IR cut-off}}
\end{itemize}
Near the cut-off at $r=r_0$, equation (\ref{0++EOM_T=0}) with $m = \tilde{m}\frac{r_0}{L^2}$ is given by:
\begin{equation}\begin{split}
\label{EOM0++_r=r0}
& q_3''(r) + \left(\frac{4 \tilde{m}^2+108}{12 {r_0}}-\frac{\left(\tilde{m}^2+9\right) (r-{r_0})}{{r_0}^2}\right) q_3'(r)\\&+q_3(r)
   \left(\frac{8 \tilde{m}^2+180}{12 {r_0}^2}-\frac{\left(32 \tilde{m}^2+360\right) (r-{r_0})}{12 {r_0}^3}\right)=0,
   \end{split}
\end{equation}
whose solution is given by:
\begin{eqnarray}
\label{0++EOM-solution}
& & q_3(r) = e^{-\frac{2 \left(4 \tilde{m}^2+45\right) r}{3 \left(\tilde{m}^2+9\right) {r_0}}} \Biggl[c_1 H_{-\frac{2 \tilde{m}^6+71 \tilde{m}^4+828
   \tilde{m}^2+2835}{9 \left(\tilde{m}^2+9\right)^3}}\left(\frac{3 \left(\tilde{m}^2+9\right)^2 r-2 \left(2 \tilde{m}^4+37
   \tilde{m}^2+153\right) {r_0}}{3 \sqrt{2} \left(\tilde{m}^2+9\right)^{3/2} {r_0}}\right)\nonumber\\
   & & +c_2 \, _1F_1\left(\frac{2 \tilde{m}^6+71
   \tilde{m}^4+828 \tilde{m}^2+2835}{18 \left(\tilde{m}^2+9\right)^3};\frac{1}{2};\frac{\left(3 \left(\tilde{m}^2+9\right)^2 r-2 \left(2
   \tilde{m}^4+37 \tilde{m}^2+153\right) {r_0}\right)^2}{18 \left(\tilde{m}^2+9\right)^3 {r_0}^2}\right)\Biggr].\nonumber\\
   & &
\end{eqnarray}
Now using Neumann boundary condition at $r=r_0$, namely $q_3^\prime(r=r_0)=0$, it can be shown numerically/graphically that for $c_1 = -0.509 c_2$, one gets, $q_3(r=r_0,\tilde{m}\approx 4.1)=0$.

We hence estimate the ground state of $0^{++}$ from metric fluctuations in M-theory to be
$4.1 \frac{r_0}{L^2}$.
\subsection{$2^{++}$ Glueball spectrum}
To study the spectrum of spin $2^{++}$ glueball, we consider the tensor type of metric perturbations where the non-zero perturbations are given as:
\begin{equation}
\begin{split}
&h_{x_2x_3}=h_{x_3x_2}=g_{x_1x_1}H(r)e^{ikx_1} \\&h_{x_2x_2}=h_{x_3x_3}=g_{x_1x_1}H(r)e^{ikx_1}.
\end{split}
\end{equation}
This time we consider the M-theory metric components corrected upto NLO in $N$.
\subsubsection{WKB approximation}
\begin{itemize}
\item{\it{Background with a black hole}}
\end{itemize}
Considering the tensor modes of metric perturbations we obtain the following second order differential equation given by,
\begin{eqnarray}
\label{2++-EOM}
& & H''(r) + A_1(r) H'(r)+A_2(r) H(r)=0,
\end{eqnarray}
where $A_1$ and $A_2$ are given upto NLO order in $N$ in (\ref{A1A2}).

The potential term in the schrodinger like equation for $2^{++}$ glueball can be obtained from (\ref{2++-EOM}) and in the IR region $(z\ll0)$ with $m=\tilde{m}\frac{\sqrt{y_h}}{L^2}$ it is given by,
\begin{equation}
\label{WKB_2++_ii}
V_{IR}(z) = e^z \left(0.52 -0.01 \tilde{m}^2\right)+\left(0.15 \tilde{m}^2-1.02\right) e^{2 z} + {\cal O}\left(\frac{g_s M^2}{N},e^{3z}\right),
\end{equation}
where like before, we have used $a=0.6 ~r_h$ and set $g_s=0.9, N\sim(g_s)^{-39}\sim100, N_f=2$ - in the MQGP limit of \cite{MQGP}.

Now the turning points are found to be at,
\begin{eqnarray}
z_1=-2.71,~~~~~~~~~~~~~ z_2=-2.53.
\end{eqnarray}
The WKB quantization with these turning points gives: $\int_{z_1}^{z_2}\sqrt{V_{IR}(z)}~dz\approx 0$.
Hence, the IR does not contribute to the $2^{++}$ glueball spectrum.

In the UV, we must consider the limit $(z\rightarrow\infty)$. Moreover, in the UV set $N_f=M=0$ to get the following potential,
\begin{equation}
\label{WKB_2++_iii}
V_{UV}(z) =1+ e^{-2 z} \left(6.56 -1.02 \tilde{m}^2\right)+\left(0.25 \tilde{m}^2-2.77\right) e^{-z} + {\cal O}\left(\frac{1}{\tilde{m}},e^{-3z}\right),
\end{equation}
where we set the same numerical values for $a$, $g_s$ and $N$ as above.

The turning points are: $\left\{z_1=\log{4.08} + {\cal O}\left(\frac{1}{\tilde{m}^2}\right),~~z_2=\infty\right\}$, giving the WKB quantization as:
\begin{eqnarray}
\label{WKB_2++_v}
\int_{z_1}^{z_2}\sqrt{V_{UV}(z)}~dz = 0.39 \tilde{m}  = \left(n + \frac{1}{2}\right)\pi,
\end{eqnarray}
%This demonstrates the isospectrality of $0^{++}$ and $2^{++}$ as in \cite{Mathur_et_al}.
implying:
\begin{equation}
\label{mn2++_Minahan_T}
m_n^{2^{++}}(T) = 8.08\left(n + \frac{1}{2}\right)\frac{r_h}{L^2}.
\end{equation}
\begin{itemize}
\item{\it{Background with an IR cut-off}}
\end{itemize}
Considering the limit $(r_h,a\rightarrow0)$ equation (\ref{2++-EOM}) is given by,
\begin{equation}
\label{2++r0}
\begin{split}
H''(r)+A_3(r)H'(r)+A_4(r)H(r)=0,
\end{split}
\end{equation}
where $A_3$ and $A_4$ are given in (\ref{A3A4}).

The `potential' term, in the IR region,  up to leading order in $N$  with $ m = \tilde{m}\frac{\sqrt{y_0}}{L^2}$ is given as:
\begin{eqnarray}
\label{V2++IR_T=0}
& & V_{IR}(z) = \frac{1}{4} \left({\tilde{m}}^2+5\right) e^{2 z}-\frac{1}{4} + {\cal O}\left(e^{3z}\right),
\end{eqnarray}

whose turning points are given as $\left\{z_1=-\log \tilde{m},~~z_2=\log(\delta^2-1)\right\}$.

Hence WKB quantization condition gives,
\begin{equation}
\label{WKB-integral-2++_T=0}
\int_{z_1}^{z_2}\sqrt{\frac{1}{4} \left({\tilde{m}}^2+5\right) e^{2 z}-\frac{1}{4} } = \frac{(\delta^2-1)}{2}\tilde{m}  - 0.78 + {\cal O}\left(\frac{1}{\tilde{m}^3}\right) = \left(n + \frac{1}{2}\right)\pi,
\end{equation}
 implying:
\begin{equation}
\label{WKB_IR}
m_n^{2^{++}}(IR) =m_n^{0^{--}}(IR).
\end{equation}
In the UV, we have:
\begin{eqnarray}
\label{V2++_UV_T=0}
& & V_{UV}(z) = \frac{1}{4} \left({\tilde{m}}^2-10\right) e^{-z}-\frac{3}{4} \left({\tilde{m}}^2-5\right) e^{-2 z}+1 + {\cal O}(e^{-3z})\nonumber\\
& & = \frac{e^{-z}}{2}\sqrt{e^z - 3}\tilde{m} + {\cal O}\left(e^{-3z},\frac{1}{\tilde{m}}\right),
\end{eqnarray}
and the turning points are: $z_1=\log{3} + {\cal O}\left(\frac{1}{\tilde{m}^2}\right),~~z_2=\infty$.

Therefore we get,
\begin{equation}
\int_{z_1}^{z_2}\sqrt{V_{UV}(z)}~dz = \frac{\pi \tilde{m}}{8\sqrt{3}}
\end{equation}
This implies,
\begin{equation}
\label{WKB2++_IR+T=0}
m_n^{2^{++}}(UV) = \left(3.46 + 6.93\right)\frac{r_0}{L^2}.
\end{equation}

Let us try to include the NLO term-in-$N$ also in the potential. Including the same in the IR region the potential with $ m = \tilde{m}\frac{\sqrt{y_0}}{L^2}$ is given by:
\begin{equation}
\begin{split}
\label{V-NLO_IR_2++_T=0}
& V^{(NLO)}_{IR}(z) =-\frac{1}{4}+\Biggl\{480   M^2 g_s \pi+128 \left(\tilde{m}^2+5\right) N \pi ^2+60 M^2 g^2_s N_f \left(12-\log {16}-\log {N}\right)\\&+ M^2 g_s\Biggl(96   \tilde{m}^2 \pi+ g_s N_f\left\{720+12 \tilde{m}^2 (6-\log {16}-\log {N})\right\}\Biggr)\log{y_0}+72 M^2 g^2_s  N_f\tilde{m}^2  \log ^2{y_0}\Biggr\},
\end{split}
\end{equation}
whose turning points are given by,
\begin{equation}\begin{split}
z_1&=\log\left\{\left(\frac{1}{\tilde{m}}\right)\left(1-\frac{ M^2 g_s \left(96 \pi +12{g_s} {N_f} \left\{6-12 \log {16}- \log {N}+6\log {y_0}\right\}
  \right)\log {y_0}}{256 \pi ^2 N}\right)\right\},\\&z_2=\log\left(\delta^2-1\right).
  \end{split}
\end{equation}
The integral $\left(\int_{z_1}^{z_2}\sqrt{V^{(NLO)}_{IR}(z)}~dz\right)$, in the large-$\tilde{m}$ limit, yields the same spectrum as $0^{--}$ up to NLO in $N$.
\subsubsection{Neumann/Diritchlet boundary condition}
\begin{itemize}
\item{\it{Background with a black hole}}
\end{itemize}
Near $r=r_h$, the solution to the above equation will be given on the same lines as {\bf 5.1} for $0^{--}$ glueballs, and the
analog of (\ref{0--vii}) is:
\begin{eqnarray}
\label{2++-solutioni}
& & \frac{T^2 \left(\frac{1.5 \sqrt{0.05 T^2-0.002 m^2}}{T}-0.11\right)+0.0002 m^2}{T \sqrt{0.05 T^2-0.002 m^2}}=-n,
\end{eqnarray}
the solutions to which are given as:
\begin{eqnarray}
\label{2++M-spectrum}
& & m_{2^{++}} = 5.086 T \nonumber\\
& & m_{2^{++}}^* = 5.269 T \nonumber\\
& & m_{2^{++}}^{**} = 5.318 T \nonumber\\
& & m_{2^{++}}^{***} = 5.338 T \nonumber\\
& & m_{2^{++}}^{****} = 5.348 T \nonumber\\
\end{eqnarray}
One can impose Dirichlet boundary condition: $H(r=r_h)=0$, and show that,

~~~~~~~~~~~~~~~~~~~~~~$m_n^{2^{++}\ {(\rm Neumann)}} = m_{n+1}^{2^{++}\ {(\rm Dirichlet)}}$, for $n=0,1,2,..$.
\begin{itemize}
\item{\it{Background with an IR cut-off}}
\end{itemize}
Up to Leading Order in $N$ near $r=r_0$, equation (\ref{2++r0}) with $m = \tilde{m}\frac{r_0}{L^2}$ is given by,
\begin{equation}
\label{2++-EOM_T=0}
H''(r)+\left(\frac{5}{{r_0}}-\frac{5 (r-{r_0})}{{r_0}^2}\right) H'(r)+H(r)
   \left(\frac{\tilde{m}^2+8}{{r_0}^2}-\frac{4 \left(\tilde{m}^2+4\right)
   (r-{r_0})}{{r_0}^3}\right)=0.
\end{equation}
The solution of (\ref{2++-EOM_T=0}) is given by,
\begin{eqnarray}
\label{solution_2++_T=0}
& &  H(r) = e^{-\frac{4 \left(\tilde{m}^2+4\right) r}{5 {r_0}}} \Biggl\{c_1 H_{\frac{1}{125} \left(16 \tilde{m}^4+53
   \tilde{m}^2+56\right)}\left(\frac{2 \left(4 \tilde{m}^2-9\right) {r_0}+25 r}{5 \sqrt{10}
   {r_0}}\right)\nonumber\\
   & & +c_2 \, _1F_1\left(\frac{1}{250} \left(-16 \tilde{m}^4-53
   \tilde{m}^2-56\right);\frac{1}{2};\frac{\left(25 r+2 \left(4 \tilde{m}^2-9\right) {r_0}\right)^2}{250
   {r_0}^2}\right)\Biggr\}.
\end{eqnarray}
The Neumann boundary condition $H'(r=r_0)=0$, numerically yields that for $c_1=-0.509 c_2$, giving the lightest $2^{++}$ glueball has a mass $1.137 \frac{r_0}{L^2}$. Similarly, by imposing Dirichlet boundary condition: $H(r=r_h)=0$, for $c_1=-0.509 c_2$, the lightest $2^{++}$ glueball has a mass $0.665 \frac{r_0}{L^2}$.
\subsection{Spin-$1^{++}$ Glueball spectrum}
Here we need to consider the vector type of metric perturbation with the non-zero components given as:
\begin{equation}
h_{ti}=h_{it}=g_{x_1x_1}G(r)e^{ikx_1}, i={x_2,x_3}.
\end{equation}
\subsubsection{WKB approximation}
\begin{itemize}
\item{\it{Background with a black hole}}
\end{itemize}
Using the above ansatz for the perturbation the differential equation in $G(r)$ is given with $k^2=-m^2$ as,
\begin{eqnarray}
\label{1++_i}
\nonumber\\
   & & \hskip -0.6in G''(r) +B_1(r)G'(r)+B_2(e)G(r)  = 0,
\end{eqnarray}
where $B_1$ and $B_2$ are given upto NLO in $N$ in (\ref{B1B2}).

The `potential' $V$ in the Schr\"{o}dinger-like equation  working with the dimensionless mass variable $\tilde{m}$ defined via: $ m = \tilde{m}\frac{r_h}{L^2}=\tilde{m}\frac{\sqrt{y_h}}{L^2}$, in the IR, can be shown to be given by:
\begin{equation}
\label{1++_iii}
V_{IR}(z)= {e^{2z}} \left(0.15 \tilde{m}^2-1.52\right)+{e^z} \left(1-0.01 \tilde{m}^2\right)-\frac{1}{4} + {\cal O}\left(e^{3z},\frac{1}{N}\right).
\end{equation}
From the above potential we find the allowed domain of integration as, $\{z_1=\log (0.07),~~z_2=-2.526\}$. Thus, in the IR:
\begin{equation}
\label{1++_iv}
\int_{z_1}^{z_2} \sqrt{{e^{2z}} \left(0.15 \tilde{m}^2-1.52\right)+{e^z} \left(1-0.01 \tilde{m}^2\right)-\frac{1}{4}}~dz\approx 0,
\end{equation}
implying a null contribution to the WKB quantization condition in the IR.

In the UV $(z\gg1)$, the potential for $a=0.6 r_h$ and $M=N_f=0$ is given by,
\begin{equation}
\label{1++_vi}
\hskip 0.35in V_{UV}(z) = e^{-2 z} \left(6.56 -1.02 \tilde{m}^2\right)+\left(0.25 \tilde{m}^2-2.77\right) e^{-z}+1 + {\cal O}(e^{-3z}).
\end{equation}
The allowed domain of integration over which the potential is positive, is: $([\log (4.08),\infty)$. Performing a large-$\tilde{m}$-expansion, one obtains:
\begin{eqnarray}
\label{1++_vii}
& & \int_{\log (4.08)}^\infty\sqrt{e^{-2 z} \left(6.56 -1.02 \tilde{m}^2\right)+\left(0.25 \tilde{m}^2-2.77\right) e^{-z}+1}~dz\nonumber\\& &=\int_{\log (4.08)}^\infty e^{-z}\sqrt{0.25 e^z - 1.02} + {\cal O}\left(\frac{1}{\tilde{m}}\right)~dz\nonumber\\
& &  = 0.389 \tilde{m} = \left(n + \frac{1}{2}\right)\pi,
\end{eqnarray}
yielding:
\begin{equation}
\label{1++_viii}
m_n^{1^{++}}(T) = 4.04\left(1 + 2n\right)\frac{r_h}{L^2}.
\end{equation}
\begin{itemize}
\item{\it{Background with an IR cut-off}}
\end{itemize}
Considering the limit of $(r_h,a)\rightarrow 0$ in equation (\ref{1++_i}), the potential obtained from the schrodinger like equation is given as,
\begin{equation}
\label{VLON_1++_T=0}
V(z) = \frac{\left(\tilde{m}^2+2\right) e^{2 z}-3 e^z+4 e^{3 z}-1}{4 \left(e^z+1\right)^3} +
{\cal O}\left(\frac{g_s M^2}{N}\right).
\end{equation}
In the IR region $(z\ll0)$ we get the potential as:
\begin{equation}
\label{V1++_T=0_IR}
V_{IR}(z) = - \frac{1}{4} + \frac{1}{4}(5 + \tilde{m}^2)e^{2z} + {\cal O}(e^{3z}),
\end{equation}
giving the turning points as, $\{z_1=-\log \tilde{m},~~z_2=\log(\delta^2-1)\}$ and the WKB quantization condition becomes:
\begin{equation}
\label{WKB_1++_T=0}
\int_{z_1}^{z_2}\sqrt{V_{IR}(z)}~dz = \frac{(\delta^2-1)}{2} \tilde{m} - \frac{\pi}{4} = \left(n + \frac{1}{2}\right)\pi.
\end{equation}
Therefore:
\begin{equation}
\label{m1++_T=0_IR}
m_n^{1^{++}}(IR,r_h=0) = m_n^{2^{++}}(IR,r_h=0) = m_n^{0^{--}}(IR,r_h=0).
\end{equation}

Further, in the UV $(z\gg1)$ the potential is,
\begin{equation}
\label{V1++_T=0_UV}
V_{UV}(z) = \frac{1}{4} \left(\tilde{m}^2-10\right) e^{-z}-\frac{3}{4} \left(\tilde{m}^2-5\right) e^{-2 z}+1 + {\cal O}\left(e^{-3z}\right),
\end{equation}
with the turning points:$\{z_1=\log{3}+ {\cal O}\left(\frac{1}{\tilde{m}^2}\right),~~z_2=\infty\}$.

 This yields the following WKB quantization condition:
\begin{equation}
\label{WKB1++_T=0}
\int_{z_1}^{z_2}\sqrt{V_{UV}(z)}~dz = \frac{\pi\tilde{m}}{4\sqrt{3}} = \left(n + \frac{1}{2}\right)\pi,
\end{equation}
which obtains:
\begin{equation}
\label{mn1++_T=0_UV}
m_n^{1^{++}}(UV,r_h=0) = \left(3.46 + 6.93 n\right)\frac{r_0}{L^2}.
\end{equation}
\subsubsection{Neumann/Diritchlet boundary condition}
\begin{itemize}
\item{\it{Background with a black hole}}
\end{itemize}
Near $r=r_h$, equation (\ref{1++_i}) up to LO in $N$, is given by:
\begin{equation}
\label{Neumann_1++_T=0}
G''(r) + \left(\frac{3.92}{r_h}\right)G'(r) + \left(\frac{2 - 0.02 \tilde{m}^2}{r_h(r-r_h)} + \frac{-1.16 + 0.57 \tilde{m}^2}{r_h^2}\right)G'(r) = 0,
\end{equation}
whose solution is given by,
{\footnotesize
\begin{eqnarray}
\label{Neumann1++_T=0-solution}
& & G(r) =  e^{\left(\frac{0.5 r \left(-2.
   \sqrt{5 -0.57 \tilde{m}^2}-3.92\right)+ {r_h} \log (r-{r_h})}{{r_h}}\right)}\nonumber\\
   & & \times\Biggl[ c_1 U\left(-\frac{-0.01 \tilde{m}^2-\sqrt{5 -0.57 \tilde{m}^2}+1}{\sqrt{5 -0.57 \tilde{m}^2}},2,\frac{2.
   \sqrt{5.0016 -0.57 \tilde{m}^2} r}{{r_h}}-2. \sqrt{5 -0.57 \tilde{m}^2}\right)\nonumber\\
    & & +c_2 L_{\frac{-1. \sqrt{5
   -0.57 \tilde{m}^2}-0.01 \tilde{m}^2+1}{\sqrt{5 -0.57 \tilde{m}^2}}}^{1}\left(\frac{2 r \sqrt{5 -0.57
   \tilde{m}^2}}{{r_h}}-2. \sqrt{5 -0.57 \tilde{m}^2}\right) \Biggr].
\end{eqnarray}}
Imposing Neumann boundary condition at $r=r_h$ and then using $\lim_{z\rightarrow0} U(p,2,z\sim0)\sim\frac{z^{-1}
\ _1F_1(p-1;0;z)}{\Gamma(p)}$, one notes that one can satisfy the Neumann boundary condition at $r=r_h$ provided $\lim_{z\rightarrow0}\ _1F_1(p-1;0;z) = \lim_{c\rightarrow0}\lim_{z\rightarrow0}\ _1F_1(p-1;c;z)$ (i.e. first set $z$ to 0 and then $c$), $p=-n\in\mathbb{Z}^-$. Hence:
\begin{eqnarray}
\label{1++_spectrum_Neumann_T}
& & m^{1^{++}}(T) = 2.6956 \pi T\nonumber\\
& & m^{1^{++*}}(T) = 2.8995 \pi T\nonumber\\
& & m^{1^{++**}}(T) = 2.9346 \pi T\nonumber\\
& & m^{1^{++***}}(T) = 2.9467 \pi T.
\end{eqnarray}
One can show that one obtains the same spectrum as (\ref{1++_spectrum_Neumann_T}) even upon imposing Dirichlet boundary condition: $G(r=r_h)=0$.
\begin{itemize}
\item{\it{Background with an IR cut-off}}
\end{itemize}
Considering the limit of $(r_h,a)\rightarrow 0$ in equation (\ref{1++_i}) up to LO in $N$ and imposing Neumann boundary condition at the IR cut-off $r=r_0$, yields isospectrality with $2^{++}$ glueball spectrum.
\section{$2^{++}$ Glueball Masses from Type IIB}
The $10$-dimensional type IIB supergravity action in the low energy limit is given by,
\begin{equation}\label{action}
\frac{1}{2k_{10}^2}\left\{\int d^{10}x~ e^{-2\phi}\sqrt{-G}\left(R-\frac{1}{2}H_3^2\right)-\frac{1}{2}\int d^{10}x
~\sqrt{-G}\left(F_1^2+\widetilde{F_3^2}+\frac{1}{2}\widetilde{F_5^2}\right)\right\},
\end{equation}
where $\phi$ is the dilaton, $G_{MN}$ is the $10$-d metric and $F_1$, $H_3$, $\widetilde{F_3}$, $\widetilde{F_5}$ are different fluxes.

The five form flux $\widetilde{F_5}$ and the three form flux $\widetilde{F_3}$ are defined as,
\begin{equation}\label{F5F3tilde}
\widetilde{F_5}=F_5+\frac{1}{2}B_2\wedge F_3 ,~~~~~~~~~~~~~~~~~~~~            \widetilde{F_3}=F_3-C_0\wedge H_3,
\end{equation}
where $F_5$ and $F_3$ are sourced by the $D_3$ and $D_5$ branes respectively. $B_2$ is the NS-NS two form and $C_0$ is the axion. The three form fluxes $\widetilde{F_3}$, $H_3$, the two form $B_2$ and the axion $C_0$ are given as \cite{metrics} - see (\ref{three-form-fluxes}). Now varying the action in (\ref{action}) with respect to the metric $g_{\mu\nu}$ one get the following equation of motion,
\begin{equation}\label{ricci scalar}
\begin{split}
R_{\mu\nu} & =\left(\frac{5}{4}\right)e^{2\phi}\widetilde{F}_{\mu p_2p_3p_4p_5}\widetilde{F}_{\nu}^{p_2p_3p_4p_5}-\left(\frac{g_{\mu\nu}}{8}\right)e^{2\phi}\widetilde{F}_{5}^2+\left(\frac{3}{2}\right)H_{\mu\alpha_2\alpha_3}
H^{\alpha_2\alpha_3}_{\nu}\\ & -\left(\frac{g_{\mu\nu}}{8}\right)H_3^2
 +\left(\frac{3}{2}\right)e^{2\phi}\widetilde{F}_{\mu\alpha_2\alpha_3}
\widetilde{F}^{\alpha_2\alpha_3}_{\nu}
-\left(\frac{g_{\mu\nu}}{8}\right)e^{2\phi}\widetilde{F}_{3}^2+\left(\frac{1}{2}\right)e^{2\phi}F_{\mu}F_{\nu}.
\end{split}
\end{equation}
 we consider the following linear perturbation of the metric,
\begin{equation}
g_{\mu\nu}=g_{\mu\nu}^{(0)}+h_{\mu\nu},
\end{equation}
Here the only non zero component according to the tensor mode of metric fluctuation is $h_{x_2x_3}$. The final equation of motion in terms of the perturbation is given as,
\begin{equation}\label{final EOM}
\begin{split}
R^{(1)}_{x_2x_3} & =\left(\frac{5}{4}\right)e^{2\phi}\left(4\widetilde{F}_{x_2 x_3p_3p_4p_5}\widetilde{F}_{x_2x_3 q_3q_4q_5}g^{p_3q_3}g^{p_4q_4}g^{p_5q_5}h^{x_2x_3}\right)-\left(\frac{h_{x_2x_3}}{8}\right)e^{2\phi}\widetilde{F}_{5}^2
\\&-\left(\frac{h_{yz}}{8}\right)H_3^2
-\left(\frac{h_{x_2x_3}}{8}\right)e^{2\phi}\widetilde{F}_{3}^2.
\end{split}
\end{equation}
Writing the perturbation $h_{x_2x_3}$ as $h_{x_2x_3}=\frac{r^2}{2 (g_{s}\pi N)^{1/2}} H(r)e^{i k x_1}$, (\ref{final EOM}) reduces to the following second order differential equation in $H(r)$,
\begin{eqnarray}
\label{EOM}
& & \hskip -0.4in H^{\prime\prime}(r)+C_1(r)H^{\prime}(r)+C_2(r)H(r)=0,
\end{eqnarray}
where $C_1$ and $C_2$ are given in (\ref{C1C2}).
\subsubsection{WKB approximation}
\begin{itemize}
\item{\it{Background with a black hole}}
\end{itemize}
The `potential' at leading order in $N$, defining $m = \tilde{m}\frac{\sqrt{y_h}}{L^2}$, yields,
\begin{equation}\begin{split}
\label{V_2++_IIB-i}
V(z) &= \frac{1}{4 \left(e^z+1\right)^3 \left(e^z+2\right)^2}\Biggl\{e^z \Biggl(3 b^2 \left(e^z+2\right) \left(-\left(\tilde{m}^2-6\right) e^z-\tilde{m}^2+3 e^{2 z}+6\right)\\& +\left(-e^z-1\right)
   \left(\left(25-3 \tilde{m}^2\right) e^z-\left(\tilde{m}^2-18\right) e^{2 z}-2 \left(\tilde{m}^2-6\right)+4 e^{3
   z}\right)\Biggr)\Biggr\} + {\cal O}\left(\frac{g_s M^2}{N}\right).
   \end{split}
\end{equation}

In the IR $(z\ll0)$, the potential with $a=0.6 r_h$ is given by:
\begin{eqnarray}
\label{V-IR}
& & V_{IR}(z) = e^z \left(\left(0.15 \tilde{m}^2-1.33\right) e^z-0.01 \tilde{m}^2+0.06\right) + {\cal O}(e^{3z}),
\end{eqnarray}
with the turning points: $\{z_1=-2.708,~~z_2=-2.526\}$, giving,
\begin{eqnarray}
\int_{z_1}^{z_2}\sqrt{V_{IR}(z)}~dz\approx0,
\end{eqnarray}
hence the IR provides no contribution to the WKB quantization.

In the UV $(z\gg1)$ one get,
\begin{equation}
\label{V_UV_T}
 V_{UV}(z) = \frac{1}{4} \left(\tilde{m}^2+9.24\right) e^{-z}-\frac{3}{4} \left(\tilde{m}^2+0.36 \left(\tilde{m}^2+9\right)+3\right) + {\cal O}\left(e^{-3z}\right),
   e^{-2 z}-1
\end{equation}
the turning points are $\left\{z_1=\log{4.08} + {\cal O}\left(\frac{1}{\tilde{m}^2}\right),~~z_2=0.25\tilde{m}^2 - 1.77\right\}$. Hence,
\begin{equation}\begin{split}
\label{WKB-2++_T_IIB}
&\int_{z_1}^{z_2}\sqrt{V_{UV}(z)}~dz\\&= \int_{z_1}^{z_2} \frac{e^{-z}}{2}\sqrt{e^z - 4.08} \tilde{m} + {\cal O}\left(\frac{1}{\tilde{m}}\right)\\
&= 0.389 \tilde{m} - 2 = \left(n + \frac{1}{2}\right)\pi,
\end{split}
\end{equation}
which obtains:
\begin{equation}
\label{mn2++_T_IIB}
m_n^{2^{++}}(T) = (9.18 + 8.08 n)\frac{r_h}{L^2}.
\end{equation}
Hence, the string theory $2^{++}$ glueball is isospectral with $0^{++}$; in the large $n$-limit of the spectrum, the M-theory and type IIB spectra coincide.
\begin{itemize}
\item{\it{Background with an IR cut-off}}
\end{itemize}
At leading order in $N$, the `potential' in the IR is given by:
\begin{equation}
\label{V1++_ii}
 V_{IR}(z) = -\frac{1}{4} + \frac{1}{4}\left(1 + \tilde{m}^2\right)e^{2z} + {\cal O}(e^{-3z}).
\end{equation}
The domain in the IR over which $V_{IR}(z)>$ is $[-\frac{1}{2}\log(5 + \tilde{m}^2),\log(\delta^2-1)]$ and the WKB quantization gives,
\begin{equation}
\label{WKB_1++_T=0}
\int_{-\frac{1}{2}\log(5 + \tilde{m}^2)}^{-2.526}\sqrt{-\frac{1}{4} + \frac{1}{4}\left(1 + \tilde{m}^2\right)e^{2z}} = \frac{(\delta^2-1)}{2}\tilde{m} - \frac{\pi}{4} = \left(n + \frac{1}{2}\right)\pi,
\end{equation}
yielding,
\begin{equation}
\label{mn1++-IR}
m_n^{2^{++}}(IR,IIB) = m_n^{2^{++}}(IR,M\ {\rm theory}).
\end{equation}

In the UV the potential reads,
\begin{equation}
\label{V1++-UV}
V_{UV}(z) = \frac{1}{4} \left(\tilde{m}^2-10\right) e^{-z}-\frac{3}{4} \left(\tilde{m}^2-5\right) e^{-2 z}+1 + {\cal O}\left(e^{-3z}\right).
\end{equation}
The domain of integration over which $V_{UV}(z)>0$ is: $[\log 3,\infty)$. Therefore:
\begin{equation}
\label{WKB1++-i}
\int_{\log 3}^\infty\sqrt{V(UV, r_h=0)} = \frac{1}{2}\int_{\log 3}^\infty e^{-z}\sqrt{e^z - 3}\tilde{m} + {\cal O}\left(\frac{1}{\tilde{m}}\right) = \frac{\tilde{m}\pi}{4\sqrt{3}} = \left(n + \frac{1}{2}\right)\pi,
\end{equation}
yielding:
\begin{equation}
\label{mn1++-ii}
m_n^{2^{++}} = (3.464 + 6.928 n)\frac{r_0}{L^2}.
\end{equation}
\section{Summary and Discussion}
The summary of all calculations is given in  Tables 4.1 (and Figure 4.2) and 4.3 - the former  table/graph having to do with a WKB quantization calculation using the coordinate/field redefinitions of \cite{Minahan} and the latter table having to do with obtaining the mass spectrum by imposing Neumann/Dirichlet boundary condition at $r_h$/IR cut-off $r_0$. Some of the salient features of the results are given as separate bullets.
It should be noted that the last two columns in Tables 4.1 and 4.3 have been prepared in the same spirit as the last columns in Table 2 of \cite{Boschi+Braga_AdS_BH_AdS_slice}.
\begin{table}[h]
\begin{tabular}{|c|c|c|c|}\hline
S. No. & {\scriptsize Glueball} & {\scriptsize $\tilde{m}$  using WKB $r_h\neq0$} & {\scriptsize
$\tilde{m}$ using WKB  $r_h=0$}   \\
&& {\scriptsize (units of $\pi T$, up to LO in $N$)} & {\scriptsize (units of $\frac{r_0}{L^2}$, up to NLO in $N$)}\\
&& {\scriptsize (large-$\tilde{m}$ limit)} & {\scriptsize (large-$\tilde{m}$ limit)}\\ \hline
1 & $0^{++}$ & (M theory) & (M theory)  \\
& {\scriptsize (Fluctuations: $h_{00,rx_1,rr}$}& $\frac{\sqrt{35 + 70 n}}{\sqrt{{\cal PL}(13.15 + 26.31 n)}}$ & {\scriptsize\rm No\ turning\ points} \\
& {\scriptsize in M-theory metric)} &  \mytriangle{red} & \\ \cline{2-4}
& $0^{++}$ & (Type IIB) & (Type IIB) \\
& {\scriptsize (Dilaton Fluctuations)}& $9.18 + 8.08 n$ \mytriangle{green} & {\scriptsize$ \frac{(4.64 + 6.28 n)}{(\delta^2-1)}\left[1 - 0.01 \frac{g_s M^2 }{N}(g_s N_f)\log N\log r_0\right]$} \\  \hline
%&& {\scriptsize UV (redefinition of \cite{WKB-i}): $5.7n$} & \\
2 & $0^{-+}$ & (Type IIA) & (Type IIA) \\
& {\scriptsize(1-form fluctuation $a_{\theta_2}$) }&  {\scriptsize  $11.12\left(n + \frac{1}{2}\right), n=0$} \mysquare{blue}& {\scriptsize $\frac{3.72 + 4.36 n}{(\delta^2-1)}, n=0$}  \\
& & {\scriptsize  $(6.22 + 4.80 n), n\in\mathbb{Z}^+$} \mytriangle{blue} & {\scriptsize $4.8\left(n + \frac{1}{2}\right), n\in\mathbb{Z}^+$} \\ \hline
3 & $0^{--}$ & (Type IIB) & (Type IIB) \\
&{\scriptsize 2-form fluctuation $A_{23}$}& {\scriptsize $= m_n^{0^{++}}({\rm dilaton},T)$} & {\scriptsize  $ \frac{6.28 n+4.71}{(\delta ^2-1)}\left(1 + \frac{0.01 {g_s}^2 {\log N} M^2 {N_f} \log ({r_0})}{ N}\right), n=0$} \\
&& \mytriangle{green} & {\scriptsize  $(7.87 + 6.93 n), n\in\mathbb{Z}^+$} \\ \hline
4 & $1^{++}$ & (M theory) & (M theory) \\
&{\scriptsize (Fluctuations: $h_{it}=h_{ti},i=x_{2,3}$}& {\scriptsize $8.08\left(n + \frac{1}{2}\right)$} & {\scriptsize $m_n^{1^{++}}(n=0,r_h=0) = m_n^{0^{--}}(n=0,r_h=0)$} \\
&{\scriptsize in M-theory metric)}& \mytriangle{purple} &  {\scriptsize $(3.46 + 6.93 n),n\in\mathbb{Z}^+$} \\ \hline
%&&& {\scriptsize $\stackrel{n\gg1}{\longrightarrow}m_n^{0^{--}}(UV,r_h=0)$} \\
5 & $2^{++}$ & (M theory) & (M theory) \\
&{\scriptsize (Fluctuations: $h_{x_2x_3}=h_{x_3x_2}, $}& {\scriptsize $8.08\left(n + \frac{1}{2}\right) = m_n^{1^{++}}(T)$} & {\scriptsize $=m_n^{1^{++}}(r_h=0)$} \\
& {\scriptsize $h_{x_2x_2}=h_{x_3x_3}$ in M-theory metric)} &\mytriangle{purple}&  \\ \cline{2-4}
& $2^{++}$& (Type IIB) & (Type IIB) \\
& {\scriptsize (Fluctuation $h_{x_2x_3}=h_{x_3x_2}$}& {\scriptsize $9.18 + 8.08 n = m_n^{0^{++}}(IIB,T)$} & {\scriptsize $=m_n^{1^{++}}(r_h=0)$} \\
& {\scriptsize in type IIB metric)}& \mytriangle{green} &
%$m_n^{2^{++}}(IR,M\ {\rm theory},LO,r_h=0)$
\\ \hline
%&&& {\scriptsize $= m_n^{2^{++}}(IR,IIB,LO,r_h=0)$} \\
%&&& {\scriptsize $= m_n^{0^{--}}(IR,IIB,LO,r_h=0)$}\\
%&&& {\scriptsize
%$= m_n^{1^{++}}(IR,M\ {\rm theory}, LO,r_h=0)$} \\
   \end{tabular}
   \caption{Summary of Glueball Spectra: $m = \tilde{m}\frac{r_h}{L^2}$  from Type IIB, IIA and M Theory using WKB quantization condition for $r_h\neq0$, and $m = \tilde{m} \frac{r_0}{L^2}$ for $r_h=0$ (equalities in the $r_h=0$ column, are valid up to NLO in $N$); the colored triangles/square in the third column correspond to the colored triangles/square that appear  in Fig. 4.2 in  the combined plot of $r_h\neq0$ supergravity calculations of glueballs}
   \end{table}

The $r_h\neq0$ glueball spectra is plotted in Figure 4.2.
\newpage
\begin{figure}
 \begin{center}
 \includegraphics[scale=0.8]{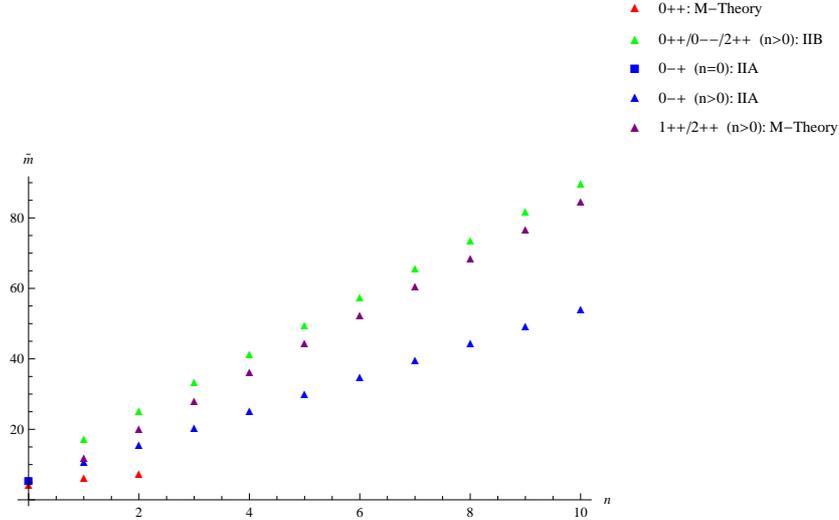}
%  \includegraphics[scale=0.8]{combinedplot_rhneq0-Karunava.jpg}
 %[height= 21cm,width=+15cm]
% \includepdf[pages=-]{combinedplot_rhneq0-Karunava.pdf}
 \end{center}
\caption{Combined Plots of BH Supergravity  Glueball Spectra}
\end{figure}
%\newpage

Some of the salient features of Table 4.1 and Figure 4.2 are presented below:
\begin{enumerate}
\item
Interestingly, via a WKB quantization condition using coordinate/field redefinitions of \cite{Minahan}, the lightest $0^{++}$ glueball spectrum for $r_h\neq0$ coming from scalar metric fluctuations in M theory compares rather well with the $N\rightarrow\infty$ lattice results of \cite{hep-lat/9804008} - refer to Table 4.2. Also, similar to \cite{Brower}, the $0^{++}$ coming from the scalar fluctuations of the M theory metric  is lighter than the $0^{++}$ coming from type IIB dilaton fluctuations. Further, interestingly, one can show that by using the coordinate and field redefinitions of \cite{WKB-i} when applied to the EOM for dilaton fluctuation to yield a WKB quantization condition, for $a=0.6 r_h$ - as in \cite{EPJC-2} - one obtains a match with the UV limit of the $0^{++}$ glueball spectrum as obtained in \cite{Minahan}. For our purpose, the method based on coordinate/field redefinitions of \cite{WKB-i}, is no good for obtaining the $0^{++}$ glueball ground state and was not used for any other glueball later on in subsequent calculations in this thesis.
\begin{table}[h]
\begin{tabular}{|c|c|c|c|}\hline
{\scriptsize State} & $N\rightarrow\infty$ {\scriptsize Entry in Table 34} of \cite{hep-lat/9804008} & {\scriptsize M-theory scalar metric perturbations}  & {\scriptsize Type IIB Dilaton  fluctuations of} \cite{Ooguri_et_al}\\
& {\scriptsize in units of square root of} & ({\bf 6.1.2} - {\scriptsize in units of} & {\scriptsize in units of reciprocal of} \\
& {\scriptsize string tension} & $\frac{r_h}{L^2}$) & {\scriptsize temporal circle's diameter} \\ \hline
$0^{++}$ & $4.065\pm0.055$ & 4.267 & 4.07 ({\scriptsize normalized to match lattice}) \\ \hline
$0^{++*}$ & $6.18\pm0.13$ & 6.251 & 7.02 \\ \hline
$0^{++**}$ & $7.99\pm0.22$ & 7.555 & 9.92 \\ \hline
$0^{++**}$ & - & 8.588 & 12.80 \\ \hline
$0^{++***}$ & - & 9.464 & 15.67 \\ \hline
\end{tabular}
\caption{Comparison of \cite{hep-lat/9804008}'s $N\rightarrow\infty$ lattice results for $0^{++}$ glueball with our supergravity results obtained  using WKB quantization condition and redefinitions of \cite{Minahan} for M theory scalar metric fluctuations}
\end{table}

\item
Also, from Table 4.1/Figure 4.2, $m_{n>0}^{2^{++}}>m_{n>0}^{0^{++}}$(scalar metric perturbations), similar to
\cite{Brower}.

\item
The higher excited states of the type IIA $0^{-+}$ glueball, for both $r_h\neq0$ and $r_h=0$, are isospectral. This is desirable because large-$n$ corresponds to the UV and that takes one away from the BH geometry, i.e., towards $r_h=0$.

\item
The non-conformal corrections up to NLO in $N$, have a semi-universal behavior of $\frac{(g_sM^2)(g_sN_f)\log r_0}{N}$ and turn out to be multiplied by a numerical pre-factor of ${\cal O}(10^{-2})$; we could disregard the same in the MQGP limit.

\item
As per a more recent lattice calculation \cite{hep-lat/0510074}\footnote{We thank P.Majumdar for bringing this reference to our attention.}, the $0^{++}$-glueball has a mass $4.16\pm0.11\pm0.04$ (in units of the reciprocal of the `hadronic scale parameter' of \cite{Sommer-r0}), which compares rather well with $m_{n=0}^{0^{++}}=4.267$ (in units of $\frac{r_h}{L^2}$) of Table 4.2 coming from scalar fluctuations of the M theory metric. Similarly, the $0^{-+}$-glueball in \cite{hep-lat/0510074} has a mass $6.25\pm0.06\pm0.06$, which matches rather nicely with  $m_{n=0}^{0^{-+}}(\delta=1.26)=6.25$ (in units of $\frac{r_0}{L^2}$) of Fig. 4.2 coming from type IIA one-form fluctuation.

\item
The ground state and the $n\gg1$ excited states of $1^{++}$ and $0^{--}$ glueballs are isospectral.

\item
The higher excited $r_h\neq0$ $2^{++}$ glueball states corresponding to metric fluctuations of the M-theory metric and the ones corresponding to fluctuations of the type IIB metric, are isospectral. The $r_h=0$ $2^{++}$ glueball states corresponding to metric fluctuations of the M-theory/type IIB string theory, are isospectral. Further, it turns out that due to internal cancellation of terms and $\frac{1}{\tilde{m}}$-suppression, a type IIB $r_h=0$ $2^{++}$ glueball spectrum, unlike an M-theoretic computation, is unable to capture the NLO-in-$N$ corrections to the LO-in-$N$ type IIB $2^{++}$ glueball spectrum.

\item
$m^{2^{++}}_n({\rm NLO},r_h=0) = m_n^{1^{++}}({\rm NLO},r_h=0)\stackrel{n\gg1}{\longrightarrow}m_n^{0^{--}}({\rm NLO},r_h=0)$, where the `NLO' implies equality with the inclusion of NLO-in-$N$ corrections.

\end{enumerate}
\newpage
\begin{table}[h]
\begin{tabular}{|c|c|c|c|}\hline
S. No. & Glueball & Spectrum Using   & Spectrum Using     \\
&& N(eumann)/D(irichlet) & N(eumann)/D(irichlet) \\
&& b.c., $r=r_h$(units of $\pi T$) & b.c., $r=r_0$(units of $\frac{r_0}{L^2}$)\\ \hline
1 & $0^{++}$ & (M theory) & (M theory)  \\
&& (N) {\scriptsize $12.25\sqrt{2+n}$} & (N) 4.1 \\
&& (D) {\scriptsize $12.25\sqrt{1+n}$} & \\ \hline
2 & $0^{-+}$ & (Type IIA) & (Type IIA) \\
& & (N/D) {\scriptsize $\frac{3.1}{\pi}\sqrt{n}$} & (N) {\scriptsize $m_{n=0}^{0^{-+}}=0, m_{n=1}^{0^{-+}}\approx 3.4, m_{n=2}^{0^{-+}}\approx 4.35$} \\
&&& (D) {\scriptsize $m_{n=0}^{0^{-+}}=0, m_{n=1}^{0^{-+}}\approx 3.06, m_{n=2}^{0^{-+}}\approx 4.81$} \\ \hline
3 & $0^{--}$ & (Type IIB) & (Type IIB) \\
&& (N/D) {\scriptsize $m_{n=0}^{0^{--}}(T)=0, m_{n=1}^{0^{--}}(T)=\frac{32.46}{\pi},$} & (large $n$)  \\
&& {\scriptsize $m_{n=2}^{0^{--}}(T)=\frac{32.88}{\pi}$} & (N/D)  \\
&&& {\scriptsize ${\scriptsize \frac{1}{2} 5^{3/4} \sqrt[4]{\frac{2 \left(\sqrt{6} \sqrt{\pi ^2 \left(16 n^2+22 n+7\right)+6}+6\right)+3 \pi ^2 (2 n+1)}{32 - 3 \pi ^2}}}$} \\ \hline
4 & $1^{++}$ & (M theory) & (M theory) \\
& & (N/D) {\scriptsize $m_{n=0}^{1^{++}}(T) = 2.6956, m_{n=1}^{1^{++}}(T)=2.8995$}  & (N) {\scriptsize $m_{n=0}^{1^{++}}(r_h=0)\approx1.137$} \\
& & {\scriptsize $m_{n=2}^{1^{++}}(T) = 2.9346$} & (D) {\scriptsize $m_{n=0}^{1^{++}}(r_h=0)\approx0.665$} \\ \hline
5 & $2^{++}$ & (M theory) & (M theory) \\
&& (N) {\scriptsize $m_{n=0}^{2^{++}}(T)=\frac{5.086}{\pi}, m_{n=1}^{2^{++}}(T)=\frac{5.269}{\pi}$} & $=m_n^{1^{++}}(r_h=0)$ \\
&& {\scriptsize $m_{n=2}^{2^{++}}(T)=\frac{5.318}{\pi}$} &  \\
&& {\scriptsize $m_{n=0}^{2^{++}}(D,T)=0, m_{n+1}^{2^{++}}(D,T)=m_n^{2^{++}}(N,T)$} & \\ \hline
   \end{tabular}
   \caption{Summary of Glueball Spectra from Type IIB, IIA and M Theory for $r_h\neq0/r_h=0$ using Neumann/Dirichlet boundary conditions at the horizon $r_h$/IR cut-off $r_0$}
   \end{table}
\newpage
Some salient features of Table 4.3 are presented below:
\begin{itemize}

\item The following is the comparison of ratios of $0^{--}$ glueball masses obtained in this work from Neumann/Dirichlet boundary conditions at the horizon,  with \cite{Ooguri_et_al}:

\begin{table}[h]
\begin{center}
\begin{tabular}{|c|c|c|}\hline
Ratio & Our Results & \cite{Ooguri_et_al}'s Results\\  \hline
$\frac{m_{0^{--}}^*}{m_{0^{--}}}$ & 1.0129 & 1.5311\\ \hline
$\frac{m_{0^{--}}^{**}}{m_{0^{--}}^*}$ & 1.0033 & 1.3244\\ \hline
$\frac{m_{0^{--}}^{***}}{m_{0^{--}}^{**}}$ & 1.0013 & 1.2393\\ \hline
$\frac{m_{0^{--}}^{****}}{m_{0^{--}}^{***}}$ & 1.0007 & 1.1588\\ \hline
 \end{tabular}
   \caption{Comparison of ratios of $0^{--}$ glueball masses obtained  from Neumann/Dirichlet boundary conditions at the horizon,  with \cite{Ooguri_et_al}}
   \end{center}
   \end{table}

Hence, for higher excited states, the ratio of masses of successive excited states approaches unity faster as per our results as compared to \cite{Ooguri_et_al}.

 \item From a comparison of results in Table 4.1/4.2 and Figure 4.2 with $N\rightarrow\infty$ lattice results, it appears that WKB quantization-based spectra are closer to $N\rightarrow\infty$ lattice results than the computations involving imposing Neumann/Dirichlet boundary conditions at the horizon/IR cut-off. In particular, it is pleasantly surprising that the WKB quantization method applied to the $0^{++}, 0^{-+}$ glueball spectra, is able to provide a good agreement (in fact for the lightest $0^{++}$ glueball spectrum, better than the classic computations of \cite{Ooguri_et_al}) with lattice results even for the ground and the lower excited states.

     \end{itemize}
%\begin{subappendices}
%\input{AppendixC}
%\end{subappendices}
%\input{empty1}
%\input{Chapter5/chap5}
\chapter{Conclusion and future outlook}
\graphicspath{{Chapter5/}{Chapter5/}}
In a `Top-down' holographic model, the precise statement of duality between the gauge theory and the dual gravity theory is known exactly.
%More precisely, the operators on each side of the of the two theories can be identified and hence the holographic dictionary can be applied efficiently.
On the other hand, in `Bottom-up' approach, one first identifies the required ingredients to study a particular phenomenon in the field theory, and incorporates them into the bulk theory. The exact duality statement in  a bottom-up approach is not significant or required.

In this thesis we have considered the `top-down' approach where the properties of strongly coupled thermal QCD-like field theories were studied from the (super)gravitational-duals-of-brane-constructs-based gauge/gravity duality. Most importantly, in top-down holographic models, considering a particular configuration of branes, one can achieve UV completion of the gauge theory. This allows the computation of the non-conformal $\frac{1}{N}$  corrections. Using the UV-complete top-down type IIB holographic dual of large-N thermal QCD as constructed in \cite{metrics}, involving a fluxed resolved warped deformed conifold, its delocalized type IIA S(trominger)-Y(au)-Z(aslow) mirror as well as its M-theory uplift constructed in \cite{MQGP}, we have evaluated the $\frac{1}{N}$ corrections to different transport coefficients, which we call as the Next-to-Leading-Order(NLO) corrections. In this thesis, we have also studied the transition between the confined and the deconfined phases of strongly coupled gauge theories and obtained a lattice-compatible deconfinement/transition temperature by tuning some dimensionless parameters in our theory.

There exists other interesting top-down type IIA IR-specific holographic models, for example, the Sakai-Sugimoto model \cite{Sakai-Sugimoto}, where the authors have calculated the spectrum of meson masses and showed confining behavior of the theory at low energy just like QCD. In this thesis, we looked upon the the spectrum of glueball masses and observed quite a nice match with the lattice data. Though not discussed in this thesis, but in \cite{Yadav+Misra+Sil}, we have also worked out a Particle-Data-Group compatible (scalar and vector) meson spectra.
%Another remarkable work using top-down holography was carried out by Karch and Katz \cite{karch} using a $D3-D7$ brane configurations. This model reflects the properties of deconfined phase of QCD, that is the aspects of QGP.

One of the future directions could be to look at the various glueball-to-meson decay modes. To that end, Performing a Kaluza-Klein reduction similar to \cite{Hashimoto-glueball}: $A_Z = \phi(Z)\pi(x^\mu), A_\mu = \psi (Z) \rho_\mu(x^\nu)$, and similar to \cite{Constable_Myers}, we can look at the following M-theory metric perturbations $h_{MN}(M,N=0,...,10;\mu=t,a, a=1,2,3)$:
\begin{eqnarray}
\label{M-theory-metric-perturbations}
& & h_{tt}(r,x^\mu) = q_1(r) G(x^\mu) G^{\cal M}_{tt}\nonumber\\
& & h_{rr}(r,x^\mu) = q_2(r)G(x^\mu) G^{\cal M}_{rr}\nonumber\\
& & h_{ra}(r,x^\mu) = g_3(r)\partial_a G(x^\mu) G^{\cal M}_{aa}\nonumber\\
& & h_{ab}(r,x^\mu) = G^{\cal M}_{ab}\left(q_4(r) + q_5 \frac{\partial_a\partial_b}{m^2}\right)G(x^\mu)\ {\rm no\ summation}\nonumber\\
& & h_{10\ 10}(r,x^\mu) = q_6(r) G(x^\mu)G^{\cal M}_{10\ 10}.
\end{eqnarray}
Using Witten's prescription of going from type IIA to M-theory
%:
%\begin{equation}
%\label{Witten_IIA_to_M+vice_versa}
%ds_{\cal M}^2 = e^{-\frac{2\phi^{IIA}}{3}}ds_{\rm IIA}^2 + e^{\frac{4\phi^{IIA}}{3}}\left(dx^{10} + A%\right)^2,
%\end{equation}
%$A$ being the type $IIA$ RR one-form,
we could work back the type IIA metric perturbations which hence would yield (in the following $\tilde{G}^{\rm IIA}_{\alpha\beta} = G^{\rm IIA}_{\alpha\beta} + h_{\alpha\beta}; \alpha,\beta=0,1...,9$ and $h_{\alpha\beta}$ being type IIA metric perturbations):
\begin{eqnarray*}
& & e^{\frac{4\phi^{\rm IIA}}{3}} = G^{\cal M}_{10\ 10} + h_{10\ 10},\nonumber\\
& & \frac{\tilde{G}^{\rm IIA}_{rr,tt}}{\sqrt{G^{\cal M}_{10\ 10} + h_{10\ 10}}} = G^{\cal M}_{rr,tt} + h_{rr,tt},\nonumber\\
& &  \frac{\tilde{G}^{\rm IIA}_{ra}}{\sqrt{G^{\cal M}_{10\ 10} + h_{10\ 10}}} = h_{ra},\nonumber\\
& & \frac{\tilde{G}^{\rm IIA}_{ab}}{\sqrt{G^{\cal M}_{10\ 10} + h_{10\ 10}}} = G^{\cal M}_{ab} + h_{ab}.
\end{eqnarray*}
Solving  the first order perturbation of the M-theory Einstein's EOM (assuming the flux term providing a cosmological constant):
%\begin{equation}
%\label{M-theory-eom-first-order-pert}
$R_{MN}^{(1)} \sim G_{PQRS}G^{PQRS}h_{MN}$,
%\end{equation}
%one can work out the perturbed type IIA metric inverse in $(\mathbb{R}^{1,3},|Z|)$-subspace is given by:
 for $q_{1,...6}$, one can obtain the glueball-meson interaction Lagrangian density (metric perturbation corresponding to glueballs and gauge field fluctuations corresponding to mesons), using which one can work out glueball decays into mesons.
\renewcommand{\chaptermark}[1]{         % Lower Case Chapter marker style
\markboth{ \thechapter.\ #1}{}} %

\appendix
%\chapter{Appendix}
%\input{Appendix}

%\appendix
%\chapter{Appendix}
%\begin{appendices}
%\begin{subappendices}
%\input{AppendixA}
\chapter{}
%\graphicspath{{AppendixA/}{AppendixA/}}
%\chapter{The Local Momentum Approximation (LMA) and its validity}
%\renewcommand{\bm}{\bibitem}
\section{Details of Exact Angular Integration in the DBI Action and Its UV Limit}
The $\theta_2$ integral in the DBI action of (\ref{SDBI-arb-mu}), is expressed in terms of elliptic integral of the first kind $F(\phi;\mu)\equiv \int_0^{\phi}\frac{d\theta}{\sqrt{1 - \mu \sin^2\theta}}$ as well as incomplete integral of the first kind $\Pi(\nu;\phi|\mu)\equiv\int_0^\phi\frac{d\theta}{\left(1 - \nu\sin^2\theta\right)\sqrt{1 - \mu\sin^2\theta}}$:
\begin{eqnarray}
\label{elliptic-fns}
& & \hskip -0.2in F\Biggl(\sin ^{-1}\left.\left(\sqrt{\frac{\frac{|\mu| }{\sqrt{|\mu| ^2-r^3}}-\frac{-7 |\mu| ^2+\sqrt{25 |\mu| ^4-104 |\mu| ^2 r^3+16 r^6}+4 r^3}{2 \left(|\mu| ^2+2
   r^3\right)}}{\frac{|\mu| }{\sqrt{|\mu| ^2-r^3}}+\frac{-7 |\mu| ^2+\sqrt{25 |\mu| ^4-104 |\mu| ^2 r^3+16 r^6}+4 r^3}{2 \left(|\mu| ^2+2
   r^3\right)}}}\right)\right|\nonumber\\
   & & \frac{\left(\frac{|\mu| }{\sqrt{|\mu| ^2-r^3}}-\frac{-7 |\mu| ^2-\sqrt{25 |\mu| ^4-104 |\mu| ^2 r^3+16 r^6}+4 r^3}{2 \left(|\mu| ^2+2 r^3\right)}\right)
   \left(-\frac{|\mu| }{\sqrt{|\mu| ^2-r^3}}-\frac{-7 |\mu| ^2+\sqrt{25 |\mu| ^4-104 |\mu| ^2 r^3+16 r^6}+4 r^3}{2 \left(|\mu| ^2+2 r^3\right)}\right)}{\left(-\frac{|\mu|
   }{\sqrt{|\mu| ^2-r^3}}-\frac{-7 |\mu| ^2-\sqrt{25 |\mu| ^4-104 |\mu| ^2 r^3+16 r^6}+4 r^3}{2 \left(|\mu| ^2+2 r^3\right)}\right) \left(\frac{|\mu| }{\sqrt{|\mu|
   ^2-r^3}}-\frac{-7 |\mu| ^2+\sqrt{25 |\mu| ^4-104 |\mu| ^2 r^3+16 r^6}+4 r^3}{2 \left(|\mu| ^2+2 r^3\right)}\right)}\Biggr);\nonumber\\
& & \hskip -0.2in F\Biggl(\sin ^{-1}\left.\left(\sqrt{\frac{-2 |\mu| ^3-7 |\mu| ^2 \sqrt{|\mu| ^2-r^3}+4 r^3 \sqrt{|\mu| ^2-r^3}+\sqrt{|\mu| ^2-r^3} \sqrt{25 |\mu| ^4-104 |\mu| ^2 r^3+16 r^6}-4 |\mu|
   r^3}{-2 |\mu| ^3+7 |\mu| ^2 \sqrt{|\mu| ^2-r^3}-4 r^3 \sqrt{|\mu| ^2-r^3}-\sqrt{|\mu| ^2-r^3} \sqrt{25 |\mu| ^4-104 |\mu| ^2 r^3+16 r^6}-4 |\mu|  r^3}}\right)\right|\nonumber\\
   & & \frac{\left(-2 |\mu|
   ^3+7 |\mu| ^2 \sqrt{|\mu| ^2-r^3}-4 r^3 \sqrt{|\mu| ^2-r^3}-\sqrt{|\mu| ^2-r^3} \sqrt{25 |\mu| ^4-104 |\mu| ^2 r^3+16 r^6}-4 |\mu|  r^3\right)}{\left(2 |\mu| ^3+7 |\mu| ^2 \sqrt{|\mu| ^2-r^3}-4
   r^3 \sqrt{|\mu| ^2-r^3}-\sqrt{|\mu| ^2-r^3} \sqrt{25 |\mu| ^4-104 |\mu| ^2 r^3+16 r^6}+4 |\mu|  r^3\right) }\nonumber\\
   & & \times\frac{ \left(2 |\mu| ^3+7 |\mu| ^2
   \sqrt{|\mu| ^2-r^3}-4 r^3 \sqrt{|\mu| ^2-r^3}+\sqrt{|\mu| ^2-r^3} \sqrt{25 |\mu| ^4-104 |\mu| ^2 r^3+16 r^6}+4 |\mu|  r^3\right)}{\left(-2 |\mu| ^3+7 |\mu| ^2 \sqrt{|\mu| ^2-r^3}-4 r^3 \sqrt{|\mu|
   ^2-r^3}+\sqrt{|\mu| ^2-r^3} \sqrt{25 |\mu| ^4-104 |\mu| ^2 r^3+16 r^6}-4 |\mu|  r^3\right)}\Biggr);
   \nonumber\\
   & & \hskip -0.2in \Pi \Biggl(\frac{\left(\frac{|\mu| }{\sqrt{|\mu| ^2-r^3}}+1\right) \left(\frac{|\mu| }{\sqrt{|\mu| ^2-r^3}}+\frac{-7 |\mu| ^2+\sqrt{25 |\mu| ^4-104 |\mu| ^2 r^3+16 r^6}+4 r^3}{2
   \left(|\mu| ^2+2 r^3\right)}\right)}{\left(1-\frac{|\mu| }{\sqrt{|\mu| ^2-r^3}}\right) \left(\frac{-7 |\mu| ^2+\sqrt{25 |\mu| ^4-104 |\mu| ^2 r^3+16 r^6}+4 r^3}{2 \left(|\mu|
   ^2+2 r^3\right)}-\frac{|\mu| }{\sqrt{|\mu| ^2-r^3}}\right)};\nonumber\\
   & & \hskip -0.2in \sin ^{-1}\left.\left(\sqrt{\frac{-2 |\mu| ^3-7 |\mu| ^2 \sqrt{|\mu| ^2-r^3}+4 r^3 \sqrt{|\mu| ^2-r^3}+\sqrt{|\mu|
   ^2-r^3} \sqrt{25 |\mu| ^4-104 |\mu| ^2 r^3+16 r^6}-4 |\mu|  r^3}{-2 |\mu| ^3+7 |\mu| ^2 \sqrt{|\mu| ^2-r^3}-4 r^3 \sqrt{|\mu| ^2-r^3}-\sqrt{|\mu| ^2-r^3} \sqrt{25 |\mu| ^4-104 |\mu|
   ^2 r^3+16 r^6}-4 |\mu|  r^3}}\right)\right|\nonumber\\
   & & \hskip -0.2in \frac{\left(-2 |\mu|
   ^3+7 |\mu| ^2 \sqrt{|\mu| ^2-r^3}-4 r^3 \sqrt{|\mu| ^2-r^3}-\sqrt{|\mu| ^2-r^3} \sqrt{25 |\mu| ^4-104 |\mu| ^2 r^3+16 r^6}-4 |\mu|  r^3\right)}{\left(2 |\mu| ^3+7 |\mu| ^2 \sqrt{|\mu| ^2-r^3}-4
   r^3 \sqrt{|\mu| ^2-r^3}-\sqrt{|\mu| ^2-r^3} \sqrt{25 |\mu| ^4-104 |\mu| ^2 r^3+16 r^6}+4 |\mu|  r^3\right) }\nonumber\\
   & & \hskip -0.2in \times\frac{ \left(2 |\mu| ^3+7 |\mu| ^2
   \sqrt{|\mu| ^2-r^3}-4 r^3 \sqrt{|\mu| ^2-r^3}+\sqrt{|\mu| ^2-r^3} \sqrt{25 |\mu| ^4-104 |\mu| ^2 r^3+16 r^6}+4 |\mu|  r^3\right)}{\left(-2 |\mu| ^3+7 |\mu| ^2 \sqrt{|\mu| ^2-r^3}-4 r^3 \sqrt{|\mu|
   ^2-r^3}+\sqrt{|\mu| ^2-r^3} \sqrt{25 |\mu| ^4-104 |\mu| ^2 r^3+16 r^6}-4 |\mu|  r^3\right)}\Biggr).\nonumber\\
   & &
\end{eqnarray}
In the large-$r$ limit of (\ref{elliptic-fns}) after angular integrations, the finite radial integrand of (\ref{SDBI-arb-mu}) is given by:
\begin{eqnarray*}
& & -\frac{1}{72 \sqrt{2} |\mu| ^3 r^6}\Biggl[\sqrt{\left({F_{rt}}^2-1\right) |\mu| ^4} \Biggl(-64 i |\mu|  \sqrt{i |\mu|  \left(\frac{1}{r}\right)^{3/2}} F\left.\left(\frac{1}{4} \left(\frac{\sqrt{2} |\mu|
   }{r^{3/2}}-4 i \sinh ^{-1}(1)\right)\right|4 i |\mu|  \left(\frac{1}{r}\right)^{3/2}-1\right) r^9\nonumber\\
   & & +32 i \sqrt{i |\mu|  \left(\frac{1}{r}\right)^{3/2}} \sqrt{|\mu| ^2-r^3}
   F\left.\left(\frac{1}{4} \left(\frac{\sqrt{2} |\mu|
   }{r^{3/2}}-4 i \sinh ^{-1}(1)\right)\right|4 i |\mu|  \left(\frac{1}{r}\right)^{3/2}-1\right) r^9\nonumber\\
   & & +64 i |\mu|  \sqrt{i |\mu|
   \left(\frac{1}{r}\right)^{3/2}} \Pi \left(i |\mu|  \left(\frac{1}{r}\right)^{3/2}-1;i \sinh ^{-1}(1)-\frac{|\mu| }{2 \sqrt{2} r^{3/2}}|4 i |\mu|
   \left(\frac{1}{r}\right)^{3/2}-1\right) r^9\nonumber\\
   & & -64 i |\mu|  \sqrt{i |\mu|  \left(\frac{1}{r}\right)^{3/2}} \Pi \left(1-3 i |\mu|  \left(\frac{1}{r}\right)^{3/2};i \sinh
   ^{-1}(1)-\frac{|\mu| }{2 \sqrt{2} r^{3/2}}|4 i |\mu|  \left(\frac{1}{r}\right)^{3/2}-1\right) r^9\nonumber\\
   & & -16 |\mu| ^2 \sqrt{i |\mu|  \left(\frac{1}{r}\right)^{3/2}}
   F\left.\left(\frac{1}{4} \left(\frac{\sqrt{2} |\mu|
   }{r^{3/2}}-4 i \sinh ^{-1}(1)\right)\right|4 i |\mu|  \left(\frac{1}{r}\right)^{3/2}-1\right)r^{15/2}\nonumber\\
   & & +136 |\mu|  \sqrt{i |\mu|
   \left(\frac{1}{r}\right)^{3/2}} \sqrt{|\mu| ^2-r^3} F\left(\frac{1}{4} \left(\frac{\sqrt{2} |\mu| }{r^{3/2}}-4 i \sinh ^{-1}(1)\right)|4 i |\mu|
   \left(\frac{1}{r}\right)^{3/2}-1\right) r^{15/2}\nonumber\\
   & & +16 |\mu| ^2 \sqrt{i |\mu|  \left(\frac{1}{r}\right)^{3/2}} \Pi \left(i |\mu|  \left(\frac{1}{r}\right)^{3/2}-1;i \sinh
   ^{-1}(1)-\frac{|\mu| }{2 \sqrt{2} r^{3/2}}|4 i |\mu|  \left(\frac{1}{r}\right)^{3/2}-1\right) r^{15/2}\nonumber\\
   & & -16 |\mu| ^2 \sqrt{i |\mu|  \left(\frac{1}{r}\right)^{3/2}} \Pi
   \left(1-3 i |\mu|  \left(\frac{1}{r}\right)^{3/2};i \sinh ^{-1}(1)-\frac{|\mu| }{2 \sqrt{2} r^{3/2}}|4 i |\mu|  \left(\frac{1}{r}\right)^{3/2}-1\right) r^{15/2}\nonumber\\
   & & +32 |\mu|
   ^3 r^6+104 i |\mu| ^3 \sqrt{i |\mu|  \left(\frac{1}{r}\right)^{3/2}} F\left.\left(\frac{1}{4} \left(\frac{\sqrt{2} |\mu|
   }{r^{3/2}}-4 i \sinh ^{-1}(1)\right)\right|4 i |\mu|  \left(\frac{1}{r}\right)^{3/2}-1\right) r^6\nonumber\\
   & & +108 i |\mu| ^2 \sqrt{i |\mu|  \left(\frac{1}{r}\right)^{3/2}} \sqrt{|\mu| ^2-r^3} F\left.\left(\frac{1}{4} \left(\frac{\sqrt{2} |\mu|
   }{r^{3/2}}-4 i \sinh ^{-1}(1)\right)\right|4 i |\mu|  \left(\frac{1}{r}\right)^{3/2}-1\right) r^6\nonumber\\
   & & +280 i |\mu| ^3 \sqrt{i |\mu|  \left(\frac{1}{r}\right)^{3/2}} \Pi
   \left(i |\mu|  \left(\frac{1}{r}\right)^{3/2}-1;i \sinh ^{-1}(1)-\frac{|\mu| }{2 \sqrt{2} r^{3/2}}|4 i |\mu|  \left(\frac{1}{r}\right)^{3/2}-1\right) r^6\nonumber\\
   & & -280 i |\mu| ^3
   \sqrt{i |\mu|  \left(\frac{1}{r}\right)^{3/2}} \Pi \left(1-3 i |\mu|  \left(\frac{1}{r}\right)^{3/2};i \sinh ^{-1}(1)-\frac{|\mu| }{2 \sqrt{2} r^{3/2}}|4 i |\mu|
   \left(\frac{1}{r}\right)^{3/2}-1\right) r^6\nonumber\\
   & & +288 i |\mu| ^3 \sqrt{i |\mu|  \left(\frac{1}{r}\right)^{3/2}} \Pi \left(1-\frac{i |\mu| }{r^{3/2}};i \sinh
   ^{-1}(1)-\frac{|\mu| }{2 \sqrt{2} r^{3/2}}|4 i |\mu|  \left(\frac{1}{r}\right)^{3/2}-1\right) r^6\nonumber\\
   & & +32 \sqrt{6} |\mu| ^2 \sqrt{|\mu| ^2-r^3} r^6+28 |\mu| ^4 \sqrt{i |\mu|
   \left(\frac{1}{r}\right)^{3/2}} F\left.\left(\frac{1}{4} \left(\frac{\sqrt{2} |\mu|
   }{r^{3/2}}-4 i \sinh ^{-1}(1)\right)\right|4 i |\mu|  \left(\frac{1}{r}\right)^{3/2}-1\right)
   r^{9/2}\nonumber\\
   & & +114 |\mu| ^3 \sqrt{i |\mu|  \left(\frac{1}{r}\right)^{3/2}} \sqrt{|\mu| ^2-r^3} F\left.\left(\frac{1}{4} \left(\frac{\sqrt{2} |\mu|
   }{r^{3/2}}-4 i \sinh ^{-1}(1)\right)\right|4 i |\mu|  \left(\frac{1}{r}\right)^{3/2}-1\right)r^{9/2}\nonumber\\
    \end{eqnarray*}
   \begin{eqnarray}
   \label{integrand_large_r}
 & & +68 |\mu| ^4 \sqrt{i |\mu|  \left(\frac{1}{r}\right)^{3/2}} \Pi \left(i |\mu|
   \left(\frac{1}{r}\right)^{3/2}-1;i \sinh ^{-1}(1)-\frac{|\mu| }{2 \sqrt{2} r^{3/2}}|4 i |\mu|  \left(\frac{1}{r}\right)^{3/2}-1\right) r^{9/2}\nonumber\\
   & & -68 |\mu| ^4 \sqrt{i |\mu|
   \left(\frac{1}{r}\right)^{3/2}} \Pi \left(1-3 i |\mu|  \left(\frac{1}{r}\right)^{3/2};i \sinh ^{-1}(1)-\frac{|\mu| }{2 \sqrt{2} r^{3/2}}|4 i |\mu|
   \left(\frac{1}{r}\right)^{3/2}-1\right) r^{9/2}\nonumber\\
   & & +72 |\mu| ^4 \sqrt{i |\mu|  \left(\frac{1}{r}\right)^{3/2}} \Pi \left(1-\frac{i |\mu| }{r^{3/2}};i \sinh
   ^{-1}(1)-\frac{|\mu| }{2 \sqrt{2} r^{3/2}}|4 i |\mu|  \left(\frac{1}{r}\right)^{3/2}-1\right) r^{9/2}\nonumber\\
   & & +\sqrt{i |\mu|  \left(\frac{1}{r}\right)^{3/2}} \sqrt{|\mu| ^2-r^3}
   \left(-32 i r^6-8 |\mu|  r^{9/2}+52 i |\mu| ^2 r^3+14 |\mu| ^3 r^{3/2}+7 i |\mu| ^4\right)\nonumber\\
   & & \times F\left.\left(\frac{1}{4} \left(\frac{\sqrt{2} |\mu|
   }{r^{3/2}}-4 i \sinh ^{-1}(1)\right)\right|4 i |\mu|  \left(\frac{1}{r}\right)^{3/2}-1\right) r^3\nonumber\\
   & & +14 i |\mu| ^5 \sqrt{i |\mu|  \left(\frac{1}{r}\right)^{3/2}} F\left.\left(\frac{1}{4} \left(\frac{\sqrt{2} |\mu|
   }{r^{3/2}}-4 i \sinh ^{-1}(1)\right)\right|4 i |\mu|  \left(\frac{1}{r}\right)^{3/2}-1\right) r^3\nonumber\\
   & & +i |\mu| ^4 \sqrt{i |\mu|
   \left(\frac{1}{r}\right)^{3/2}} \sqrt{|\mu| ^2-r^3} F\left.\left(\frac{1}{4} \left(\frac{\sqrt{2} |\mu|
   }{r^{3/2}}-4 i \sinh ^{-1}(1)\right)\right|4 i |\mu|  \left(\frac{1}{r}\right)^{3/2}-1\right) r^3\nonumber\\
   & & +34 i |\mu| ^5 \sqrt{i |\mu|  \left(\frac{1}{r}\right)^{3/2}} \Pi \left(i |\mu|  \left(\frac{1}{r}\right)^{3/2}-1;i \sinh
   ^{-1}(1)-\frac{|\mu| }{2 \sqrt{2} r^{3/2}}|4 i |\mu|  \left(\frac{1}{r}\right)^{3/2}-1\right) r^3\nonumber\\
   & & -34 i |\mu| ^5 \sqrt{i |\mu|  \left(\frac{1}{r}\right)^{3/2}} \Pi
   \left(1-3 i |\mu|  \left(\frac{1}{r}\right)^{3/2};i \sinh ^{-1}(1)-\frac{|\mu| }{2 \sqrt{2} r^{3/2}}|4 i |\mu|  \left(\frac{1}{r}\right)^{3/2}-1\right) r^3\nonumber\\
   & & +36 i |\mu| ^5
   \sqrt{i |\mu|  \left(\frac{1}{r}\right)^{3/2}} \Pi \left(1-\frac{i |\mu| }{r^{3/2}};i \sinh ^{-1}(1)-\frac{|\mu| }{2 \sqrt{2} r^{3/2}}|4 i |\mu|
   \left(\frac{1}{r}\right)^{3/2}-1\right) r^3\nonumber\\
   & & +8 |\mu| ^5 \sqrt{i |\mu|  \left(\frac{1}{r}\right)^{3/2}} \sqrt{|\mu| ^2-r^3} F\left.\left(\frac{1}{4} \left(\frac{\sqrt{2} |\mu|
   }{r^{3/2}}-4 i \sinh ^{-1}(1)\right)\right|4 i |\mu|  \left(\frac{1}{r}\right)^{3/2}-1\right) r^{3/2}\nonumber\\
   & & -2 |\mu|  \sqrt{i |\mu|  \left(\frac{1}{r}\right)^{3/2}} \sqrt{|\mu|
   ^2-r^3} \left(32 r^{15/2}+8 i |\mu|  r^6+24 |\mu| ^2 r^{9/2}+6 i |\mu| ^3 r^3+4 |\mu| ^4 r^{3/2}+i |\mu| ^5\right)\nonumber\\
   & & \times E\left.\left(\frac{1}{4} \left(\frac{\sqrt{2} |\mu|
   }{r^{3/2}}-4 i \sinh ^{-1}(1)\right)\right|4 i |\mu|  \left(\frac{1}{r}\right)^{3/2}-1\right)\Biggr)\Biggr]\nonumber\\
   & & \approx \sqrt{|\mu|}\sqrt{1-F_{rt}^2}r^{\frac{9}{4}} + {\cal O}(r^{\frac{3}{2}}).
\end{eqnarray}
\section{EOMs involving $N_f=2$ Gauge Field Fluctuations, Solution and On-Shell Action}
Choosing the momentum four-vector in $\mathbb{R}^{1,3}$ as $q^\mu = (w, q, 0, 0)$, and  writing the fluctuation as: $A_\mu^a(x,u)=\int d^4q e^{- i w t + i q x}A^a_\mu(q,u)$ (flavor index $a=1,2,3$), the equation of motion for the same follows in a straightforward way from (\ref{eom}) and that in momentum space is given as:
\begin{equation}\begin{split}\label{Ay}
& A_y^a{''}+\frac{\partial_u(\sqrt{{\rm det}\ G}G^{uu}G^{yy})}{\sqrt{{\rm det}\ G}G^{uu}G^{yy})}A_y^a{'}-w^2\frac{G^{tt}}{G^{uu}}A^a_y-\frac{(iw)}{2}\frac{G^{tt}}{G^{uu}}\frac{r^2_h}{2\pi{\alpha^\prime}}\tilde{A}^3_0f^{ab3}A^b_y\\
&+(iw)\frac{G^{tt}}{G^{uu}}\frac{r^2_h}{2\pi{\alpha^\prime}}\tilde{A}^3_0f^{ab3}A^b_y-\frac{1}{2}(\frac{r^2_h}{2\pi{\alpha^\prime}}\tilde{A}^3_0)^2
\frac{G^{tt}}{G^{uu}}f^{ab3}f^{bc3}A^c_y=0.
\end{split}
\end{equation}
Setting $q=0$ and defining $A^a_y=\frac{1}{\omega}E^a_T$ the above equation can be rewritten as:
\begin{equation}\begin{split}\label{ET}
& E_T^a{''}+\frac{\partial_u(\sqrt{{\rm det}\ G}G^{uu}G^{yy})}{\sqrt{{\rm det}\ G}G^{uu}G^{yy})}E_T^a{'}-w^2\frac{G^{tt}}{G^{uu}}E^a_T+\frac{iw}{2}\frac{G^{tt}}{G^{uu}}\frac{r^2_h}{2\pi{\alpha^\prime}}
\tilde{A}^3_0f^{ab3}E^b_T\\&-\frac{1}{2}(\frac{r^2_h}{2\pi{\alpha^\prime}}\tilde{A}^3_0)^2\frac{G^{tt}}{G^{uu}}f^{ab3}f^{bc3}
E^c_T=0.
\end{split}
\end{equation}
Now for $a=1$ and $b=2$ equation (\ref{ET}) gives:
\begin{equation}\begin{split}
\label{E^1}
& E_T^1{''}+\frac{\partial_u(\sqrt{{\rm det}\ G}G^{uu}G^{yy})}{\sqrt{{\rm det}\ G}G^{uu}G^{yy})}E_T^1{'}-\frac{G^{tt}}{G^{uu}}\Biggl(w^2-\frac{1}{2}(\frac{r^2_h}{2\pi{\alpha^\prime}}\tilde{A}^3_0)^2\Biggr)E^1_T\\&+
 \frac{iw}{2}\frac{G^{tt}}{G^{uu}}\frac{r^2_h}{2\pi{\alpha^\prime}}\tilde{A}^3_0f^{ab3}E^2_T=0.
 \end{split}
\end{equation}
For $a=2,b=1$ equation (\ref{ET}) gives:
\begin{equation}\begin{split}
\label{E^2}
&E_T^2{''}+\frac{\partial_u(\sqrt{{\rm det}\ G}G^{uu}G^{yy})}{\sqrt{{\rm det}\ G}G^{uu}G^{yy})}E_T^2{'}-\frac{G^{tt}}{G^{uu}}\Biggl(w^2-\frac{1}{2}(\frac{r^2_h}{2\pi{\alpha^\prime}}
 \tilde{A}^3_0)^2\Biggr)E^2_T\\&-\frac{iw}{2}\frac{G^{tt}}{G^{uu}}\frac{r^2_h}{2\pi{\alpha^\prime}}\tilde{A}^3_0f^{ab3}E^1_T=0.
 \end{split}
\end{equation}
Finally for $a=3$ the same equation (\ref{ET}) gives:
\begin{equation}
\label{E^3}
E_T^3{''}+\frac{\partial_u(\sqrt{{\rm det}\ G}G^{uu}G^{yy})}{\sqrt{{\rm det}\ G}G^{uu}G^{yy})}E_T^3{'}-w^2\frac{G^{tt}}{G^{uu}}E_T^3=0.
\end{equation}
Defining $X=E^1+ i E^2$, $Y=E^1- i  E^2$ and $\overline{A^3_0}\equiv\frac{r_h}{2\pi\alpha^\prime}\tilde{A}^3_0$, the $SU(2)$  equations of motion (\ref{E^1}) - (\ref{E^3}) can be rewritten as:
\begin{equation}\begin{split}
\label{SU2_EOMs}
& \frac{\partial^2 X}{\partial u^2}+\Sigma(u)\frac{\partial X}{\partial u}+(w-\overline{A^3_0})^2\Lambda(u)X=0,\\&
\frac{\partial^2 Y}{\partial u^2}+\Sigma(u)\frac{\partial Y}{\partial u}+(w+\overline{A^3_0})^2\Lambda(u)X=0,\\&
\frac{\partial^2 E^3}{\partial u^2}+\Sigma(u)\frac{\partial X}{\partial u}+w^2\Lambda(u)X=0,
\end{split}
\end{equation}
where $\Sigma(u)$ and $\Lambda (u)$ are given as:
\begin{equation}\begin{split}
&\Sigma(u)=\left(\frac{1}{4(u^4-1)\sqrt{\frac{r_h}{u}}(r_h^4\sqrt{\frac{r_h}{u}}+C^2 e^{2\phi}u^{5})^3}\right)
\left(16 C^6 e^{6\phi}\sqrt{\frac{r_h}{u}}
u^{14}(2u^4-1)\right.\\& \left. +6C^2e^{2\phi}r_h^{9}\sqrt{\frac{r_h}{u}}u^{5}(13u^4-5)
+r^{14}_h(23u^4-7)+3C^4e^{4\phi}r_h^5u^9(29u^4-13)\right);\\&
\Lambda(u)=\frac{1}{\pi^2 T^2 (u^4-1)^2}
\end{split}
\end{equation}
The $U(1)$ EOM corresponding to gauge-invariant variable $E(u)$ is similar to the one with $E^3$ in (\ref{SU2_EOMs}).
The EOM for $Z(u)\equiv E^3(u)$ or $E(u)$ can be written as:
\begin{equation}
\label{Z}
(u-1)^2\frac{d^2Z(u)}{du^2} + \frac{(u-1)\Sigma(u)}{(u+1)(u^2+1)}\frac{dZ(u)}{du} + \frac{w_3^2 Z(u)}{(u+1)^2(u^2+1)^2} = 0.
\end{equation}
One realizes that $u=1$ is a regular singular point with solutions to the indicial equation given by: $\pm i\frac{w_3}{4}$ and we choose the minus sign for incoming-wave solutions:
$Z(u)=(1-u)^{-\frac{iw_3}{4}}{\cal Z}(u)$. Using a perturbative ansatz:
\begin{equation}
\label{pert}
{\cal Z}(u) = {\cal Z}^{(0)}(u) + w_3 {\cal Z}^{(1)}(u) + {\cal O}(w_3^2),
\end{equation}
one finds (\ref{Z}) splits up into the following system of differential equations:
\begin{equation}\begin{split}
\label{systemdiffeqs}
(u-1)^2 \frac{d^2{\cal Z}^{(0)}}{du^2}  + \frac{(u-1)\Sigma(u)}{(u+1)(u^2+1)}\frac{d{\cal Z}^{(0)}}{du} &= 0;\\
(u-1)^2\frac{d{\cal Z}^{(1)}}{du^2} + \frac{(u-1)\Sigma(u)}{(u+1)(u^2+1)}\frac{d{\cal Z}^{(1)}}{du}& = \frac{i}{4}\left\{-1 + \frac{\Sigma(u)}{4(u+1)(u^2+1)}\right\}{\cal Z}^{(0)}(u) \\&+ \frac{i}{2}(u-1)\frac{d{\cal Z}^{(0)}}{du},
\end{split}
\end{equation}
with the following solutions to (\ref{systemdiffeqs}):
\begin{equation}\begin{split}
\label{sol1}
{\cal Z}^{(0)}(u) &= \frac{2 c_1 \left(-21  u(1-2 u)^{1/4} \ _2F_1\left(\frac{1}{4},\frac{1}{4};\frac{5}{4};2 u\right)+6 u^2+u-2\right)}{3 u^{3/4}
(2 u-1)^{1/4}}+c_2\\& = (-1)^{3/4}\left(\frac{ 4}{3 u^{3/4}}+ 14 u^{1/4}\right)c_1+c_2+{\cal O}\left(u^{5/4}\right)\\ & \equiv \frac{\alpha_0 }{u^{\frac{3}{4}}}c_1 + c_2 + {\cal O}\left(u^{1/4}\right);\\
{\cal Z}^{(1)}(u) &= c_3+\left(\frac{1}{1008 u^{3/4}
   (2 u-1)^{1/4}}\right)
\left(-14112 c_2 u(1-2 u)^{1/4}  \ _2F_1\left(\frac{1}{4},\frac{1}{4};\frac{5}{4};2 u\right)\right. \\& \left. +672 c_2 \left(6 u^2+u-2\right)+68(1+i) \sqrt{2} c_1 (2 u-1)^{1/4}\right)\\&
=\frac{17(1+i)}{126\sqrt{2}u^{3/4}}c_1+(-1)^{3/4}\left(\frac{ 4}{3 u^{3/4}}+14 u^{1/4}\right)c_2+c_3\\&  \equiv
\frac{\alpha_1}{u^{3/4}}c_1+\frac{\beta_1}{u^{3/4}}c_2+c_3+ {\cal O}\left(u^{1/4}\right),
   \end{split}
\end{equation}
where $c_{1,2}\in\mathbb{R}$ and it is understood that $u\rightarrow0$ as $u\rightarrow\delta\rightarrow0$ and $c_{1,2,3}\rightarrow\delta^{\frac{3}{4}}:\frac{c_1}{c_2}$ is finite, to ensure finite gauge field perturbations ${\cal Z}^{(0),(1)}(u\rightarrow0)$ in (\ref{sol1}) and finite electrical conductivity (\ref{sigma-DC}). From (\ref{sol1}), we obtain the following:
\begin{equation}\begin{split}
\label{sol2}
Z(u)&= \frac{\alpha_0 c_1 + w_3\left[\alpha_1c_1 + \beta_1c_2\right]}{u^{3/4}}
+ c_2 + c_3 w_3 + c_1\gamma_0 u^{1/4} \\&+ \left(\frac{i}{4}\alpha_0c_1 + c_2\gamma_0\right)w_3u^{1/4} + \frac{i c_2}{4}w_3 u + ......;\\
\frac{dZ(u)}{du} &= \frac{1}{u^{7/4}}\left\{-\frac{3\alpha_0c_1}{4} - \frac{\left(3\alpha_1c_1+ 3 \beta_1c_2\right)}{4}
w_3 + \frac{c_1\gamma_0}{4}u + \left(\frac{i}{16}\alpha_0c_1 + \frac{\gamma_0}{4}c_2\right)w_3u + ...\right\}.
\end{split}
\end{equation}
We notice that the only distinction between the $SU(2)$ and $U(1)$ EOMs is the shift in the roots of the indicial equation corresponding to the horizon being a regular singular point; the incoming plane-wave root of the former (in $\alpha^\prime=\frac{1}{2\pi}$-units) is given by:
\begin{equation}
-\frac{i}{4}\left(w_3 + \overline{A^3_0}(u=1)\right) = -\frac{i}{4}\left(w_3 + \left\{\frac{2^{4/9} \Gamma \left(\frac{5}{18}\right) \Gamma \left(\frac{11}{9}\right)}{\pi^{1/18} \left(\frac{g_{s} N_{f} \log{\mu }-2
   \pi }{C g_{s}}\right)^{4/9}} - 1\right\}r_h\right).
\end{equation}
We will not say more about this in this thesis.

Let us work out the on-shell action to calculate the DC conductivity. For $\sigma=u$ the LHS of equation (\ref{eom}) simplifies to:
\begin{equation}
\partial_t\left(\sqrt{{\rm det}\ G}\left(2G^{tt}G^{uu}-2G^{ut}G^{ut}\right)\widehat{F^a_{ut}}\right)+\partial_x\left(\sqrt{{\rm det}\ G}\left(2G^{xx}G^{uu}\right)\widehat{F^a_{ux}}\right).
\end{equation}
Similarly the RHS simplifies to:
\begin{equation}\begin{split}
&\sqrt{{\rm det}\ G}\frac{r^2_h}{2\pi{\alpha^\prime}}
\tilde{A}^3_0f^{ab3}\left(G^{\nu t}G^{u\mu}-G^{\nu u}G^{t\mu}\right)\widehat{F^b_{\mu\nu}}\\ &= \sqrt{{\rm det}\ G}\frac{r^2_h}{2\pi{\alpha^\prime}}
\tilde{A}^3_0f^{ab3}\biggl[\left(G^{tt}G^{uu}-G^{tu}G^{tu}\right)\widehat{F^b_{ut}}+\left(G^{ut}G^{ut}-G^{uu}G^{tt}\right)\widehat{F^b_{tu}}\biggr]
\\&= \sqrt{{\rm det}\ G}\frac{r^2_h}{2\pi{\alpha^\prime}}
\tilde{A}^3_0f^{ab3}\biggl[2G^{tt}G^{uu}-2G^{ut}G^{ut}\biggr]\widehat{F^b_{ut}}.
\end{split}
\end{equation}
Now, working in the gauge $A^a_u=0$ which implies
\begin{equation}\begin{split}
 \partial_t\widehat{F^a_{ut}}&=2\left(-iw\right)\partial_u A^a_t\\
\partial_x\widehat{F^a_{ux}}&=2\left(iq\right)\partial_u A^a_x,
\end{split}
\end{equation}
we get the EOM:
\begin{equation}\begin{split}
 &\left(-iw\right)\left(G^{tt}G^{uu}-G^{ut}G^{ut}\right)\partial_u A^a_t+\left(iq\right)\left(G^{xx}G^{uu}\right)\partial_u A^a_x\\& = \frac{r^2_h}{2\pi{\alpha^\prime}}
\tilde{A}^3_0f^{ab3}\left(G^{tt}G^{uu}-G^{ut}G^{ut}\right)\partial_u A^b_t,
\end{split}
\end{equation}
which implies,
\begin{equation}\begin{split}
\partial_uA^a_x=\frac{\left(G^{tt}G^{uu}-G^{ut}G^{ut}\right)}{\left(iq\right)\left(G^{xx}G^{uu}\right)}\biggl[\frac{r^2_h}{2\pi{\alpha^\prime}}
\tilde{A}^3_0f^{ab3}\left(\partial_uA^b_t\right)+iw\left(\partial_uA^a_t\right)\biggr]
\end{split}
\end{equation}
Now, as shown in \cite{Erdmenger_et_al}, the on-shell action is given by:
\begin{equation}
\label{S-on-shell-1}
S_{\rm on-shell}\sim T_r T_{D7}\int d^4x\sqrt{{\rm det}\ G}\left.\left(G^{\nu u}G^{\nu'\mu}-G^{\nu \nu'}G^{u \mu}\right)A^a_{\nu'}\widehat{F^a_{\mu\nu}}\right|_{u=0},
\end{equation}
wherein:
\begin{equation}\begin{split}
\label{S-on-shell-2}
& \sqrt{{\rm det}\ G}\left(G^{\nu u}G^{\nu'\mu}-G^{\nu \nu'}G^{u \mu}\right)A^a_{\nu'}\widehat{F^a_{\mu\nu}}\\
 &= \sqrt{{\rm det}\ G}\biggl[\left(G^{uu}G^{tt}-G^{ut}G^{ut}\right)A^a_{t}\widehat{F^a_{tu}}+\left(G^{tu}G^{tu}-G^{tt}G^{uu}\right)A^a_{t}
\widehat{F^a_{ut}} +\left(G^{uu}G^{xx}\right)A^a_x\widehat{F^a_{xu}}\\&+\left(G^{tu}G^{xx}\right)A^a_x\widehat{F^a_{xt}} +\left(-G^{uu}G^{xx}\right)A^a_{x}\widehat{F^a_{ux}}+\left(-G^{ut}G^{xx}\right)A^a_{x}\widehat{F^a_{tx}}
+\left(G^{uu}G^{\alpha\alpha}\right)A^a_{\alpha}\widehat{F^a_{\alpha u}}\\&+\left(G^{tu}G^{\alpha\alpha}\right)A^a_{\alpha}\widehat{F^a_{\alpha t}} +\left(-G^{uu}G^{\alpha\alpha}\right)A^a_{\alpha}\widehat{F^a_{u \alpha}}+\left(-G^{ut}G^{\alpha\alpha}\right)A^a_{\alpha}\widehat{F^a_{t \alpha}}\biggr]\\&= \sqrt{{\rm det}\ G}\biggl[\left(2G^{ut}G^{ut}-2G^{uu}G^{tt}\right)A^a_{t}\widehat{F^a_{ut}}-\left(2G^{uu}G^{xx}\right)A^a_{x}\widehat{F^a_{ux}}-\left(2G^{uu}G^{\alpha\alpha}\right)A^a_{\alpha}\widehat{F^a_{u \alpha}}\biggr]\\
& =\sqrt{{\rm det}\ G}\biggl[4\left(G^{ut}G^{ut}-G^{uu}G^{tt}\right)A^a_{t}\left(\partial_uA^a_t\right)-4\left(G^{uu}G^{xx}\right)A^a_{x}\left(\partial_uA^a_x\right)
\\&-4\left(G^{uu}G^{\alpha \alpha}\right)A^a_{\alpha}\left(\partial_u A^a_{\alpha}\right)\biggr].
\end{split}
\end{equation}
In equation (\ref{S-on-shell-2}), the first term as an example can be simplified to:
\begin{equation}\begin{split}
& 4\sqrt{{\rm det}\ G}\left[\left(G^{ut}G^{ut}-G^{uu}G^{tt}\right)A^a_{t}\left(\partial_uA^a_t\right)\right] -\frac{A^a_x}{iq}\Biggl[\left(G^{uu}G^{tt}-G^{ut}G^{ut}\right)\\&\left(\frac{r^2_h}{2\pi{\alpha^\prime}}
\tilde{A}^3_0f^{ab3}\left(\partial_uA^b_t\right)+iw\left(\partial_uA^a_t\right)\right)-\left(G^{uu}G^{\alpha\alpha}\right)
A^a_{\alpha}\left(\partial_uA^a_{\alpha}\right)\Biggr]
\\& =4\sqrt{{\rm det} G}\left\{\left(G^{ut}G^{ut}-G^{uu}G^{tt}\right)\left(\partial_uA^a_t\right)\left(A^a_{t}+\frac{w}{q}A^a_{x}\right)
+\left(G^{ut}G^{ut}-G^{uu}G^{tt}\right)\right. \\& \left. \left(\frac{r^2_h}{2\pi{\alpha^\prime}}
\tilde{A}^3_0f^{ab3}\right)
\left(\frac{A^a_{x}}{iq}\right)\left(\partial_u A^b_t\right)-\left(G^{uu}G^{\alpha\alpha}\right)A^a_{\alpha}\left(\partial_u A^a_{\alpha}
\right)
\right\}.
\end{split}
\end{equation}
Let us work with the gauge-invariant electric field components $E^a_x=q A_t + w A^a_x$ and $E^a_{\alpha} = wA^a_{\alpha}, \alpha = y, z$. Differentiating we get
\begin{equation}\begin{split}
\label{duEax}
 \partial_uE^a_x &=q\partial_uA^a_t+wA^a_x\\
& = q\partial_uA^a_t+w\frac{w\left(G^{uu}G^{tt}-G^{ut}G^{ut}\right)}{\left(iq\right)\left(G^{uu}G^{xx}\right)}\left(\frac{r^2_h}{2\pi{\alpha^\prime}}
\tilde{A}^3_0f^{ab3}\partial_uA^b_t\right)\\& + \frac{w^2}{q}\frac{\left(G^{uu}G^{tt}-G^{ut}G^{ut}\right)}{G^{xx}G^{uu}}\left(\partial_uA^a_t\right).
\end{split}
\end{equation}
Now the terms in the on-shell action have to write in terms of $\partial_u E^a_x$. Assuming one will be interested in evaluation of flavor-diagonal two-point correlation functions for simplicity, we will disregard the flavor anti-symmetric terms and therefore obtain:
\begin{eqnarray}
\partial_uA^a_x=\frac{w}{q}\frac{\left(G^{tt}G^{uu}-G^{ut}G^{ut}\right)}{\left(G^{xx}G^{uu}\right)}\left(\partial_uA^a_t\right).
\end{eqnarray}
Substituting for $\partial_uA^a_x$, the action (\ref{S-on-shell-2}) then simplifies to:
\begin{eqnarray}
& & 4\sqrt{{\rm det}\ G}\left[\left(G^{ut}G^{ut}-G^{uu}G^{tt}\right)\left(A^a_{t}+\frac{w}{q}A^a_x\right)\left(\partial_uA^a_t\right)-4\left(G^{uu}G^{\alpha\alpha}\right)E^a_{\alpha}\left(\partial_uE^a_{\alpha}\right)\right]\nonumber\\
& & = \sqrt{{\rm det}\ G}\left[\frac{4}{q}\left(G^{ut}G^{ut}-G^{uu}G^{tt}\right)E^a_{x}\left(\partial_uA^a_t\right)-\frac{4}{w^2}\left(G^{uu}G^{\alpha\alpha}\right)
E^a_{\alpha}\left(\partial_uE^a_{\alpha}\right)\right].
\end{eqnarray}
Again disregarding the flavor-antisymmetric factor the expression for $\partial_uE^a_x$ in equation (\ref{duEax}), one gets: \begin{equation}
\partial_uE^a_x=q\partial_uA^a_t+wA^a_x
=q\partial_uA^a_t+\frac{w^2}{q}\frac{\left(G^{uu}G^{tt}-G^{ut}G^{ut}\right)}{G^{xx}G^{uu}}(\partial_uA^a_t),
\end{equation}
using which one obtains the following on-shell action's integrand:
\begin{equation}\begin{split}
\label{DBI-EOM-ii}
&\sqrt{{\rm det}\ G}\left(\frac{4G^{uu}G^{xx}(G^{ut}G^{ut}-G^{uu}G^{tt})}{q^2(G^{uu}G^{xx})+w^2(G^{tt}G^{uu}-G^{ut}G^{ut})}E^a_x(\partial_uE^a_x)-\frac{4}{w^2}G^{uu}G^{\alpha \alpha}E^a_{\alpha}(\partial_uE^a_{\alpha})+...\right)_{u=0}\\
&= \frac{4u^2(u^4-1)}{w^2\left(\frac{1}{\sqrt{\frac{r_h}{u}}+c^2e^{2\phi}u^4}\right)^{1/2}}\Biggl(E^a_x(\partial_uE^a_x) + E^a_{\alpha}(\partial_uE^a_{\alpha})+...
\Biggr)_{u=0}\\
&\sim \left.\frac{r_h^{1/4}u^{7/4}}{w^2}\left(E^a_x(\partial_uE^a_x) + E^a_{\alpha}(\partial_uE^a_{\alpha})\right)+....\right|_{u\rightarrow0},
\end{split}
\end{equation}
where the dots include the flavor anti-symmetric terms.
%\end{subappendices}
%\end{appendices}
%\input{empty1}
%\begin{appendices}
%\begin{subappendices}
%\input{AppendixB}
\chapter{}

\section{EOMs for (Vector Mode) Metric and Gauge Fluctuations, and Their Solutions near $u=0$}
The EOM involving double derivative only on $H_{ty}$ is given as:
\begin{equation}\begin{split}
\label{EOM-Hty}
&\frac{ e^{ i (q x-t w)}r^4_{h}}{2u^6L^7g_{s}^{5/3}}\left(u(1-u^4)r_{h}^2H_{ty}^{\prime\prime}-3(1-u^4)r_{h}^2H_{ty}^{\prime}-L^4 q^2 u H_{ty}-L^4 q w u
H_{xy}\right)\\&+\frac{e^{2 i (q x-t w)} \sqrt{c^2 e^{2 \Phi }+\left(\frac{r_h}{u}\right)^{9/2}}\sqrt{\mu } r_h}{36\left(u^4-1\right)\sqrt{c^2 e^{2 \Phi }
   u^4+\left(\frac{r_h^9}{u}\right)^{1/2}}}\left(\sqrt{c^2 e^{2 \Phi }+\left(\frac{r_h}{u}\right)^{9/2}}H_{ty}-c
   e^{\Phi } u^4 \phi^{\prime}+c e^{\Phi } \phi^{\prime} \right)=0
\end{split}
\end{equation}
The EOM involving double derivative only on $H_{xy}$ is given as:
\begin{equation}\begin{split}
\label{EOM-Hxy}
&\Biggl[L^4 \left(\sqrt{\mu } r_{h}^3 (1-u^4) e^{i q x} \sqrt{\frac{r_{h}}{u}} \sqrt{\frac{r_{h}^4}{c^2 e^{2 \Phi } u^4+\left(\frac{r_h^9}{u}\right)^{1/2}}}\right.\\&
\left.+18 u^4 w^2 e^{i t w} \sqrt{\frac{r_h^8}{L^6 u^{10}
   g_s^{10/3}}}\right)H_{xy} +18 e^{i t w}\frac{r_h^4 L}{ u g_s^{5/3}  q w }H_{ty}\\&+18e^{i t w} \frac{r_h^6 }{L^3 u^2 g_s^{5/3}}(1-u^4) \Biggl(u (1-u^4) H_{xy}^{\prime\prime}-\left(3+u^4\right)H_{xy}^{\prime}\Biggr)\Biggr]=0.
\end{split}
\end{equation}
The EOM involving double derivative only on $\phi$ is given as:
{\footnotesize
\begin{equation}\begin{split}
\label{EOM-phi}
&\left(c^6 e^{6 \Phi} u^{13}+3 c^4 e^{4 \Phi } r_h^4 u^9 \sqrt{\frac{r_{h}}{u}}+3 c^2 e^{2 \Phi } r_{h}^9 u^4+r_{h}^{13}
   \sqrt{\frac{r_{h}}{u}}\right)\\&\times\Biggl[\left(\frac{\sqrt{\mu } r_{h}^6 \left(u^4-1\right) \left(\frac{r_{h}}{u}\right)^{7/4} e^{2 i (q x-t w)}}{36 \left(c^2 e^{2 \Phi
   } u^5+r_{h}^4 u \sqrt{\frac{r_{h}}{u}}\right)^4 \left(\frac{r_{h}^5}{c^2 e^{2 \Phi } u^5
   \sqrt{\frac{r_{h}}{u}}+r_{h}^5}\right)^{3/2}}\right)\phi^{\prime\prime}
-\left(\frac{c \sqrt{\mu } r_{h}^6 \left(\frac{r_{h}}{u}\right)^{7/4} \sqrt{c^2 e^{2 \phi }+\left(\frac{r_{h}}{u}\right)^{9/2}} e^{\Phi +2 i q x-2 i t w}}{36 \left(c^2 e^{2 \Phi } u^4+r_{h}^4 \sqrt{\frac{r_{h}}{u}}\right)^5
   \left(\frac{r_{h}^5}{c^2 e^{2 \Phi } u^5 \sqrt{\frac{r_{h}}{u}}+r_{h}^5}\right)^{3/2}}\right)H^{\prime}_{ty}
\\& +\left(\frac{c \sqrt{\mu } r_h^5 \left(\frac{r_{h}}{u}\right)^{11/4} \sqrt{c^2 e^{2 \Phi }+\left(\frac{r_{h}}{u}\right)^{9/2}} e^{\Phi +2 i q x-2 i t w}}{18 \left(c^2 e^{2 \Phi } u^4+r_{h}^4 \sqrt{\frac{r_{h}}{u}}\right)^5
   \left(\frac{r_h^5}{c^2 e^{2 \Phi } u^5 \sqrt{\frac{r_{h}}{u}}+r_{h}^5}\right)^{3/2}}\right)H_{ty}\Biggr]
\\& -\Biggl[\frac{\sqrt{\mu } r_{h}^6 \left(\frac{r_{h}}{u}\right)^{7/4}e^{2 i (q x-t w)}}{144 u^5 \left(c^2 e^{2 \Phi } u^4+r_{h}^4 \sqrt{\frac{r_{h}}{u}}\right)^4
\left(\frac{r_{h}^5}{c^2 e^{2 \Phi } u^5 \sqrt{\frac{r_{h}}{u}}+r_{h}^5}\right)^{3/2}}
\left(-8 c^6 e^{6 \Phi } u^{13} \left(u^4+1\right)\right.\\&\left.-3 c^4 e^{4 \Phi }  u^9
\left(5 u^4+11\right) \sqrt{\frac{r_h^9}{u}}-6c^2e^{2 \Phi }r_{h}^9 u^4 \left(u^4+7\right)+ \left(u^4-17\right)
\sqrt{\frac{r_h^{27}}{u}}\right)\Biggr]\phi^{\prime}
 \\&+\Biggl(\frac{\pi  {g_s} \sqrt{\mu } N u \left(\frac{r_{h}}{u}\right)^{3/4} \sqrt{\frac{r_{h}^5}{c^2 e^{2 \Phi } u^5
   \sqrt{\frac{r_{h}}{u}}+r_{h}^5}} e^{2 i (q x-t w)}}{9 r_{h}^6 \left(u^4-1\right)}\Biggr)
    \Biggl(-w^2 \left(c^2 e^{2 \Phi}u^4+r_{h}^4
   \sqrt{\frac{r_{h}}{u}}\right)-q^2 r_{h}^4 \left(u^4-1\right) \sqrt{\frac{r_{h}}{u}}\Biggr)\phi = 0.
   \end{split}
\end{equation}}
Equation (\ref{EOM-Hty}), setting $q=0$ and near $u=0$ is given as:
\begin{equation}
\label{Htyq0u0}
\frac{r_{h}^6u}{2g_{s}^{5/3}L^7}H_{ty}^{\prime\prime}-3\frac{r_{h}^6}{2g_{s}^{5/3}L^7}H_{ty}^{\prime}-\frac{1}{36}\sqrt{\mu } r_{h}^{13/4} u^{7/4} H_{ty}=0,
\end{equation}
whose solution is given by:
\begin{equation}\begin{split}
\label{Htyq0u0solution}
 H_{ty} &= \frac{4 2^{2/11} c_1 {g_s}^{40/33} L^{56/11} \mu ^{4/11} u^2 \Gamma \left(-\frac{5}{11}\right) I_{-\frac{16}{11}}\left(\frac{4 \sqrt{2} g_{s}^{5/6}L^{7/2} \mu^{1/4}  u^{11/8}}{33 r_{h}^{11/8}}\right)}{33 33^{5/11} r_{h}^2}\\
   &  -\frac{64 (-1)^{5/11} 2^{2/11} c_2 g_{s}^{40/33} L^{56/11} \mu
   ^{4/11} u^2 \Gamma \left(\frac{16}{11}\right) I_{\frac{16}{11}}\left(\frac{4 \sqrt{2} g_{s}^{5/6} L^{7/2} \mu^{1/4} u^{11/8}}{33
   r_{h}^{11/8}}\right)}{363 33^{5/11} r_{h}^2}\\
    & = \kappa_1 + \kappa_2 u^{\frac{11}{4}} + \gamma\kappa_2 u^4 + ....,
   \end{split}
\end{equation}
where:
\begin{eqnarray}
\label{Htynearu0}
& & \kappa_1 \equiv c_1;\nonumber\\
& & \kappa_2 \equiv -\frac{64 \sqrt{2} \pi ^{7/4} c_1 {g_s}^{41/12} \sqrt{\mu } N^{7/4}}{495 r_{h}^{11/4}};\nonumber\\
& & \gamma \equiv \frac{320 (-1)^{5/11} 2^{21/22} \pi ^{35/44} c_2 {g_s}^{205/132} \mu ^{5/22} N^{35/44} \Gamma \left(\frac{16}{11}\right)}{121 33^{10/11} c_1 r_{h}^{5/4}
   \Gamma \left(\frac{27}{11}\right)}.
\end{eqnarray}
Equation (\ref{EOM-Hxy}) near $u=0$ is given as:
\begin{eqnarray}
\label{HxyEOM-i}
& & 18 r_h^2 \left(u {H_{xy}}''(u)-3 {H_{xy}}'(u)\right) \sqrt{\frac{r_h^8}{L^6 g_s^{10/3}}}+L^4 \sqrt{\mu } r_{h}^{13/4} u^{7/4} {H_{xy}}(u)=0,
\end{eqnarray}
whose solution is given by:
\begin{eqnarray}
& & H_{xy}=\frac{64 2^{2/11} c_2 L^{56/11} \mu ^{4/11} r_{h}^{26/11} u^2 \Gamma \left(\frac{16}{11}\right) g_s^{40/33} J_{\frac{16}{11}}\left(\frac{4 \sqrt{2} L^{7/2}
   \sqrt[4]{\mu } r_{h}^{13/8} u^{11/8} g_s^{5/6}}{33 r_h^3}\right)}{363 33^{5/11} r_h^{48/11}}\nonumber\\
   & & +\frac{4 2^{2/11} c_1 L^{56/11} \mu ^{4/11} r_{h}^{26/11}
   u^2 \Gamma \left(-\frac{5}{11}\right) g_s^{40/33} J_{-\frac{16}{11}}\left(\frac{4 \sqrt{2} L^{7/2} \sqrt[4]{\mu } r_{h}^{13/8} u^{11/8} g_s^{5/6}}{33
   r_h^3}\right)}{33 33^{5/11} r_h^{48/11}}.
\end{eqnarray}
Substituting (\ref{Htyq0u0solution}), the $\phi(u)$ EOM near $u=0$ can be approximated by:
\begin{equation}\begin{split}
\label{phi EOM i}
& -\frac{4 r^{25/4}_{h}}{u^{5/4}} \phi^{\prime\prime}+\frac{17 r^{25/4}_{h}}{u^{9/4}} \phi^{\prime}-4 c {g_s} r^4_{h} u H^{\prime}_{ty}\\&+8 c g_{s} r^4_{h} H_{ty}+16 \pi  g_{s} i N r^{9/4}_{h} u^{11/4} \left(q^2-w^2\right) \phi=0,
   \end{split}
\end{equation}
whose solution is given by:whose solution is given by:{\footnotesize
\begin{eqnarray}
\label{phiu0solution_i}
& &\phi(u) = -\frac{1}{3024 \pi
   ^{9/16} {r_h}^{9/4} \left(-w^2\right)^{23/16} \left(\frac{u^3 \sqrt{-{g_s} i N w^2}}{{r_h}^2}\right)^{7/8} \Gamma \left(\frac{2}{3}\right) \Gamma
   \left(\frac{4}{3}\right) \Gamma \left(\frac{37}{24}\right) \Gamma \left(\frac{53}{24}\right)}\nonumber\\
   & & \times\Biggl\{{g_s}^{7/16} u^{13/4} \Biggl(\frac{1}{{r_h}^4}
   \Biggl\{u^{19/8} \Gamma \left(\frac{2}{3}\right) \Biggl[192 3^{7/8} c \gamma  {g_s}^{9/16} {k_2} \sqrt[8]{\pi }
   {r_h}^4 u^{13/8} \left(-w^2\right)^{23/16} I_{\frac{7}{8}}\left(\frac{2 \sqrt{\pi } u^3 \sqrt{-{g_s} i N w^2}}{3 {r_h}^2}\right)\nonumber\\&& \Gamma
   \left(\frac{1}{3}\right) \Gamma \left(\frac{37}{24}\right) \Gamma \left(\frac{15}{8}\right) \Gamma \left(\frac{53}{24}\right)  \,
   _1F_2\left(\frac{1}{3};\frac{1}{8},\frac{4}{3};-\frac{{g_s} i N \pi  u^6 w^2}{9 {r_h}^4}\right)-\frac{1}{\sqrt[4]{\frac{u^3 \sqrt{-{g_s} i N w^2}}{{r_h}^2}}}
   \Biggl\{w^2 \Gamma \left(\frac{4}{3}\right)
   \nonumber\\
   & &\Biggl(64
   \sqrt[8]{3} c {g_s}^{25/16} i {k_1} N \pi  \left(-w^2\right)^{23/16} I_{-\frac{7}{8}}\left(\frac{2 \sqrt{\pi } u^3 \sqrt{-{g_s} i N w^2}}{3
   {r_h}^2}\right) \Gamma \left(\frac{1}{8}\right) \Gamma \left(\frac{13}{24}\right) \Gamma \left(\frac{53}{24}\right)\nonumber\\
   & & \times \,
   _1F_2\left(\frac{13}{24};\frac{37}{24},\frac{15}{8};-\frac{{g_s} i N \pi  u^6 w^2}{9 {r_h}^4}\right) u^{29/8}\nonumber\\
   & &+\Gamma \left(\frac{37}{24}\right)
   \Biggl[9 {r_h}^{5/2} \Gamma \left(\frac{53}{24}\right) \Biggl(\frac{1}{u^3}\Biggl\{\left(-w^2\right)^{7/16} I_{-\frac{7}{8}}\left(\frac{2 \sqrt{\pi } u^3 \sqrt{-{g_s}
   i N w^2}}{3 {r_h}^2}\right) \Gamma \left(\frac{1}{8}\right)\nonumber\\
    & & \times\Biggl[21 \sqrt[8]{3} c {g_s}^{9/16} {k_2} u^{27/8}-24 c {g_s}^{9/16} {k_2}
   \sqrt[16]{\pi } \sqrt[8]{\frac{\sqrt{{g_s}} \sqrt{i} \sqrt{N} u^3 \sqrt{-w^2}}{{r_h}^2}} I_{-\frac{1}{8}}\left(\frac{2 \sqrt{{g_s}} \sqrt{i}
   \sqrt{N} \sqrt{\pi } u^3 \sqrt{-w^2}}{3 {r_h}^2}\right)\nonumber\\
   & &\Gamma \left(\frac{15}{8}\right) u^{27/8}-112 \sqrt[8]{3} i^{7/16} N^{7/16} \pi  \sqrt{{r_h}}
   \left(-w^2\right)^{7/16} \left(\frac{u^3 \sqrt{-{g_s} i N w^2}}{{r_h}^2}\right)^{9/8} c_1\Biggr] {r_h}^{3/2}\Biggr\}
   \nonumber\\
   & & +8 \sqrt[8]{{g_s}}
   \sqrt[8]{3 \pi } \sqrt[8]{\frac{u^3 \sqrt{-{g_s} i N w^2}}{{r_h}^2}} I_{\frac{7}{8}}\left(\frac{2 \sqrt{\pi } u^3 \sqrt{-{g_s} i N w^2}}{3
   {r_h}^2}\right)\Gamma \left(\frac{15}{8}\right)  \left(14 i^{9/16} N^{9/16} (-\pi )^{7/8} \left(-{g_s} i N w^2\right)^{3/8} c_2 w^2\right.\nonumber\\
   && \left.+3^{3/4} c
   {g_s}^{7/16} {k_2} {r_h}^{3/2} u^{3/8} \left(-w^2\right)^{7/16} \sqrt[8]{\frac{u^3 \sqrt{-{g_s} i N w^2}}{{r_h}^2}}\Gamma
   \left(\frac{1}{8}\right) \left(;\frac{9}{8};-\frac{{g_s} i N \pi  u^6 w^2}{9 {r_h}^4}\right)\right)\Biggr) \nonumber\\
   &&-64 \sqrt[8]{3} c \gamma  {g_s}^{25/16}
   i {k_2} N \pi  u^{61/8} \left(-w^2\right)^{23/16} I_{-\frac{7}{8}}\left(\frac{2 \sqrt{\pi } u^3 \sqrt{-{g_s} i N w^2}}{3 {r_h}^2}\right) \nonumber\\
   & & \Gamma
   \left(\frac{1}{8}\right) \Gamma \left(\frac{29}{24}\right) \, _1F_2\left(\frac{29}{24};\frac{15}{8},\frac{53}{24};-\frac{{g_s} i N \pi  u^6 w^2}{9
   {r_h}^4}\right)\Biggr]
   \Biggr)\Biggr\}\Biggr]\Biggr\} \nonumber\\
   && -192 3^{7/8} c {g_s}^{9/16} {k_1}
   \sqrt[8]{\pi } \left(-w^2\right)^{23/16} I_{\frac{7}{8}}\left(\frac{2 \sqrt{\pi } u^3 \sqrt{-{g_s} i N w^2}}{3 {r_h}^2}\right)\nonumber\\
& & \Gamma
   \left(-\frac{1}{3}\right) \Gamma \left(\frac{4}{3}\right) \Gamma \left(\frac{37}{24}\right) \Gamma \left(\frac{15}{8}\right) \Gamma
   \left(\frac{53}{24}\right) \, _1F_2\left(-\frac{1}{3};\frac{1}{8},\frac{2}{3};-\frac{{g_s} i N \pi  u^6 w^2}{9 {r_h}^4}\right)\Biggr)\Biggr\}.
       \end{eqnarray}}
\section{Frobenius Solution  of  EOM of Gauge-Invariant $Z_s(u)$ for Scalar Modes of Metric Fluctuations for ($\alpha^\prime=1$) $r:\log r<\log N$}
The $Z_s(u)$ EOM can be rewritten as:
\begin{equation}
(u-1)^2Z_s^{\prime\prime}(u) + (u-1)P(u-1) Z_s^\prime(u) + Q(u-1) Z_s(u) = 0,
\end{equation}
in which $P(u-1) = \sum_{n=0}^\infty p_n(u-1)^n$ and $Q(u-1) = \sum_{m=0}^\infty q_n (u-1)^n$  wherein, up to ${\cal O}\left(\frac{1}{N}\right)$:
{\footnotesize
\begin{equation}\begin{split}
\label{pn+qn_up_to_2nd_order}
&  p_0 = 1,\\
& p_1 = \frac{3 g_{s}^2 M^2 {N_f} \log (N) \left(28 q_{3}^4+36 q_{3}^2 \omega_{3}^2-81 \omega_{3}^4\right)}{64 \pi ^2 N \left(2 q_{3}^2-3
   \omega_{3}^2\right)^2}+\frac{10 q_{3}^2+9 \omega_{3}^2}{4 q_{3}^2-6 \omega_{3}^2},\\
    & p_2 = \frac{3 g_{s}^2 M^2 {N_f} \log (N) \left(712 q_{3}^6-948 q_{3}^4 \omega_{3}^2-162 q_{3}^2 \omega_{3}^4+405 \omega_{3}^6\right)}{64 \pi ^2 N \left(2
   q_{3}^2-3 \omega_{3}^2\right)^3}+\frac{364 q_{3}^4-420 q_{3}^2 \omega_{3}^2+99 \omega_{3}^4}{4 \left(2 q_{3}^2-3 \omega_{3}^2\right)^2};\\
    & q_0 = \frac{3 g_{s}^2 M^2 {N_f} \log (N) \left(\left(\omega_{3}^2+4\right) \left(27 \omega_{3}^2-10 q_{3}^2\right)-8 q_{3}^2 \omega_{3}^2 \log
   {N}\right)}{4096 \pi ^2 N q_{3}^2}+\frac{\omega_{3}^2}{16},\\
    & q_1 = \frac{1}{4096 \pi ^2 N q_{3}^2 \left(2 q_{3}^2-3 \omega_{3}^2\right)^2}\\
    & \times\Biggl\{3 g_{s}^2 M^2 {N_f} \log (N) \biggl(-8 \log (N) \left(4 q_{3}^2-3 \omega_{3}^2\right) \left(2 q_{3}^3-3 q_{3} \omega_{3}^2\right)^2 \\
    & -96
   q_{3}^8+8 q_{3}^6 \left(51 \omega_{3}^2-52\right)-36 q_{3}^4 \omega_{3}^2 \left(9 \omega_{3}^2+52\right)-54 q_{3}^2 \omega_{3}^4 \left(9
   \omega_{3}^2-28\right)+81 \omega_{3}^6 \left(7 \omega_{3}^2+20\right)\biggr)\Biggr\}\\
   & +\frac{8
   q_{3}^4-2 q_{3}^2 \left(9 \omega_{3}^2+32\right)+9 \omega_{3}^4}{32 q_{3}^2-48 \omega_{3}^2},\\
    & q_2 = \frac{-96 q_{3}^6+52 q_{3}^4 \left(7 \omega_{3}^2-64\right)+q_{3}^2 \left(3456 \omega_{3}^2-444 \omega_{3}^4\right)+171 \omega_{3}^6}{64 \left(2
   q_{3}^2-3 \omega_{3}^2\right)^2} -\frac{3 g_{s}^2 M^2 {N_f} \log (N) }{16384 \pi ^2 N q_{3}^2 \left(2 q_{3}^2-3 \omega_{3}^2\right)^3}\\
   & \times \Biggl[-8 q_{3}^2 \log (N) \left(24 q_{3}^2-19 \omega_{3}^2\right) \left(2
   q_{3}^2-3 \omega_{3}^2\right)^3+1920 q_{3}^{10}-16 q_{3}^8 \left(157 \omega_{3}^2-6268\right)-96 q_{3}^6 \omega_{3}^2 \left(351
   \omega_{3}^2+1444\right)\\&+216 q_{3}^4 \omega_{3}^4 \left(559 \omega_{3}^2+76\right)-864 q_{3}^2 \omega_{3}^6 \left(173 \omega_{3}^2+53\right)+243 \omega_{3}^8
   \left(265 \omega_{3}^2+308\right)\Biggr].
\end{split}
\end{equation}}
The Frobenius method then dictates that the solution is given by:
\begin{equation}
\label{solution-i}
Z_s(u) = \left(1 - u \right)^{\frac{3 g_{s}^2 M^2 {N_f} q_{3} \omega_{3}^2 \log (N) \left(8 q_{3}^2 \omega_{3}^2 \log (N)+\left(\omega_{3}^2+4\right) \left(10 q_{3}^2-27
   \omega_{3}^2\right)\right)}{2048 \pi ^2 N \left(-q_{3}^2 \omega_{3}^2\right)^{3/2}}-\frac{i \omega_{3}}{4}}\left(1 + \sum_{m=1}a_m (u - 1)^m\right),
\end{equation}
where
{\footnotesize
\begin{eqnarray}
\label{a1a2-sound}
& & a_1 = \frac{8 i q_{3}^4+2 q_{3}^2 \left(-9 i \omega_{3}^2+10 \omega_{3}-32 i\right)+9 (2+i \omega_{3}) \omega_{3}^3}{8 (\omega_{3}+2 i) \left(3 \omega_{3}^2-2
   q_{3}^2\right)}\nonumber\\
   & & -\frac{1}{4096 \pi ^2 N
   q_{3}^2 \omega_{3} (\omega_{3}+2 i)^2 \left(2 q_{3}^2-3 \omega_{3}^2\right)^2}\nonumber\\
   & & \times\Biggl\{3 i g_{s}^2 M^2 {N_f} \log (N) \biggl((\omega_{3}+2 i) \biggl(-32 q_{3}^8 (\omega_{3}+10 i)-8 q_{3}^6 \left(27
   \omega_{3}^3-146 i \omega_{3}^2+364 \omega_{3}-520 i\right)\nonumber\\
   & & +12 q_{3}^4 \omega_{3}^2 \left(141 \omega_{3}^3-486 i \omega_{3}^2+356 \omega_{3}-1336 i\right)-54
   q_{3}^2 \omega_{3}^4 \left(59 \omega_{3}^3-130 i \omega_{3}^2+44 \omega_{3}-200 i\right)\nonumber\\
   & & +81 \omega_{3}^6 \left(23 \omega_{3}^3-18 i \omega_{3}^2+4 \omega_{3}+72
   i\right)\biggr)-8 q_{3}^2 \omega_{3} \log (N) \biggl[16 q_{3}^6 (\omega_{3}+4 i)-4 q_{3}^4 \omega_{3} \left(15 \omega_{3}^2+60 i \omega_{3}-52\right)\nonumber\\
   & & +24
   q_{3}^2 \omega_{3}^3 \left(3 \omega_{3}^2+12 i \omega_{3}-10\right)-27 \omega_{3}^5 \left(\omega_{3}^2+4 i \omega_{3}+4\right)\biggr]\biggr)\Biggr\},\nonumber\\
   & & a_2 = \frac{1}{128 (\omega_{3}+2 i) (\omega_{3}+4 i) \left(3 \omega_{3}^2-2 q_{3}^2\right)}\Biggl\{32 q_{3}^6-32 q_{3}^4 \omega_{3} (3 \omega_{3}+8 i)\nonumber\\
   & & +2 q_{3}^2 \left(45 \omega_{3}^4+98 i \omega_{3}^3+624 \omega_{3}^2+32 i \omega_{3}+3072\right)-3
   \omega_{3}^3 \left(9 \omega_{3}^3+2 i \omega_{3}^2+48 \omega_{3}+32 i\right)\Biggr\}\nonumber\\
   & & -\frac{1}{32768 \pi ^2 N q_{3}^2 \omega_{3} (\omega_{3}+2 i)^2
   (\omega_{3}+4 i)^2 \left(2 q_{3}^2-3 \omega_{3}^2\right)^2}\Biggl\{3
   g_{s}^2 M^2 {N_f} \log {N}\nonumber\\
   & &\times \Biggl[(-\omega_{3}-2 i) \biggl(128 q_{3}^{10} \left(\omega_{3}^2+19 i \omega_{3}-30\right)+256 q_{3}^8 \left(3
   \omega_{3}^4-18 i \omega_{3}^3+33 \omega_{3}^2-76 i \omega_{3}+160\right)\nonumber\\
   & & -8 q_{3}^6 \left(927 \omega_{3}^6-104 i \omega_{3}^5+17784 \omega_{3}^4+11808 i
   \omega_{3}^3+70256 \omega_{3}^2+42368 i \omega_{3}+93440\right)\nonumber\\
   & & +12 q_{3}^4 \omega_{3}^2 \left(1485 \omega_{3}^6+128 i \omega_{3}^5+31912 \omega_{3}^4+14176 i
   \omega_{3}^3+96656 \omega_{3}^2-68992 i \omega_{3}+262912\right)\nonumber\\
   & & -18 q_{3}^2 \omega_{3}^4 \left(945 \omega_{3}^6-1388 i \omega_{3}^5+27920 \omega_{3}^4+13664 i
   \omega_{3}^3+54736 \omega_{3}^2-80768 i \omega_{3}+171776\right)\nonumber\\
   & & +81 \omega_{3}^6 \left(69 \omega_{3}^6-416 i \omega_{3}^5+3960 \omega_{3}^4+4128 i \omega_{3}^3+5520
   \omega_{3}^2+4736 i \omega_{3}+768\right)\biggr)-8 q_{3}^2 \omega_{3} \log (N)\nonumber\\
    & & \times\biggl(64 q_{3}^8 \left(\omega_{3}^2+9 i \omega_{3}-16\right)-32 q_{3}^6
   \omega_{3} \left(9 \omega_{3}^3+89 i \omega_{3}^2-240 \omega_{3}-192 i\right)\nonumber\\
   & & +4 q_{3}^4 \omega_{3} \left(117 \omega_{3}^5+1198 i \omega_{3}^4-3612 \omega_{3}^3-1624
   i \omega_{3}^2-8064 \omega_{3}-9344 i\right)\nonumber\\
   & & -12 q_{3}^2 \omega_{3}^3 (\omega_{3}+4 i)^2 \left(27 \omega_{3}^3+52 i \omega_{3}^2+116 \omega_{3}+296 i\right)\nonumber\\
   & & +9
   \omega_{3}^5 (\omega_{3}+2 i)^2 \left(9 \omega_{3}^3+46 i \omega_{3}^2+64 \omega_{3}+32 i\right)\biggr)\Biggr]\Biggr\}.
   \end{eqnarray}}
As stated in {\bf 5.3.1}, imposing Dirichlet boundary condition $Z_s(u=0)=0$ and going up to second order in powers of $(u-1)$ in (\ref{solution-i}) and considering in the hydrodynamical limit $\omega_3^nq_3^m:m+n=2$ one obtains:
\begin{equation}
\label{vs+Gammas-second_order}
\omega_3 = -\frac{2 q_{3}}{\sqrt{3}}-\frac{9 i q_{3}^2}{32},
\end{equation}
which yields a result for the speed of sound similar to (\ref{v_s 1}) for $n=0,1$.
To get the LO or conformal result for the speed of sound $v_s = \frac{1}{\sqrt{3}}$, let us go to the fourth order in (\ref{solution-i}). For this, up to ${\cal O}\left(\frac{1}{N}\right)$, we will need:
{\footnotesize
\begin{eqnarray}
\label{pn+qn_up_to_fourth_order}
& & p_3 = \frac{3 g_{s}^2 M^2 {N_f} \log (N) \left(6256 q_{3}^8-9600 q_{3}^6 \omega_{3}^2-4104 q_{3}^4 \omega_{3}^4+12960 q_{3}^2 \omega_{3}^6-5589
   \omega_{3}^8\right)}{64 \pi ^2 N \left(2 q_{3}^2-3 \omega_{3}^2\right)^4}\nonumber\\
   & & +\frac{3880 q_{3}^6-4788 q_{3}^4 \omega_{3}^2+270 q_{3}^2 \omega_{3}^4+729
   \omega_{3}^6}{8 \left(2 q_{3}^2-3 \omega_{3}^2\right)^3},\nonumber\\
   & & p_4 = \frac{3 g_{s}^2 M^2 {N_f} \log (N) }{64 \pi ^2 N \left(2 q_{3}^2-3 \omega_{3}^2\right)^5}\nonumber\\
   & & \times \left(53536 q_{3}^{10}-110256 q_{3}^8 \omega_{3}^2-2736 q_{3}^6 \omega_{3}^4+168264 q_{3}^4
   \omega_{3}^6-156006 q_{3}^2 \omega_{3}^8+47385 \omega_{3}^{10}\right)\nonumber\\
   & &+\frac{3 \left(17360
   q_{3}^8-32992 q_{3}^6 \omega_{3}^2+19320 q_{3}^4 \omega_{3}^4-5112 q_{3}^2 \omega_{3}^6+1485 \omega_{3}^8\right)}{16 \left(2 q_{3}^2-3
   \omega_{3}^2\right)^4};\nonumber\\
   & & q_3 = \frac{3 g_{s}^2 M^2 {N_f} \log (N) }{4096 \pi ^2 N q_{3}^2 \left(2 q_{3}^2-3
   \omega_{3}^2\right)^4}\nonumber\\
   & & \times \biggl[-40 q_{3}^2 \log (N) \left(2 q_{3}^2-3 \omega_{3}^2\right)^4 \left(q_{3}^2-\omega_{3}^2\right)-3552
   q_{3}^{12}+416 q_{3}^{10} \left(31 \omega_{3}^2-648\right)+96 q_{3}^8 \omega_{3}^2 \left(466 \omega_{3}^2+5695\right)
   \nonumber\\& & -288 q_{3}^6 \omega_{3}^4
   \left(1107 \omega_{3}^2+818\right)+54 q_{3}^4 \omega_{3}^6 \left(12319 \omega_{3}^2+840\right)-162 q_{3}^2 \omega_{3}^8 \left(3779
   \omega_{3}^2+1632\right)\nonumber\\
   & & +729 \omega_{3}^{10} \left(293 \omega_{3}^2+250\right)\biggr]\nonumber\\
   & &+\frac{40 q_{3}^8-4 q_{3}^6 \left(55 \omega_{3}^2+1488\right)+6 q_{3}^4 \omega_{3}^2 \left(75 \omega_{3}^2+1696\right)-9
   q_{3}^2 \omega_{3}^4 \left(45 \omega_{3}^2+464\right)+135 {\omega _3}^8}{16 \left(2 q_{3}^2-3 \omega_{3}^2\right)^3},\nonumber\\
   & & q_4 =\frac{1}{256 \left(2 q_{3}^2-3
   \omega_{3}^2\right)^4}\Biggl\{-640 q_{3}^{10}+48 q_{3}^8 \left(99 \omega_{3}^2-12352\right)-288 q_{3}^6 \omega_{3}^2 \left(49 \omega_{3}^2-4400\right)\nonumber\\
   & &+216 q_{3}^4 \omega_{3}^4
   \left(97 \omega_{3}^2-3680\right)-1728 q_{3}^2 \omega_{3}^6 \left(9 \omega_{3}^2-70\right)+4617 \omega_{3}^{10}\Biggr\}\nonumber\\
   & & -\frac{1}{65536 \pi ^2 N q_{3}^2 \left(2 q_{3}^2-3 \omega_{3}^2\right)^5}\Biggl\{3 g_{s}^2 M^2 {N_f} \log (N) \nonumber\\
   & & \biggl(3 \biggl[112128 q_{3}^{14}-64 q_{3}^{12} \left(6133 \omega_{3}^2-190428\right)-192
   q_{3}^{10} \omega_{3}^2 \left(15391 \omega_{3}^2+151076\right)\nonumber\\
   & & +720 q_{3}^8 \omega_{3}^4 \left(30785 \omega_{3}^2+17332\right)-28800 q_{3}^6 \omega_{3}^6
   \left(2079 \omega_{3}^2-443\right)+324 q_{3}^4 \omega_{3}^8 \left(252677 \omega_{3}^2+420\right)\nonumber\\
   & & -972 q_{3}^2 \omega_{3}^{10} \left(58571
   \omega_{3}^2+17828\right)+729 \omega_{3}^{12} \left(22027 \omega_{3}^2+12156\right)\biggr]\nonumber\\
   & &-8 q_{3}^2 \log (N) \left(40 q_{3}^2-57 \omega_{3}^2\right) \left(2
   q_{3}^2-3 \omega_{3}^2\right)^5\biggr)\Biggr\}.
\end{eqnarray}}
We will not quote the expressions for $a_3$ and $a_4$ because they are too cumbersome. Substituting the expressions for $a_{1,2,3,4}$ into $Z(u)$ and implementing the Dirichlet boundary condition: $Z_s(u=0)=0$, in the hydrodynamical limit, going up to ${\cal O}(\omega_3^4)$ one sees that one can write the Dirichlet boundary condition as a quartic: $a \omega_3^4 + b \omega_3^3 + c \omega_3^2 + f \omega_3 + g = 0$ where:
\begin{eqnarray}
\label{a+b+c+f+g}
& & a = -\frac{(17978967982080+432082299912192 i) g_{s}^2 M^2 {N_f} \log (N)}{N},\nonumber\\
& & b = -16384 q_{3}^2 \left(55717134336 \pi ^2-\frac{(8065585152-2189804544 i) g_{s}^2 M^2 {N_f} \log (N)}{N}\right),\nonumber\\
& & c = \frac{(6351753314304+163465918414848 i) g_{s}^2 M^2 {N_f} q_{3}^2 \log (N)}{N},\nonumber\\
& & f = 196608 q_{3}^4 \left(981467136 \pi ^2-\frac{(25958400-36690432 i) g_{s}^2 M^2 {N_f} \log (N)}{N}\right),\nonumber\\
& & g = -\frac{(842551787520+22613002813440 i) g_{s}^2 M^2 {N_f} q_{3}^4 \log (N)}{N}.
\end{eqnarray}

\section{Frobenius Solution of  EOM of Gauge-Invariant $Z_s(u)$  for Scalar Modes of Metric Fluctuations for $ (\alpha^\prime=1)\ r: \log r\sim\log N$}
Constructing a $Z_s(u)$ which is invariant under diffeomorphisms: $h_{\mu\nu}\rightarrow h_{\mu\nu} - \nabla_{(\mu}\xi_{\nu)}$, one sees one obtain the following equation of motion for $Z_s(u)$:
\begin{eqnarray}
\label{Z-EOM}
& & Z_s^{\prime\prime}(u) = \Biggl[\frac{q_3^2 \left(7 u^8-8 u^4+9\right)-3 \left(u^4+3\right) \omega_3^2}{u \left(u^4-1\right) \left(q_3^2 \left(u^4-3\right)+3
   \omega_3^2\right)}\nonumber\\
   & &   -\frac{1}{64 \pi ^2 N u \left(u^8-4 u^4+3\right)
   \left(q_3^2 \left(u^4-3\right)+3 \omega_3^2\right)^2}\Biggl\{-3 g_{s}^2 M^2 {N_f} \log N\nonumber\\
   & &  \times \Biggl(q_3^4 \left(5 u^{16}-98 u^{12}+372 u^8-414 u^4+135\right)+2
   q_3^2 \left(32 u^{12}-183 u^8+306 u^4-135\right) \omega_3^2\nonumber\\
   & & +3 \left(u^8-66 u^4+45\right) \omega_3^4\Biggr)\Biggr\}\Biggr]Z_s^\prime(u)\nonumber\\
& & + \Biggl[\frac{1}{128 \pi ^2 N
   q_3^2 \left(u^4-3\right) \left(u^4-1\right)^3 \left(q_3^2 \left(u^4-3\right)+3 \omega_3^2\right)^2}\Biggl\{-3 g_{s}^2 M^2 {N_f} \log N\nonumber\\
   & & \times \Biggl(30 q_3^6 u^{22}-542 q_3^6 u^{18}-7 q_3^6 u^{16} \omega_3^2+2540
   q_3^6 u^{14}+46 q_3^6 u^{12} \omega_3^2-4764 q_3^6 u^{10}-84 q_3^6 u^8 \omega_3^2\nonumber\\
   & & +4086 q_3^6 u^6+18 q_3^6 u^4
   \omega_3^2-1350 q_3^6 u^2+27 q_3^6 \omega_3^2+318 q_3^4 u^{18} \omega_3^2-2464 q_3^4 u^{14} \omega_3^2\nonumber\\
   & & -49 q_3^4 u^{12}
   \omega_3^4+6972 q_3^4 u^{10} \omega_3^2+189 q_3^4 u^8 \omega_3^4-8496 q_3^4 u^6 \omega_3^2-99 q_3^4 u^4 \omega_3^4+3510
   q_3^4 u^2 \omega_3^2\nonumber\\
   & & -81 q_3^4 \omega_3^4+114 q_3^2 u^{14} \omega_3^4-2262 q_3^2 u^{10} \omega_3^4-105 q_3^2 u^8
   \omega_3^6+5598 q_3^2 u^6 \omega_3^4+144 q_3^2 u^4 \omega_3^6\nonumber\\
   & &-2970 q_3^2 u^2 \omega_3^4+81 q_3^2 \omega_3^6 -8 \left(u^8-4
   u^4+3\right) \left(q_3^2 \left(u^4-1\right)+\omega_3^2\right)\nonumber\\
   & & \left(q_3^3 \left(u^4-3\right)+3 q_3 \omega_3^2\right)^2 \log
   \left(\frac{r_{h}}{u}\right)+18 u^{10} \omega_3^6-1188 u^6 \omega_3^6-63 u^4 \omega_3^8+810 u^2 \omega_3^6-27 \omega_3^8\Biggr)\Biggr\}\nonumber\\
   & & -\frac{q_3^4 \left(u^8-4 u^4+3\right)+2
   q_3^2 \left(8 u^{10}-8 u^6+2 u^4 \omega_3^2-3 \omega_3^2\right)+3 \omega_3^4}{\left(u^4-1\right)^2 \left(q_3^2 \left(u^4-3\right)+3
   \omega_3^2\right)}\Biggr]Z_s(u).
\end{eqnarray}
The horizon $u=1$ due to inclusion of the non-conformal corrections to the metric, ceases to be an irregular singular point. One then tries the ansatz: $Z_s(u) = e^{S(u)}$ near $u=1$. Assuming that $\left(S^{\prime}\right)^2\gg S^{\prime\prime}(u)$ near $u=1$ the differential equation (\ref{Z-EOM}), which could written as $Z_s^{\prime\prime}(u) = m(u)Z_s^\prime + l(u) Z_s(u)$
can be approximated by:
\begin{equation}
\label{S_EOM}
\left(S^\prime\right)^2 - m(u) S^\prime(u) - l(u) \approx 0.
\end{equation}
A solution to (\ref{S_EOM}) is:
\begin{eqnarray}
\label{S-solution}
& & S(u) = \frac{1}{2}\left(m(u) - \sqrt{m^2(u) + 4 l(u)}\right)\nonumber\\
& & = -\frac{\sqrt{\frac{15}{2}} \sqrt{\frac{g_{s}^2 M^2 {N_f} \omega_{3}^2 \left(\omega_{3}^2+4\right) \log
   \left(\frac{1}{N}\right)}{N q_{3}^2}}}{64 \pi  (u-1)^{3/2}} + \frac{\frac{15 g_{s}^2 M^2 {N_f} \omega_{3}^2 \log \left(\frac{1}{N}\right)}{256 \pi ^2 N \left(2 q_{3}^2-3
   \omega_{3}^2\right)}-\frac{1}{2}}{u-1}\nonumber\\
   & & +\frac{3 g_{s}^2 M^2 {N_f} \log \left(\frac{1}{N}\right) \left(112 q_{3}^4+214
   q_{3}^2 \omega_{3}^2-369 \omega_{3}^4\right)+128 \pi ^2 N \left(-20 q_{3}^4+12 q_{3}^2 \omega_{3}^2+27 \omega_{3}^4\right)}{512 \pi ^2 N \left(2
   q_{3}^2-3 \omega_{3}^2\right)^2}\nonumber\\
   & & +\frac{1}{1024 \sqrt{30} \pi ^3
   N^2 q_{3}^2 \sqrt{u-1} \left(2 q_{3}^2-3 \omega_{3}^2\right)^2 \sqrt{\frac{g_{s}^2 M^2 {N_f} \omega_{3}^2 \left(\omega_{3}^2+4\right) \log
   \left(\frac{1}{N}\right)}{N q_{3}^2}}}\nonumber\\
   & & \times\Biggl\{-225 g_{s}^4 M^4 {N_f}^2 q_{3}^2 \omega_{3}^4 \log ^2\left(\frac{1}{N}\right)+12 \pi ^2 g_{s}^2 M^2
   N {N_f} \log \left(\frac{1}{N}\right) \left(2 q_{3}^2-3 \omega_{3}^2\right)\nonumber\\
   & & \times \Biggl[-80 q_{3}^4 \left(\omega_{3}^2-4\right)+2 q_{3}^2 \omega_{3}^2
   \left(57 \omega_{3}^2-412\right)+64 \left(2 q_{3}^4 \omega_{3}^2-3 q_{3}^2 \omega_{3}^4\right) \log (r_{h})\nonumber\\
   & &+9 \omega_{3}^4
   \left(\omega_{3}^2+164\right)\Biggr]+4096 \pi ^4 N^2 q_{3}^2 \left(\omega_{3}^2-4\right) \left(2 q_{3}^2-3 \omega_{3}^2\right)^2\Biggr\}  + {\cal O}\left(\sqrt{u-1}\right).
\end{eqnarray}
Taking first the MQGP limit, the first term in the RHS of (\ref{S-solution}) can be dropped. After integrating with respect to $u$, the solution (\ref{S-solution}) to equation (\ref{S_EOM}) will reflect the singular nature of $Z(u)$'s equation of motion (\ref{S-solution}) via
\begin{equation}
\label{pole-soln-Z}
Z_s(u)\sim \left(1 - u \right)^{-\frac{1}{2} + \frac{15 g_{s}^2 M^2 {N_f} \omega_{3}^2 \log \left(\frac{1}{N}\right)}{256 \pi ^2 N \left(2 q_{3}^2-3\omega_{3}^2\right)}}F(u),
\end{equation}
 where $F(u)$ is regular in $u$ and its equation of motion, around $u=0$, is given by:
 {\footnotesize
\begin{eqnarray}
\label{F-EOM}
& & 256 F''(u)+\frac{F'(u) \left(\frac{60 g_{s}^2 M^2 {N_f} \log \left(\frac{1}{N}\right)}{\pi ^2 N}-768\right)}{u}\nonumber\\
& & -\frac{3 F(u) \left(64 \pi ^2 N-5
   g_{s}^2 M^2 {N_f} \log \left(\frac{1}{N}\right)\right) \left(15 g_{s}^2 M^2 {N_f} \omega_{3}^2 \log (N)+128 \pi ^2 N \left(2 q_{3}^2-3
   \omega_{3}^2\right)\right)}{64 \pi ^4 N^2 u \left(2 q_{3}^2-3 \omega_{3}^2\right)} = 0.\nonumber\\
   & &
\end{eqnarray}}
The solution to (\ref{F-EOM}) is given by:
{\footnotesize
\begin{eqnarray}
\label{F-eom-solution}
& & F(u) = 2^{-\frac{105 g_{s}^2 M^2 {N_f} \log (N)}{64 \pi ^2 N}-28} 3^{\frac{15 g_{s}^2 M^2 {N_f} \log (N)}{128 \pi ^2 N}+2} N^{-\frac{15 g_{s}^2 M^2
   {N_f} \log (N)}{64 \pi ^2 N}-4} \pi^{-\frac{15 g_{s}^2 M^2 {N_f} \log (N)}{32 \pi ^2 N}-8}\nonumber\\
  & & \times  \left(5 g_{s}^2 M^2 {N_f} \log (N)+64 \pi ^2
   N\right)^{2-\frac{15 g_{s}^2 M^2 {N_f} \log (N)}{128 \pi ^2 N}} u^{\frac{15 g_{s}^2 M^2 {N_f} \log (N)}{128 \pi ^2 N}+2} \left(2 q_{3}^2-3
   \omega_{3}^2\right)^{-\frac{15 g_{s}^2 M^2 {N_f} \log (N)}{128 \pi ^2 N}-2}\nonumber\\
   & & \times \left(15 g_{s}^2 M^2 {N_f} \omega_{3}^2 \log (N)+128 \pi ^2 N \left(2
   q_{3}^2-3 \omega_{3}^2\right)\right)^{2-\frac{15 g_{s}^2 M^2 {N_f} \log (N)}{128 \pi ^2 N}}\nonumber\\
 & & \times   \Biggl(c_1 \left(5 g_{s}^2 M^2 {N_f} \log (N)+64
   \pi ^2 N\right)^{\frac{15 g_{s}^2 M^2 {N_f} \log (N)}{64 \pi ^2 N}} \Gamma \left(-\frac{15 g_{s}^2 {N_f} \log (N) M^2}{64 N \pi ^2}-3\right)\nonumber\\
& & \times   \left(15 g_{s}^2 M^2 {N_f} \omega_{3}^2 \log (N)+128 \pi ^2 N \left(2 q_{3}^2-3 \omega_{3}^2\right)\right)^{\frac{15 g_{s}^2 M^2 {N_f} \log
   (N)}{64 \pi ^2 N}}\nonumber\\
   & & \times I_{-\frac{15 g_{s}^2 {N_f} \log (N) M^2}{64 N \pi ^2}-4}\left(\frac{\sqrt{3} \sqrt{u} \sqrt{\left(5 g_{s}^2 {N_f} \log (N)
   M^2+64 N \pi ^2\right) \left(15 g_{s}^2 M^2 {N_f} \log (N) \omega_{3}^2+128 N \pi ^2 \left(2 q_{3}^2-3 \omega_{3}^2\right)\right)}}{64 \pi ^2
   \sqrt{N^2 \left(2 q_{3}^2-3 \omega_{3}^2\right)}}\right)\nonumber\\
   & & +c_2 N^{\frac{15 g_{s}^2 M^2 {N_f} (2 \log (N)+i \pi )}{64 \pi ^2 N}} \Gamma \left(\frac{15
   g_{s}^2 {N_f} \log (N) M^2}{64 N \pi ^2}+5\right) \left(2 q_{3}^2-3 \omega_{3}^2\right)^{\frac{15 g_{s}^2 M^2 {N_f} \log (N)}{64 \pi ^2 N}}\nonumber\\
& & \times   \left(\left(5 g_{s}^2 M^2 {N_f} \log (N)+64 \pi ^2 N\right) \left(15 g_{s}^2 M^2 {N_f} \omega_{3}^2 \log (N)+128 \pi ^2 N \left(2 q_{3}^2-3
   \omega_{3}^2\right)\right)\right)^{\frac{15 g_{s}^2 M^2 {N_f} \log (N)}{64 \pi ^2 N}}\nonumber\\
   & & \times \left(N^2 \left(2 q_{3}^2-3
   \omega_{3}^2\right)\right)^{-\frac{15 g_{s}^2 M^2 {N_f} \log (N)}{64 \pi ^2 N}}\nonumber\\
   & & \times I_{\frac{15 g_{s}^2 {N_f} \log (N) M^2}{64 N \pi
   ^2}+4}\left(\frac{\sqrt{3} \sqrt{u} \sqrt{\left(5 g_{s}^2 {N_f} \log (N) M^2+64 N \pi ^2\right) \left(15 g_{s}^2 M^2 {N_f} \log (N)
   \omega_{3}^2+128 N \pi ^2 \left(2 q_{3}^2-3 \omega_{3}^2\right)\right)}}{64 \pi ^2 \sqrt{N^2 \left(2 q_{3}^2-3 \omega_{3}^2\right)}}\right)\Biggr).\nonumber\\
   & &
\end{eqnarray}}
One notes from (\ref{F-eom-solution}) that $F(u\sim0) = c_1$. This needs to be improved upon by including the sub-leading terms in $u$ in $F'(u)$ in (\ref{F-EOM}), implying that we should look at:
{\footnotesize
\begin{eqnarray}
\label{improved_F_EOM}
& &  256 F''(u)+F'(u) \Biggl(\frac{120 g_{s}^2 M^2 {N_f} \log N\left(2 q_{3}^2-3 \omega_{3}^2\right)+30 g_{s}^2 M^2
   {N_f} \omega_{3}^2 \log (N)+1792 \pi ^2 N \left(2 q_{3}^2-3 \omega_{3}^2\right)}{\pi ^2 N \left(2 q_{3}^2-3
   \omega_{3}^2\right)}\nonumber\\
& & +\frac{\frac{-60 g_{s}^2 M^2 {N_f} \log N}{\pi ^2 N}-768}{u}\Biggr)\nonumber\\
   & & -\frac{3 F(u) \left(64 \pi ^2 N+5
   g_{s}^2 M^2 {N_f} \log N\right) \left(15 g_{s}^2 M^2 {N_f} \omega_{3}^2 \log (N)+128 \pi ^2 N \left(2
   q_{3}^2-3 \omega_{3}^2\right)\right)}{64 \pi ^4 N^2 u \left(2 q_{3}^2-3 \omega_{3}^2\right)} = 0.
   \end{eqnarray}}
   The solution to (\ref{improved_F_EOM}) near $u=0$, is given as under:
   \begin{eqnarray}
   \label{solution-improved-F-EOM}
   & & N^{\frac{15 g_{s}^2 M^2 {N_f} u \left(11 \omega_{3}^2-8 q_{3}^2\right)}{128 \pi ^2 N \left(2 q_{3}^2-3 \omega_{3}^2\right)}}
   \nonumber\\
   & & \times\Biggl[u^4
   \Biggl(\frac{c_1 \Gamma \left(-\frac{15 g_{s}^2 {N_f} \log (N) M^2}{64 N \pi ^2}-4\right)}{\Gamma \left(-\frac{3 \left(5 g_{s}^2 {N_f}
   \log (N) M^2+64 N \pi ^2\right) \left(15 g_{s}^2 {N_f} \left(16 q_{3}^2-23 \omega_{3}^2\right) \log (N) M^2+1664 N \pi ^2 \left(2
   q_{3}^2-3 \omega_{3}^2\right)\right)}{128 N \pi ^2 \left(15 g_{s}^2 {N_f} \left(8 q_{3}^2-11 \omega_{3}^2\right) \log (N) M^2+896 N \pi
   ^2 \left(2 q_{3}^2-3 \omega_{3}^2\right)\right)}\right)}\nonumber\\
   & & +c_2 L_{-\frac{225 g_{s}^4 M^4 {N_f}^2 \omega_{3}^2 \log ^2(N)+4800 \pi ^2
   g_{s}^2 M^2 N {N_f} \log (N) \left(4 q_{3}^2-5 \omega_{3}^2\right)+139264 \pi ^4 N^2 \left(2 q_{3}^2-3 \omega_{3}^2\right)}{128 \pi ^2 N
   \left(15 g_{s}^2 M^2 {N_f} \log (N) \left(8 q_{3}^2-11 \omega_{3}^2\right)+896 \pi ^2 N \left(2 q_{3}^2-3
   \omega_{3}^2\right)\right)}}^{\frac{15 g_{s}^2 M^2 {N_f} \log (N)}{64 \pi ^2 N}+4}(0)\Biggr)\nonumber\\
   & & +\frac{1}{\Gamma \left(\frac{225 g_{s}^4 {N_f}^2 \omega_{3}^2 \log ^2(N) M^4+4800 g_{s}^2 N
   {N_f} \pi ^2 \left(4 q_{3}^2-5 \omega_{3}^2\right) \log (N) M^2+139264 N^2 \pi ^4 \left(2 q_{3}^2-3 \omega_{3}^2\right)}{128 N \pi ^2
   \left(15 g_{s}^2 {N_f} \left(8 q_{3}^2-11 \omega_{3}^2\right) \log (N) M^2+896 N \pi ^2 \left(2 q_{3}^2-3
   \omega_{3}^2\right)\right)}\right)}\nonumber\\
   & & \times\Biggl\{c_1 2^{\frac{105 g_{s}^2 M^2 {N_f}
   \log (N)}{64 \pi ^2 N}+28} \pi ^{\frac{15 g_{s}^2 M^2 {N_f} \log (N)}{32 \pi ^2 N}+8} u^{-\frac{15 g_{s}^2 M^2 {N_f} \log (N)}{64 \pi
   ^2 N}} \Gamma \left(\frac{15 g_{s}^2 {N_f} \log (N) M^2}{64 N \pi ^2}+4\right)\nonumber\\
   & &  \left(\frac{15 g_{s}^2 M^2 {N_f} \log (N) \left(8
   q_{3}^2-11 \omega_{3}^2\right)+896 \pi ^2 N \left(2 q_{3}^2-3 \omega_{3}^2\right)}{N \left(2 q_{3}^2-3 \omega_{3}^2\right)}\right)^{-\frac{15
   g_{s}^2 M^2 {N_f} \log (N)}{64 \pi ^2 N}-4}\Biggr\}\Biggr].\nonumber\\
   & &
   \end{eqnarray}
\section{Gauge Transformations Preserving $h_{m\ \mu}=0$,  Pole Structure of $\Omega(\omega_3,q_3)$ and Solutions to ${\cal H}_{ab}(u)$}
\subsection{Gauge Transformations Preserving $h_{m\ \mu}=0$}
There are three gauge transformations that preserve $h_{\mu u}=0$, for the black $M3$-brane metric having integrated out the $M_6$ in the (asymptotic) $AdS_5\times M_6$ in the MQGP limit. They are given below:

{\bf Set I}: The Gauge transformations are generated by
\begin{eqnarray}
&& \xi_{{x}}= \frac{C_{{x}}(t,{x})}{u^2}+\xi^{(1)}_{{x}}(u,t,{x})\nonumber\\
&& \xi_{t}=\xi^{(1)}_{t}(u,t,{x})
\end{eqnarray}
The Gauge Solutions for the above kind of  transformations are given as:
\begin{eqnarray}
\label{GTI}
&& H^{\rm Gauge(I)}_{tt}=\frac{g^{\frac{2}{3}}_{s}}{i L^2}\left[\frac{2 \omega_3 u^2}{g_1}\tilde{\xi}^{(1)}_{t}-q_3 \tilde{C}_{{x}} H_{t t}-\frac{ 2 \omega_3 \tilde{C}_{{x}}}{g_1} H_{t {x}}\right]\nonumber\\
&& H^{\rm Gauge(I)}_{{x} {x}}=\frac{g^{\frac{2}{3}}_{s}}{i L^2}\left[ q_3 \tilde{C}_{{x}} H_{{x} {x}}-2 q_3 u^2 \tilde{\xi}^{(1)}_{{x}}- { 2 q_3 \tilde{C}_{{x}}} \right]\nonumber\\
&&H^{\rm Gauge(I)}_{t {x}}=\frac{g^{\frac{2}{3}}_{s}}{i L^2}\left[\omega_3 \tilde{C}_{{x}}+ u^2 \omega_3 \tilde{\xi}^{(1)}_{x} - u^2 q_3 \tilde{\xi}^{(1)}_{t}-\omega_3 \tilde{C}_{{x}}H_{{x} {x}}\right]\nonumber\\
&&  H^{\rm gauge (I)}_{a a}= \frac{g^{\frac{2}{3}}_{s}}{i L^2}\left[ -q_3 \tilde{C}_{{x}} H_{aa}\right]
\end{eqnarray}
where $H_{aa}=H_{{y} {y}}+H_{{z} {z}}, \tilde{C}_{{x}}\equiv\frac{C_{{x}}}{\pi T}, \tilde{\xi}^{(1)}_t\equiv\frac{\xi^{(1)}_t}{\pi T}$.

{\bf Set II}: The Gauge transformations are generated by
\begin{eqnarray}
&& \xi_{t}=- \frac{ g_1 C_t(t,{x})}{u^2}+\xi^{(1)}_{t}(u,t,{x})\nonumber\\
&& \xi_{{x}}=\xi^{(1)}_{{x}}(u,t,{x})
\end{eqnarray}
The Gauge Solutions for the above kind of  transformations are given as:
\begin{eqnarray}
\label{GTII}
&& H^{\rm Gauge(II)}_{tt}=\frac{g^{\frac{2}{3}}_{s}}{i L^2}\left[-2 \omega_3 \tilde{C}_{t}+\frac{2 \omega_3 u^2}{g_1}\tilde{\xi}^{(1)}_{t}-\omega_3 \tilde{C}_{t} H_{tt} \right]\nonumber\\
&& H^{\rm Gauge(II)}_{{x} {x}}=\frac{g^{\frac{2}{3}}_{s}}{i L^2}\left[ -2 q_3 u^2 \tilde{\xi}^{(1)}_{{x}}+ 2 q_3 \tilde{C}_{t} H_{t {x}}+ \omega_3   \tilde{C}_t H_{{x} {x}} \right]\nonumber\\
&&H^{\rm Gauge(II)}_{t {x}}=\frac{g^{\frac{2}{3}}_{s}}{i L^2}\left[q_3 g_1 \tilde{C}_{t}- u^2 q_3 \tilde{\xi}^{(1)}_{t}+   u^2 \omega_3 \tilde{\xi}^{(1)}_{{x}}+ q_3 g_1 \tilde{C}_{t}H_{tt}\right]\nonumber\\
&&  H^{\rm gauge (II)}_{a a}= \frac{g^{\frac{2}{3}}_{s}}{i L^2}\left[ \omega_3 \tilde{C}_{t} H_{aa}\right]
\end{eqnarray}
where $H_{aa}=H_{{y} {y}}+H_{{z} {z}}, \tilde{\xi}^{(1)}_{{x}}\equiv\frac{\xi^{(1)}_{{x}}}{\pi T}$.

{\bf Set III}: Writing $\xi_u^{(0)}=\frac{C_u(t,{x})}{u\sqrt{g}}, \xi^{(0)}_t = - \partial_t C_u(t,{x})\psi(u), \xi^{(0)}_{{x}} = - \partial_{{x}}C_u(t,{x})\chi(u)$, and demanding the solutions to be well behave at $u=0$, one obtains:
\begin{eqnarray}
\label{GT-III}
& & \xi_u^{(0)} = \frac{C_u(t,{x})}{u\sqrt{g}};\nonumber\\
& & \xi^{(0)}_t = - \left(\frac{1}{2} - \frac{u^4}{3}\right)\sqrt{g} \partial_t C_u(t,{x});
\nonumber\\
& & \xi^{(0)}_{{x}} = - \partial_{{x}}C_u(t,{x}) \frac{F(\sin^{-1}u|1)}{u}  = - \partial_{{x}}C_u(t,{x})\left(1 + \frac{u^4}{10} + {\cal O}(u^8)\right).
\end{eqnarray}
This yields the following:
\begin{eqnarray}
\label{GTIII}
& & H^{\rm Gauge(III)}_{tt}=2\frac{g_s^{\frac{2}{3}}u^2\omega_3^2C_u(t,{x})}{L^2 \sqrt{1-u^4}}\left(\frac{1}{2} - \frac{u^4}{3}\right) - 2\frac{g_s^{\frac{2}{3}}}{L^2\sqrt{1-u^4}}C_u(t,{x})(1+u^4) \nonumber\\
& & + \frac{\omega_3q_3g_s^{\frac{2}{3}}}{L^2} H_{tt} C_u F\left(\sin^{-1}u|1\right)u - \omega_3^2g_s^{\frac{2}{3}}H_{tt}\frac{\left(\frac{1}{2} - \frac{u^4}{3}\right)}{\sqrt{1-u^4}} + \frac{g_s^{\frac{2}{3}}C_u\left(u H_{t{x}}' - 2 H_{t{x}}\right)}{2L^2\sqrt{1-u^4}};\nonumber\\
& & H^{\rm Gauge(III)}_{{x}t}=-\frac{\omega_3q_3g_s^{\frac{2}{3}}}{L^2}\sqrt{1-u^4}C_u u^2\left(\frac{1}{2} - \frac{7u^4}{30}\right);\nonumber\\
& & H^{\rm Gauge(III)}_{{x} {x}}=-2\frac{q_3^2g_s^{\frac{2}{3}}}{L^2}
C_u\left(1+\frac{u^4}{10}\right) + 2\frac{\sqrt{g}g_s^{\frac{2}{3}}}{L^2}C_u(1+u^4);
\nonumber\\
& & H^{\rm Gauge(III)}_{{y}{y}}=\frac{g_s^{\frac{2}{3}}\sqrt{1-u^4}}{L^2}\left(-H_{{x}{x}} + u H_{{x}{x}}'\right).
\end{eqnarray}
\subsection{Pole Structure of $\Omega(\omega_3,q_3)$}
The equation (\ref{pole-speed_s}) can be solved for $\omega_3$ and the solution is given by:
\begin{eqnarray}
\label{pole-speed_s_ii}
& & \omega_3 =  -\frac{2 \left({\alpha_{yy}^{(1,0)}}+\sqrt{{\alpha_{yy}^{(1,0)}}^2-{\alpha_{yy}^{(0,0)}} (4 {C_{1yy}^{(0,2)}}+4
   {C_{2yy}^{(0,2)}}+i {\Sigma_{2yy}^{(0,1)}})}\right)}{4 {C_{1yy}^{(0,2)}}+4 {C_{2yy}^{(0,2)}}+i {\Sigma_{2\ yy}^{(0,1)}}} \nonumber\\
   & & + \frac{q_{3} \left(\frac{{\alpha_{yy}^{(1,0)}} \left(72 {C_{1yy}^{(0,2)}}+8 e^3 {C_{1yy}^{(1,1)}}+72 {C_{2yy}^{(0,2)}}-36 {C_{2yy}^{(1,1)}}+18 i {\Sigma_{2\ yy}^{(0,1)}}+9
   i\right)}{\sqrt{{\alpha_{yy}^{(1,0)}}^2-{\alpha_{yy}^{(0,0)}} (4 {C_{1yy}^{(0,2)}}+4 {C_{2yy}^{02}}+i
   {\Sigma_{2\ yy}^{(0,1)}})}}+8 e^3 {C_{1yy}^{(1,1)}}-36 {C_{2yy}^{(1,1)}}+9 i\right)}{18
   (4 {C_{1yy}^{(0,2)}}+4 {C_{2yy}^{(0,2)}}+i {\Sigma_{2\ yy}^{(0,1)}})}
   \nonumber\\
   & & + {\cal O}(q_3^2);\nonumber\\
   & & -\frac{2 \left({\alpha_{yy}^{(1,0)}}-\sqrt{{\alpha_{yy}^{(1,0)}}^2-{\alpha_{yy}^{(0,0)}} (4 {C_{1yy}^{(0,2)}}+4
   {C_{2yy}^{(0,2)}}+i {\Sigma_{2yy}^{(0,1)}})}\right)}{4 {C_{1yy}^{(0,2)}}+4 {C_{2yy}^{(0,2)}}+i {\Sigma_{2\ yy}^{(0,1)}}} \nonumber\\
   & & + \frac{q_{3} \left(-\frac{{\alpha_{yy}^{(1,0)}} \left(72 {C_{1yy}^{(0,2)}}+8 e^3 {C_{1yy}^{(1,1)}}+72 {C_{2yy}^{(0,2)}}-36 {C_{2yy}^{(1,1)}}+18 i {\Sigma_{2\ yy}^{(0,1)}}+9
   i\right)}{\sqrt{{\alpha_{yy}^{(1,0)}}^2-{\alpha_{yy}^{(0,0)}} (4 {C_{1yy}^{(0,2)}}+4 {C_{2yy}^{02}}+i
   {\Sigma_{2\ yy}^{(0,1)}})}}+8 e^3 {C_{1yy}^{(1,1)}}-36 {C_{2yy}^{(1,1)}}+9 i\right)}{18
   (4 {C_{1yy}^{(0,2)}}+4 {C_{2yy}^{(0,2)}}+i {\Sigma_{2\ yy}^{(0,1)}})}
   \nonumber\\
   & & + {\cal O}(q_3^2)
   \end{eqnarray}
 Assuming $\alpha_{yy}^{(0,0)}\ll1, |\Sigma_{2\ yy}^{(0,1)}|\gg1(i \Sigma_{2\ yy}^{(0,1)}\in\mathbb{R}): \alpha_{yy}^{(0,0)}\Sigma_{2\ yy}^{(0,1)}<1; \alpha_{yy}^{(1,0)} = - |\alpha_{yy}^{(1,0)}|$, consistent with the constraints (\ref{constraints_I}),  (\ref{pole-speed_s_ii}) implies the following.
\noindent\underline{Root 1}:
 {\footnotesize
 \begin{eqnarray}
 \label{root1-i}
 & & \omega_3 = \left(-\frac{2 \left(\sqrt{{\alpha_{yy}^{(1,0)}}^2}+{\alpha_{yy}^{(1,0)}}\right)}{4 {C_{1yy}^{(0,2)}}+4 {C_{2yy}^{(0,2)}}+i {\Sigma_{2yy}^{(0,1)}}}-\frac{{\alpha_{yy}^{(0,0)}}
   \sqrt{{\alpha_{yy}^{(1,0)}}^2} (-4 {C_{1yy}^{(0,2)}}-4 {C_{2yy}^{(0,2)}}-i {\Sigma_{2yy}^{(0,1)}})}{{\alpha_{yy}^{(1,0)}}^2 (4 {C_{1yy}^{(0,2)}}+4 {C_{2yy}^{(0,2)}}+i
   {\Sigma_{2yy}^{(0,1)}})}+O\left({\alpha_{yy}^{(0,0)}}^2\right)\right)+\nonumber\\
   & & q_3 \Biggl[\frac{\sqrt{{\alpha_{yy}^{(1,0)}}^2} \left(8 e^3 {C_{1yy}^{(1,1)}}-36
   {C_{2yy}^{(1,1)}}+9 i\right)+{\alpha_{yy}^{(1,0)}} \left(72 {C_{1yy}^{(0,2)}}+72 {C_{2yy}^{(0,2)}}-36 {C_{2yy}^{(1,1)}}+18 i {\Sigma_{2yy}^{(0,1)}}+8 {C_{1yy}^{(1,1)}} e^3+9
   i\right)}{18 \sqrt{{\alpha_{yy}^{(1,0)}}^2} (4 {C_{1yy}^{(0,2)}}+4 {C_{2yy}^{(0,2)}}+i {\Sigma_{2yy}^{(0,1)}})}+\nonumber\\
   & &\frac{1}{18 (4 {C_{1yy}^{(0,2)}}+4 {C_{2yy}^{(0,2)}}+i
   {\Sigma_{2yy}^{(0,1)}})} \Biggl[\frac{\left(8 e^3 {C_{1yy}^{(1,1)}}-36
   {C_{2yy}^{(1,1)}}+9 i\right) (-4 {C_{1yy}^{(0,2)}}-4 {C_{2yy}^{(0,2)}}-i {\Sigma_{2yy}^{(0,1)}})}{2 {\alpha_{yy}^{(1,0)}}^2}\nonumber\\
   & &  -\frac{\left(\sqrt{{\alpha_{yy}^{(1,0)}}^2}
   \left(8 e^3 {C_{1yy}^{(1,1)}}-36 {C_{2yy}^{(1,1)}}+9 i\right)+{\alpha_{yy}^{(1,0)}} \left(72 {C_{1yy}^{(0,2)}}+72 {C_{2yy}^{(0,2)}}-36 {C_{2yy}^{(1,1)}}\right)\right)}{2 {\alpha_{yy}^{(1,0)}}^2
   \sqrt{{\alpha_{yy}^{(1,0)}}^2}}\nonumber\\
   & &-\frac{\alpha_{yy}^{(1,0)}(18i \Sigma^{(0,1)}_{2yy}+8C^{1,1}_{1yy}e^3+9i)}{2 {\alpha_{yy}^{(1,0)}}^2
   \sqrt{{\alpha_{yy}^{(1,0)}}^2}}\Biggr]  \times  (-4 {C_{1yy}^{(0,2)}}-4 {C_{2yy}^{(0,2)}}-i {\Sigma_{2yy}^{(0,1)}}){\alpha_{yy}^{(0,0)}} +O\left({\alpha_{yy}^{(0,0)}}^2\right)\Biggr]+O\left(q_3^2\right).
 \end{eqnarray}}
 The expansion (\ref{root1-i}) implies:
 {
\begin{eqnarray}
\label{root1-ii}
& & \omega_3= -\frac{{\alpha_{yy}^{(0,0)}}}{{\alpha_{yy}^{(1,0)}}} + \nonumber\\
& & q_3\Biggl[ -\frac{72 {\alpha_{yy}^{(0,0)}} {C_{1yy}^{(0,2)}}+8 e^3 {\alpha_{yy}^{(0,0)}} {C_{1yy}^{(1,1)}}+72 {\alpha_{yy}^{(0,0)}} {C_{2yy}^{(0,2)}}-36 {\alpha_{yy}^{(0,0)}} {C_{2yy}^{(1,1)}}}{36 {\alpha_{yy}^{(1,0)}}^2}\nonumber\\
   & & -\frac{q_3(18
   i {\alpha_{yy}^{(0,0)}} {\Sigma_{2yy}^{(0,1)}}+9 i {\alpha_{yy}^{(0,0)}}+36 {\alpha_{yy}^{(1,0)}}^2)}{36 {\alpha_{yy}^{(1,0)}}^2}\Biggr]\nonumber\\
   & & \approx - q_3\left(1 + i \frac{\alpha_{yy}^{(00)}\Sigma_{2\ yy}^{(0,1)}}{2\left(\alpha_{yy}^{(1,0)}\right)^2}\right).
\end{eqnarray} }
\noindent\underline{Root 2}:
 {\footnotesize
 \begin{eqnarray}
 \label{root2-i}
  && \hskip -0.6in \omega_3 = \left(\frac{2 \left(\sqrt{{\alpha_{yy}^{(1,0)}}^2}-{\alpha_{yy}^{(1,0)}}\right)}{4 {C_{1yy}^{(0,2)}}+4 {C_{2yy}^{(0,2)}}+i {\Sigma_{2yy}^{(0,1)}}}-\frac{{\alpha_{yy}^{(0,0)}}
   \sqrt{{\alpha_{yy}^{(1,0)}}^2}}{{\alpha_{yy}^{(1,0)}}^2}+O\left({\alpha_{yy}^{(0,0)}}^2\right)\right)
   + \nonumber\\
& & \hskip -0.6in   q_3 \left(\frac{\sqrt{{\alpha_{yy}^{(1,0)}}^2} \left(8 e^3
   {C_{1yy}^{(1,1)}}-36 {C_{2yy}^{(1,1)}}+9 i\right)-{\alpha_{yy}^{(1,0)}} \left(72 {C_{1yy}^{(0,2)}}+72 {C_{2yy}^{(0,2)}}-36 {C_{2yy}^{(1,1)}}+18 i {\Sigma_{2yy}^{(0,1)}}+8
   {C_{1yy}^{(1,1)}} e^3+9 i\right)}{18 \sqrt{{\alpha_{yy}^{(1,0)}}^2} (4 {C_{1yy}^{(0,2)}}+4 {C_{2yy}^{(0,2)}}+i
   {\Sigma_{2yy}^{(0,1)}})}\right.\nonumber\\
   & & \hskip -0.6in \left.-\frac{\left(\sqrt{{\alpha_{yy}^{(1,0)}}^2} \left(72 {C_{1yy}^{(0,2)}}+72 {C_{2yy}^{(0,2)}}-36 {C_{2yy}^{(1,1)}}+18 i {\Sigma_{2yy}^{(0,1)}}+8
   {C_{1yy}^{(1,1)}} e^3+9 i\right)\right) {\alpha_{yy}^{(0,0)}}}{36 {\alpha_{yy}^{(1,0)}}^3}\right.\nonumber\\
   & &\hskip -0.6in \left.+O\left({\alpha_{yy}^{(0,0)}}^2\right)\right)+O\left(q_3^2\right).
 \end{eqnarray}}
 The expansion (\ref{root2-i}) implies:
 {\footnotesize
 \begin{eqnarray}
 \label{root2-ii}
 & &  \omega_3 =  \sqrt{{\alpha_{yy}^{(1,0)}}^2} \left(-\frac{{\alpha_{yy}^{(0,0)}}}{{\alpha_{yy}^{(1,0)}}^2}+\frac{4}{4 {C_{1yy}^{(0,2)}}+4 {C_{2yy}^{(0,2)}}+i {\Sigma_{2yy}^{(0,1)}}}\right)\nonumber\\
 & & + q_3\Biggl[\frac{ \left(36 {C_{1yy}^{(0,2)}}+8 e^3 {C_{1yy}^{(1,1)}}+36 {C_{2yy}^{(0,2)}}-36 {C_{2yy}^{(1,1)}}+9 i {\Sigma_{2yy}^{(0,1)}}+9 i\right)}{9 (4 {C_{1yy}^{(0,2)}}+4
   {C_{2yy}^{(0,2)}}+i {\Sigma_{2yy}^{(0,1)}})} \nonumber\\
   & &  + \frac{\alpha_{yy}^{(0,0)} \left(72 {C_{1yy}^{(0,2)}}+8 e^3 {C_{1yy}^{(1,1)}}+72 {C_{2yy}^{(0,2)}}-36 {C_{2yy}^{(1,1)}}+18 i {\Sigma_{2yy}^{(0,1)}}+9 i\right)}{36
   {\alpha_{yy}^{(1,0)}}^2} \Biggr]\nonumber\\
   & &  \approx q_3\left(1 + i \frac{\alpha_{yy}^{(00)}\Sigma_{2\ yy}^{(0,1)}}{2\left(\alpha_{yy}^{(1,0)}\right)^2}\right).
 \end{eqnarray}}
\subsection{Solutions to ${\cal H}_{ab}(u)$}
Making double perturbative ansatze:
${\cal H}_{ab}(u) = \sum_{m=0}^\infty\sum_{n=0}^\infty {\cal H}_{ab}^{(m,n)}(u)q_3^m\omega_3^n$, one obtains near $u=0$ the solutions to the scalar modes' EOMs (\ref{7scalar_EOMs}):
\begin{eqnarray*}
& & {\cal H}_{yy}^{(0,0)}(u) = \alpha_{yy}^{(0,0)} + \beta_{yy}^{(0,0)} u^4,\nonumber\\
& & {\cal H}_{yy}^{(1,0)}(u) = \alpha_{yy}^{(1,0)} + \beta_{yy}^{(1,0)} u^4,\nonumber\\
& & {\cal H}_{yy}^{(0,1)}(u) = \Sigma_{2\ yy}^{(0,1)} + \frac{i}{16} \alpha_{yy}^{(0,0)}u^2\nonumber\\
& & {\rm where:}\Sigma_{2\ yy}^{(0,1)}\equiv \frac{1}{8}\left(- 195 i - 60 \pi - 35\beta_{yy}^{(0,0)}(13 i + 4\pi) + 8 C_{2yy}^{(0,1)}\right),\nonumber\\
& & {\cal H}_{yy}^{(1,1)} = -\frac{i}{4} - \frac{2}{9} e^3 C_{1yy}^{(1,1)} + C_{2yy}^{(1,1)} +
i \frac{u}{4},\nonumber\\
& & {\cal H}_{yy}^{(2,0)}(u) = C_{1yy}^{(2,0)} + C_{2yy}^{(2,0)} u^4,\nonumber\\
& & {\cal H}_{yy}^{(0,2)}(u) = i \frac{\Sigma_{2\ yy}^{(0,1)}}{4} + C_{1yy}^{(0,2)} + C_{2yy}^{(0,2)} - \frac{i}{4} \Sigma_{2\ yy}^{(0,1)} u;\nonumber\\
& & {\cal H}_{xt}^{(0,0)}(u) = \alpha_{xt}{(0,0)},\nonumber\\
& & {\cal H}_{xt}^{(1,0)}(u) = \alpha_{xt}^{(1,0)} + \beta_{xt}^{(1,0)} u^4,\nonumber\\
& & {\cal H}_{xt}^{(0,1)}(u) = C_{2xt}^{(0,1)} + \frac{i}{4} \alpha_{xt}^{(0,0)} u,\nonumber\\
& & {\cal H}_{xt}^{(1,1)}(u) = C_{2xt}^{(1,1)} + \frac{1}{96} \left[12 i \left(\alpha_{xt}^{(1,1)} + 5 \beta_{xt}^{(1,1)}\right) u +
     6 \left(3 i \alpha_{xt}^{(1,1)} + 4 \alpha_{yy} + 5 i \beta_{xt}^{(1,1)} - 4 \beta_{yy}\right) u^2\right],\nonumber\\
& & {\cal H}_{xt}^{(2,0)}(u) = \alpha_{xt}^{(2,0)} + \beta_{xt}^{(2,0)} u^4,\nonumber\\
& & {\cal H}_{xt}^{(0,2)}(u) = \frac{i}{4} C_{2xt}^{(0,1)} u + \frac{1}{4} u^4 C_{1xt}^{(0,2)} + C_{2xt}^{(0,2)};\nonumber\\
& & {\cal H}_{tt}^{(0,0)}(u) = \left(\frac{4}{3} \beta_{yy}^{(0,0)} -
    i C_{1tt}^{(0,0)}\right) + \left(-4 \beta_{yy}^{(0,0)}/3 - \frac{i}{2} C_{1tt}^{(0,0)}\right) u^4 - \frac{3i}{8}C_{1tt}^{(0,0)}u^8,\nonumber\\
& & {\cal H}_{tt}^{(0,1)}(u) =
 \alpha_{tt}^{(0,1)} + \frac{i}{12} \left(6 \alpha_{yy}^{(0,0)} + 4 \beta_{yy}^{(0,0)} - 3 i C_{1tt}^{(0,0)}\right) u +
  \frac{i}{24} \left(3 \alpha + 4 \beta - 3 i C_{tt}^{(0,0)}\right) u^2,\nonumber\\
& & {\cal H}_{tt}^{(1,0)}(u) = \left(\frac{4}{3} \beta_{yy}^{(1,0)} -
    i C_{1tt}^{(1,0)}\right) + \left(-\frac{4}{3} \beta_{yy}^{(1,0)} - (i/2) C_{1yy}^{(1,0)}\right) u^4,\nonumber\\
    & & {\cal H}_{tt}^{(1,1)}(u) =
 C_{1tt}^{(1,1)} + 1/12 i \left(-6 + 6 \alpha_{yy}^{(1,0)} + 4 \beta_{yy}^{(1,0)} - 3 i C1tt^{(1,0)}\right) u\nonumber\\
& & +
  1/24 i \left(6 \alpha_{yy}^{(1,0)} + 4 \beta_{yy}^{(1,0)} - 3 i C_{1tt}^{(1,0)}\right) u^2,\nonumber\\
& & {\cal H}_{tt}^{(0,2)}(u) =
 \frac{i}{192} \left(12 \alpha_{tt}^{(0,1)} \pi + 12 \alpha_{xt}^{(1,0)} \pi +
     6 i \alpha_{yy}^{(0,0)} \pi + 4 i \beta_{yy}^{(0,0)} \pi  + 3 C_{1tt}^{(0,0)} \pi -
     24 \pi \Sigma_{2\ yy}^{(0,1)} - 192 C_{1tt}^{(0,2)}\right)\nonumber\\
     & &  -
  \frac{i}{8} \left(2 \alpha_{tt}^{(0,1)} + 2 \alpha_{xt}^{(1,0)} - \alpha_{xt}^{(1,1)} - 5 \beta_{xt}^{(1,1)}-
     12 \Sigma_{2\ yy}^{(0,1)}\right) u \nonumber\\
  & &
  -\frac{i}{96} \left(12 \alpha_{tt}^{(0,1)} + 12 \alpha_{xt}^{(1,0)} + 6 i \alpha_{yy}^{(0,0)} +
     4 i \beta_{yy}^{(0,0)} + 3 C_{1tt}^{(0,0)} - 24 \Sigma_{2\ yy}^{(0,1)}\right) u^2,\nonumber\\
& & {\cal H}_{tt}^{(2,0)}(u) = \left(\frac{4}{3} C_{2yy}^{(2,0)} -
    i C_{1tt}^{(2,0)}\right) + \left(-\frac{4}{3} C_{2yy}^{(2,0)} - 1/2 i C_{1tt}^{(2,0)}\right) u^4;\nonumber\\
    \end{eqnarray*}
 \begin{eqnarray}
 \label{hab}
    && {\cal H}_s^{(0,0)}(u) = \frac{C_{1s}^{(0,0)}}{2}u^2  + C_{2s}^{(0,0)},\nonumber\\
& & {\cal H}_s^{(1,0)}(u) = \frac{C_{1s}^{(1,0)}}{2} u^2 + C_{2s}^{(1,0)},\nonumber\\
& & {\cal H}_s^{(0,1)}(u) = -\frac{(2 + 2 i) C_{1s}^{(0,1)}}{\pi},\nonumber\\
& & {\cal H}_s^{(2,0)}(u) = \frac{C_{1s}^{(2,0)}}{2} u^2 + C_{2s}^{(2,0)},\nonumber\\
& & {\cal H}_s^{(0,2)}(u) = \Sigma_{s}^{(0,2)} u + \frac{C_{1s}^{(0,2)}}{2} u^2 + C_{2s}^{(0,2)}.
\end{eqnarray}
such that:
\begin{eqnarray}
\label{constraints_I}
& & 171 i + 2 i \alpha_{yy}^{(0,0)} + 319 i  \beta_{yy}^{(0,0)} + 24 C_{1yy}^{(0,0)} = 0;\nonumber\\
& & 3 \alpha_{yy}^{(0,0)} + 4 \beta_{yy}^{(0,0)} - 3 i C_{1tt}^{(0,0)} - 3 C_{2s}^{(0,0)} = 0
\end{eqnarray}
For consistency checks, we have ensured that (\ref{hab}) obtained from the fourth, fifth and the sixth equations of (\ref{7scalar_EOMs}), also solve the first, second, third and seventh equations near $u=0$ and up to ${\cal O}(q_3^m\omega_3^n):m+n=2$ by imposing suitable additional constraints on the constants appearing in (\ref{hab}).
\section{Frobenius Solution to EOM of Gauge-Invariant $Z_v(u)$ for Vector Modes of Metric Fluctuations}
The equations of motion for the vector perturbation modes up next-to-leading order in $N$, can be reduced to the following single equation of motion in terms of a gauge-invariant variable $Z_v(u)$:
\begin{equation}
\label{vector-modes-ZEOM}
Z_v^{\prime\prime}(u) - m(u) Z_v^\prime(u) - l(u) Z_v(u) = 0,
\end{equation}
where
\begin{eqnarray}
\label{m+l-vec-definitions}
& & m(u)\equiv\frac{15 g_{s}^2 M^2 {N_f} \left(u^4-1\right) \log (N) \left(q_{3}^2 \left(u^4-1\right)+\omega_{3}^2\right)+64 \pi ^2 N \left(3 q_{3}^2
   \left(u^4-1\right)^2-\left(u^4+3\right) \omega_{3}^2\right)}{64 \pi ^2 N u \left(u^4-1\right) \left(q_{3}^2 \left(u^4-1\right)+\omega_{3}^2\right)},\nonumber\\
   & & l(u)\equiv -\frac{\left(q_{3}^2 \left(u^4-1\right)+\omega_{3}^2\right) \left(32 \pi ^2 N-3 g_{s}^2 M^2 {N_f} \log ^2(N)\right)}{32 \pi ^2 N \left(u^4-1\right)^2}.
\end{eqnarray}
The horizon $u=1$ is a regular singular point of (\ref{vector-modes-ZEOM}) and the root of the indicial equation corresponding to the incoming-wave solution is given by:
\begin{equation}
\label{root-solution-incoming-wave}
-\frac{i \omega_{3}}{4} + \frac{3 i g_{s}^2 M^2 {N_f} \omega_{3} \log ^2(N)}{256 \pi ^2 N}.
\end{equation}
(a) Using the Frobenius method, taking the solution about $u=1$ to be:
\begin{equation}
\label{solution1}
Z_v(u) = (1 - u)^{-\frac{i \omega_{3}}{4} + \frac{3 i g_{s}^2 M^2 {N_f} \omega_{3} \log ^2(N)}{256 \pi ^2 N}}\left(1 + \sum_{n=1}^\infty a_n (u - 1)^n\right),
\end{equation}
by truncating the infinite series in (\ref{solution1}) to ${\cal O}((u-1)^2)$ one obtains:
{\footnotesize
\begin{eqnarray}
\label{a1a2}
& & \hskip -0.8in a_1 = \frac{1}{512 \pi ^2 N \omega_{3} (\omega_{3}+2 i)^2}\Biggl\{3 i g_{s}^2 M^2 {N_f} \log (N)\left(20 (2-i \omega_{3}) \omega_{3}^2-\log (N) \left(3 \omega_{3}^2 \left(\omega_{3}^2+4 i \omega_{3}+4\right)-4
   q_{3}^2 \left(\omega_{3}^2+4 i \omega_{3}-8\right)\right)\right)\Biggr\}\nonumber\\
   & & \hskip -0.8in +\frac{4 q_{3}^2 (4-i \omega_{3})+3 (2+i
   \omega_{3}) \omega_{3}^2}{8 \omega_{3} (\omega_{3}+2 i)} + {\cal O}\left(\frac{1}{N^2}\right),
   \nonumber\\
& & \hskip -0.8in  a_2 = -\frac{1}{\Sigma}\Biggl\{4 \Biggl(405 i g_{s}^8 M^8 {N_f}^4 \omega_{3}^5 \log ^7(N)+8640 i \pi ^2 g_{s}^6 M^6 N {N_f}^3 \omega_{3}^2 \log ^5(N) \left(16 q_{3}^2
   (\omega_{3}+2 i)+(-13 \omega_{3}+4 i) \omega_{3}^2\right)\nonumber\\
   & &\hskip -0.8in +368640 \pi ^4 g_{s}^4 M^4 N^2 {N_f}^2 \omega_{3} \log ^3(N) \left(4 q_{3}^2 \left(-3 i
   \omega_{3}^2+12 \omega_{3}+16 i\right)+i \omega_{3}^2 \left(9 \omega_{3}^2+8 i \omega_{3}+16\right)\right)\nonumber\\
   & &\hskip -0.5in +7864320 \pi ^6 g_{s}^2 M^2 N^3 {N_f} \omega_{3} \log
   (N) \left(4 i q_{3}^2 \left(\omega_{3}^2+6 i \omega_{3}-16\right)+\omega_{3}^2 \left(-3 i \omega_{3}^2+4 \omega_{3}-16 i\right)\right)
   \nonumber\\
   & &\hskip -0.8in -49152 \pi ^4 g_{s}^2
   M^2 N^2 {N_f} \omega_{3} \log ^2(N) \Biggl(75 g_{s}^2 M^2 {N_f} (\omega_{3}+4 i) \omega_{3}^2+8 \pi ^2 N \Biggl[32 q_{3}^4 (\omega_{3}+6 i)-48
   q_{3}^2 \left(\omega_{3}^3+12 \omega_{3}+16 i\right)\nonumber\\
   & &\hskip -0.8in +\omega_{3}^2 \left(18 \omega_{3}^3-111 i \omega_{3}^2+200 \omega_{3}+16 i\right)\Biggr]\Biggr)\nonumber\\
   & & \hskip -0.8in+108 g_{s}^6
   M^6 {N_f}^3 \omega_{3} \log ^6(N) \left(-75 g_{s}^2 M^2 {N_f} \omega_{3}^3+2 \pi ^2 N \left(128 i q_{3}^4-24 q_{3}^2 \omega_{3}^2 (\omega_{3}+4
   i)+\omega_{3}^4 (19 \omega_{3}+22 i)\right)\right)\nonumber\\
   & &\hskip -0.8in +2304 \pi ^2 g_{s}^4 M^4 N {N_f}^2 \log ^4(N)\nonumber\\
   & &\hskip -0.8in \times \left(150 g_{s}^2 M^2 {N_f} (\omega_{3}+2 i)
   \omega_{3}^3+\pi ^2 N \left(256 q_{3}^4 \left(\omega_{3}^2+3 i \omega_{3}+4\right)-24 q_{3}^2 \omega_{3}^2 \left(15 \omega_{3}^2-8 i
   \omega_{3}+96\right)+\omega_{3}^4 \left(125 \omega_{3}^2-636 i \omega_{3}+208\right)\right)\right)\nonumber\\
   & &\hskip -0.8in +4194304 \pi ^8 N^4 \omega_{3} \left(16 q_{3}^4 (\omega_{3}+8
   i)-24 q_{3}^2 \left(\omega_{3}^3+24 \omega_{3}+64 i\right)+\omega_{3}^2 \left(9 \omega_{3}^3-74 i \omega_{3}^2+200 \omega_{3}+32
   i\right)\right)\Biggr)\Biggr\} + {\cal O}\left(\frac{1}{N^2}\right),
\end{eqnarray}
where:
\begin{eqnarray}
& & \Sigma \equiv \omega_{3}^2 \Biggl[-81 g_{s}^8 M^8 {N_f}^4 \omega_{3}^4 \log ^8(N)+41472 i \pi ^2 g_{s}^6 M^6 N {N_f}^3 \omega_{3}^3 \log
   ^6(N)+294912 \pi ^4 g_{s}^4 M^4 N^2 {N_f}^2 \nonumber\\
   & & \times(26-3 i \omega_{3}) \omega_{3}^2 \log ^4(N)-201326592 \pi ^6 g_{s}^2 M^2 N^3 N_{f} \omega_{3}
   (\omega_{3}+3 i) \log ^2(N)+2147483648 \pi ^8 N^4 \left(\omega_{3}^2+6 i \omega_{3}-8\right)\Biggr].
   \end{eqnarray}}
The Dirichlet boundary condition $Z_v(u=0)=0$ in the hydrodynamical limit retaining therefore terms only up to ${\cal O}(\omega_3^mq_3^n):\ m+n=4$, reduces to:
$a \omega_3^4 + b \omega_3^3 + c \omega_3^2 + f \omega_3 + g = 0$ where:
\begin{eqnarray}
\label{a b c d f g-i}
& & a = 3\left(96\pi^2 + \frac{13 g_s^2 M^2 N_f (\log N)^2}{N}\right),\nonumber\\
& & b = 2 i \left(1664\pi^2 + 39 g_s^2 M^2 N_f \frac{(\log N)^2}{N}\right),\nonumber\\
& & c = 128 \pi^2\left(-70 + 3 q_3^2\right) + 78 g_s^2 M^2 N_f\left(-2 + q_3^2\right)\frac{(\log N)^2}{N},\nonumber\\
& & f = 8 i\left(64\pi^2(-16 + 7 q_3^2) - 6 g_s^2 M^2 N_f \frac{(\log N)^2}{N}q_3^2\right),\nonumber\\
& & g = 16 q_3^2\left(64\pi^2(-4 + q_3^2) + 3 g_s^2 M^2 N_f(4 - 3 q_3^2)\frac{(\log N)^2}{N}\right),
\end{eqnarray}
One of the four roots of  $Z(u=0)=0$ is:
\begin{equation}
\label{root-at-second-order}
\omega_3 = -8.18 i + \frac{0.14 i g_s^2 M^2 N_f(\log N)^2}{N} + \left(-0.005 i - \frac{0.002 i g_s^2 M^2 N_f (\log N)^2}{N}\right)q_3^2 + {\cal O}(q_3^3).
\end{equation}

(b) Using the Frobenius method and going up to ${\cal O}((u-1)^3)$ in (\ref{solution1}), one obtains:
{\footnotesize
\begin{eqnarray}
\label{NLOa3}
& & \hskip -0.6in a_3 = \frac{1}{65536 \pi ^2 N \omega_{3}^2 (\omega_{3}+2 i)^2 (\omega_{3}+4 i)^2 (\omega_{3}+6 i)^2}\Biggl\{i g_{s}^2 M^2 {N_f} \log (N) \Biggl(20 i \omega_{3} \left(\omega_{3}^3+12 i \omega_{3}^2-44 \omega_{3}-48 i\right)\nonumber\\
& & \hskip -0.6in \times \Biggl[48 q_{3}^4
   \left(\omega_{3}^2+12 i \omega_{3}-48\right)-8 q_{3}^2 \left(9 \omega_{3}^4+48 i \omega_{3}^3+60 \omega_{3}^2+1472 i \omega_{3}-3840\right)\nonumber\\
   & & \hskip -0.6in+\omega_{3}^2 \left(27\omega_{3}^4-42 i \omega_{3}^3+1288 \omega_{3}^2+2464 i \omega_{3}-2048\right)\Biggr]\Biggr) -\log (N) \Biggl[64 q_{3}^6 \omega_{3} \left(3 \omega_{3}^4+72 i
   \omega_{3}^3-652 \omega_{3}^2-2400 i \omega_{3}+2880\right)\nonumber\\
   & & \hskip -0.6in -48 q_{3}^4 \left(9 \omega_{3}^7+156 i \omega_{3}^6-668 \omega_{3}^5+3072 i \omega_{3}^4-37024
   \omega_{3}^3-124416 i \omega_{3}^2+160768 \omega_{3}+49152 i\right)\nonumber\\
   & & \hskip -0.6in +4 q_{3}^2 \omega_{3} \left(81 \omega_{3}^8+852 i \omega_{3}^7+4324 \omega_{3}^6+85824 i
   \omega_{3}^5-444320 \omega_{3}^4-1143552 i \omega_{3}^3+1270784 \omega_{3}^2-454656 i \omega_{3}+1769472\right)\nonumber\\
   & & \hskip -0.6in -\omega_{3}^3 \left(81 \omega_{3}^8+288 i
   \omega_{3}^7+13136 \omega_{3}^6+103296 i \omega_{3}^5-183440 \omega_{3}^4+289152 i \omega_{3}^3-925696 \omega_{3}^2-436224 i
   \omega_{3}+221184\right)\Biggr]\Biggr)\Biggr\}\nonumber\\
   & & \hskip -0.6in +\frac{1}{3072 \omega_{3}^2 \left(\omega_{3}^3+12 i \omega_{3}^2-44 \omega_{3}-48 i\right)}\Biggl\{64 i q_{3}^6 \omega_{3}
   (\omega_{3}+12 i)\nonumber\\
   & &  \hskip -0.6in +48 q_{3}^4 \left(-3 i \omega_{3}^4+6 \omega_{3}^3-208 i \omega_{3}^2+960 \omega_{3}+512 i\right)+4 q_{3}^2 \omega_{3} \left(27 i
   \omega_{3}^5+222 \omega_{3}^4+2272 i \omega_{3}^3-7200 \omega_{3}^2+4736 i \omega_{3}-36864\right)\nonumber\\
   & & \hskip -0.6in +\omega_{3}^3 \left(-27 i \omega_{3}^5-504 \omega_{3}^4-932 i
   \omega_{3}^3-5424 \omega_{3}^2-4544 i \omega_{3}+4608\right)\Biggr\} + {\cal O}\left(\frac{1}{N^2}\right).
\end{eqnarray}}
The Dirichlet condition $Z_v(u=0)=0$ reduces to $a \omega_3^4 + b \omega_3^3 + c \omega_3^2 + f \omega_3 + g = 0$
where
\begin{eqnarray}
\label{a b c d f g-ii}
& & a = -\frac{957 g_{s}^2 M^2 {N_f} \log ^2(N)}{N}-63264 \pi ^2,\nonumber\\
& & b = -48 i \left(\frac{27 g_{s}^2 M^2 {N_f} \log ^2(N)}{N}+2240 \pi ^2\right),\nonumber\\
& & c = 8 \left(\frac{15 g_{s}^2 M^2 {N_f} q_{3}^2 \log ^2(N)}{N}+32 \pi ^2 \left(127 q_{3}^2+288\right)\right)\nonumber\\
& & f = 576 i q_{3}^2 \left(64 \pi ^2-\frac{3 g_{s}^2 M^2 {N_f} \log ^2(N)}{N}\right),\nonumber\\
& & g = 384 q_{3}^4 \left(32 \pi ^2-\frac{3 g_{s}^2 M^2 {N_f} \log ^2(N)}{N}\right).
\end{eqnarray}
One of the four roots of the quartic in $\omega_3$ is:
\begin{equation}
\label{root-at-third-order}
\omega_3 =  \left(- 0.73 i + \frac{0.003 i g_s^2 M^2 N_f (\log N)^2}{N}\right)q_3^2 + {\cal O}(q_3^3).
\end{equation}
The leading order coefficient of $q_3^2$ is not terribly far off the correct value $-\frac{i}{4}$ already at the third order in the infinite series (\ref{solution1}).

(c) Let us look at (\ref{solution1}) up to the fourth order. One finds:
{\footnotesize
\begin{eqnarray}
\label{a_4}
& & a_4 = \frac{1}{98304 \omega_{3}^4 \left(\omega_{3}^4+20 i \omega_{3}^3-140 \omega_{3}^2-400 i
   \omega_{3}+384\right)}\nonumber\\
   & & \times\Biggl\{256 q_{3}^8 \omega_{3}^3 (\omega_{3}+16 i)-768 q_{3}^6 \left(\omega_{3}^6+4 i \omega_{3}^5+136 \omega_{3}^4+832 i \omega_{3}^3+256 \omega_{3}^2+7168 i
   \omega_{3}-12288\right)\nonumber\\
   & &  +32 q_{3}^4 \omega_{3}^2 \left(27 \omega_{3}^6-222 i \omega_{3}^5+4880 \omega_{3}^4+18176 i \omega_{3}^3+110464 \omega_{3}^2+652288 i
   \omega_{3}-675840\right)\nonumber\\
   & & -16 q_{3}^2 \omega_{3}^3 \left(27 \omega_{3}^7-558 i \omega_{3}^6+3320 \omega_{3}^5-9232 i \omega_{3}^4+198656 \omega_{3}^3+888320 i
   \omega_{3}^2-774144 \omega_{3}+589824 i\right)\nonumber\\
   & & +3 \omega_{3}^5 \left(27 \omega_{3}^7-900 i \omega_{3}^6-1316 \omega_{3}^5-53104 i \omega_{3}^4+108800 \omega_{3}^3+147200
   i \omega_{3}^2-487424 \omega_{3}-344064 i\right)\Biggr\}\nonumber\\
   & & -\frac{1}{524288 \pi ^2 N \omega_{3}^4
   \left(\omega_{3}^2+6 i \omega_{3}-8\right)^2 \left(\omega_{3}^2+14 i \omega_{3}-48\right)^2}\nonumber\\
   & & \times\Biggl\{-g_{s}^2 M^2 {N_f} \log N \biggl(-\log N \Biggl[256 q_{3}^8 \omega_{3}^3
   \left(\omega_{3}^5+37 i \omega_{3}^4-530 \omega_{3}^3-3500 i \omega_{3}^2+10368 \omega_{3}+10752 i\right)\nonumber\\
   & & -768 q_{3}^6 \Biggl(\omega_{3}^{10}+28 i \omega_{3}^9-222
   \omega_{3}^8+848 i \omega_{3}^7-24192 \omega_{3}^6-153184 i \omega_{3}^5+399360 \omega_{3}^4\nonumber\\
   & & +133120 i \omega_{3}^3+1531904 \omega_{3}^2+3293184 i
   \omega_{3}-2359296\Biggr)\nonumber\\
   & & +16 q_{3}^4 \omega_{3}^2 \Biggl(54 \omega_{3}^{10}+1017 i \omega_{3}^9+2420 \omega_{3}^8+195388 i \omega_{3}^7-1954848 \omega_{3}^6-8216832
   i \omega_{3}^5+5373440 \omega_{3}^4\nonumber\\
   & & -87731200 i \omega_{3}^3+345751552 \omega_{3}^2+510885888 i \omega_{3}-259522560\Biggr)\nonumber\\
   & & -8 q_{3}^2 \omega_{3}^3 \Biggl(54
   \omega_{3}^{11}+513 i \omega_{3}^{10}+14300 \omega_{3}^9+252484 i \omega_{3}^8-1373088 \omega_{3}^7-588832 i \omega_{3}^6\nonumber\\
   & & -30598656 \omega_{3}^5-183382016 i
   \omega_{3}^4+519692288 \omega_{3}^3+707788800 i \omega_{3}^2-297271296 \omega_{3}+113246208 i\Biggr)\nonumber\\
   & & +3 \omega_{3}^5 \Biggl(27 \omega_{3}^{11}+11672 \omega_{3}^9+105584 i
   \omega_{3}^8+196016 \omega_{3}^7+6136320 i \omega_{3}^6-29371904 \omega_{3}^5-60586752 i \omega_{3}^4\nonumber\\
   & & +67778560 \omega_{3}^3+79093760 i \omega_{3}^2-93585408
   \omega_{3}-33030144 i\Biggr)\Biggr]+20 i \left(\omega_{3}^4+20 i \omega_{3}^3-140 \omega_{3}^2-400 i \omega_{3}+384\right)\nonumber\\
   & & \times \omega_{3}^2 \biggl[64 q_{3}^6 \omega_{3}
   \left(\omega_{3}^2+18 i \omega_{3}-96\right)-16 q_{3}^4 \left(9 \omega_{3}^5+84 i \omega_{3}^4+192 \omega_{3}^3+6496 i \omega_{3}^2-23296 \omega_{3}-9216 i\right)\nonumber\\
   & & +4
   q_{3}^2 \omega_{3} \left(27 \omega_{3}^6+12 i \omega_{3}^5+2756 \omega_{3}^4+22208 i \omega_{3}^3-71680 \omega_{3}^2+27136 i \omega_{3}-270336\right)\nonumber\\
   & & +\omega_{3}^4
   \left(-27 \omega_{3}^5+234 i \omega_{3}^4-3704 \omega_{3}^3-4224 i \omega_{3}^2+1408 \omega_{3}+52224 i\right)\biggr]\biggr)\Biggr\}
   + {\cal O}\left(\frac{1}{N^2}\right).
\end{eqnarray}}
In the hydrodynamical limit the Dirichlet boundary condition $Z_v(u=0)=0$ reduces to $a \omega_3^4 + b \omega_3^3 + c \omega_3^2 + f \omega_3 + g = 0$
where
\begin{eqnarray}
\label{a b c d f g-iii}
& & a = 9849372385059274752 i \pi ^2 + {\cal O}\left(q_3^2\right),\nonumber\\
& & b = \frac{19237055439568896 q_{3}^2 \left(3 g_{s}^2 {\log N} (2 {\log N}+5) M^2 {N_f}-128 \pi ^2 N\right)}{N},\nonumber\\
& & c = {\cal O}\left(q_3^4\right),\nonumber\\
& & f = {\cal O}\left(q_3^6\right),\nonumber\\
& & f = {\cal O}\left(q_3^6\right).
\end{eqnarray}
\section{$Z_t(u)$  from Tensor Mode of Metric Fluctuations}
The EOM for the tensor metric perturbation mode $Z_t(u)$, inclusive of the non-conformal corrections was written out in equation (\ref{EOM-tensor}). Realizing that $u=1$ is a regular singular point of (\ref{EOM-tensor}), using the Frobenius method we made a double perturbative ansatz (\ref{ansatz-solution-tensor}) for the analytic part of the solution. Substituting (\ref{ansatz-solution-tensor}) into (\ref{EOM-tensor}), setting the coefficient of $\omega_3$ to zero one gets:
\begin{eqnarray}
\label{w3}
& & {z_{00}}(u) \left(-6 g_{s}^2 M^2 {N_f} \log (N) \log r_{h}-3 g_{s}^2 M^2 {N_f} \log ^2(N)+64 \pi ^2
   N\right)\nonumber\\
    & & \times\left(64 \pi ^2 N \left(u^2+2 u+3\right)+15 g_{s}^2 M^2 {N_f} \left(u^3+u^2+u+1\right) \log \left({N}\right)\right)\nonumber\\
    & & -128 i \pi ^2 N
   \Biggl[2 \Biggl(z_{01}'(u) \left(-15 g_{s}^2 M^2 {N_f} \left(u^4-1\right) \log \left({N}\right)+64 \pi ^2 N \left(u^4+3\right)\right)\nonumber\\
   & &+64 \pi ^2
   N u \left(u^4-1\right) z_{01}''(u)\Biggr)\nonumber\\
   & & -i u \left(u^3+u^2+u+1\right) {z_{00}}'(u) \left(-6 g_{s}^2 M^2 {N_f} \log (N) \log r_{h}-3 g_{s}^2 M^2 {N_f} \log ^2(N)+64 \pi ^2 N\right)\Biggr].\nonumber\\
   & &
   \end{eqnarray}
By setting the coefficient of $q_3$ to zero:
   \begin{eqnarray}
   \label{q3}
   {z_{10}}'(u) \left(-15 g_{s}^2 M^2 {N_f} \left(u^4-1\right) \log \left({N}\right)+64 \pi ^2 N \left(u^4+3\right)\right)+64 \pi ^2 N u
   \left(u^4-1\right) {z_{10}}''(u) = 0,\nonumber\\
   & &
\end{eqnarray}
which solves to yield:
\begin{eqnarray}
\label{h10solution}
& & z_{10}(u) = c_2-\frac{1}{\left(64 \pi ^2 N+15 g_{s}^2 M^2 {N_f} \log \left({N}\right)\right) \left(128 \pi ^2 N+15
   g_{s}^2 M^2 {N_f} \log \left({N}\right)\right)}\nonumber\\
   & & \times\Biggl\{16 \pi ^2 c_1 N u^{1+\frac{15 g_{s}^2 M^2 {N_f} \log {N}}{64 \pi ^2 N}} \Biggl(2 u \left(64 \pi ^2 N+15 g_{s}^2 M^2
   {N_f} \log \left({N}\right)\right)\nonumber\\
    & & \times \, _2F_1\left(1,1+\frac{15 g_{s}^2 M^2 {N_f} \log \left({N}\right)}{128 N \pi ^2};2+\frac{15
   g_{s}^2 M^2 {N_f} \log \left({N}\right)}{128 N \pi ^2};-u^2\right)+\left(128 \pi ^2 N+15 g_{s}^2 M^2 {N_f} \log
   \left({N}\right)\right)\nonumber\\
    & & \times\, _2F_1\left(1,1+\frac{15 g_{s}^2 M^2 {N_f} \log \left({N}\right)}{64 N \pi ^2};2+\frac{15 g_{s}^2 M^2
   {N_f} \log \left({N}\right)}{64 N \pi ^2};-u\right)\nonumber\\
   & & -\left(15 g_{s}^2 M^2 {N_f} \log \left({N}\right)-128 \pi ^2 N\right) \,
   _2F_1\left(1,1+\frac{15 g_{s}^2 M^2 {N_f} \log \left({N}\right)}{64 N \pi ^2};2+\frac{15 g_{s}^2 M^2 {N_f} \log
   \left({N}\right)}{64 N \pi ^2};u\right)\Biggr)\Biggr\}\nonumber\\
   & &  = u^{\frac{15 g_{s}^2 M^2 {N_f} \log \left({N}\right)}{64 \pi ^2 N}} \Bigg(\frac{64 N \pi ^2 c_1 u^4}{256 N \pi ^2+15 g_{s}^2 M^2 {N_f} \log
   \left({N}\right)}\nonumber\\
   & & +\frac{64 N \pi ^2 c_1 u^8}{512 N \pi ^2+15 g_{s}^2 M^2 {N_f} \log \left({N}\right)}+\frac{64 N \pi ^2 c_1 u^{12}}{768 N
   \pi ^2+15 g_{s}^2 M^2 {N_f} \log \left({N}\right)}+O\left(u^{13}\right)\Biggr)+c_2\nonumber\\
\end{eqnarray}
Setting $c_1=0$ for convenience, one obtains:
\begin{eqnarray}
\label{c10i}
& & z_{01}(u)= c_4+\frac{1}{3072}\Biggl\{u \Biggl[\frac{6 i c_2 g_{s}^2 M^2 {N_f} \left(3 u^3+4 u^2+6 u+12\right) \log (N) \log r_{h}}{\pi ^2
   N}\nonumber\\
   & & +\frac{3 i c_2 g_{s}^2 M^2 {N_f} \left(3 u^3+4 u^2+6 u+12\right) \log ^2(N)}{\pi ^2 N}+64 \Biggl(\frac{48 c_3 u^{3+\frac{15 g_{s}^2 M^2 {N_f}
   \log \left({N}\right)}{64 \pi ^2 N}}}{4+\frac{15 g_{s}^2 M^2 {N_f} \log \left({N}\right)}{64 \pi ^2 N}}\nonumber\\
   & & -i c_2 \left(3 u^3+4 u^2+6
   u+12\right)\Biggr)\Biggr]\Biggr\}\nonumber\\
   & & = - \left(\frac{c_5 u^4}{4}+c_3\right)\frac{15 \left(c_5 g_{s}^2 M^2 {N_f} u^4 \log \left({N}\right) (4 \log (u)-1)\right)}{1024 \pi ^2
   N}+ {\cal O}\left(\frac{1}{N^2}\right)\nonumber\\
   & & = u^3 \left(\frac{i c_2 g_{s}^2 M^2 {N_f} \log \left({N}\right) \left(\log \left({N}\right)+2 \log \left(2 \pi ^{3/2} \sqrt{g_{s}}
   T\right)\right)}{256 \pi ^2 N}-\frac{i c_2}{12}\right)\nonumber\\
   & & +u^2 \left(\frac{3 i c_2 g_{s}^2 M^2 {N_f} \log {N}\left(\log
   {N}+2 \log r_{h}\right)}{512 \pi ^2 N}-\frac{i c_2}{8}\right)\nonumber\\
   & & +u \left(\frac{3 i c_2 g_{s}^2 M^2
   {N_f} \log \left({N}\right) \left(\log {N}+2 \log r_{h}\right)}{256 \pi ^2 N}-\frac{i
   c_2}{4}\right)+c_4 + {\cal O}\left(\frac{u^4}{N},\frac{1}{N^2}\right).\nonumber\\
   & &
\end{eqnarray}
Similarly,
\begin{eqnarray}
\label{phi10}
& & z_{10}(u) = \frac{15 c_5 g_{s}^2 M^2 {N_f} u^4 \log \left({N}\right) (4 \log (u)-1)}{1024 \pi ^2 N}+\frac{c_5 u^4}{4}+c_3.
\end{eqnarray}
The constant (in $\omega_3, q_3$) yields:
\begin{eqnarray}
\label{constant}
& & \frac{{z_{00}}'(u) \left(-15 g_{s}^2 M^2 {N_f} \left(u^4-1\right) \log \left({N}\right)+64 \pi ^2 N \left(u^4+3\right)\right)}{64 \pi ^2 N u
   \left(u^4-1\right)}+{z_{00}}''(u) = 0,
\end{eqnarray}
which is identical in form to the EOM of $z_{10}(u)$.
Setting $q_3=0$ in (\ref{Phi}), one  obtains:
\begin{eqnarray}
\label{Phi'overPhi}
& &\hskip -0.6in \frac{Z_t^\prime(u)}{Z_t(u)} = \left(\frac{i c_4 \omega_{3}^2}{4 \left(c_4 \omega_{3}+c_2\right)}+\frac{3 i c_4 g_{s}^2 M^2 {N_f} \omega_{3}^2 \log \left({N}\right) \left(\log
   \left({N}\right)+2 \log r_{h}\right)}{256 \pi ^2 N \left(c_4
   \omega_{3}+c_2\right)}+{\cal O}\left(\frac{1}{N^2}\right)\right)\nonumber\\
   & & \hskip -0.6in +u \Biggl(\frac{\omega_{3}^2 \left(c_2^2+4 i c_4 c_2+4 i \omega_{3} c_4^2\right)}{16
   \left(c_2+\omega_{3} c_4\right){}^2}\nonumber\\
   & & \hskip -0.6in -\frac{3 i g_{s}^2 M^2 {N_f} \omega_{3}^2 \left(-i c_2^2+2 c_4 c_2+2 \omega_{3} c_4^2\right) \log
   \left({N}\right) \left(\log \left({N}\right)+2 \log r_{h}\right)}{512 \pi ^2 \left(c_2+\omega_{3}
   c_4\right){}^2 N}\nonumber\\
   & &\hskip -0.6in  + {\cal O}\left(\frac{1}{N^2}\right)\Biggr)+u^2 \Biggl\{\frac{\omega_{3}^2 \left((i \omega_{3}+6) c_2^3+2 (3 \omega_{3}+8 i) c_4 c_2^2+32 i
   \omega_{3} c_4^2 c_2+16 i \omega_{3}^2 c_4^3\right)}{64 \left(c_2+\omega_{3} c_4\right){}^3}\nonumber\\
   & &\hskip -0.6in  -\frac{3 i g_{s}^2 M^2 {N_f} \omega_{3}^2 \left(3 (\omega_{3}-4 i)
   c_2^3+4 (4-3 i \omega_{3}) c_4 c_2^2+32 \omega_{3} c_4^2 c_2+16 \omega_{3}^2 c_4^3\right) \log \left({N}\right) \left(\log \left({N}\right)+2 \log
   r_{h}\right)}{4096 \pi ^2 \left(c_2+\omega_{3} c_4\right){}^3 N}\nonumber\\
   & &\hskip -0.6in  +{\cal O}\left(\frac{1}{N^2}\right)\Biggr\}+u^3\nonumber\\
   & & \hskip -0.6in \times
   \Biggl(\frac{1}{768 \left(c_2+\omega_{3} c_4\right){}^4}\Biggl\{\omega_{3} \Biggl[\left(-3 \omega_{3}^3+24 i \omega_{3}^2+88 \omega_{3}+192 i\right) c_2^4+8 \omega_{3} \left(3 i \omega_{3}^2+22 \omega_{3}+96 i\right) c_4
   c_2^3\nonumber\\
   & & \hskip -0.6in+8 \omega_{3}^2 (11 \omega_{3}+144 i) c_4^2 c_2^2+768 i \omega_{3}^3 c_4^3 c_2+192 i \omega_{3}^4 c_4^4\Biggr]\Biggr\}\nonumber\\
   & & \hskip -0.6in -\frac{1}{4096 \pi ^2 \left(c_2+\omega_{3} c_4\right){}^4
   N}\Biggl\{i
   g_{s}^2 M^2 {N_f} \omega_{3}\Biggl[\left(3 i \omega_{3}^3+18 \omega_{3}^2-44 i \omega_{3}+48\right) c_2^4+2 \omega_{3} \left(9 \omega_{3}^2-44 i
   \omega_{3}+96\right) c_4 c_2^3\nonumber\\
   & & \hskip -0.6in+4 (72-11 i \omega_{3}) \omega_{3}^2 c_4^2 c_2^2+192 \omega_{3}^3 c_4^3 c_2+48 \omega_{3}^4 c_4^4\Biggr] \log \left({N}\right)
   \left(\log \left({N}\right)+2 \log r_{h}\right)\Biggr\}\nonumber\\
   & & \hskip -0.6in +{\cal O}\left(\frac{1}{N^2}\right)\Biggr)+{\cal O}\left(u^4\right).\nonumber\\
   & &
   \end{eqnarray}
%\end{subappendices}
%\end{appendices}
%\input{empty1}
%\begin{appendices}
%\begin{subappendices}
%\input{AppendixC}
\chapter{}
\section{Equation of motion for different Glueballs}
\subsection{Type IIB Dilaton wave equation for $0^{++}$ glueball}
\begin{itemize}
\item{\it{Background with a black hole}}
\end{itemize}
The EOM for the dilaton is given as,
\begin{equation}
\partial_{z}(E_{z}\partial_{z}\tilde{\phi})+y^2_{h}F_{z}m^2\tilde{\phi}=0,
\end{equation}
 where the Coefficients $E_z$ and $F_z$ in the above are given below,
\begin{equation}
\label{EF}
\begin{split}
E_{z}(r)&=\frac{\left(e^z+2\right) y_h^2}{128 \pi ^2 g_s^2 L^5}\Biggl(8 \pi+4 g_s N_f \log{4}+2 N_f g_s \log{N}-3g_s N_f  \log \left\{y_h\left(e^z+1\right) \right\}\Biggr)
\\&\Biggl( 6 a^2 \left(4 \pi+ 2 g_s N_f  \log{4}-6g_s N_f \right)+2\left(e^z+1\right) y_h \left(8 \pi+4g_s N_f\log {4}\right)
\\&+2 g_s N_f \log{N} \left(3 a^2+2\left(e^z+1\right)y_h  \right)-3g_s N_f  \left(3 a^2+2 \left(e^z+1\right) y_h\right) \log \left( y_h\left(e^z+1\right)
  \right)\Biggr)\\\\
F_{z}&=\frac{1}{128 \pi ^2 g_s^2 L y_h \left(e^z+1\right)}e^z\left(4 \pi+ g_s N_f \log {16}+g_s N_f \log {N}-\frac{3}{2}g_s N_f  \log \left[\left(e^z+1\right) y_h\right]\right)
\\&\Biggl[\left(e^z+1\right) y_h \Biggl(8 \pi+2 g_s N_f \log {N}+g_s N_f \log {256} -3g_s N_f  \log \left[\left(e^z+1\right) y_h\right] \Biggr)
\\&-3 a^2 \left(4 \pi+6g_s N_f +g_s N_f  \log {N}+g_s N_f  \log {16} -\frac{3}{2}g_s N_f  \log \left[\left(e^z+1\right) y_h\right]\right)\Biggr]
\end{split}
\end{equation}
\begin{itemize}
\item{\it{Background with an IR cut-off}}
\end{itemize}
The dilaton EOM is given as,
\begin{equation}
\partial_{z}(C_{z}\partial_{z}\tilde{\phi})+y^2_{h}D_{z}m^2\tilde{\phi}=0,
\end{equation}
with,
\begin{equation}\label{CD}\begin{split}
C_{z}&=\frac{e^{-z} \left(e^z+1\right)^3~y_0^3}{32768 \sqrt{2} ~\pi ^{21/4} N^{9/4} g_s^{13/4}} \Biggl(8 \pi+2 g_s N_f \log {N}+4 g_s N_f  \log {4} -3 g_s N_f  \log \left[y_0 \left(e^z+1\right)\right]\Biggr)^2
\\&\Biggl(128 \pi ^2 N+15 g_s  M^2 \left\{-8 \pi+g_s N_f
   (\log {16}-6) +g_s N_f  \log {N} \right\}\log \left[ \left(e^z+1\right)y_0\right]\\&-90 M^2 g_s^2 N_f  \log \left[ \left(e^z+1\right)y_0\right]^2\Biggr)\\\\
D_{z}&=\frac{e^{z}}{32768 \sqrt{2}~ \pi^{17/4} N^{5/4} g_s^{9/4}}\Biggl(8 \pi+2 g_s N_f \log {N}+4 g_s N_f  \log {4} -3 g_s N_f  \log \left[y_0 \left(e^z+1\right)\right]\Biggr)^2
\\&\Biggl(128 \pi ^2 N+3 g_s  M^2 \left\{-8 \pi+g_s N_f
   (\log {16}-6) +g_s N_f  \log {N} \right\}\log \left[ \left(e^z+1\right)y_0\right]\\&-18 M^2 g_s^2 N_f  \log \left[ \left(e^z+1\right)y_0\right]^2\Biggr)
\end{split}
\end{equation}
\subsection{$0^{--}$ glueball EOM}
\begin{itemize}
\item{\it{Background with a black hole}}
\end{itemize}
\begin{eqnarray}
\label{D1D2}
 & &  D_1(r)=\Biggl(\frac{3 g_{s} M^2 (g_{s} {N_f} ({\log  N}-6+\log (16))-24 g_{s} {N_f} \log (r)-8 \pi )}{64 \pi ^2 N r}\nonumber\\
 & & -\frac{75.
   r_{h}^2 \left(4. g_{s} M^2 \log (r_{h})+4. g_{s} M^2+0.6 N\right)^2}{N^2 r^3}+\frac{5 r^4-r_{h}^4}{r^5-r r_{h}^4}\Biggr)\nonumber\\
   & & D_2(r)= -\frac{g_{s}m^2}{4 \pi  r^2 \left(r^4-r_{h}^4\right)} \left(\frac{3 r_{h}^2 \left(4. g_{s} M^2 \log (r_{h})+4. g_{s} M^2+0.6 N\right)^2}{N^2}-r^2\right)\nonumber\\
   & & \times \biggl[36 g_{s}^2 M^2 {N_f}
   \log ^2(r)-3 g_{s} M^2 \log (r) (g_{s} {N_f} ({\log  N}-6+\log (16))-8 \pi )+16 \pi ^2 N\biggr]\nonumber\\
   & &
\end{eqnarray}
\subsection{$0^{--}$ glueball EOM near $r=r_h$}
{\footnotesize
\begin{equation}\begin{split}
\label{b1b2a2}
 &b_1=\frac{g_{s} M^2 (g_{s} (0.005 {\log  N}-0.015) {N_f}+(-0.114 g_{s} {N_f}-360.) \log (r_{h})-360.119)}{N r_{h}}-\frac{24.5}{r_{h}},\\&
 a_2=\frac{0.02 g_{s}^2 m^2 M^2 \left(\log (r_{h}) (g_{s} (0.24 {\log  N}-0.775) {N_f}-2279.99)-2.88 g_{s} {N_f} \log
   ^2(r_{h})-2273.96\right)}{r_{h}^3}\\&-\frac{0.251 g_{s} m^2 N}{r_{h}^3},\\
   &  b_2=\frac{0.04 g_{s}^2 m^2 M^2 }{r_{h}^4}\\&\Biggl(\log (r_{h}) (g_{s} {N_f} (8.158 -3.42 {\log  N})+4065.38)+g_{s} (0.12 {\log  N}-0.387) {N_f}+41.04
   g_{s} {N_f} \log ^2(r_{h})+3976.41\Biggr)\\
   & +\frac{7.163 g_{s} m^2 N}{r_{h}^4}.
   \end{split}
\end{equation}}
\subsection{$0^{++}$ glueball EOM from M-theory}
\begin{itemize}
\item{\it{Background with a black hole}}
\end{itemize}
{\scriptsize
\begin{eqnarray}
\label{GH}
& & \hskip -.6in G(r)=\frac{\left(r^2 \left(16 \pi ^2 g_{s}^2 m^2 N^2 r^2+12 \pi  g_{s} N \left(9 r^4-r_{h}^4\right)-3 r \left(r^4-r_{h}^4\right)^2\right)-3
   a^2 \left(16 \pi ^2 g_{s}^2 m^2 N^2 r^2+36 \pi  g_{s} N \left(r^4-r_{h}^4\right)+3 r \left(r^4-r_{h}^4\right)^2\right)\right)}{12 \pi  g_{s}
   N r^3 \left(r^4-r_{h}^4\right)}\nonumber\\
   & & \hskip -.6 in H(r)=\frac{ \left(r^2 \left(32 \pi ^2 g_{s}^2 m^2 N^2 r^2+12 \pi  g_{s} N \left(15 r^4+r_{h}^4\right)-3 r
   \left(5 r^8-6 r^4 r_{h}^4+r_{h}^8\right)\right)-36 a^2 \left(4 \pi ^2 g_{s}^2 m^2 N^2 r^2+\pi  g_{s} N \left(11 r^4+r_{h}^4\right)+r^9-r^5
   r_{h}^4\right)\right)}{12 \pi  g_{s} N r^4 \left(r^4-r_{h}^4\right)}\nonumber\\
   & & \hskip -1in.
\end{eqnarray}}
\subsection{$2^{++}$ glueball EOM from M-theory}
\begin{itemize}
\item{\it{Background with a black hole}}
\end{itemize}
\begin{equation}\begin{split}
\label{A1A2}
& A_1(r)= -\frac{3 a^2}{r^3}+\frac{15 g_{s} M^2 (g_{s} {N_f} \log (N)-24 g_{s} {N_f} \log (r)-6 g_{s}
   {N_f}+g_{s} {N_f} \log (16)-8 \pi )}{64 \pi ^2 N r} \\&+\frac{5 r^4-r_{h}^4}{r^5-r r_{h}^4}\\
 &  A_2(r)=\frac{1}{4\pi r^{4}\left( r^{4}-rh^{4}\right) } \Biggl[ 8\pi \Biggl\{3 a^2 \left(-2 \pi  g_{s} N m^2 r^2-r^4+r_{h}^4\right)+2 \pi  g_{s} N m^2 r^4+4 r^6\Biggr\}\\
   & -
3 g_{s}^2 M^2 m^2 r^2 \left(r^2-3 a^2\right) \log (r) \Biggl\{g_{s} {N_f} \log (N)+g_{s} {N_f} (\log (16)-6)-8 \pi \Biggr\}\\
& +36 g_{s}^3 M^2 {N_f} m^2 r^2 \left(r^2-3 a^2\right)  \log ^2(r)\Biggr]
\end{split}
\end{equation}
\begin{itemize}
\item{\it{Background with an IR cut-off}}
\end{itemize}
\begin{equation}
\begin{split}
\label{A3A4}
&A3=\frac{5 \left(3 M^2 g_s \left(-24 N_f g_s \log (r)-6 N_f g_s+N_f g_s \log (N)+\log (16) N_f g_s-8 \pi \right)+64 \pi ^2 N\right)}{64 \pi ^2 N r}\\&A4=\frac{1}{4 \pi  r^4}36 m^2 M^2 N_f g_s^3 \log ^2(r)-3 m^2 M^2 g_s^2 \log (r) \left(N_f g_s \log (N)+(\log (16)-6) N_f g_s-8 \pi \right)\\&+16 \pi  \left(\pi  m^2 N g_s+2
   r^2\right)
\end{split}
\end{equation}
\subsection{$1^{++}$ glueball EOM from M-theory}
\begin{itemize}
\item{\it{Background with a black hole}}
\end{itemize}
{\footnotesize
\begin{eqnarray}
\label{B1B2}
\nonumber\\
& &B_1= -\frac{3 a^2}{r^3}-\frac{15 g_{s} M^2 (-g_{s} {N_f} \log (N)+24
g_{s} {N_f} \log (r)+6 g_{s} {N_f}-2 g_{s} {N_f} \log (4)+8 \pi )}{64 \pi ^2 N r}+\frac{5}{r}
\nonumber\\
& & \hskip -0.8in B_2= \frac{1}{4 \pi  r^4 \left(r^4-r_{h}^4\right)}\Biggl\{ \Biggl(36 g_{s}^3 M^2 {N_f} m^2 r^2 \left(r^2-3 a^2\right) \log ^2(r)-3 g_{s}^2 M^2 m^2 r^2 \left(r^2-3 a^2\right) \log (r) (g_{s}
{N_f} \log (N)+g_{s} {N_f} (\log (16)-6)-8 \pi )\nonumber\\
& & \hskip -0.8in +8 \pi  \left(3 a^2 \left(-2 \pi  g_{s} N m^2 r^2-r^4+r_{h}^4\right)+2 \pi   g_{s} N m^2
r^4+4 r^6\right)\Biggr)\Biggr\}.
\end{eqnarray}}
\subsection{$1^{++}$ glueball EOM from M-theory}
\begin{itemize}
\item{\it{Background with a black hole}}
\end{itemize}
{\scriptsize
\begin{eqnarray}
\label{C1C2}
& & \hskip -0.4in C_1(r)=\Biggl(\frac{5 r^4-r_{h}^4}{r \left(r^4-r_{h}^4\right)}-\frac{9 a^2}{r^3}+\Biggl\{\frac{3}{256 \pi ^2 N^{2/5} r^3}\Biggl[-54 a^2 g_{s}^2 M^2 {N_f}-72 \pi  a^2 g_{s} M^2+768 \pi ^2 a^2+12 g_{s}^2 M^2 {N_f} r^2+\nonumber\\
& & \hskip -0.4in  9 a^2 g_{s}^2 M^2 {N_f} \log (16)-2 g_{s}^2 M^2 {N_f} r^2 \log (16)+16 \pi  g_{s} M^2 r^2+g_{s}^2 M^2 {N_f} \left(9 a^2-2 r^2\right) \log (N)-24 g_{s}^2 M^2 {N_f}\nonumber\\
& & \hskip -0.4in \left(9 a^2-2 r^2\right) \log (r)\Biggr]\Biggr\}\Biggr)
\\& &C_2(r)=\Biggl(\frac{1}{4 \pi  r^4 \left(r^4-r_{h}^4\right)}\Biggr\{8 \pi  \left(a^2 \left(6 \pi  g_{s} N q^2 r^2-9 r^4+9 r_{h}^4\right)-2 \pi  g_{s} N q^2 r^4+4 r^6\right)\nonumber\\
& &\hskip -0.4in  +3 g_{s}^2 M^2 q^2 r^2 \left(r^2-3 a^2\right) \log (r) (g_{s} {N_f} \log (16 N)-6 g_{s} {N_f}-8 \pi )-36 g_{s}^3 M^2 {N_f} q^2 r^2 \left(r^2-3 a^2\right) \log ^2(r)\Bigg\}-\nonumber\\
& &\hskip -0.4in  \frac{g_{s}^{2}}{512\pi^{3}}\Biggl\{\frac{34992 a^2 g_{s} M^2 \left(\sqrt[5]{N}+3\right) {N_f}^2 \log (r)}{r^3}+9 a^2 g_{s} {N_f} \left(\frac{7831552 \pi ^5}{\left(r^4-r_{h}^4\right) (g_{s} {N_f} \log (16 N)-3 g_{s} {N_f} \log (r)+4 \pi )^3}-\frac{81 M^2
  \left(7 \sqrt[5]{N}-1\right) {N_f}}{r^4}\right)\nonumber\\
& & \hskip -0.4in + \frac{2 \left(243 g_{s} M^2 \left(\sqrt[5]{N}+1\right) {N_f}^2+\frac{3915776 \pi ^5 r^4}{\left(r^4-r_{h}^4\right) (g_{s} {N_f} \log (16 N)-3 g_{s}
   {N_f} \log (r)+4 \pi )^2}\right)}{r^2}\Biggr\}\Biggr).
\end{eqnarray}}
\section{Flux generated cosmological constant term}
The flux generated cosmological constant term is given as,
\begin{equation}
G_{MNPQ} G^{MNPQ}=\frac{\mathcal{A}(r)}{N^{7/10}},
\end{equation}
with
\begin{equation}\begin{split}
\label{cc}
\mathcal{A}(r)&=\left(\frac{288\ 6^{5/6} b^4 M^3 g_s r_h^4}{125 \pi ^{4/3} N_f^{2/3} \log ^{\frac{17}{3}}(N)}\right)\\&\left(\frac{68260644 \log (10) \left(r_h^4-10000\right) \left(54 b^2 r_h^2+5\right)}{\left(100-3 b^2 r_h^2\right){}^4}-\frac{30876125 \log (9)
   \left(r_h^4-6561\right) \left(12 b^2 r_h^2+1\right)}{9 \left(b^2 r_h^2-27\right){}^4}\right)
\end{split}
\end{equation}
%\end{subappendices}
%\end{appendices}
%\input{green}
%\input{redgreen}
%\input{seven}
%\input{six}

\clearpage
\addcontentsline{toc}{chapter}{Bibliography}

\begin{spacing}{1}
\end{spacing}

%\begin{spacing}{1}
%\bibliographystyle{unsrt}
%\cleardoublepage
%\addcontentsline{toc}{chapter}{Bibliography}
%\bibliography{sarita_23911}
%\end{spacing}

\end{document}